%% file: Dissertation_arxiv_upload.tex
\documentclass[12pt,a4paper,DIV12,fleqn,BCOR2mm]{scrbook}
\usepackage[utf8]{inputenc}
\usepackage{amsmath,amsfonts,amssymb,amscd,braket}
\usepackage{etex} % wichtig: Ohne das gibt es errors bei zu vielen Paketen! (pre:2015)
\usepackage{mathtools} %mapsto
\usepackage{graphicx}
\usepackage{epsfig} %eps bilder, für feynmp
\usepackage{cite}
\usepackage{mathrsfs}
\usepackage{bbm}
\usepackage{pifont}
\usepackage{marvosym}
\usepackage{units}
\usepackage{bibentry} %Display bibliography entries in the text.
\nobibliography* %Setting options for bibentry package
\usepackage[format=plain]{caption}
\usepackage[labelformat=simple,format=plain]{subcaption}

\usepackage{xcolor}
\usepackage{stmaryrd} %zusätzliche symbole, widerspruchsblitz
\usepackage[all]{xy} 
\usepackage{wrapfig} 
\usepackage{feynmp} %Feynman diagramme
\usepackage{multirow} %tabellen mit mehrzeiligen spalten
\usepackage{rotating} %vertikale schrift in tabelle
\usepackage{arydshln} %dashed lines in a table (cdashline)
\usepackage{slashed} %Feynman slash
\usepackage{euscript} %ParityP and TimeT
\usepackage{xspace} %variable space after macros for example for \Dff delta 54
\usepackage{enumitem} %\enmuerate w\ roman numerals
\usepackage[hang,flushmargin]{footmisc} % fussnoten linksbündig
\usepackage{chngcntr} %let footnotes run continuously, not restart at every chapter
\counterwithout{footnote}{chapter}

\usepackage[debug=false,colorlinks,linkcolor=black,filecolor=black,citecolor=black,pdftex]{hyperref} %for online version only

\usepackage[left=2.75cm, top=2.5cm, bottom=4.5cm, right=2.25cm]{geometry}
\addtokomafont{disposition}{\boldmath}

\newcommand{\changefont}[3]{
\fontfamily{#1} \fontseries{#2} \fontshape{#3} \selectfont}

\newcommand{\Eqref}[1]{equation~\eqref{#1}}
\newcommand{\Figref}[1]{figure~\ref{#1}}

\newcommand{\Tabref}[1]{table~\ref{#1}}
\newcommand{\Secref}[1]{section~\ref{#1}}
\newcommand{\Appref}[1]{appendix~\ref{#1}}

\newcommand{\SemiDirect}[0]{\ensuremath{\rtimes}}

\newcommand{\eVdist}{\kern-0.06em}
\newcommand{\Ev}{\text{e\eVdist V}}     % solely as unit

\newcommand{\Gev}{\text{Ge\eVdist V}}

\DeclareMathOperator{\im}{Im}
\DeclareMathOperator{\tr}{tr}
\DeclareMathOperator{\diag}{diag}

\newcommand{\D}{\mathrm{d}}
\newcommand{\I}{\mathrm{i}}
\newcommand{\e}{\mathrm{e}}

\newcommand{\SO}[1]{\ensuremath{\mathrm{SO}(#1)}}
\newcommand{\SU}[1]{\ensuremath{\mathrm{SU}(#1)}}
\newcommand{\U}[1]{\ensuremath{\mathrm{U}(#1)}}
\newcommand{\Z}[1]{\ensuremath{\mathbbm{Z}_{#1}}} % Z_N ->\Z{N}
\newcommand{\A}[1]{\ensuremath{\mathrm{A}_{#1}}}

\newcommand*{\rep}[2][]{\ensuremath{{\boldsymbol{#2}#1}}} 

\renewcommand{\bar}[1]{\overline{#1}}
\newcommand{\UU}[2][]{\ensuremath{U_{\rep[#1]{#2}}}}
\newcommand{\WW}[2][]{\ensuremath{W_{\rep[#1]{#2}}}}
\newcommand{\SIGMA}[2][]{\ensuremath{\Sigma_{\rep[#1]{#2}}}}
\newcommand{\rhoR}[1]{\ensuremath{\rho_{\rep[#1]{r}}}}
\newcommand{\chiR}[1]{\ensuremath{\chi_{\rep[#1]{r}}}}
\newcommand{\elm}[1]{\mathsf{#1}}

\newcommand{\ord}[1]{\mathcal O({#1})}
\newcommand{\op}[1]{\boldsymbol{\mathsf{#1}}}
\newcommand{\Dff}[0]{\ensuremath{\Delta(54)}\xspace}
\newcommand{\Dts}[0]{\ensuremath{\Delta(27)}\xspace}
\newcommand{\Sfour}{\ensuremath{\mathrm{S}_4}\xspace}
\newcommand{\Tprime}{\ensuremath{\mathrm{T}'}\xspace}

\newcommand{\DiscreteGroup}{\ensuremath{G}}
\newcommand{\FSI}{\ensuremath{\mathrm{FS}_u}\xspace}
\newcommand{\TFS}[1]{\ensuremath{\mathrm{FS}_{#1}}}
\DeclareMathOperator{\Out}{Out}
\newcommand*{\Equi}[0]{\ensuremath{E}\xspace}
\newcommand{\Id}{\ensuremath{\mathbbm{1}}}

\newcommand{\vecp}{\ensuremath{{\vec{p}}}}
\newcommand{\parityP}{\ensuremath{\EuScript P}}
\newcommand{\timeT}{\ensuremath{\EuScript T}}

\newcommand{\vev}{\ensuremath{\phi}\xspace}
\newcommand{\VEV}{\ensuremath{\Phi}\xspace}

\newcommand{\HI}{\ensuremath{H_\mathrm{I}}\xspace}
\newcommand{\HY}{\ensuremath{H_\mathrm{Y}}\xspace}
\newcommand{\EEp}{\ensuremath{E^{1}_\mathrm{+}}\xspace}
\newcommand{\EEm}{\ensuremath{E^{1}_\mathrm{-}}\xspace}
\newcommand{\ETp}{\ensuremath{E^{\theta}_\mathrm{+}}\xspace}
\newcommand{\ETm}{\ensuremath{E^{\theta}_\mathrm{-}}\xspace}
\newcommand{\EZp}{\ensuremath{E^{2}_\mathrm{+}}\xspace}
\newcommand{\EZm}{\ensuremath{E^{2}_\mathrm{-}}\xspace}
\newcommand{\EEpm}{\ensuremath{E^{1}_\mathrm{\pm}}\xspace}
\newcommand{\ETpm}{\ensuremath{E^{\theta}_\mathrm{\pm}}\xspace}
\newcommand{\EZpm}{\ensuremath{E^{2}_\mathrm{\pm}}\xspace}

\newcommand{\HRule}{\rule{\linewidth}{0.5mm}}

\def\myvec{\empty}

\begin{document}
 
\frontmatter

\include{Titel_arxiv}

%%%%%%%%%%%%%%%%%%%%%%%%%%%%%%%%%%%%%%%%%%%%%%%%%%%%%%%%%%%%%%%%%%%%%%%%%%%%%%%%%%%%%%%%%%%%%%%%%%%%%%%%%%%%%%%
\cleardoublepage

\thispagestyle{empty}

\vspace*{\fill}
\begin{center}
 {\large\changefont{pzc}{m}{n} As far as I see, all a priori statements in physics have their origin in symmetry.}
\end{center}
\begin{flushright}
 \small\textsc{Hermann Weyl, 1952} \\
\end{flushright}
\vspace*{\fill}
\nocite{Weyl:1952}
%%%%%%%%%%%%%%%%%%%%%%%%%%%%%%%%%%%%%%%%%%%%%%%%%%%%%%%%%%%%%%%%%%%%%%%%%%%%%%%%%%%%%%%%%%%%%%%%%%%%%%%%%%%%%%%

\cleardoublepage
\thispagestyle{empty}

\enlargethispage{2cm}
\begin{center}
\textsc{\Large Abstract}\\[0.5cm]
\end{center}
\small
This work is devoted to the study of outer automorphisms of symmetries (``symmetries of symmetries'') in relativistic quantum field theories (QFTs).
Prominent examples of physically relevant outer automorphisms are the discrete transformations of charge conjugation (C), space--reflection (P) , and time--reversal (T). 
After an introduction to the Standard Model (SM) flavor puzzle, CP violation in the SM, and the group theory of outer automorphisms,
it is discussed how CP transformations can be viewed as special outer automorphisms of the global, local, and space--time symmetries of a model.
Special emphasis is put on the study of finite (discrete) groups. Based on their outer automorphism properties, finite groups are classified into three categories.
It is shown that groups from one of these categories generally allow for a prediction of CP violating complex phases with fixed geometrical values, also referred to as explicit geometrical CP violation.
The remainder of this thesis pioneers the study of outer automorphisms which are not related to C, P, or T. 
It is shown how outer automorphisms, in general, give rise to relations between symmetry invariant operators.
This allows to identify physically degenerate regions in the parameter space of models. Furthermore, in QFTs with spontaneous symmetry breaking, 
outer automorphisms imply relations between distinct vacuum expectation values (VEVs) and give rise to emergent symmetries.
An example model with a discrete symmetry and three copies of the SM Higgs field is discussed in which the rich outer automorphism structure
completely fixes the Higgs VEVs in their field space direction, including relative phases. 
This underlies the prediction of spontaneously CP violating complex phases with fixed geometrical values, also referred to as spontaneous geometrical CP violation.
It is concluded with an outlook, highlighting the possible physical relevance of outer automorphisms for a wide field of future studies.
\vspace{1cm}
\begin{center}
\textsc{\Large Zusammenfassung}\\[0.5cm]
\end{center}
Diese Arbeit befasst sich mit äußeren Automorphismen von Symmetriegruppen (``Symmetrien von Symmetrien'') in relativistischen Quantenfeldtheorien.
Bekannte Beispiele für physikalisch relevante äußere Automorphismen sind die diskreten Transformationen der Ladungskonjugation (C), Raumspiegelung (P),
sowie der Zeitumkehr (T). Nach einer Einführung in das Flavor Puzzle des Standardmodells der Elementarteilchenphysik (SM), in die CP Verletzung im SM und in die Gruppentheorie 
von äußeren Automorphismen, wird dargelegt wie CP Transformationen als spezielle äußere Automorphismen von globalen, lokalen und raum--zeit Symmetrien aufgefasst werden können.
Im Fokus stehen insbesondere endliche (diskrete) Gruppen, welche aufgrund der Eigenschaften ihrer äußeren Automorphismen in drei Kategorien klassifiziert werden.
Es wird gezeigt, dass Gruppen aus einer dieser Klassen im Allgemeinen vorhersagekräftig sind im Bezug auf die Werte von CP verletzenden komplexen Phasen.
Die so erzeugte CP Verletzung wird auch als explizite geometrische CP Verletung bezeichnet.
Weiterhin werden erstmals äußere Automorphismen untersucht die nichts mit C, P oder T zu tun haben. Es wird gezeigt, 
dass äußere Automorphismen im Allgemeinen Relationen zwischen symmetrieinvarianten Operatoren herstellen. Diese erlauben physikalisch äquivalente Regionen 
im Parameterraum von Theorien zu identifizieren. In Theorien mit spontaner Symmetriebrechung stellen äußere Automorphismen Relationen zwischen unterschiedlichen Vakuumerwartungswerten her und führen zu emergenten Symmetrien.
Als Beispiel wird ein Drei--Higgs--Modell diskutiert in welchem die relativen Werte und komplexen Phasen 
der Higgs Vakuumerwartungswerte durch die reichhaltige Struktur der äußeren Automorphismen gänzlich festlegt werden. 
Der zugrundeliegende Mechanismus erklärt somit das Auftreten von spontaner CP Verletzung durch fixe geometrische komplexe Phasen. 
Ein abschliessender Ausblick betont die mögliche Relevanz von äußeren Automorphismen für viele weitere Anwendungsbereiche.

\normalsize
\cleardoublepage
\thispagestyle{empty}

\null
\vfill

\begin{center}
\textsc{Publications within the context of this dissertation}\\[0.5cm]
\end{center}
\begin{NoHyper}
\begin{itemize}
\item\bibentry{Chen:2013dpa}
\item\bibentry{Chen:2014tpa}
\item\bibentry{Fallbacher:2015rea}
\item\bibentry{Chen:2015aba}
\item\bibentry{Chen:2015dka}
\end{itemize}
\end{NoHyper}

\tableofcontents

\mainmatter
\setcounter{page}{13}
\chapter{Introduction}

Understanding the asymmetry between matter and anti--matter in the observable universe is one of the great unsolved questions in physics.
If not due to arcane initial conditions, the observed asymmetry should have a natural explanation which manifests itself also on the level of fundamental
interactions of particles and anti--particles. Indeed, it is well--known that microscopic violation of the discrete symmetries of charge conjugation and parity (CP)
is a necessary condition for the creation of a macroscopic baryon asymmetry in standard scenarios of the early universe \cite{Sakharov:1967dj}.\footnote{%
There are alternative scenarios for the creation of a baryon asymmetry which circumvent Sakharov's conditions \cite{Cohen:1987vi,Hook:2015foa}.}

Astonishingly, the Standard Model (SM) of particle physics features a source of explicit CP violation (CPV) \cite{Kobayashi:1973fv,NPP:2008}, which, however, does not suffice 
to explain the observed baryon asymmetry \cite{Kuzmin:1985mm} (cf.\ \cite{Riotto:1998bt,Bernreuther:2002uj} for reviews).
In fact, CPV in the SM only arises with a minimum of three generations of matter fields (cf.\ e.g.\ \cite{Branco:1999fs}),
while the QCD $\theta$--term, as alternative source of CPV in the SM, is absent by observation \cite{Baker:2006ts, Agashe:2014kda}.
The origin of CPV in the SM, therefore, is intimately related to the flavor puzzle and the strong CP problem.
Many ideas have been put forward in the endeavor to understand these puzzles but there is presently no commonly 
accepted theory of flavor (cf.\ e.g.\ \cite{Weinberg:1977hb, Peccei:1997mz, Fritzsch:1999ee, Xing:2014sja, Feruglio:2015jfa} for reviews). 
Arguably, \emph{the} theory of flavor should also be \emph{the} theory of CP violation, as it must  simultaneously explain the origin of both 
phenomena in consistency with observations. Understanding the origin of CPV, therefore, could give invaluable directions also for a solution to the flavor puzzle and the strong CP problem.

\enlargethispage{0.5cm}
In this thesis, CPV is studied from the bottom up, starting with a review of the SM flavor puzzle and the strong CP problem. 
Facilitated by a pedagogical introduction to outer automorphism transformations (``symmetries of symmetries''), 
the discrete transformations of C, P, and T are identified as outer automorphism transformations of space--time \cite{Buchbinder:2000cq}, 
gauge \cite{Grimus:1995zi}, and additional global symmetries. 
This allows for a novel and very general definition of CP as a complex conjugation outer automorphism which maps 
all present symmetry representations to their respective complex conjugate representations. Subsequently, CP transformations are studied 
in models with discrete symmetries \cite{Holthausen:2012dk} and it is found that CP outer automorphisms are not allowed in 
certain models based on certain discrete groups \cite{Chen:2014tpa}. 
Necessary and sufficient conditions are found for the appearance of explicit (``geometrical'') CP violation by calculable complex phases \cite{Chen:2009gf,Branco:2015hea}. 
These complex phases are understood to originate from the complex Clebsch--Gordan (CG) coefficients of certain groups \cite{Chen:2014tpa}. 
Therefore, there are settings in which explicit CPV is understood to originate from the requirement of other symmetries. 
Also, certain settings of spontaneous CPV \cite{Lee:1973iz} are studied in which the CP violating phases likewise originate from 
complex CGs of certain groups \cite{Fallbacher:2015rea} and, therefore, are calculable \cite{Branco:1983tn}. 
This is called spontaneous geometrical CP violation \cite{Branco:1983tn}.

In general, CP transformations are only a special subset of all possible outer automorphisms, meaning that there can be others. 
The generality of the concept indeed suggests that outer automorphism transformations also play a role in many other situations.
In this work, it is shown that outer automorphisms correspond to mappings in the parameter space of a model 
and that stationary points of potentials always appear in representations of the group of all available outer automorphisms \cite{Fallbacher:2015rea}. 
These findings are demonstrated based on a three Higgs doublet (3HDM) example model with $\Delta(54)$ symmetry \cite{Branco:1983tn}.
Here, outer automorphisms give rise to emergent symmetries and thereby explain the origin of spontaneous geometrical CP violation.

The results presented in this thesis have to some extend already been covered in the publications \cite{Chen:2014tpa,Fallbacher:2015rea}.
Nevertheless, some results are new. This includes clarifying remarks on the relation of so--called generalized CP transformations in a horizontal space, to
CP transformations which are outer automorphisms of a symmetry acting in such a horizontal space. 
Furthermore, it is firstly remarked that discrete groups of the so--called type~II~B necessarily give rise to so--called half--odd \cite{Ivanov:2015mwl} 
or even more exotic CP eigenstates. 
Also, it is noted that outer automorphisms can give rise to emergent symmetries in settings with spontaneous symmetry breaking (SSB).
Finally, also the very general definition of CP as a complex conjugation (outer) automorphism is firstly published, 
while it is remarked that this is merely a generalization of the findings in \cite{Grimus:1995zi,Buchbinder:2000cq}.

This work is held in the style of a review article and, in this sense, should serve as a coherent and self--contained introduction
for students and researchers interested in the origin of CPV, its possible relation to the flavor puzzle, 
and the topic of outer automorphisms in general. 
The current knowledge on these topics is summarized and interesting future directions are highlighted.
Basic knowledge of quantum field theory, group theory, and the structure of the SM is assumed.
For brevity, many technical details have to be skipped but an effort is made to highlight the crucial points in a coherent manner. 
References to the original literature are provided throughout, to facilitate further reading.
Advanced readers might be familiar with the content covered in the introductory parts as well as with the machinery of outer automorphisms.
For them, it is recommended to skip directly to the respective point of interest.
Whoever is mainly interested in CPV from finite groups, the classification of finite groups according to their CP properties, or the 
conditions for explicit geometrical CPV should skip to chapter \ref{sec:CPAndFiniteGroups}. 
The topic of spontaneous geometrical CPV is touched in \ref{sec:SSBandCPV} and treated in detail in \ref{sec:3HDMExample}. 
Who just wants to learn about symmetries of symmetries themselves should consider the introduction in \ref{sec:OutIntro} and then skip directly 
to chapter \ref{sec:OtherOuts} where their fascinating power is revealed in a calculation of stationary points with emergent symmetries.

\chapter{The Standard Model and CP violation in Nature}
\section{The flavor puzzle}
\subsection{Repetition of families; masses and mixings}

The Standard Model of particle physics is the best theory of Nature known to man. 
Being a relativistic quantum field theory in $3+1$ space--time dimensions, it successfully describes 
the forces of electromagnetism, weak, and strong interactions by gauge symmetries.
The corresponding spin--$1$ gauge vector bosons arise as mediators of local transformations of
spin--$\nicefrac12$ fermion matter fields. 
The gauge symmetry of the Standard Model is
\begin{equation}
G_\mathrm{SM}~=~\SU3_\mathrm{c}\times \SU2_\mathrm{L}\times \U1_\mathrm{Y}\;,
\label{eq:GSM}
\end{equation}
where the first factor corresponds to the strong interaction, called color, 
and the last two factors are the gauge symmetry of the electroweak (EW) interaction.
In addition to matter fermions and gauge vector bosons, the SM contains a third species, the so--called Higgs field, 
whose vacuum expectation value (VEV) spontaneously breaks $\SU2_\mathrm{L}\times \U1_\mathrm{Y}\rightarrow \U1_\mathrm{EM}$.
The spontaneous breaking of the EW symmetry explains why we -- living in the non--symmetric ground state of EW interactions -- 
do not observe the complete $\SU2_\mathrm{L}\times \U1_\mathrm{Y}$ symmetry, but only electromagnetism with a massless photon, 
massive matter fermions and weak interaction with massive gauge bosons. 
In addition, the mechanism of SSB predicts the presence of the neutral Brout--Englert--Higgs scalar boson corresponding to excitations
around the VEV. With the much--anticipated discovery of the Brout--Englert--Higgs boson at the LHC in 2012 \cite{Aad:2012tfa,Chatrchyan:2012xdj}, the whole particle content of the 
SM is now experimentally accessible, behaving in complete consistency with the SM predictions. 
The only known phenomenon persisting a fully consistent gauge theory description is gravity. 

\renewcommand{\arraystretch}{1.55}
\begin{table}[t]
\begin{center}
\begin{tabular}{|rr|lc|c|l|c|}
\cline{3-7}
\multicolumn{2}{l}{}&\multicolumn{2}{|c|}{Names} & Fields & $~G_\mathrm{SM}$ & $\U1_\mathrm{EM}$ \\  
\cline{3-7}
\hline
\multirow{5}{9pt}{\begin{sideways}Matter\end{sideways}} & \multirow{5}{9pt}{\begin{sideways} (spin $\nicefrac 12 $) \end{sideways}} 
 & \multirow{3}{*}{Quarks \mbox{($\times 3$ families)}} & $Q$ & $(u_\mathrm{L}\>\>\>d_\mathrm{L})$ & $({\rep 3},{\rep2})_{\nicefrac 16}$ & $\nicefrac23\>\>\>-\!\nicefrac13$\\
&& & $\bar U$ & $u^\dagger_\mathrm{R}$ & $({\rep{\bar3}},{\rep 1})_{\nicefrac{-2}{3}}$  &  $\!\!\!\!-\nicefrac23$ \\ 
&&	& $\bar D$ & $d^\dagger_\mathrm{R}$ & $({\rep{\bar3}},{\rep 1})_{\nicefrac 13}$ &	$\nicefrac13$ \\  
\cline{3-7}
&& \multirow{3}{*}{Leptons \mbox{($\times 3$ families)}}& $L$ & $(\nu_\mathrm{L}\>\>\>\>e_\mathrm{L})$ & $({\rep  1},{\rep 2})_{\nicefrac {-1}2}$ & $0\>\>\>-\!1$ \\
&& & $\bar E$& $e^\dagger_\mathrm{R}$ & $({\rep  1},{\rep  1})_1$ & $1$ \\
\cdashline{3-7}
&& & $\bar N$& $\nu^\dagger_\mathrm{R}$ & $({\rep  1},{\rep 1})_0$ & $0$ \\
\hline
\hline
\multirow{2}{9pt}{\begin{sideways}\vspace{-40pt}Higgs\end{sideways}} & \multirow{2}{9pt}{\begin{sideways} (spin $0 $) \end{sideways}} 
 &\multirow{2}{9pt}{Higgs}& \multirow{2}{*}{$H$} & \multirow{2}{*}{$(H^+\>\>\>H^0 )$}& \multirow{2}{*}{$({\rep  1},{\rep 2}\>)_{\nicefrac 12}$} & \multirow{2}{*}{$1\>\>\>0$} \\ 
 &&							&&&&\\
\hline
\hline
\multirow{3}{9pt}{\begin{sideways}Gauge\end{sideways}} & \multirow{3}{9pt}{\begin{sideways} (spin $1$) \end{sideways}} 
&Gluon & & $g$ & $({\rep  8},{\rep  1})_0$ & 0 \\
&&W bosons& & $W^\pm\>\>\> W^0$ & $({\rep  1},{\rep  3})_0$ & $\!\!\!\!\pm1\>\>\> 0$ \\
&&B boson & &$B^0$ & $({\rep  1},{\rep  1})_0$ & $0$\\
\hline
\end{tabular}
\caption{The Standard Model fields and their gauge group embedding. A hypothetical right--handed neutrino has been added to generate neutrino masses.}
\label{tab:SM}
\end{center}
\end{table}
\renewcommand{\arraystretch}{1.0}%

As a curiosity -- since not required by any means of theoretical consistency but only due to observation -- all matter fields of the SM appear in three identical copies
called families or generations. The gauge representations of one generation of fermions is given by
\begin{equation}\label{SMgeneration}
\text{generation}~=~(\rep{3} ,\rep{2} )_{\nicefrac 16}+(\bar{\rep{3}},\rep1)_{\nicefrac {-2}3}+(\bar{\rep3},\rep1)_{\nicefrac 13}+
                    (\rep1,\rep 2)_{\nicefrac{-1}2}+(\rep1,\rep1)_1+(\rep1,\rep1)_0\;,
\end{equation}
and the complete SM field content is displayed in \Tabref{tab:SM}.
The convention used here is such that all fields are introduced as left--handed Weyl spinors emphasizing that the SM is a chiral theory.\footnote{%
In this work a mixed notation of two and four component spinors is used where 
$\Psi_\mathrm{L/R}:=\mathrm{P}_\mathrm{L/R}\Psi=\frac12\left(1\mp\gamma_5\right)\Psi$ is a Weyl spinor which can be treated as
a Dirac spinor for notational convenience.}
The classical formulation of the SM cannot explain the experimentally observed family mixing in the lepton sector and its most plausible 
interpretation in the form of non--zero neutrino masses \cite{NPP:2015}.

Arguably, the most straightforward way to reconcile neutrino oscillations with the SM is to introduce three gauge singlet fermions, 
typically referred to as right--handed neutrinos. These give rise to Dirac and possibly also (lepton number violating) Majorana mass terms for the neutrinos,
thereby also allowing for the observed lepton mixing. 
For semantics, note that when referring to the SM in the following it is meant the classical SM extended by three right--handed neutrinos, 
and these states have already been included in \eqref{SMgeneration}.

As a result of the field content and gauge symmetries, the SM exhibits two accidental global $\U1$ symmetries called Baryon (B) and Lepton number (L).
The latter is broken in case neutrinos acquire Majorana masses. A possible charge assignment for these symmetries is $q_\mathrm{B}=+1/3$
for all quarks, or $q_\mathrm{L}=+1$ for all leptons, respectively, while all other fields remain neutral.
Taken individually, both B and L are anomalous, i.e.\ violated by quantum effects, while the combination $B-L$ is anomaly free.

The flavor puzzle of the SM has many facets, with the very starting point being the repetition of fermion families. 
The gauge kinetic terms of the SM admit the large global flavor symmetry
\begin{equation}
G_\mathrm{F}~=~\U3_Q \times \U3_U \times \U3_D \times \U3_L \times \U3_E \times \U3_N \;.
\end{equation}
That is, taken aside Yukawa couplings, there is no differentiation between the multiple copies of each fermion,
meaning that the gauge couplings are ``flavor blind''.
However, taking into account the Yukawa couplings between Higgs field and fermions
\begin{equation}\label{eq:SMYukawas}
 -\mathscr L_\mathrm{Yuk.}~=~\bar{Q}^{i}\widetilde{H}\,y^{ij}_u\,u_\mathrm{R}^{j}+\bar{Q}^{i}H\,y^{ij}_d\,d_\mathrm{R}^{j}+
 \bar{L}^{i}H\,y^{ij}_e\,e_\mathrm{R}^{j}+\bar{L}^{i}\widetilde{H}\,y^{ij}_\nu\,\nu_\mathrm{R}^{j}+\mathrm{h.c.}\;,
\end{equation}
where $\widetilde{H}:=\varepsilon H^*$ and it is implicitly summed over the flavor indices $(i,j=1,2,3)$, the flavor symmetry is 
explicitly broken as $G_\mathrm{F}\rightarrow \U1_\mathrm{B} \times \U1_\mathrm{L}$. In this sense, 
smallness of the Yukawa couplings $y_f$ $(f=u,d,e,\nu)$ is
technically natural and has to be expected. 

In general, $y_f$ are complex $3\times3$ matrices in flavor space, which, however, feature many redundant parameters.
The number of independent physical parameters shall be counted in the following.
By singular value decomposition, also called bi--unitary diagonalization, any of the matrices $y_f$ can be written in the form
\begin{equation}
y_f~=~V^{f}_\mathrm{L}\,\lambda_f\,V^{f\dagger}_\mathrm{R}\;,\quad\text{where}\quad\lambda_f~=~\mathrm{diag}(\lambda_{f,i},\dots)\;,
\end{equation}
with real and positive singular values $\lambda_{f,i}$, and unitary matrices $V^{f}_\mathrm{L}$ and $V^{f}_\mathrm{R}$. 
By using this form for all the $y_f$ in \eqref{eq:SMYukawas}, it is straightforward
to perform appropriate basis transformations in flavor space to eliminate redundant degrees of freedom from the Lagrangian.
Thus, working with the redefined fields (flavor indices are suppressed in the following)
\begin{align}\label{eq:MassBasis}
Q'~&=~V^{u\dagger}_\mathrm{L}\,Q\,,&                            L'~&=~V^{e\dagger}_\mathrm{L}\,L\,,& \\
u_\mathrm{R}'~&=~V^{u\dagger}_\mathrm{R}\,u_\mathrm{R}\,,&      e_\mathrm{R}'~&=~V^{e\dagger}_\mathrm{R}\,e_\mathrm{R}\,,&\\
d_\mathrm{R}'~&=~V^{d\dagger}_\mathrm{R}\,d_\mathrm{R}\,,&      \nu_\mathrm{R}'~&=~V^{\nu\dagger}_\mathrm{R}\,\nu_\mathrm{R}\,,& 
\end{align}
the Lagrangian changes its form to
\begin{equation}\label{eq:SMYukawasMassBasis}
 -\mathscr L_\mathrm{Yuk.}~=~\bar{Q}'\widetilde{H}\,\lambda_u\,u_\mathrm{R}'+\bar{Q}'H\left(V^{u\dagger}_\mathrm{L} V^{d}_\mathrm{L}\right)\lambda_d\,d_\mathrm{R}'+
 \bar{L}'H\,\lambda_e\,e_\mathrm{R}'+\bar{L}'\widetilde{H}\left(V^{e\dagger}_\mathrm{L} V^{\nu}_\mathrm{L}\right)\lambda_\nu\,\nu_\mathrm{R}'+\mathrm{h.c.}\;.
\end{equation}
Here, the $\lambda_f$ are diagonal, real, and positive matrices and the primes will be dropped in the following.
Inspecting \eqref{eq:SMYukawasMassBasis}, it makes sense to define the unitary CKM \cite{Kobayashi:1973fv,NPP:2008} and PMNS \cite{Maki:1962mu} matrices
\begin{equation}
V_\mathrm{CKM}~:=~V^{u\dagger}_\mathrm{L} V^{d}_\mathrm{L}\;, \quad\mathrm{and}\quad U_\mathrm{PMNS}~:=~V^{e\dagger}_\mathrm{L} V^{\nu}_\mathrm{L}\;.
\end{equation}
For $n$ families of quarks, the CKM matrix is a $n\times n$ unitary matrix which generally has $n^2$ real parameters. Besides the already 
performed basis transformations it is in addition possible to rephase the fields $Q^{i}_\mathrm{L}$, $u^{i}_\mathrm{R}$, and $d^{i}_\mathrm{R}$
by which one can remove $(2n-1)$ unphysical\footnote{%
These phases are not entirely unphysical but are shifted to the
$\theta$ parameter of QCD, as will be discussed in detail below.} 
phases from $V_\mathrm{CKM}$. Analogously, a rephasing of $L^{i}$, $e^{i}_\mathrm{R}$, and $\nu^{i}_\mathrm{R}$ 
could remove $(2n-1)$ unphysical phases from $U_\mathrm{PMNS}$.
Unphysical phase rotations of $\nu^{i}_\mathrm{R}$, however, are possible if and only if there are no Majorana mass terms for $\nu_\mathrm{R}$. 
If there were such terms then only $n$ phases of $U_\mathrm{PMNS}$ are unphysical and there are $n-1$ additional physical ``Majorana'' phases. 

In the SM with $n=3$ families, a standard parametrization for the CKM matrix is given by \cite{Chau:1984fp, Agashe:2014kda}%
\begin{equation}\label{eq:CKM}
V_\mathrm{CKM}~=~\mathrm{diag}\left(\e^{\I \delta_{u}},\e^{\I \delta_{c}},\e^{\I \delta_{t}} \right)\,V\!\!\left(\theta^{q}_{12},\theta^{q}_{23},\theta^{q}_{13},\delta_\mathrm{CKM}\right)\,
\mathrm{diag}\left(1,\e^{\I \delta_{s}},\e^{\I \delta_{b}} \right)\;,
\end{equation}
where the phases that can be absorbed by rephasing of the quark fields are explicitly displayed for later convenience and $V\!\!\left(\theta_{12},\theta_{23},\theta_{13},\delta\right)$ is given by
\begin{equation}
V~=~
\begin{pmatrix}
c_{12}\,c_{13} & s_{12}\,c_{13} & s_{13}\,\e^{-\I\delta} \\ 
-s_{12}\,c_{23}-c_{12}\,s_{23}\,s_{13}\,\e^{\I\delta} & c_{12}\,c_{23}-s_{12}\,s_{23}\,s_{13}\,\e^{\I\delta} & s_{23}\,c_{13}\\
s_{12}\,s_{23}-c_{12}\,c_{23}\,s_{13}\,\e^{\I\delta} & -c_{12}\,s_{23}-s_{12}\,c_{23}\,s_{13}\,\e^{\I\delta} & c_{23}\,c_{13}
\end{pmatrix}\;,
\end{equation}
with the abbreviations $s_{ij}=\sin(\theta_{ij})$ and $c_{ij}=\cos(\theta_{ij})$.
The angles can be chosen in the range $\theta_{ij}\in[0,\pi/2]$ such that $s_{ij}, c_{ij}\geq0$, and $\delta\in[0,2\pi]$. 
The fact that there remains a physical complex phase implies that generally $V_\mathrm{CKM}^*\neq V_\mathrm{CKM}$.
This is an unambiguous sign of CPV in flavor changing processes as will be elucidated below.

In complete analogy it is possible to parametrize the PMNS matrix by 
\begin{equation}
U_\mathrm{PMNS}~=~%\mathrm{diag}\left(\e^{\I \delta_{\nu_e}},\e^{\I \delta_{\nu_\mu}},\e^{\I \delta_{\nu_\tau}} \right)\,
V\!\!\left(\theta^{\ell}_{12},\theta^{\ell}_{23},\theta^{\ell}_{13},\delta_\mathrm{PMNS}\right)\,
\mathrm{diag}\left(1,\e^{\I \alpha/2},\e^{\I \beta/2} \right)\;,
\end{equation}
where, in contrast to the CKM matrix above, only the ``right--handed'' phases $\alpha,\beta\in[0,2\pi]$ have been explicitly displayed.
In the Dirac neutrino case $\alpha$ and $\beta$ can be absorbed by a rephasing of the right--handed neutrino fields,
in contrast to the Majorana neutrino case where $\alpha$ and $\beta$ are physical parameters. 

The gauge symmetry of the SM generally prohibits mass terms for fermions. However, the Higgs EW doublet field acquires a VEV
\begin{equation}
\bra{0}H\ket{0}~=~\frac{1}{\sqrt2}\begin{pmatrix} 0 \\ v \end{pmatrix}\;,
\end{equation}
with $v\approx246\,\Gev$, thereby breaking \mbox{$\SU2_\mathrm{L}\times \U1_\mathrm{Y}\rightarrow \U1_\mathrm{EM}$}. 
This mechanism of SSB simultaneously explains the appearance of $W^\pm$ and $Z$ boson masses and their ratio, as well 
as the appearance of fermion mass terms. Plugging the Higgs VEV into \eqref{eq:SMYukawasMassBasis} gives rise to Dirac fermion mass terms 
of the form
\begin{equation}
M_{f}\left(f_\mathrm{L}^\dagger\,f_\mathrm{R} + \mathrm{h.c.}\right)\;, 
\end{equation} 
where $M_{f}$ has real and positive eigenvalues
\begin{equation}
m_{f,i}~=~\lambda_{f,i}\,\frac{v}{\sqrt{2}}\;.
\end{equation}
Note that the down type quark masses as well as the neutrino masses are not diagonal in \eqref{eq:SMYukawasMassBasis}.
After the EW symmetry is broken, however, it is possible to diagonalize the respective mass terms by rotating $d_\mathrm{L}$, 
as well as $\nu_\mathrm{L}$, independently of their $\SU2_\mathrm{L}$ doublet partners. In the basis
\begin{align}
d_\mathrm{L}'~=&~V_\mathrm{CKM}^\dagger\,d_\mathrm{L}\;,\\
\nu_\mathrm{L}'~=&~U_\mathrm{PMNS}^\dagger\,\nu_\mathrm{L}\;,
\end{align}
all mass terms are finally diagonal. Note, however, that these rotations change the gauge interaction terms with the $W$ bosons,
\begin{align}
\frac{g}{\sqrt{2}}W^+_\mu\left(\bar{u}_\mathrm{L}\,\gamma^\mu\,d_\mathrm{L}\right)+\mathrm{h.c.}
~&=~
\frac{g}{\sqrt{2}}W^+_\mu\left(\bar{u}_\mathrm{L}\,\gamma^\mu\,V_\mathrm{CKM}\,d_\mathrm{L}'\right)+\mathrm{h.c.}\,,\quad \mathrm{and} &\\
\frac{g}{\sqrt{2}}W^+_\mu\left(\bar{\nu}_\mathrm{L}\,\gamma^\mu\,e_\mathrm{L}\right)+\mathrm{h.c.}
~&=~
\frac{g}{\sqrt{2}}W^+_\mu\left(\bar{\nu}_\mathrm{L}'\,\gamma^\mu\,U_\mathrm{PMNS}^\dagger\,e_\mathrm{L}\right)+\mathrm{h.c.}\;.&
\end{align}
This implies the presence of flavor changing interactions in both, quark and lepton sectors, which are, in this basis, mediated by the $W$ bosons.

In summary the threefold repetition of fermion generations in the SM leads to a rich phenomenological structure that can be interpreted 
in terms of $4\times3$ fermion masses, $2\times3$ flavor changing mixing angles and two or, in the case of Majorana neutrinos, four
CP violating phases. Up to date best fit values for the experimentally determined parameters can be found in the PDG review \cite{Agashe:2014kda} 
or from the global fits \cite{Bona:2006ah,Baak:2014ora,Charles:2015gya,Gonzalez-Garcia:2014bfa}. 
In summary, all quark and charged lepton masses have been determined and it is well established that they 
exhibit a strong hierarchy spanning about six orders of magnitude from the top quark to the electron. 
In contrast, the absolute neutrino mass scale is currently unknown. Nevertheless, there are stringent upper limits from 
cosmology pointing to the sub--$\Ev$ regime \cite{Thomas:2009ae, Riemer-Sorensen:2013jsa, Ade:2015xua}, putting 
the neutrino mass scale down at least by another six orders of magnitude compared to the charged leptons. 
In consistency with a low neutrino mass scale, neutrino oscillation experiments have determined the neutrino mass squared differences 
$\Delta m^2_{21}=7.50_{-0.17}^{+0.19}\times10^{-5}\,\Ev^2$ and  $\Delta m^2_{31(32)}={2.457_{-0.047}^{+0.047}(-2.449_{-0.047}^{+0.048})\times 10^{-3}\,\Ev^2}$
for normal (inverted) ordering of the neutrino masses \cite{Gonzalez-Garcia:2014bfa}, also implying that at least two neutrinos are massive.

The quark mixing angles have been pinned down to an enormous precision and 
show a hierarchical pattern descending by an order of magnitude each from $\theta^q_{12}\approx0.23$,
over $\theta^q_{23}$, down to $\theta^q_{13}$ corresponding to a CKM matrix with an almost unit matrix structure.
The complex phase of the quark sector has been determined as $\delta_\mathrm{CKM}=69.4\pm3.4^\circ$ \cite{Bona:2006ah}.
This proves that CP is violated in Nature. Altogether, the experimental data suggests that the CKM mechanism, most likely, 
is the dominant source of the observed CPV in the quark sector \cite{Nir15}.

The lepton mixing angles are known to the precision of about a degree by now, 
except for $\theta^\ell_{23}$ whose best fit value is either $\approx\!42^\circ$ or 
$\approx\!50^\circ$, discriminating between normal and inverted neutrino mass ordering. In contrast to the quark sector, 
the PMNS matrix shows an almost anarchical structure with all entries being approximately of the same size.
If neutrinos are Dirac particles, then the only currently unknown parameter (besides the overall mass scale) 
of the right--handed neutrino extended SM is the phase of the PMNS matrix with 
a current best fit value of $\delta_\mathrm{PMNS}=306_{-70}^{+39}(254_{-62}^{+63})$ \cite{Gonzalez-Garcia:2014bfa}.
It is very likely that this parameter will be known to an acceptable precision within the next decade.
If neutrinos are Majorana particles, then there are two additional phases $\alpha$ and $\beta$,
which, albeit difficult, could in principle be measured as well, cf.\ \cite{Mohapatra:1998rq,Delepine:2009qg,Simkovic:2012hq,Xing:2013woa,Minakata:2014jba} and references therein. 

In summary, as an experimental fact there is a clear pattern amongst the flavor parameters. 
On the one hand, the SM allows to consistently describe this pattern. On the other hand, however,
the reason for family repetition, for mass hierarchies, for the hierarchical quark and anarchical lepton mixing, as well as the
origin of CP violation is not known at present. Despite many possible approaches for explanations, 
cf.\ e.g.\ \cite{Weinberg:1977hb, Peccei:1997mz, Fritzsch:1999ee, Xing:2014sja, Feruglio:2015jfa} for reviews, 
there is up to date no solution to the flavor puzzle.

\subsection{The strong CP problem}

The SM flavor puzzle has yet another aspect, commonly referred to as the strong CP problem.
Note that $\SU3_\mathrm{c}$ gauge invariance allows for the presence of a so--called $\theta$--term 
\begin{equation}\label{eq:Ltheta}
\mathscr L_\theta~=~\theta\,\frac{g^2}{32\pi^2}\,G^a_{\mu\nu}\,\widetilde{G}^{\mu\nu,a}\;,
\end{equation}
where $G^a_{\mu\nu}$ is the gluon field strength and $\widetilde{G}^{\mu\nu,a}:=\frac{1}{2}\varepsilon^{\mu\nu\rho\sigma}G^a_{\rho\sigma}$ its dual.
This term is odd under parity or time--reversal transformations and, therefore, violates CP.
Consider now a chiral fermion $\Psi_{\mathrm{L}}=\mathrm{P_L}\Psi$ transforming in the fundamental representation of $\SU3_\mathrm{c}$.
It can be shown that a chiral \U1 rotation of such a fermion
\begin{equation}
\Psi_{\mathrm{L}}~\rightarrow~\e^{\I\alpha}\Psi_{\mathrm{L}}\;,
\end{equation}
induces an anomalous transformation of the path integral measure \cite{Fujikawa:1979ay,Fujikawa:1980eg}%
\begin{equation}
 \mathcal{D}\Psi\,\mathcal{D}\overline{\Psi} ~\to~ 
 \mathcal{D}\Psi\,\mathcal{D}\overline{\Psi}\,\exp\left\lbrace -\I\,\int\!\D^4x\,\frac{\alpha\,g^2}{32\pi^2}\,G^a_{\mu\nu}\,\widetilde{G}^{\mu\nu,a}\right\rbrace\;.
\end{equation} 
Thereby, the ``bare'' parameter $\theta$ is shifted to $\theta-\alpha$, implying that $\theta$ by itself is not a reparametrization invariant parameter.

Note that the rephasing transformations of quarks performed in order to remove phases from the CKM matrix,
for example the phases $\delta_{u,c,t}$ in \eqref{eq:CKM}, are exactly of the chiral type discussed above.
Hence, they induce shifts of $\theta$. It can be shown by splitting the unitary matrices $V^u_\mathrm{L,R}$ and $V^d_\mathrm{L,R}$ into 
their determinant (which is a complex phase) and an \SU3 matrix, that general (chiral) basis rotations of the type \eqref{eq:MassBasis} leave the quantity
\begin{equation}
\bar{\theta}~:=~\theta~+~\arg\det{y_u\,y_d}\;,
\end{equation}
invariant. The parameter $\bar\theta$, hence, is a reparametrization invariant physical parameter.
As it is not constrained by any otherwise unbroken symmetry, naively one would expect $|\bar\theta|\sim\mathcal O(1)$. Experimental upper bounds on the 
electric dipole moment (EDM) of the neutron, however, imply that $|\bar\theta|\lesssim 10^{-10}$ \cite{Baker:2006ts, Agashe:2014kda}.
While CP violation is well established in flavor changing processes of quarks (i.e.\ involving the CKM matrix) and 
most likely also present for leptons, flavor conserving CP violating processes like the neutron EDM are highly suppressed, if existing at all.
The reason for this suppression is unknown and commonly referred to as the strong CP problem, which is an integral part of the flavor puzzle. 

It should be remarked that a term similar to \eqref{eq:Ltheta} does not appear for the $\U1_\mathrm{Y}$ and $\SU2_\mathrm{L}$ gauge factors of the SM for the following reasons.
The $\theta$ term \eqref{eq:Ltheta} can be rewritten as a total derivative, which distinguishes gauge field configurations by a winding number when integrated
over the infinite volume boundary surface. For Abelian gauge fields, however, all such configurations are equivalent, meaning that there is
no difference in winding number, and $\theta_{\U1}$ does not exist, cf.\ e.g.\ \cite{Srednicki_2007}.
%\ToDo{This view has recently been challenged see \cite{Hsu:2010jm}}
In contrast, for $\SU2$ gauge groups there could, in principle, exist a non--vanishing $\theta$ parameter. In the SM, however, there is a possible 
anomalous (and therefore necessarily chiral) global symmetry transformation, for example $B+L$, which allows to absorb $\theta_{\SU2}$ in field redefinitions without changing 
any other parameter of the Lagrangian. This demonstrates that the EW $\theta$ angle is unphysical in the SM.
This argument works whenever there is an anomalous global symmetry rotating fermions that are charged under the gauge group which exhibits the $\theta$ term.
Therefore, this argument does not hold upon introducing B \textit{and} L violating terms, in which case the EW $\theta$ angle would become physical \cite{Perez:2014fja}.

\section{Standard definition of C, P, and T}
\label{sec:CPTDefinition}

In the previous section it has already been remarked that CP is violated in the SM. 
This section serves to formally introduce the standard C, P, and T transformations and investigate their implications.
 
The continuous transformations of boosts, rotations, and translations are forming the ``proper orthochronous'' part of the Poincar\'e group.
The representation theory of the Poincar\'e group is based solely on these proper orthochronous transformation, meaning that representation matrices 
are continuously connected to the identity and, therefore, have $\mathrm{det}=+1$.
In addition, however, there are discrete transformations, acting as automorphisms (which are in this case sometimes also called isometries) of the Poincar\'e group which 
are represented by matrices with $\mathrm{det}=-1$. The fact that these elements are not continuously connected to the identity implies 
that the corresponding automorphisms are outer \cite{Buchbinder:2000cq}. Outer automorphisms will formally be introduced in \Secref{sec:OutIntro},
and their action on representations of the Poincar\'e group will be discussed in \Secref{sec:OutPoincare}.

Two well--known transformations that form outer automorphisms of the proper orthochronous Poincar\'e group are parity and time--reversal
which act on the space--time coordinates as
\begin{align}\label{eq:PT}
\mathrm{P}\,:\,(t,\vec{x})\mapsto(t,-\vec{x})\; \qquad \mathrm{and} \qquad \mathrm{T}\,:\,(t,\vec{x})\mapsto(-t,\vec{x})\;.
\end{align}
Explicit matrices for the transformation $x^\mu\mapsto{\Lambda^\mu}_{\nu} x^\nu$ are given by
\begin{equation}\label{eq:explicitPT}
 \parityP^\mu_{~\nu}~=~\begin{pmatrix} 1 & & & \\  & -1 & & \\ & & -1 & \\ & & & -1 \end{pmatrix} \quad \mathrm{and} 
 \quad \timeT^\mu_{~\nu}~=~\begin{pmatrix} -1 & & & \\  & 1 & & \\ & & 1 & \\ & & & 1  \end{pmatrix}\;,
\end{equation}
from which one immediately reads off that $\parityP^{-1}=\parityP$ and $\timeT^{-1}=\timeT$. 
Hence, the corresponding automorphisms are involutory, meaning that they square to the identity. 
In the following $\parityP\,x\equiv(t,-\vec{x})$ and $\timeT x\equiv(-t,\vec{x})$ are used to denote 
parity or time--reversal transformed coordinates, respectively.

One should note that multiple ways of implementing time--reversal have been pursued in the literature, 
cf.\ e.g.\ \cite{Bell:1996nh, Buchbinder:2000cq} and references therein. In this work, time--reversal refers
to a transformation in the sense of Wigner \cite{Wigner1932}. This operation is implemented as an anti--unitary 
operator on the Hilbert space, such that it does \textit{not} flip the sign of the Hamiltonian \cite{Srednicki_2007, Weinberg:1995mt2}. 
This type of time reversal inverts the direction of momentum and reverses spin, while conserving charge 
and handedness of all particles. Physically, this corresponds to the classical intuition of ``motion reversal''.

If a theory features complex representations then there is another possible 
discrete (outer) automorphism transformation called charge conjugation. 
For additional internal symmetries, such as the automatically present global \U1 phase rotations of Dirac spinor fields,
this transformation corresponds to complex conjugation and flips the sign of all charges.
For general, possibly non--Abelian, internal symmetries charge conjugation corresponds to mapping symmetry representations 
to their complex conjugate representations, which will in detail be discussed in \Secref{sec:OutPoincare}. 
For now the focus is on the action of C, P, and T in the presence of just Abelian internal symmetries,
such as for example in the theory of Quantum Electrodynamics (QED). 

\renewcommand{\arraystretch}{1.2}
\begin{table}
\begin{center}
\begin{tabular}{rcrrr}
 & & \multicolumn{1}{c}{P} & \multicolumn{1}{c}{T} & \multicolumn{1}{c}{C} \\
\hline
$\varphi(x)$ & $\mapsto$  & $\pm\varphi(\parityP\,x)$ 														& $\pm\varphi(\timeT\,x)$ 																	
& $\varphi(x)$ \\
$\phi(x)$  & $\mapsto$    & $\eta_\mathsf{P}\,\phi(\parityP\,x)$ 									& $\eta_\mathsf{T}\,\phi(\timeT\,x)$ 												
& $\eta_\mathsf{C}\,\phi^*(x)$ \\
$A_\mu(x)$ 	& $\mapsto$ 	& $\varepsilon(\mu)\,A_\mu(\parityP\,x)$ 								& $\varepsilon(\mu)\,A_\mu(\timeT\,x)$ 											
& $-A_\mu(x)$ \\
$\Psi(x)$   & $\mapsto$   & $\eta_\mathsf{P}\,\mathcal{\beta}\,\Psi(\parityP\,x)$ & $\eta_\mathsf{T}\,\gamma_5\,\mathcal{C}\,\Psi(\timeT\,x)$ 
%& $\eta_\mathsf{C}\,\mathcal{C}\,{\bar{\Psi}}^\mathrm{T}(x)$ \\
& $\eta_\mathsf{C}\,\mathcal{C}\,\mathcal{\beta}\,\Psi^*(x)$ \\
%\hline
\end{tabular}
\caption{Action of the discrete transformations P, T, and C on real (pseudo--)scalar $\varphi$, complex scalar $\phi$, (gauge) vector $A_\mu$, and 
Dirac spinor fields $\Psi$. See text for an explanation of the symbols.}
\label{tab:classicalCPT}
\end{center}
\end{table}
\renewcommand{\arraystretch}{1.0}%

The action of C, P, and T transformations on
real and complex scalar, vector, and spinor fields is summarized in \Tabref{tab:classicalCPT}. Several remarks are in order:%
\begin{itemize}
\item  In order for a theory to be invariant under the action of P or T the complete Lagrangian density must transform as
\begin{align}
 \op{P}^{-1}\,\mathscr{L}(x)\,\op{P}~=&~+\mathscr{L}(\parityP\,x)\;, \quad\mathrm{or}\\
 \op{T}^{-1}\,\mathscr{L}(x)\,\op{T}~=&~+\mathscr{L}(\timeT\,x)\;,
\end{align}
such that the action $S=\int \mathrm{d}^4x \, \mathscr{L}(x)$ can be shown to be invariant by a change of integration variables.
\item There are free complex phases $\eta_\mathsf{P}$, $\eta_\mathsf{T}$, and $\eta_\mathsf{C}$
in the transformation of complex fields due to the ubiquitous rephasing freedom.
\item The function $\varepsilon(\mu)$ is defined as
\begin{equation}
  \varepsilon(\mu)~:=~\left\{\begin{array}{ll}
    +1\;, & \mu=0\;,\\
	-1\;, & \mu=1,2,3\;.
  \end{array}\right.
 \end{equation}
\item The transformation behavior of $A_\mu(x)$ under time--reversal may appear strange regarding \eqref{eq:explicitPT}.
However, recalling that $A_\mu(x)$ is a (gauge--)vector potential it follows directly from the physical requirement 
that $\vec{E}=\partial_0\vec{A}\mapsto\vec{E}$ under physical time--reversal, 
where the transformation behavior of $\partial_\mu$ directly follows from the transformation of the coordinate $x_\mu$. 
This is one manifestation of the Wigner time--reversal transformation as ``motion reversal''. 
\item The objects $\beta$ and $\mathcal{C}$ are $4\times4$ complex matrices which fulfill
\begin{align}
\beta^{-1}\,\gamma^\mu\,\beta~=&~{\gamma^\mu}^{\dagger}\;, \\ \label{eq:Ccondition}
\mathcal{C}^{-1}\,\gamma^\mu\,\mathcal{C}~=&~-\left({\gamma^\mu}\right)^\mathrm{T}\;,
\end{align}
and
\begin{align}
&\beta^\mathrm{T}~=~\beta^\dagger~=~\beta^{-1}~=~\beta\;,& \\
&\mathcal{C}^\mathrm{T}~=~\mathcal{C}^\dagger~=~\mathcal{C}^{-1}~=~-\mathcal{C}\;,\quad \mathrm{with} \quad\left[\beta,\mathcal{C}\right]~=~0\;.
\end{align}
In general, $\beta$ and $\mathcal{C}$ have a different index structure\footnote{%
A good reference to become acquainted with the details of two-- and four--component spinor notation is \cite{Srednicki_2007}, whose
notation also has loosely been followed here. Another highly recommended resource is \cite{Dreiner:2008tw}.} 
than the Dirac matrices $\gamma^\mu$, but similar to $\gamma_5:=\I\gamma^0\gamma^1\gamma^2\gamma^3$.
Nevertheless, correct numerical solutions are obtained by the identification
\begin{equation}
\beta~=~\gamma^0\,, \quad \text{and}\quad \mathcal{C}~=~\I\,\gamma^2\,\gamma^0\;,
\end{equation}
which holds in the Weyl (chiral) basis and in the Dirac basis for the gamma matrices.
With this solution one also has $\gamma_5\mathcal{C}=\gamma^1\gamma^3$ for the time--reversal transformation, 
and recovers the usual definition of $\bar{\Psi}:=\Psi^\dagger\beta=\Psi^\dagger\gamma^0$. 
For completeness, the explicit form of all matrices is given in \Appref{app:DiracMatrices}.
\item Under the combined transformation CP a Dirac spinor transforms like
\begin{equation}\label{eq:DiracCP}
\Psi(x)~\mapsto~\eta_\mathsf{CP}\,\mathcal{C}\,\Psi^*(\parityP\,x)\;.
\end{equation}
\end{itemize}
\pagebreak
The physical implications of the C, P, and T transformations can be understood after investigating the mode expansion of the 
Dirac field operator
\begin{equation}
\hat{\Psi}(x)~=~\sum_{s=\pm}\int \frac{{\mathrm d}^3p}{(2\pi)^3\sqrt{2p_0}} \left\{ \hat{c}_s(\vecp)\,u_s(\vecp)\,\e^{-\I\,p\,x} + \hat{d}^\dagger_s(\vecp)\,v_s(\vecp)\,\e^{\I\,p\,x}  \right\}\;.
\end{equation}
Here and $\hat{c}_s(\vecp)$ and $\hat{d}^\dagger_s(\vecp)$ are the particle annihilation and anti--particle creation operators, respectively, 
and $u_s(\vecp)$ and $v_s(\vecp)$ denote orthogonal basis spinors. 
To be explicit, Fock space operators have been denoted by a hat which will be dropped in the following. 
The previously stated transformation behavior holds if and only if the Fock space creation and annihilation operators transform as 
stated in \Tabref{tab:FockCPT}. 

\renewcommand{\arraystretch}{1.2}
\begin{table}
\begin{center}
\begin{tabular}{rcrrr}
 & & \multicolumn{1}{c}{P} & \multicolumn{1}{c}{T} & \multicolumn{1}{c}{C} \\
\hline
$c_s(\vecp)$  & $\mapsto$ & $\eta_{\mathsf{P}}\,c_s(-\vecp)$ 														& $s\,\eta_{\mathsf{T}}\,c_{-s}(-\vecp)$ 																	
& $\eta_{\mathsf{C}}\,d_{s}(\vecp)$ \\
$d_s^\dagger(\vecp)$ & $\mapsto$  & $-\eta_{\mathsf{P}}\,d_s^\dagger(-\vecp)$ 									& $s\,\eta_{\mathsf{T}}\,d_{-s}^\dagger(-\vecp)$ 												
& $\eta_{\mathsf{C}}\,c_{s}^\dagger(\vecp)$ \\
%\hline
\end{tabular}
\caption{Action of the discrete transformations P, T, and C on the particle annihilation and anti--particle creation operators $c_s(\vecp)$ and $d_s^\dagger(\vecp)$, respectively.}
\label{tab:FockCPT}
\end{center}
\end{table}
\renewcommand{\arraystretch}{1.0}%

The transformations of creation an annihilation operators in \Tabref{tab:FockCPT} explicitly show that:%
\begin{itemize}
\item The parity operation reverses the direction of momentum and exchanges left-- and right--handed spinors, thus, intuitively corresponds to a spacial reflection.
\item The time--reversal operation reverses the direction of momentum and reverses the spin, thus, intuitively corresponds to a reversal of all dynamics.
\item The charge conjugation operation exchanges creation and annihilation operators of particles and anti--particles.
\end{itemize}

Finally, under a sequential application of all three transformations one finds
\begin{equation}
 \left(\op{C\,P\,T}\right)^{-1}\,\Psi(x)\,\op{C\,P\,T}~=~-\eta_{\mathsf{CPT}}\,\gamma_5\,\Psi^*(-x)\;.
\end{equation}
This is the involution that in the most general sense provides the connection between particles and anti--particles, 
equating their masses and decay rates \cite{Weinberg:1995mt2}. It can be shown
under very general assumptions that CPT is a symmetry of any Lorentz invariant local QFT \cite{Streater:1989vi}.
The CPT theorem can easily be understood in the following way. 
Inspecting the complete basis of possible Hermitean fermion bilinear operators listed in \Tabref{tab:CPTbilinears},
one notes that any possible Lorentz invariant contraction of a bilinear with other fermion bilinears, the derivative, or the gauge vector field also conserves CPT.
Therefore, CPT is automatically conserved if a theory is Lorentz invariant.

Nevertheless, note that in a more common language already the CP conjugate states are referred to as anti--particles.
This is because a CP transformation maps fields, and in particular Weyl spinors, to their \textit{own} complex conjugate,
thereby providing a relation between particles with opposite charge (e.g.\ lepton or baryon number) without referring to any 
additional degrees of freedom or the need to invoke motion reversal. It is also the CP conjugate states which can annihilate
each other to a neutral gauge boson.

\renewcommand{\arraystretch}{1.2}
\begin{table}%[!h!]
\centerline{\begin{tabular}{c|rrrrr|rr}
  & \multicolumn{1}{c}{$\overline{\Psi}\Psi$} & \multicolumn{1}{c}{$\overline{\Psi}\gamma_5\Psi$} & \multicolumn{1}{c}{$\overline{\Psi}\gamma^\mu\Psi$} & 
  \multicolumn{1}{c}{$\overline{\Psi}\gamma^\mu\gamma_5\Psi$} & \multicolumn{1}{c}{$\overline{\Psi}\sigma^{\mu\nu}\Psi$} & \multicolumn{1}{|c}{$\partial_\mu$} & \multicolumn{1}{c}{$A_\mu$} \\
 \hline
 P & 	$+1$ & $-1$ & $\varepsilon(\mu)$ & $-\varepsilon(\mu)$ & $\varepsilon(\mu)\,\varepsilon(\nu)$ & $\varepsilon(\mu)$  & $\varepsilon(\mu)$ \\
 T & 	$+1$ & $-1$ & $\varepsilon(\mu)$ & $\varepsilon(\mu)$ & $-\varepsilon(\mu)\,\varepsilon(\nu)$ & $-\varepsilon(\mu)$ & $\varepsilon(\mu)$ \\
 C & 	$+1$ & $+1$ & $-1$ & $+1$ & $-1$ & $+1$ & $-1$ \\
 CPT &  $+1$ & $+1$ & $-1$ & $-1$ & $+1$ & $-1$ & $-1$ \\ 
\end{tabular}}
\caption{Transformation of (pseudo--)scalar, (pseudo--)vector and tensor fermion bilinears as well as the partial derivative and the gauge vector field under C, P, and T.}
\label{tab:CPTbilinears}
\end{table}
\renewcommand{\arraystretch}{1.0}%

\section{C, P, and CP violation in the Standard Model}
So far, the discrete transformations have been discussed based on a Dirac spinor field.
The SM, however, is a chiral theory in the sense that individual left-- and right--handed Weyl fermions
carry gauge quantum numbers in such a way that they cannot be paired up into Dirac fermion representations.

The transformations C and P as discussed before, however, necessitate the exchange of left-- and right--handed 
components within a single Dirac spinor representation. That is, these transformations are well defined transformations if and only if all
Weyl fermions can be paired up into Dirac spinors without conflicting other quantum numbers. Since this is by construction not the case for chiral theories, 
both, C and P transformations, are broken explicitly and ``maximally'' in the SM.

CP or T transformations, on the contrary, map a single Weyl fermion onto its own complex conjugate or to itself, respectively, and are, therefore, well defined transformations 
irrespective of whether a theory is chiral or not.

Both of the preceding statements have a very clear formulation in the group theoretical language introduced in \Secref{sec:OutIntro}, where C, P, and T are
understood as outer automorphisms of all symmetries of a theory. Then it will also be possible to uniquely assign a clear and well defined meaning
to the term of ``maximal'' violation of a possible symmetry. Namely, when it is broken by the field content, i.e.\ the symmetry representations, of a model. 

The fact that CP or T are well defined transformations of the SM does, of course, not automatically imply that they are symmetries. 
A Lagrangian that gives rise to a real action is schematically given by
\begin{equation}
\mathscr{L}~=~ c \, \mathcal{O}(x) + c^{\ast} \, \mathcal{O}^{\dagger}(x) \;,
\end{equation}
with some operator $\mathcal{O}$ and coupling $c$. By mapping each field to its complex conjugate, also all operators in the 
Lagrangian $\mathcal{O}$ are mapped to their respective Hermitian conjugate operators $\mathcal{O}\mapsto\mathcal{O}^\dagger$. 
This, however, is a symmetry operation if and only if the couplings $c$ fulfill certain relations, typically constraining their complex phases.

For example, in the SM in the basis of \eqref{eq:SMYukawasMassBasis} and neglecting phases which can be absorbed into $\bar{\theta}$, the only complex parameters
are the phases of the CKM and PMNS matrix. In this basis, performing a CP transformation on all fields corresponds to a mapping $V_\mathrm{CKM}\mapsto V_\mathrm{CKM}^*$ 
and $U_\mathrm{PMNS}\mapsto U_\mathrm{PMNS}^*$ (and $\bar\theta\mapsto-\bar\theta$). Therefore, CP is a symmetry of the SM if and only if $V_\mathrm{CKM}$ and $U_\mathrm{PMNS}$ are 
real (and $\bar{\theta}=0$). The question of whether or not CP is violated in the SM, thus, can only be answered experimentally. 
CP violation in the quark sector has been experimentally observed \cite{Christenson:1964fg, NPP:1980} 
in decays and oscillations of K and B mesons and is in broad consistency with the SM CKM mechanism \cite{Agashe:2014kda}.
Complementary to CPV, the CPT theorem implies the violation of T which has also been experimentally verified \cite{Lees:2012}. 
Even though CPV is a necessary condition for baryogenesis \cite{Sakharov:1967dj}, 
the observed amount of CPV does not suffice to explain the observed matter--anti matter asymmetry of the universe \cite{Kuzmin:1985mm}, 
cf.\ e.g.\ the reviews \cite{Riotto:1998bt,Bernreuther:2002uj}. 
The other possible sources of CPV in the SM are either experimentally known to be highly suppressed ($\bar\theta$) 
or not yet experimentally accessible ($\delta_\mathrm{PMNS}$). 

An important point to note is that the discussion so far has been based on a specific parametrization, i.e.\ a specific basis choice.
Physics, of course, cannot depend on the chosen mathematical formulation and has to be basis independent. 
Because of that it is very useful to define basis invariant quantities. A basis invariant measure of the quark sector CPV
is the so--called Jarlskog invariant \cite{Jarlskog:1985ht} (see also the earlier \cite{Gronau:1984nm}) which can be expressed as
\begin{equation}\label{eq:J}
J~=~\frac{1}{\I}\,\det\left[y_u\,y_u^\dagger,y_d\,y_d^\dagger\right]\;.
\end{equation} 
In the mass basis, as obtained above, this takes the form
\begin{equation}\label{eq:JinMB}
\begin{split}
J~=&~\frac{1}{\I}\,\det\left[V_\mathrm{CKM}^\dagger\,\lambda_u\,\lambda_u^\dagger\,V_\mathrm{CKM},\lambda_d\,\lambda_d^\dagger\right]~=~
2\,c_{12}\,c^2_{13}\,c_{23}\,s_{12}\,s_{13}\,s_{23}\,\sin\left(\delta_\mathrm{CKM}\right)\times \\
~&~\times\,\left(m_t^2-m_c^2\right)\left(m_t^2-m_u^2\right)\left(m_c^2-m_u^2\right)\left(m_b^2-m_s^2\right)\left(m_b^2-m_d^2\right)\left(m_s^2-m_d^2\right)\;.
\end{split}
\end{equation}
Under CP transformations $\delta_\mathrm{CKM}\mapsto-\delta_\mathrm{CKM}$, for what reason also $J$ changes its sign, 
which is just the statement that $J$ is CP odd. Therefore, the experimentally found non--vanishing value of $J$ is an 
unambiguous and basis independent sign of CPV. Indeed, as $J$ is the only CP odd basis invariant quantity in the classical SM (neglecting $\bar{\theta}$), 
a vanishing of $J$ would be a necessary and sufficient condition for CP conservation \cite{Bernabeu:1986fc}.
As manifest in the expression \eqref{eq:JinMB} for $J$, sufficient conditions for CP conservation are (i) $\delta_\mathrm{CKM}=0,\pi$; but also 
(ii) there is a pair of mass degenerate quarks in either the up-- or down--sector, or (iii) the sine or cosine of any mixing angle vanishes.
The fact that $J$ is the only basis invariant CP odd quantity implies that, in the classical SM, all rates for CP violating processes are proportional to $J$. 
If the classical SM is amended by three right--handed neutrinos, a CP odd invariant analogue to $J$ appears in the lepton sector. 
In general, the vanishing of individual CP odd basis invariants is only a necessary condition for CP conservation \cite{Bernabeu:1986fc, Botella:1994cs}. 
A sufficient condition for CP conservation is that all CP odd basis invariants vanish.

\chapter{Group theoretical introduction to outer automorphisms}
\label{sec:OutIntro}

After discussing the classical definitions of C, P, and T, the following sections will pave the way to understand these transformations 
in a possibly more formal, yet certainly more vivid,
group theoretical language.
An effort is made to use a physical language and unnecessary mathematical details will be skipped whenever possible.
Knowledge of basic group theory is assumed, and the reader is reminded of the classical introductory literature to the 
subject of group theory in physics, cf.\ e.g.\ \cite{Cornwell:1985xs,Georgi:1999jb,Fuchs:1997jv,Ramond:2010zz}. 
The crucial parts of group theory related to (outer) automorphisms are typically not entirely satisfactory covered in the standard literature
and, therefore, will briefly be introduced in this section.
The formal discussion will mostly be focused on the case of finite (discrete) groups which will serve as benchmark throughout this work.
As an explicit example, the complete automorphism structure of the discrete group $\Delta(54)$ will be investigated in detail.
In addition, outer automorphisms of semisimple and compact Lie--algebras will briefly be discussed, illustrated on the basis of the example \SU3.
A more detailed treatment of outer automorphisms of semisimple and compact Lie--algebras can be found in \cite{Grimus:1995zi}.
The discussion of outer automorphisms of the Poincaré (including the Lorentz) group
will mostly be reviewed and not performed in every detail. A more detailed treatment of this case can be found in \cite{Buchbinder:2000cq}.
Most of the formalities discussed for finite groups straightforwardly adapt also to the other cases and the analogies will be 
pointed out at the appropriate places.

\section{Definitions}
\label{sec:Definitions}
For clarity the possibly not so well--known necessary terms are defined. 
For the definitions of other group theoretical terms see any book on group theory, for example \cite{Cornwell:1985xs, Fuchs:1997jv, Ramond:2010zz}.
The focus is on finite groups.

\paragraph{Group homomorphism.}
Given two groups, $G$ with multiplication $\bullet$ and $H$ with multiplication $\circ$, a group homomorphism is a map $h:G\rightarrow H$ such that for all $\elm{g_1,g_2}\in G$
\begin{equation}
h(\elm g_1\bullet\elm g_2)~=~h(\elm g_1)\circ h(\elm g_2)\;.
\end{equation} 
A direct consequence of the definition is that the identity elements of $G$ and $H$ are identified,
and that inverse elements in $G$ are mapped to inverse elements in $H$.
Therefore, a group homomorphism preserves the group structure.

\paragraph{Automorphism group.}
A bijective group homomorphism $G\rightarrow G$ is called an automorphism. 
All the automorphisms of a group themselves form a group (under composition) called the automorphism group of $G$, $\mathrm{Aut}(G)$.
Note that the automorphism group $\mathrm{Aut}(G)$ contains \textit{all} possible maps of a group $G$ to itself, that is,
it describes the symmetry properties of $G$.

\paragraph{Inner automorphism group.}
For each group element $\elm g$, the conjugation map $\mathrm{conj}_{\elm g}:\elm h\rightarrow\elm{g\,h\,g}^{-1}$ for all $\elm h\in G$ is an automorphism 
of $G$. Together, all automorphisms that can be represented by such a conjugation map form the inner automorphism group $\mathrm{Inn}(G)$,
which is a subgroup of $\mathrm{Aut}(G)$.

\paragraph{Conjugacy classes.}
The set of elements of a group which are related by the conjugation with a group element, that is by an inner automorphism, form a so--called conjugacy class.

\paragraph{Normal subgroup.}
A normal subgroup $N\subseteq G$ is a subgroup of $G$ that is invariant under all inner automorphisms. 

\paragraph{Quotient group.}
Let $N$ be a normal subgroup of $G$. The set of all (left) cosets of $N$ in $G$, that is 
\begin{equation}
G/N~:=~\{\elm gN:\elm g\in G\}\;,
\end{equation}
where $\elm gN$ stands for the set of products of a group element $\elm g$ with \textit{all} elements of $N$, forms a group on its own.
$G/N$ is called the quotient group of $G$ by $N$. An easy way to visualize the quotient group $G/N$ is to simply identify all elements of
$N$ with the identity.

\paragraph{Center.} 
The center of a group $\mathrm{Z}(G)$ is the set of elements in $G$ which commute with every other element. 
$\mathrm{Z}(G)$ always is a normal subgroup of $G$. 
Due to the fact that conjugation with a group element $\elm g\in G$ leads to the trivial 
automorphism if and only if $\elm g\in\mathrm{Z}(G)$ it is clear that there is an isomorphism
\begin{equation}
\mathrm {Inn}(G)\,\cong\,{G}\,/\,\mathrm{Z}(G)\,.
\label{eq:inner_aut}
\end{equation}

\paragraph{Outer automorphism group.}
Any automorphism in $\mathrm{Aut}(G)$ which is not inner, that is, which cannot be represented by the conjugation with a group element, is an outer automorphism.
Due to the fact that inner automorphisms $\mathrm{Inn}(G)$ form a normal subgroup of $\mathrm{Aut}(G)$ the outer automorphism group can be constructed via
\begin{equation}
\mathrm{Out}(G)~:=~\mathrm{Aut}(G)/\mathrm{Inn}(G)\;.
\end{equation}
Note that outer automorphisms are strictly speaking not automorphisms but equivalence classes of automorphisms. 
Colloquially speaking this means that each outer automorphism contains \textit{all} inner automorphisms. 

Inner automorphisms, being represented by conjugation operations with group elements themselves, always leave the conjugacy classes 
of a group invariant. In contrast, outer automorphisms typically (but not necessarily) interchange different conjugacy classes.
For explicit (matrix) representations of a group this implies that outer automorphisms may interchange inequivalent representations,
a fact that will later be explained in much more detail.

A very useful, albeit mathematically not completely correct, way to think about outer automorphisms $A_{\elm a}:G\rightarrow G$ 
is in form of a conjugation operation
\begin{equation}
A_{\elm a}(\elm g)~=~\elm {a\,g\,a^{-1}} ~~~\forall \elm g\in G\;,
\end{equation}
where $\elm a\notin G$. This is not entirely correct because there is formally no multiplication defined between $\elm a$ and $\elm g$.

\paragraph{Direct product.}
It is always possible to combine two groups to a larger group $F$ via the Cartesian product of the elements of $G$ and $G'$ 
(i.e. ordered pairs $(\elm g,\elm g')$) and the group multiplication law in $F$ defined by
\begin{equation}
(\elm {g,g'})\,(\elm {h,h'})~:=~(\elm {g\,h,g'\,h'})\;,
\end{equation}
where $\elm {g,h}\in G$ and $\elm {g',h'}\in G'$. $F\cong G\times G'$ is called the direct product of $G$ and $G'$.

\paragraph{Semidirect product.}
A more involved way of combining two groups is the semidirect product. Let $N$ and $H$ be groups, and $f:H\rightarrow\mathrm{Aut}(N)$ a group homomorphism 
from $H$ to the automorphism group of $N$ with $f(\elm h)\equiv f_{\elm h}$ and $f_{\elm h}(\elm n)=\elm {h\,n\,h^{-1}}~\forall\elm h\in H,~\elm n\in N$. 
The multiplication law of the semidirect product group $G\cong N\rtimes_fH$ can then be defined via the Cartesian product 
\begin{equation}
(\elm n_1,\elm h_1)\,(\elm n_2,\elm h_2)~:=~(\elm n_1\,f_{\elm h_1}(\elm n_2),\elm h_1\,\elm h_2)\;.
\end{equation}
In the following the subscript $f$ will be dropped whenever the corresponding homomorphism is obvious from the context. 
The elements of $G$ are uniquely given by $\elm {nh}$ where $\elm n\in N$ and $\elm h\in H$. Note that $N$ is a normal subgroup of the semidirect product group.

\paragraph{Group presentation.}
A very intuitive way to define groups and understand the formation of direct and semidirect products of groups is via so--called group presentations. 
A group presentation is given by a set of generators $\mathcal G(G)$ and a set of relations $\mathcal R(G)$ on them. A group $G$ then can be defined by
\begin{equation}
G~:=~\Braket{\mathcal G(G)\,|\,\mathcal R(G)}\;.
\end{equation}
For example, an Abelian group of order $n$ can be presented by a single generator $\elm a$ that fulfills the relation $\elm a^n=\elm e$ ($\elm e$ denotes the identity) and therefore,
\begin{equation}
\Z{n}~:=~\Braket{\elm a \,|\, \elm a^n~=~\elm e}\;.
\end{equation}
Note that group presentations are typically not unique, not even in the number of generators or relations.
Nevertheless, there are so--called minimal generating sets which contain a minimal number of generators and relations. 
For the combination of two groups $N$ and $H$, both the direct and the semidirect product group $G$ can be presented by the union of both sets of generators 
$\mathcal G(G) = \mathcal{G}(N)\,\cup\,\mathcal G(H)$, and $\mathcal R(G)=\mathcal R(N)\,\cup\,\mathcal R(H)+\mathcal R_{\mathrm{new}}$ relations.
Here, $\mathcal R_{\mathrm{new}}$ are $|\mathcal R_{\mathrm{new}}|=|\mathcal{G}(N)|\times|\mathcal G(H)|$ new additional relations between the generators.

In case of a direct product, the generators of both groups commute by assumption, implying that the additional 
new relations are trivial, i.e.\ of the type $\elm{h}\,\elm{n}\,\elm{h}^{-1}=\elm n,~\forall\elm{h}\in H~\text{and}~\forall\elm n\in N$. 

In contrast, for the case of a semidirect product there are new and non--trivial relations of the type $\elm h\,\elm n\,\elm h^{-1}=f_{\elm{h}}(\elm n)$.
This allows one to understand why $f_{\elm{h}}(\elm n)$ must be a mapping into the automorphism group of $N$.
If it were not, the new relations would be in contradiction with the, of course, still present relations $\mathcal R(N)$ of $N$, thereby invalidating 
the whole construction.

\section{Representation matrices of outer automorphisms}

The action of outer automorphisms on group representations shall briefly be discussed in the following.
The notation is such that $\rho_{\rep r}(\mathsf{g})$ denotes the unitary matrix representation of an abstract group element $\elm g$ in the representation $\rep{r}$.

From the preceding subsection it is clear that automorphisms are transformations 
that leave the structure of a group, i.e.\ the group algebra, invariant. Furthermore, it has also been noted 
that outer automorphisms may induce non--trivial permutations among the conjugacy classes of a group.
As a consequence also class functions, such as the characters of a representation, 
may be non--trivially permuted under the action of outer automorphisms.
However, the characters uniquely determine a representation (up to equivalence, that is up to similarity transformations) \cite{Cornwell:1985xs}. 
Therefore, whenever an outer automorphism induces a permutation of the characters, it will also induce a permutation 
of inequivalent representations. 

In general, for an (outer) automorphism that acts as $u: \mathsf{g}\mapsto u(\mathsf{g})$ and maps a representation $\rep r$ to a representation $\rep{r'}$, 
the explicit representation matrix $U$ of $u$ is given by the solution to
\begin{equation}\label{eq:consistencyEquation}
U\,\rho_{\rep{r'}}(\mathsf{g})\,U^{-1}~=~\rho_{\rep{r}}(u(\mathsf{g}))\;,\qquad\forall \mathsf{g}\in G\;.
\end{equation}
This definition equally holds for inner and outer automorphisms, where for inner automorphisms $\rep{r}\equiv\rep{r'}$ is automatically implied.
Furthermore, note that the matrices $U$ are always defined only up to a phase and up to an element representing the center of $G$. 
Equation \eqref{eq:consistencyEquation} is a consistency condition in the sense that one can find a non--trivial solution for $U$ 
if and only if there exists an appropriate automorphism $u(\mathsf{g})$. This statement is proven in \Appref{app:CCProof}.

\section{Outer automorphisms of finite groups}

\subsection{Explicit example: \texorpdfstring{$\Delta(54)$}{}}
\label{Sec:Delta54}

In order to become acquainted with the just introduced definitions,  
this section explicitly demonstrates the step by step construction of the outer automorphism group and its implications for the finite group
$\Delta(54)$. Even though all of the computations can be performed manually, the reader be reminded of \textsc{GAP} \cite{GAP4}, a powerful 
computer code for the work with finite groups, and the \textsc{Discrete} \cite{Holthausen:2011vd} package which provides a \textsc{GAP}--\textsc{Mathematica} interface.
In the SmallGroup catalogue of \textsc{GAP}, $\Delta(54)$
is included as $\mathrm{SG}(54,8)$.

A presentation for the group \Dff is given by
\begin{equation}\label{eq:DffPrsentation}
\Dff~=~\Braket{\elm{A,B,C}\,|\,
\mathsf{A}^3\,=\,\mathsf{B}^3\,=\,\mathsf{C}^2\,=\,\left(\mathsf{A}\,\mathsf{B}\right)^3\,=\,\left(\mathsf{A}\,\mathsf{C}\right)^2\,=\,\left(\mathsf{B}\,\mathsf{C}\right)^2\,=\,\mathsf{e}}\;.
\end{equation}
The group has $54$ elements and its conjugacy classes are given by
\begin{align}
C_{1a}& : \mathsf{\{e \}\;,}               &  \nonumber\\
C_{3a}& : \mathsf{\{A, A^2, BAB^2, B^2AB, BA^2B^2, B^2A^2B \}\;,} &\nonumber\\
C_{3b}& : \mathsf{\{B, B^2, ABA^2, A^2BA, AB^2A^2, A^2B^2A \}\;,} &\nonumber\\
C_{3c}& : \mathsf{\{AB^2, A^2B, BA^2, B^2A, ABA, BAB \}\;, }      &\nonumber\\
C_{3d}& : \mathsf{\{AB, BA, A^2B^2, B^2A^2, AB^2A, A^2BA^2 \}\;,} &\nonumber\\
C_{2a}& : \mathsf{\{C, AC, A^2C, BC, B^2C, ABAC, BABC, A^2BA^2C, AB^2AC \}\;,} &\nonumber\\
C_{6a}& : \mathsf{\{BAC, A^2BC, AB^2C, B^2A^2C, B^2ABC, BA^2B^2C, ABA^2C, A^2B^2AC, AB^2ABAC \}\;,} &\nonumber\\
C_{6b}& : \mathsf{\{ABC, BA^2C, B^2AC, A^2B^2C, A^2BAC, BAB^2C, AB^2A^2C, B^2A^2BC, BA^2BABC \}\;,} &\nonumber\\
C_{3e}& : \mathsf{\{AB^2ABA \}\;,}\qquad C_{3f}: \mathsf{\{BA^2BAB\}\;,}&
\label{eq:ccs}
\end{align}
and have been labeled by the order of their elements and a letter.
The non--trivial irreducible representations (irreps) of the group are the real representations $\rep[_1]{1}$ and $\rep[_i]{2}$ ($i=1,2,3,4$),
as well as the complex representations $\rep[_1]{3}$ and $\rep[_2]{3}$ and their respective conjugates.  The character table is
shown in \Tabref{tab:Delta54char}.

\renewcommand{\arraystretch}{1.}
\begin{table}[t!]
\centering
\resizebox{\textwidth}{!}{\begin{tabular}{c|rrrrrrrrrr|}
                      &  $C_{1a}$ & $C_{3a}$ & $C_{3b}$ & $C_{3c}$ & $C_{3d}$ & $C_{2a}$ & $C_{6a}$ & $C_{6b}$ & $C_{3e}$ & $C_{3f}$  \\
                      &  1 &  6 &  6 &  6 &  6 &  9 &  9 &  9 &   1 &  1  \\
\Dff         & $\mathsf{e}$ & $\mathsf{A}$  & $\mathsf{B}$  & $\mathsf{ABA}$ & $\mathsf{AB}$ & $\mathsf{C}$  & $\mathsf{ABC}$       & $\mathsf{BAC}$   & $\mathsf{AB^2ABA}$ & $\mathsf{BA^2BAB}$ \\
\hline
 $\rep[_0]{1}$       & $1$ & $1$  & $1$  & $1$  & $1$  & $1$  & $1$         & $1$         & $1$         & $1$        \\
 $\rep[_1]{1}$       & $1$ & $1$  & $1$  & $1$  & $1$  & $-1$ & $-1$        & $-1$        & $1$         & $1$        \\
 $\rep[_1]{2}$       & $2$ & $2$  & $-1$ & $-1$ & $-1$ & $0$  & $0$         & $0$         & $2$         & $2$        \\
 $\rep[_2]{2}$       & $2$ & $-1$ & $2$  & $-1$ & $-1$ & $0$  & $0$         & $0$         & $2$         & $2$        \\
 $\rep[_3]{2}$       & $2$ & $-1$ & $-1$ & $2$  & $-1$ & $0$  & $0$         & $0$         & $2$         & $2$        \\
 $\rep[_4]{2}$       & $2$ & $-1$ & $-1$ & $-1$ & $2$  & $0$  & $0$         & $0$         & $2$         & $2$        \\
 $\rep[_1]{3}$       & $3$ & $0$  & $0$  & $0$  & $0$  & $1$  & $\omega^2\hspace{-5pt}$  & $\omega$    & $3\omega$ & $3\omega^2\hspace{-5pt}$  \\
 $\rep[_1]{\bar{3}}$ & $3$ & $0$  & $0$  & $0$  & $0$  & $1$  & $\omega$    & $\omega^2\hspace{-5pt}$  & $3\omega^2\hspace{-5pt}$ & $3\omega$  \\
 $\rep[_2]{3}$       & $3$ & $0$  & $0$  & $0$  & $0$  & $-1$ & $-\omega^2\hspace{-5pt}$ & $-\omega$   & $3\omega$ & $3\omega^2\hspace{-5pt}$  \\
 $\rep[_2]{\bar{3}}$ & $3$ & $0$  & $0$  & $0$  & $0$  & $-1$ & $-\omega$   & $-\omega^2\hspace{-5pt}$ & $3\omega^2\hspace{-5pt}$ & $3\omega$  \\
\hline
\end{tabular}}
\caption{Character table of \Dff. The definition $\omega:=\mathrm{e}^{2\pi\,\I/3}$ is used.
The second line gives the cardinality of the conjugacy class (c.c.)\ and
the third line gives a representative of the corresponding c.c.\ in the presentation specified
in \eqref{eq:DffPrsentation}.}
\label{tab:Delta54char}
\end{table}

The outer automorphism group of \Dff can be found via the construction outlined in the last section.\footnote{%
For manual computations in finite groups it is extremely useful and highly recommended to firstly identify useful identities 
from the group algebra. For \Dff those relations are  $\elm{CA}=\elm A^2\elm C$; $\elm{CB}=\elm B^2\elm C$; 
$\elm{BAB}=\elm A^2\elm B^2\elm A^2$; $\elm{ABA}=\elm B^2\elm A^2\elm B^2$; $\elm B\elm A^2\elm B=\elm A\elm B^2\elm A$; $\elm A^2\elm B\elm A^2=\elm B^2\elm A\elm B^2$.}
The starting point is the automorphism group. 
In a brute force way the automorphism group can be obtained 
by successively mapping every generator to every other element of the same order, while checking 
whether the group structure is preserved. All maps that preserve the group structure are automorphisms. 
The outer automorphism group is then given by the 
quotient of the automorphism group with respect to the inner automorphism group.
The inner automorphism group can be found by taking the quotient group of \Dff with respect to its center, while
the center of \Dff can straightforwardly be found by checking commutation properties of group elements.
For \Dff it is given by the subgroup \Z3 which, in this presentation, is generated by the element $\mathsf{AB^2ABA}$.
Since performing these computations is very tedious, in practice it is much more convenient to use computer codes such as \textsc{GAP}.
A computer code which performs the computation of the outer automorphism group in \textsc{GAP} is given in \Appref{app:Out}.

The automorphism structure of \Dff can be summarized as
\begin{align}
\mathrm{Z}(\Dff)&~=~\Z3\;,& \mathrm{Aut}(\Dff)&~=~\mathrm{SG}(432,734)\;,&\\
\mathrm{Inn}(\Dff)&~=~(\Z3\times\Z3)\rtimes \Z2\;,&  \mathrm{Out}(\Dff)&~=~\Sfour\;.&
\end{align}
The outer automorphism group of \Dff turns out to be $\Sfour$, the permutation group of four elements. 
A minimal generating set for $\Sfour$ has only two elements, and the group can be presented via
\begin{equation}\label{eq:S4algebra}
\Sfour~=~\Braket{\elm{S,T}\,|\,\mathsf{S}^2\,=\,\mathsf{T}^3\,=\,\left(\mathsf{T}^2\,\mathsf{S}\right)^4\,=\,\mathsf{e}}\;.
\end{equation}
A possible choice for the action of the outer automorphisms $s$ and $t$ on the group elements of \Dff is given by
\begin{equation}\label{eq:SfourAction}
s:~(\mathsf{A,B,C})\mapsto(\elm A\,\elm B^2\,\elm A,\elm B, \elm C)\quad\text{and}\quad t:~(\mathsf{A,B,C})\mapsto(\mathsf{A,A\,B\,A,C})\,.
\end{equation}
Stating the explicit action of $s$ and $t$ it is important to keep in mind that the outer automorphism group 
\Sfour in this construction is not a group of automorphisms but of cosets of automorphisms.
This implies that an element of \Sfour is not a single automorphism, but an outer automorphism that additionally 
contains all inner automorphisms. Nevertheless, it is possible and useful for practical computations to choose one 
particular representative, that is one particular inner automorphism, of each coset. This has been done 
in stating the explicit action \eqref{eq:SfourAction}.
The results of the computations will not depend on the particular choice as all other elements of the coset
can be obtained by taking into account inner automorphisms. 

As another consequence of this, note that acting with the identity outer automorphism of \Sfour on the group 
elements of \Dff does not necessarily refer to a conjugation with the identity element of \Dff.
Instead, the trivial outer automorphism $\mathsf{e}$ of \Sfour refers to every inner automorphism of \Dff. 
The chosen presentation of \Sfour actually suffers from this degeneracy because the composition 
$\left(t^2\circ s\right)^4$ only closes to an inner automorphism of \Dff which corresponds to
the conjugation with the group element $\mathsf{C}$.

Physically most relevant is the action of outer automorphisms on the representations of \Dff which shall be derived in the following.
Knowing the explicit action of the outer automorphisms on the group elements, \Eqref{eq:SfourAction}, one can also track the action on
the characters of \Dff, cf.\ \Tabref{tab:Delta54char}. Comparing the sequence of characters after application of the outer automorphism 
to the original character table, it is possible to sort out the permutation of representations under the action of the outer automorphism. 
For example, for the outer automorphism $s$ one can easily check that the last four columns of the character table are permuted in such a way
that $\rep[_1]{3}\leftrightarrow\rep[_1]{\bar{3}}$ and $\rep[_2]{3}\leftrightarrow\rep[_2]{\bar{3}}$. Therefore, $s$ corresponds to a 
complex conjugation outer automorphism for these representations. 
Note that by the outlined procedure one has identified a symmetry of the character table under simultaneous permutation of rows and columns.
It is in general true that outer automorphisms of finite groups correspond to some symmetry of the character table \cite{Holthausen:2012dk}.
Nevertheless, this is generally not a one--to--one relation as there exist class--preserving outer automorphisms, which do not permute characters.\footnote{%
An example for a group that has a class--preserving outer automorphism is $\mathrm{SG}(32,43)$ \cite{Fallus:2015}.
}
 
In order to obtain the explicit action of outer automorphisms on group elements (operators) and states (fields) it is -- for the first time in this computation -- 
necessary to specify an explicit basis for the representations. A possible choice of representation matrices for the triplet representation $\rep[_{1}]{3}$ is\footnote{%
The representation matrices of $\rep[_{2}]{3}$ can be chosen as $A$, $B$, and $-C$.}
\begin{equation}\label{eq:delta54gens}
 A~=~\begin{pmatrix} 0 & 1 & 0 \\ 0 & 0 & 1 \\ 1 & 0 & 0 \end{pmatrix}\;, 
 \qquad B\,=\,\begin{pmatrix} 1 & 0 & 0 \\ 0 & \omega & 0 \\ 0 & 0 & \omega^2 \end{pmatrix}\;,
 \qquad C\,=\,\begin{pmatrix} 1 & 0 & 0 \\ 0 & 0 & 1 \\ 0 & 1 & 0 \end{pmatrix}\;,
\end{equation}
where here and in the following $\omega$ is defined as $\omega:=\mathrm{e}^{2\pi\,\I/3}$. 
Even though not required, it is highly recommended and very convenient to chose for $\rep[_{1}]{\bar{3}}$ the 
respective complex conjugate matrices.
An explicit representation matrix of the outer automorphism then can be obtained by solving the consistency condition \eqref{eq:consistencyEquation}.
For the action of the outer automorphism $s$ the consistency condition takes the form 
\begin{equation}\label{eq:ExplicitConsistencyEquation}
U_s\,\rho_{\rep[_1]{\bar{3}}}(\mathsf{g})\,U_s^{-1}~=~\rho_{\rep[_1]{3}}(s(\mathsf{g}))\;,\qquad\forall \mathsf{g}\in \Dff\;.
\end{equation}
It is sufficient to solve this equations for the generators of \Dff to fix $U_s$. Therefore, $U_s$ is given by the simultaneous solution to 
\begin{equation}\label{eq:}
U_s\,A^*\,U_s^{-1}~=~A\,B^2\,A\;,\quad U_s\,B^*\,U_s^{-1}~=~B\;, \quad\text{and}\quad U_s\,C^*\,U_s^{-1}~=~C\;.
\end{equation}
Completely analogous, it is possible to find $U_t$ as the explicit representation of the automorphism $t$ which, however, maps all triplet representations 
to themselves. Altogether one finds
\begin{equation}\label{eq:Delta54Aut}
U_{s}~=~\begin{pmatrix} \omega^2 & 0 & 0 \\ 0 & 0 & 1  \\ 0 & 1 & 0 \end{pmatrix} \quad \text{and}\quad
U_{t}~=~\frac{\I}{\sqrt{3}}\begin{pmatrix} 1 & \omega^2 & \omega^2 \\ \omega^2 & 1 & \omega^2  \\ \omega^2 & \omega^2 & 1 \end{pmatrix}\;,
\end{equation}
with the explicit action on the triplet representations
\begin{equation}\label{eq:Delta54AutTriplet}
 s:~\rep[_i]{3}\mapsto U_{s}\,\rep{3}^*_i\;,\quad\text{and}\quad t:~\rep[_i]{3}\mapsto U_{t}\,\rep[_i]{3}\;.
\end{equation}
As mentioned earlier, $U_s$ and $U_t$ are only fixed up to a complex phase. 
Nevertheless, for convenient computations the phase here is fixed by the requirement that $U_s$ and $U_t$ fulfill the group algebra of \Sfour \eqref{eq:S4algebra}
(up to inner automorphisms). Interestingly, all odd permutations in \Sfour correspond to transformations that interchange complex conjugate triplet representations, 
whereas all even permutations map the triplets to themselves.

In general, note that every transformation which fulfills the consistency condition, i.e.\ is consistent with the original group algebra, can be used to 
enlarge the group by a non--trivial semidirect product to a bigger group. An explicit construction of how to construct this bigger group is outlined in the following.
Again the outer automorphism $s$ of \Dff will be used as an example. From the preceding paragraphs it is clear that $s$ maps $\rep[_1]{3}\leftrightarrow\rep[_1]{\bar{3}}$.
Consequently, if the corresponding transformation should be added to the symmetry transformations of the group, it is clear that the bigger group will have a 
six--dimensional representation that unifies $\rep[_1]{3}$ and $\rep[_1]{\bar{3}}$. That is, the larger group $H$ will have some representation $\rep{6}$ that branches 
as $\rep{6}\rightarrow\rep[_1]{3}\oplus\rep[_1]{\bar{3}}$ in the group $G\subset H$. 

The construction of the group $H$ as a matrix group will be discussed in the following. 
A somewhat more mathematical treatment of this can be found in \cite{Holthausen:2012dk}. 
The basic idea is to start with a reducible representation of $G$ containing $\rep[_1]{3}\oplus\rep[_1]{\bar{3}}$
and then add the explicit action of the outer automorphism group.
The main point is that elements of $G$ only act on the triplets separately, while elements of $H$ which are not in $G$
will interrelate the triplets. This is manifest in the explicit form of the representation matrices. 
The representation matrices of $\rep[_1]{3}\oplus\rep[_1]{\bar{3}}$ in $G$ are given by
\begin{equation}\label{eq:Rep6}
  A_{\rep{6}}  ~=~ \begin{pmatrix} A & \mathbf{0}\\ \mathbf{0} & A^*\end{pmatrix}\,, \qquad 
  B_{\rep{6}}~=~\begin{pmatrix} B & \mathbf{0}\\ \mathbf{0} & B^*\end{pmatrix}\,,\quad \text{and}\quad 
  C_{\rep{6}}~=~\begin{pmatrix} C & \mathbf{0}\\ \mathbf{0} & C^*\end{pmatrix}\,. 
\end{equation}
Together with the new matrix 
\begin{equation}
 S_{\rep{6}}  ~=~ \begin{pmatrix} \mathbf{0} & U_s \\ U_s^* & \mathbf{0}\end{pmatrix}\;,
\end{equation}
these matrices define the group $H$. By putting the matrices into \textsc{GAP}, one finds that $H=\mathrm{SG}(108,17)$ and
one confirms that this group has the corresponding $\rep{6}$--plet representation.
For completeness, note that the analogous construction for the extension of \Dff by $\mathsf{T}$ results in the group $\mathrm{SG}(162,10)$.

Finally, the behavior of the doublet representations of \Dff under outer automorphisms shall be discussed. 
A set of possible representation matrices for the doublets is given in \Tabref{tab:DffDoublets}, using the matrices
\begin{equation}\label{eq:DoubletDefs}
  \mathbbm{1}_{\rep{2}}~=~ \begin{pmatrix} 1 & 0\\ 0 & 1\end{pmatrix}\,, \qquad 
  \Omega_{\rep{2}}~=~ \begin{pmatrix} \omega^2 & 0\\ 0 & \omega\end{pmatrix}\,,\quad \text{and}\quad 
  S_{\rep{2}}~=~\begin{pmatrix} 0 & 1\\ 1 & 0\end{pmatrix}\,.
\end{equation}
\renewcommand{\arraystretch}{1.2}
\begin{table}%[!h!]
\centerline{\begin{tabular}{c|cccc}
   & \rep[_1]{2} & \rep[_2]{2} & \rep[_3]{2} & \rep[_4]{2} \\
 \hline
 $A_{\rep[_i]{2}}$ & $\mathbbm{1}_{\rep{2}}$ & $\Omega_{\rep{2}}$ & $\Omega_{\rep{2}}$ & $\Omega_{\rep{2}}$ \\
 $B_{\rep[_i]{2}}$ & $\Omega_{\rep{2}}$ & $\mathbbm{1}_{\rep{2}}$ & $\Omega_{\rep{2}}$ & $\Omega_{\rep{2}}^*$ \\
 $C_{\rep[_i]{2}}$ & $S_{\rep{2}}$ & $S_{\rep{2}}$ & $S_{\rep{2}}$ & $S_{\rep{2}}$ \\
\end{tabular}}
\caption{Explicit matrices for the doublet representations of \Dff, see \eqref{eq:DoubletDefs} for a definition of the matrices $\Omega_{\rep{2}}$ and $S_{\rep{2}}$.}
\label{tab:DffDoublets}
\end{table}\renewcommand{\arraystretch}{1.0}%
The action on the doublet representations then is found to be
\begin{equation}\label{eq:STActionDoublets}
s:~\begin{pmatrix} \rep[_1]{2} \\ \rep[_2]{2} \\ \rep[_3]{2} \\ \rep[_4]{2} \end{pmatrix} \mapsto 
 \begin{pmatrix} S_{\rep2}\,\rep[_4]{2} \\ S_{\rep2}\,\rep[_2]{2} \\ \phantom{S_{\rep2}\,}\rep[_3]{2} \\ S_{\rep2}\,\rep[_1]{2} \end{pmatrix}\;,\quad\text{and}\quad
 t:~\begin{pmatrix} \rep[_1]{2} \\ \rep[_2]{2} \\ \rep[_3]{2} \\ \rep[_4]{2} \end{pmatrix} \mapsto 
 \begin{pmatrix} \rep[_1]{2} \\ \rep[_4]{2} \\ \rep[_2]{2} \\ \rep[_3]{2} \end{pmatrix}\;.
\end{equation}
Consequently, the permutations of \Dff doublets under the outer automorphism group (generated by $s$ and $t$) correspond to all possible 
permutations of four elements. That is, the four doublets form a \rep{4}--plet under the outer automorphism group \Sfour.

Analogous to the construction of the \rep{6}--plet above it is possible to construct representations which contain multiple copies of 
the two dimensional representations and the corresponding groups by amending \Dff by outer automorphism.

Starting from a given group, it may be extended by one or multiple of its outer automorphisms.
The thereby resulting group will in general have new outer automorphisms which were not present on the level of the original group.
Conversely, it is also true that subgroups sometimes have outer automorphisms which are neither part of the 
supergroup nor part of any of its outer automorphisms. 
Altogether, thus, moving in a ``stack'' of supergroups and subgroups, outer automorphisms may appear and disappear at any level.
This unpredictability of the appearance and disappearance of outer automorphisms seems to 
be closely related to the so--called extension problem of finite groups, which is an essential obstacle 
in the systematic classification of finite groups. This question will not be discussed here any further.

The action of outer automorphisms on representations of finite groups can be summarized as follows. In general, outer automorphisms 
act as a permutation of representations of the same dimensionality. Whether or not such a permutation is possible
is entirely fixed by the structure of the group.
Representations which are permuted by a specific outer automorphism are merged to larger representations in the extended group, 
which is obtained as the semidirect product of the original group with the corresponding outer automorphism. 
Representations which are not permuted under the action of the outer automorphism, in general, are present in the extended group as well -- but there appear 
altogether $n$ copies of them, where $n$ is the order of the corresponding outer automorphism. 

\section{Outer automorphisms of continuous groups}

In the last section some feeling for outer automorphism has been gained from the considerations of finite groups. 
To obtain a complete picture including also the case of gauge and space--time symmetries, 
outer automorphisms will briefly be discussed for the case of continuous groups in the following. 
The focus will be on intuitive and picturesque arguments and mathematical details will be skipped where they are unnecessary.
A more thorough treatment of some of the formalities can, for example, be found in \cite{Gantmacher:1939, Cornwell:1985xt, Grimus:1995zi, Fuchs:1997jv, Hall:2000, Hall:2015}.

Considering a Lie algebra $\mathfrak{L}$ with elements $x$ and $y$, an automorphism of the Lie algebra is given by a linear
mapping $\psi:\mathfrak{L}\to\mathfrak{L}$ that respects the structure of the Lie algebra as
\begin{equation}\label{eq:LieAutStructure}
 \psi(\left[x,y\right])~=~\left[\psi(x),\psi(y)\right]\;\quad\forall x,y\in\mathfrak{L}\;.
\end{equation}
Choosing an explicit basis $\left\{x_a\right\}$ for the generators, the automorphism acts on the generators as 
\begin{equation}
 \psi_R:~x_a~\mapsto~R_{ab}\,x_b\;.
\end{equation}
For compact Lie algebras\footnote{%
A semisimple Lie group $G$ is compact if and only if its Lie algebra has a negative definite Killing form, cf.\ e.g.\ \cite{Grimus:1995zi}.
The Lie algebras of compact Lie groups are called compact.
}
one has the well--known
\begin{equation}
 \left[x_a,x_b\right]~=~\I\,f_{abc}\,x_c\;.
\end{equation}
Therefore, from condition \eqref{eq:LieAutStructure} one finds that 
\begin{equation}\label{eq:ConditionAnonON}
 R_{aa'}\,R_{bb'}\,f_{a'b'c}~=~f_{abc'}\,R_{c'c}\;
\end{equation}
must be fulfilled by $R$ in order for it to be an automorphism.

By choosing an orthonormal basis $\left\{x_a\right\}$, which obeys the normalization $\tr\left(x_ax_b\right)=k\delta_{ab}$,
one finds that $R$ must be an orthogonal matrix in order to conserve the norm and \eqref{eq:ConditionAnonON} can be
written as 
\begin{equation}\label{eq:ConditionA}
 R_{aa'}\,R_{bb'}\,R_{cc'}\,f_{a'b'c'}~=~f_{abc}\;.
\end{equation}

Defining the adjoint map w.r.t.\ a Lie algebra element $x\in\mathfrak{L}$ as
\begin{equation}
 \mathrm{ad}_x:~y~\mapsto~\mathrm{ad}_x(y)~:=~\left[x,y\right]\quad\forall y\in\mathfrak{L}\;,
\end{equation}
one can show that it fulfills \eqref{eq:LieAutStructure}.
The adjoint map defines an automorphism of $\mathfrak{L}$ w.r.t.\ to the element $x$ via the mapping \cite{Grimus:1995zi}%
\begin{equation}
 \psi_x(y)~:=~\e^{\mathrm{ad}_x}(y)~=~\e^x\,y\,\e^{-x}\;.
\end{equation}
Here the exponential of the adjoint map is meant as the power series in composition of maps, and the last equality holds in case $x$ and $y$ are
taken as explicit matrices, which is always possible \cite{Jacobson:1979} (cf.\ also \cite[exercise 3.14]{Hall:2000}).
Automorphisms which can be written in this way are called inner automorphisms of a Lie algebra, whereas all other automorphisms are called outer.

Completely analogous to the conjugation map for finite groups, 
also for Lie groups $L$ there is the usual conjugation map
\begin{equation}\label{eq:InnerAutL}
L~\mapsto~\mathrm{Ad}_{A}(L)~:=~A\,L\,A^{-1}\;,
\end{equation}
with respect to any element $A\in L$, which corresponds to an inner automorphism of $L$. One can show that \eqref{eq:InnerAutL}
also corresponds to an inner automorphism of the corresponding Lie algebra $\mathfrak{L}$ of $L$ (cf.\ e.g.\ \cite[exercise 3.13]{Hall:2015}).
In particular, $\mathrm{Ad}_{A}(x)=A\,x\,A^{-1}\in\mathfrak{L}$ for all $x\in\mathfrak{L}$.
Again, every automorphism that can be written in the form \eqref{eq:InnerAutL} is called an inner automorphism,
whereas an automorphism is called outer if this is not the case.

Note that the consistency condition \eqref{eq:consistencyEquation} has a straightforward translation to 
the elements of a Lie algebra. That is, for generators of a given explicit representation $T^{(\rep{r})}_a$ one can find matrices
$U$ and $R$ such that 
\begin{equation}\label{eq:consistencyEquationLie}
U\,T^{(\rep{r'})}_a\,U^{-1}~=~R_{ab}\,T^{(\rep{r})}_b\;,\qquad\forall a\;,
\end{equation}
if and only if there is an automorphism mapping $\rep{r}\mapsto\rep{r'}$ with $\dim(\rep{r})=\dim(\rep{r'})$.
 
For the adjoint representation, which is unique in its dimension and, therefore, always mapped to itself, \Eqref{eq:consistencyEquationLie} 
nicely merges to \eqref{eq:ConditionAnonON}. Consequently, all possible automorphisms can readily be classified from the possible 
non--trivial mappings of the adjoint to itself. Since the roots of a Lie algebra are the weights of the adjoint representation, the 
root system of a Lie algebra completely reflects this symmetry. That is, the possible automorphisms of a given Lie algebra 
can be obtained from the symmetries of the root system.

\begin{figure}[t]
\centerline{\includegraphics[width=0.6\textwidth]{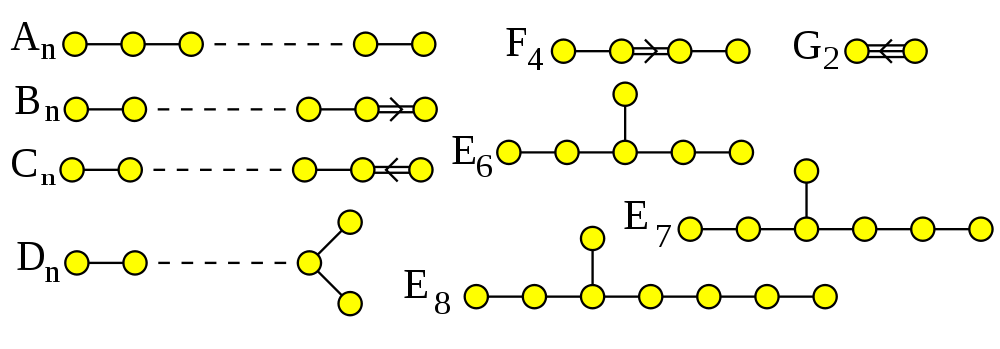}}
  \caption{Dynkin diagrams of all simple Lie algebras. The outer automorphism group of a Lie algebra is isomorphic to the symmetry of
its Dynkin diagram. Figure taken from \cite{Dynkin:2016} under the \textit{Creative Commons Attribution-Share Alike 3.0 Unported} license.}
  \label{fig:Dynkin} 
\end{figure}

Of highest interest for gauge theories are (semi)simple Lie groups. 
The corresponding simple Lie algebras can be classified in terms of their root system $\rho$, which is commonly done in the form of Dynkin diagrams.
As argued above, the complete root system has an automorphism group $\mathrm{Aut}(\rho)$ 
which is isomorphic to the automorphism group of the corresponding Lie algebra $\mathrm{Aut}(\mathfrak{L})$.
The normal subgroup of inner automorphisms $\mathrm{Inn}(\mathfrak{L})$ corresponds to the so--called Weyl symmetry of the root system $W$, which 
consists of all possible reflections of roots on hyperplanes perpendicular to each of the roots. In contrast, 
the outer automorphism group $\mathrm{Out}(\mathfrak{L})$ corresponds to ambiguities in the ordering of simple roots. Therefore,
it is isomorphic to the symmetry of the corresponding Dynkin diagram $S_{\mathrm{Dyn.}}$. In summary,
\begin{equation}
\mathrm{Out}(\mathfrak{L})~\cong~\mathrm{Aut}(\mathfrak{L})/\mathrm{Inn}(\mathfrak{L})~\cong~\mathrm{Aut}(\rho)/W~\cong~S_{\mathrm{Dyn.}}\;.
\end{equation}
The Dynkin diagrams of all simple Lie algebras are shown in \Figref{fig:Dynkin} and the corresponding outer automorphism groups are readily obtained from them.
In \Tabref{tab:SimpleAuts} the simple Lie algebras and the corresponding compact groups are summarized together with their outer automorphism groups
and the corresponding action on the representations.
\renewcommand{\arraystretch}{1.2}
\begin{table}[t!]
\centerline{
\begin{tabular}{lccc}
 ~Algebra~ & ~Group~ & ~Out~ & ~Action on irreps~   \\
 \hline  
    $\mathrm{A}_{n>1}$ & $\SU{n+1}$  & $\Z2$ & $\rep{r}\,~\mapsto~\rep{r}^*$  \\
    $\mathrm{D}_{n=4}$ & $\SO{8}$  & $\mathrm{S}_3$ & $\rep{r}_i~\mapsto~\rep{r}_j$ \\
    $\mathrm{D}_{n>4}$ & $\SO{2n}$ & $\Z2$ & $\rep{r}\,~\mapsto~\rep{r}^*$ \\
    $\mathrm{E}_6$     & $\mathrm{E}_6$     & $\Z2$ & $\rep{r}\,~\mapsto~\rep{r}^*$ \\ 
    all others    &  & $\elm e$  & $\rep{r}\,~\mapsto~\rep{r}^{\phantom{*}}$  \\
\end{tabular}}
\caption{List of all simple Lie algebras which have a non--trivial outer automorphism group, together with their compact real forms. The last column lists 
the action on the irreps of the corresponding group. All other simple Lie algebras (cf.\ \Figref{fig:Dynkin}) only have the trivial outer automorphism 
group.}
\label{tab:SimpleAuts}
\end{table}
\renewcommand{\arraystretch}{1.0}%
While finite groups generally feature very rich structures of outer automorphisms, there are only very few simple Lie groups
with non--trivial outer automorphisms.

\subsection{Explicit example: \SU3}
\label{sec:SU3example}

As an example, the Lie algebra $\mathfrak{su}(3)$ of the compact simple Lie group \SU3 shall be investigated with respect to its automorphism structure.
It is convenient to work in a basis with non--Hermitian generators in order to emphasize the connection to 
the symmetries of the root system. The relation to the usual basis of Gell--Mann matrices, as
well as further details, are given in \Appref{app:SU3}. 

\begin{figure}[t]
\centerline{\includegraphics[width=0.4\textwidth]{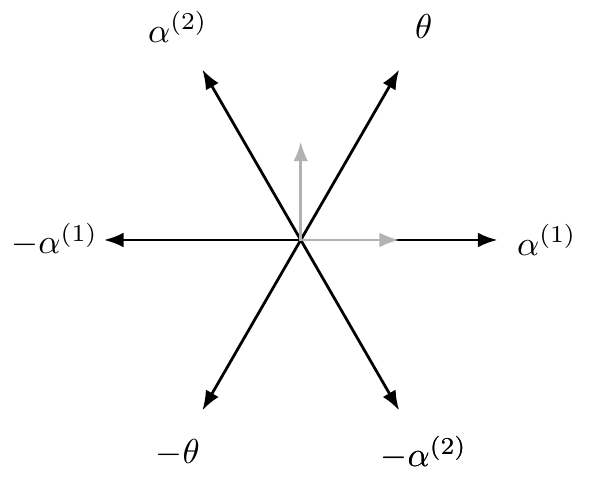}}
  \caption{Root system of $\mathfrak{su}(3)$ in the basis $(H_\mathrm{I},H_\mathrm{Y})$.}
  \label{fig:Su3Roots} 
\end{figure}

The generators of the fundamental representation are given by
\begin{equation}
\begin{split}\label{eq:SU3Gens}
H_\mathrm{I}~&=~\frac12\begin{pmatrix} 1 & 0 & 0 \\ 0 & -1 & 0  \\ 0 & 0 & 0 \end{pmatrix},&
H_\mathrm{Y}~&=~\frac{1}{2\sqrt{3}}\begin{pmatrix} 1 & 0 & 0 \\ 0 & 1 & 0  \\ 0 & 0 & -2 \end{pmatrix},\\
E^{1}_\mathrm{+}~&=~\frac{1}{\sqrt{2}}\begin{pmatrix} 0 & 1 & 0 \\ 0 & 0 & 0  \\ 0 & 0 & 0 \end{pmatrix},&
E^{\theta}_\mathrm{+}~&=~\frac{1}{\sqrt{2}}\begin{pmatrix} 0 & 0 & 1 \\ 0 & 0 & 0  \\ 0 & 0 & 0 \end{pmatrix},&
\!\!\!E^{2}_\mathrm{+}~&=~\frac{1}{\sqrt{2}}\begin{pmatrix} 0 & 0 & 0 \\ 0 & 0 & 1  \\ 0 & 0 & 0 \end{pmatrix}, \\
E^{1}_\mathrm{-}~&=~\frac{1}{\sqrt{2}}\begin{pmatrix} 0 & 0 & 0 \\ 1 & 0 & 0  \\ 0 & 0 & 0 \end{pmatrix},&
E^{\theta}_\mathrm{-}~&=~\frac{1}{\sqrt{2}}\begin{pmatrix} 0 & 0 & 0 \\ 0 & 0 & 0  \\ 1 & 0 & 0 \end{pmatrix},&
\!\!\!E^{2}_\mathrm{-}~&=~\frac{1}{\sqrt{2}}\begin{pmatrix} 0 & 0 & 0 \\ 0 & 0 & 0  \\ 0 & 1 & 0 \end{pmatrix}.
\end{split}
\end{equation}
The generators of the maximally commuting (Cartan) subalgebra $\vec{H}=\left(\HI,\HY\right)$ obey the commutation relations
\begin{equation}
\begin{split}\label{eq:SU3Comms}
[\vec{H},\EEpm]~&=~\pm\left(1,0\right)^\mathrm{T}\,\EEpm\;,& 
[\vec{H},\EZpm]~&=~\pm\frac12\left(-1,\sqrt{3}\right)^\mathrm{T}\,\EZpm\;,\\
[\vec{H},\ETpm]~&=~\pm\frac12\left(1,\sqrt{3}\right)^\mathrm{T}\,\ETpm\;.
\end{split}
\end{equation}
One can read off the roots as (cf.\ e.g.\ \cite{Fuchs:1997jv}) 
\begin{equation}
\alpha^{(1)}~=~\left(1,0\right)^\mathrm{T}\,,\quad \alpha^{(2)}~=~\frac12\left(-1,\sqrt{3}\right)^\mathrm{T}\,,\quad \text{and}\quad \theta~=~\frac12\left(1,\sqrt{3}\right)^\mathrm{T}\,.
\end{equation}
They are shown in \Figref{fig:Su3Roots}. 
The symmetry group of the root system shall be analyzed in the following, in order 
to find the complete automorphism group of $\mathfrak{su}(3)$. 
As mentioned above, inner automorphisms correspond to reflections of the root system on hyperplanes perpendicular to any of the roots.
With the above relations between generators and roots it is straightforward to obtain the action of a given root reflection on the generators.

For example, the reflection on a plane perpendicular to $\alpha^{(1)}$ corresponds to a mapping of the generators as
\begin{equation}\label{eq:SU3Inns1}
u_1:~\EEp~\leftrightarrow~\EEm\,,\quad\ETp~\leftrightarrow~\EZp\,,\quad\EZm~\leftrightarrow~\ETm\,,\quad\HI~\mapsto~-\HI\,,\quad\HY~\mapsto~\HY.
\end{equation}
Numbering the generators as $\vec{T}:=(\EEp,\EEm,\HI,\ETp,\ETm,\EZp,\EZm,\HY)$, the corresponding transformation matrix in the 
adjoint space is given by
\begin{equation}
R_{u_1}~=~\begin{pmatrix} 
0 & 1 & 0 & 0 & 0 & 0 & 0 & 0 \\ 
1 & 0 & 0 & 0 & 0 & 0 & 0 & 0 \\
0 & 0 &-1 & 0 & 0 & 0 & 0 & 0 \\
0 & 0 & 0 & 0 & 0 & 1 & 0 & 0 \\
0 & 0 & 0 & 0 & 0 & 0 & 1 & 0 \\
0 & 0 & 0 & 1 & 0 & 0 & 0 & 0 \\
0 & 0 & 0 & 0 & 1 & 0 & 0 & 0 \\
0 & 0 & 0 & 0 & 0 & 0 & 0 & 1 
\end{pmatrix}\;.
\end{equation}
Given the structure constants in this basis of the generators (cf.\ \Appref{app:SU3}) it is straightforward to check that \eqref{eq:ConditionA} is fulfilled.\footnote{%
One should note that the basis choice \eqref{eq:SU3Gens} is \emph{not} an orthonormal basis w.r.t.\ $\tr(T_aT_b)$. The matrices $R$, therefore, are not guaranteed to be orthogonal
in this basis and only the condition \eqref{eq:ConditionAnonON} must be fulfilled for any automorphism.
}
Furthermore, the consistency condition \eqref{eq:consistencyEquationLie} here takes the form
\begin{equation}
U\,T_a\,U^{-1}~=~\left(R_{u_1}\right)_{ab}T_b\;,\qquad\forall a\;,
\end{equation}
and is solved by 
\begin{equation}
U_{u_1}~=~\begin{pmatrix} 
0 & -1 & 0 \\ 
-1 & 0 & 0 \\ 
0 & 0 & -1 \\ 
\end{pmatrix}\;.
\end{equation}
As usual, $U_{u_1}$ is only defined up to a phase which has been chosen as $-1$ here such as to make $\det(U_{u_1})=1$. 
Nevertheless, any other phase choice would, in principle, be admissible.
The decisive criterion by which one can tell that $u_1$ is an inner automorphism 
is that $u_1:~\rep{r}\mapsto\rep{r}$ and $\det(R_{u_1})=1$, 
i.e.\ representations are mapped to themselves and $R$ is contained in the adjoint representation.

Completely analogous, the reflection on a plane perpendicular to the root $\theta$ corresponds to the mapping 
\begin{equation}\label{eq:SU3Inns2}
\begin{split}
u_\theta:~\EEp~&\leftrightarrow~\EZm\,,\quad\EEm~\leftrightarrow~\EZp\,,\quad\ETp~\leftrightarrow~\ETm\,,\\
\HI~&\mapsto~\frac12\left(\HI-\sqrt{3}\,\HY\right)\,,\quad\HY~\mapsto~-\frac12\left(\sqrt{3}\,\HI+\HY\right)\,,
\end{split}
\end{equation}
or equivalently 
\begin{equation}
R_\theta~=~\begin{pmatrix} 
0 & 0 & 0 & 0 & 0 & 0 & 1 & 0 \\ 
0 & 0 & 0 & 0 & 0 & 1 & 0 & 0 \\
0 & 0 & \frac12 & 0 & 0 & 0 & 0 & -\frac{\sqrt{3}}{2} \\
0 & 0 & 0 & 0 & 1 & 0 & 0 & 0 \\
0 & 0 & 0 & 1 & 0 & 0 & 0 & 0 \\
0 & 1 & 0 & 0 & 0 & 0 & 0 & 0 \\
1 & 0 & 0 & 0 & 0 & 0 & 0 & 0 \\
0 & 0 & -\frac{\sqrt{3}}{2} & 0 & 0 & 0 & 0 & -\frac12 
\end{pmatrix}\;.
\end{equation}
Again \eqref{eq:ConditionA} is fulfilled and the consistency condition \eqref{eq:consistencyEquationLie} is solved by
\begin{equation}
U_{u_\theta}~=~\begin{pmatrix} 
0 & 0 & -1 \\ 
0 & -1 & 0 \\ 
-1 & 0 & 0 \\ 
\end{pmatrix}\;,
\end{equation}
where the free phase has been fixed to obtain $\det(U_{u_\theta})=1$. 
Also $u_\theta$ is inner by the observation that $u_\theta:~\rep{r}\mapsto\rep{r}$ and $\det(R_{u_\theta})=1$.

Taken together, $U_{u_1}$ and $U_{u_\theta}$ generate the complete inner automorphism group $\mathrm{S}_3$ of the fundamental (\rep{3}) space 
of \SU3 which is equivalent to all possible Weyl reflections of the root system. 
The matrices $R_{u_1}$ and $R_\theta$ generate the same group in the adjoint space.

There are, however, more possible symmetries of the root system which obey \eqref{eq:ConditionA}.
In particular, take the point reflection through the origin, which maps all roots to their negative. Clearly, this is not a Weyl reflection.
This transformation is also called the contragredient automorphism \cite{Grimus:1995zi}.
The corresponding action on the generators of the fundamental representation is given by
\begin{equation}\label{eq:SU3Out}
\begin{split}
u_\Delta:~\EEp~&\leftrightarrow~-\EEm\,,\quad\ETp~\leftrightarrow~-\ETm\,,\quad\EZp~\leftrightarrow~-\EZm\,,\\
\HI~&\mapsto~-\HI\,,\quad\HY~\mapsto~-\HY\,.
\end{split}
\end{equation}
Therefore, the mapping in the adjoint space is given by
\begin{equation}
R_\Delta~=~\begin{pmatrix} 
0 &-1 & 0 & 0 & 0 & 0 & 0 & 0 \\ 
-1& 0 & 0 & 0 & 0 & 0 & 0 & 0 \\
0 & 0 &-1 & 0 & 0 & 0 & 0 & 0 \\
0 & 0 & 0 & 0 &-1 & 0 & 0 & 0 \\
0 & 0 & 0 &-1 & 0 & 0 & 0 & 0 \\
0 & 0 & 0 & 0 & 0 & 0 &-1 & 0 \\
0 & 0 & 0 & 0 & 0 &-1 & 0 & 0 \\
0 & 0 & 0 & 0 & 0 & 0 & 0 &-1 
\end{pmatrix}\;.
\end{equation}
The corresponding consistency condition \eqref{eq:consistencyEquationLie} can not be solved for $\rep{3}\mapsto\rep{3}$.
However, one can solve the condition if one takes $u_\Delta$ as a mapping $\rep{3}\mapsto\rep{3}^*$. The
corresponding consistency condition reads
\begin{equation}
U\,(-T^\mathrm{T}_a)\,U^{-1}~=~\left(R_\Delta\right)_{ab}T_b\;,\qquad\forall a\;,
\end{equation}
and it is simply solved for $U=\mathbbm{1}$. This shows that $u_\Delta$ is an outer automorphism, as expected.
One should not be confused by the fact that $\det(U)=1$, as there is again a free phase in $U$.
Rather, by observing that $\det(R_\Delta)=-1$, it is clear that $R_\Delta$ cannot be part of the adjoint 
space and the automorphism cannot be inner. That \eqref{eq:SU3Out} is indeed the complex conjugation automorphism is 
not obvious in the chosen basis \eqref{eq:SU3Gens}. However, rotating the adjoint space to the standard 
Gell--Mann basis one confirms that $u_\Delta$ is indeed the usual complex conjugation which transforms the Gell--Mann matrices \cite{GellMann:1962xb} as
$\lambda_{2,5,7}\mapsto\lambda_{2,5,7}$ and $\lambda_{1,3,4,6,8}\mapsto-\lambda_{1,3,4,6,8}$ (cf.\ \Appref{app:SU3}).

Together, $u_1$, $u_\theta$, and $u_\Delta$ generate the complete automorphism group of $\mathfrak{su}(3)$, isomorphic to the symmetry of its root system and also
isomorphic to the automorphism group of $\SU3$. The group is the dihedral group $D_{12}$ which is, without surprise, also the symmetry group of a regular hexagon.

\section{Outer automorphisms of the Poincar\'e group}
\label{sec:OutPoincare}

Completely analogous to the above cases of finite groups and compact Lie groups also the proper orthochronous Poincar\'e group\footnote{%
The terms ``proper'' and ``orthochronous'' will be dropped in the following but they are implicit in any mentioning of both, the 
Poincar\'e and Lorentz group.}
has outer automorphisms. The Poincar\'e group is a Lie group which is non--compact, for what reason the discussion has to be led separately from the preceding section.
For the scope of this thesis, only a concise and mostly informal treatment will be provided.
A more thorough treatment including detailed derivations of most of the presented results can be found in \cite{Buchbinder:2000cq} and references therein.
Practical application of the Poincar\'e group outer automorphisms C, P, and T has already been discussed in \Secref{sec:CPTDefinition}. 
This section now serves to help interpret these transformations as outer automorphisms, and furthermore, 
sets the stage for a more general definition of what determines a physical CP transformation in \Secref{sec:CPIsOut}.

The Poincar\'e group is a semidirect product of the (proper orthochronous) Lorentz group $\mathrm{SO}_\mathrm{+}(3,1)$, containing boosts and rotations in
Minkowski space--time, and the group of four--dimensional translations.
Outer automorphisms of the translational part are generated by dilatations $x^\mu\mapsto R\,x^\mu (R\neq0,1)$ \cite{Buchbinder:2000cq}.
This type of outer automorphisms certainly bears a lot of interest on its own, but it is beyond the scope of this work to discuss it.
Anyways, note that the scaling outer automorphism is involutory (meaning that it squares to the identity) only for $R=-1$.
In this case it actually corresponds to a combined application of the P and T transformations given in \eqref{eq:PT}.
Indeed, there is a one--to--one correspondence between involutory outer automorphisms of the Poincar\'e group 
and reflections in Minkowski space \cite{Buchbinder:2000cq}. Fortunately, the possible reflections in Minkoswki space are already 
exhausted by the transformations P and T. However, upon introducing functions on the Poincar\'e group (i.e.\ representations) also complex functions 
(i.e.\ complex representations) arise. Complex representations are then forming two invariant subspaces under the 
Poincar\'e group including the transformations P and T. These invariant subspaces 
can be mapped onto each other by complex conjugation of the corresponding function, i.e.\ a C transformation \cite{Buchbinder:2000cq}.

Let us now focus on the Lorentz part of the Poincar\'e group. The Lorentz group is not simply connected for what reason
its Lie algebra is also the Lie algebra of a bigger group. 
This bigger group is $\mathrm{SL}(2,\mathbbm{C})$ which is the double covering group of the Lorentz group. 
Taking into account spinorial representations, the representations of the Lorentz group are actually representations of $\mathrm{SL}(2,\mathbbm{C})$.
The corresponding Lie algebra $\mathfrak{sl}(2,\mathbbm{C})$ is the complexification of the Algebra $\mathfrak{su}(2)$. 
Therefore, there is an isomorphism $\mathfrak{sl}(2,\mathbbm{C})\cong\mathfrak{su}(2)\oplus\mathfrak{su}(2)$.
The representations of $\mathrm{SL}(2,\mathbbm{C})$, hence, can conveniently be discussed as the simultaneous representations of two \SU2's.
It is well--known that the irreps of the usual spin group \SU2 can be labeled by half integers, characterizing the spin of the representation.
The representations of the Lorentz group (more precisely of $\mathrm{SL}(2,\mathbbm{C})$), thus, are conventionally labeled as a 
pair of two half integers $(j,k)$ corresponding to representations under the two \SU2 groups.

For definiteness, the two lowest non--trivial representations and their interplay under outer automorphisms shall be discussed. 
This is the ``left--handed'' Weyl spinor $\chi$ in the representation $(\frac12,0)$ 
and the ``right--handed'' Weyl spinor $\xi^\dagger$ in the representation $(0,\frac12)$.\footnote{%
The explicit spinor indices are omitted because they are not needed for this discussion. The corresponding indices would be restored as $\chi_a$ and $\xi^{\dagger\dot{a}}$.
}
Each of these fields describes two real degrees of freedom.
The dagger for $\xi^\dagger$ is part of the name. One should be very careful in noting that the respective Hermitian conjugated fields
are of the opposite ``handedness'' than the original fields. For example, $\chi\mapsto\chi^\dagger$ is right--handed, whereas
$\xi^\dagger\mapsto\xi$ is left--handed. This explains our notation: left--handed fields are simply the ones without a dagger.
The two real degrees of freedom per field, thus, correspond to two possible helicity states. 
One should be careful however, because the two helicity states cannot necessarily be turned into one another by a spin flip.
This is the case if the states are additionally charged under another conserved symmetry which could, for example, be a \U1 lepton number. 
Then the two helicity states will have exactly opposite charge under this \U1, and, therefore, cannot be turned into one another. 
Therefore, one should really think of the four real degrees of freedom as $\{\chi,\chi^\dagger;\xi^\dagger,\xi\}$.

The analogy of the Poincar\'e group outer automorphism transformations C, P, and T 
to the outer automorphism transformations of finite and compact Lie groups shall be emphasized in the following.
The Dirac spinor $\Psi$ has already been discussed in \Secref{sec:CPTDefinition}. It transforms in the 
(reducible) representation $(\frac12,0)\oplus(0,\frac12)$. 
Therefore, $\Psi$ can be constructed out of the two fields $\chi$ and $\xi^\dagger$ as
\begin{equation}
\Psi~:=~\begin{pmatrix} \chi \\ \xi^\dagger \end{pmatrix}\;.
\end{equation}
The action of C, P, and T on the Dirac spinor representation $\Psi$ has been summarized in \Tabref{tab:classicalCPT}.
The explicit transformation matrices of the outer automorphisms in this representation can be read off as 
$\mathcal{C}\mathcal{\beta}$, $\mathcal{\beta}$, and $\gamma_5\mathcal{C}$ for C, P, and T transformations, respectively 
(cf.\ \Appref{app:DiracMatrices}). 
Furthermore, from \eqref{eq:DiracCP} the explicit representation matrix for the CP transformation is read off as $\mathcal{C}$.
The appearing free phases $\eta_{\mathsf{C,P,T,CP}}$ are understood from the fact that representation matrices of outer automorphisms are,
as always, only defined up to a phase. To make the analogy to the preceding discussion manifest, note how the $\gamma$--matrices 
transform under the respective operations:\footnote{%
The $\gamma$--matrices are, of course, not the generators of the Dirac spinor representation. However, the translation to the 
actual generators of this representation, $S^{\mu\nu}=\frac{\I}{4}\left[\gamma^\mu,\gamma^\nu\right]$, is straightforward.
}
\begin{align}
\mathrm{C}&:&\quad \mathcal{(C\,\beta)}\,(-{\gamma^\mu})^\mathrm{T}\,\mathcal{(C\,\beta)}^{-1}~&=~\parityP^\mu_{~\nu}\,{\gamma^\nu}\;,&\\
\mathrm{P}&:&\quad \mathcal{\beta}\,{\gamma^\mu}\,\mathcal{\beta}^{-1}~&=~\parityP^\mu_{~\nu}\,{\gamma^\nu}\;,&\\\label{eq:LorentzT}
\mathrm{T}&:&\quad (\gamma_5\,\mathcal{C})\,{(-\gamma^\mu)^*}\,{(\gamma_5\,\mathcal{C})}^{-1}~&=~\timeT^\mu_{~\nu}\,{\gamma^\nu}\;,&\\ \label{eq:LorentzCP}
\mathrm{CP}&:&\quad\mathcal{C}\,(-{\gamma^\mu})^\mathrm{T}\,\mathcal{C}^{-1}~&=~{\gamma^\mu}\;,&
\end{align}
Note the striking similarity to the consistency conditions \eqref{eq:consistencyEquation} and \eqref{eq:consistencyEquationLie}.

\begin{figure}[t]
\centerline{\includegraphics[width=0.5\textwidth]{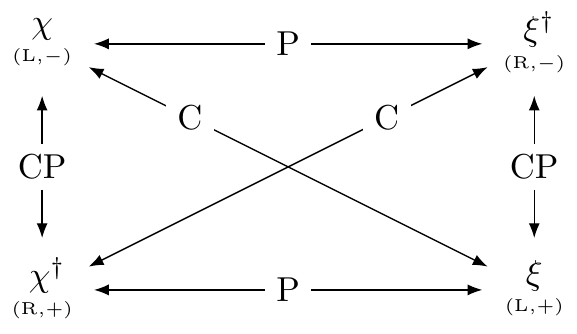}}
  \caption{C, P, and CP transformation for the chiral Weyl fermions $\chi$ and $\xi^\dagger$.
	The undersets denote the handedness and potential \U1 charge of the corresponding field.}
  \label{fig:LorentzCP} 
\end{figure}

Consider now the action of outer automorphisms on the four individual states $\{\chi,\chi^\dagger;\xi^\dagger,\xi\}$. 
As seen before, outer automorphisms can -- but do not have to -- exchange representations of the same dimensionality.
Here, the transformation T does not permute any of the fields, but as discussed in \Secref{sec:CPTDefinition}, merely corresponds to motion reversal
(the anti--unitarity of the T operation explains why \eqref{eq:LorentzT} involves an additional conjugation).
In contrast, the two order--two transformations C and P correspond to all possible pairwise permutations of the four fields.
The transformations are summarized in \Figref{fig:LorentzCP}.

Note that C or P transformations by themselves only make sense for a complete Dirac spinor. 
This is because these transformations are interrelating representations which are not mutually complex conjugate.
If there is only a single chiral Weyl fermion without the corresponding counterpart, then C and P transformations are broken explicitly and maximally, i.e.\ by
the absence of representations. This is the case in the SM. 
It is emphasized that \emph{the only possible} outer automorphism transformation for a single
chiral Weyl fermion is CP. Under CP each state is mapped to \emph{its own} respective complex conjugate state. This
is also true for the Dirac spinor representation itself, as is clear already from \Eqref{eq:DiracCP}, 
and it can also be inferred for the generators of the Dirac representation from \eqref{eq:LorentzCP}. 

In this sense, note that charge conjugation is not the complex conjugation outer automorphism of the Lorentz group.
In contrast, the transformation which maps each representation to its own complex conjugate representation is CP.
This observation will be picked up shortly. First, however, this chapter shall be concluded by a remark advocating the title of this work.

\section{(Outer) automorphisms are symmetries of a symmetry}

It has been understood that automorphisms are all possible ways to map a certain symmetry, 
i.e.\ the abstract generators of a group or the elements of the abstract group algebra, 
to itself without changing the structure of the group.
This justifies the term ``symmetry of a symmetry'' for automorphisms. 

In this group of automorphism transformations, the so--called outer automorphisms play a special role.
This is because, in contrast to inner automorphisms, outer automorphisms cannot be represented by elements of the original symmetry group.
Outer automorphisms, therefore, are truly the non--trivial ways to map a symmetry to itself.

\chapter{CP as a symmetry of symmetries}
\label{sec:CPasOuts}

\section{CP as an (outer) automorphism of space--time and gauge symmetries}

After having discussed CP transformations and outer automorphisms largely separately, it is only a short stroll to 
interpret the former as a special class of the latter. Let us start doing this for the case of 
a gauge theory. Many technical details of this discussion can be found in \cite{Grimus:1995zi}.

Consider a gauge theory with a compact semisimple non--Abelian gauge group.
The gauge part of the Lagrangian is given by 
\begin{equation}\label{eq:LGauge}
\mathscr L_\mathrm{G}~=~-\frac14\,G^a_{\mu\nu}\,G^{\mu\nu,a}\;,
\end{equation}
with the field strength tensor
\begin{equation}
G^a_{\mu\nu}~=~\partial_\mu\,W^a_\nu-\partial_\nu\,W^a_\mu+g\,f^{abc}\,W^b_\mu\,W^c_\nu\;.
\end{equation}
Furthermore, assume that there is a left--handed Weyl fermion $\Psi_\mathrm{L}$ charged under the gauge group and transforming 
in a representation generated by matrices $\{T_a\}$. The gauge--kinetic part of the fermion Lagrangian, hence, is given by
\begin{equation}
\mathscr L_\mathrm{F}~=~\I\,\bar{\Psi}_\mathrm{L}\,\gamma^\mu\left(\partial_\mu-\I\,g\,T_a\,W^a_\mu \right) \Psi_\mathrm{L}\;.
\end{equation}
The most general possible CP transformation then acts on the gauge and fermion fields as (cf.\ \Tabref{tab:classicalCPT} and \eqref{eq:DiracCP})
\begin{equation}
\begin{split}\label{eq:generalCP}
W^a_\mu(x)~&\mapsto~\varepsilon(\mu)\,R^{ab}\,W^b_\mu(\parityP\,x)\;, \\
\Psi_\mathrm{L}(x)~&\mapsto~\eta_\mathsf{CP}\,U\,\mathcal{C}\,\Psi^*_\mathrm{L}(\parityP\,x)\;.
\end{split}
\end{equation}
To be as general as possible, the gauge fields are allowed to rotate in the adjoint space of the gauge group parametrized by $R$,
and the fermions are allowed to rotate in their representation space of the gauge group parametrized by $U$, respectively. 
Furthermore, possible rotations in the Dirac representation space of the Lorentz group are parametrized by $\mathcal{C}$. 
Here, $\mathcal{C}$ and $U$ are general unitary matrices, while $R$ can be chosen real due to the reality of the gauge fields.

It is straightforward to show that the most general CP transformation \eqref{eq:generalCP} is a conserved symmetry of the action if and only if
\begin{align}\label{eq:CPasOut1}
(\mathrm{i})~&:~&R_{aa'}\,R_{bb'}\,f_{a'b'c}~&=~f_{abc'}\,R_{c'c}\;,& \\ \label{eq:CPasOut2}
(\mathrm{ii})~&:~&U\,(-T^\mathrm{T}_a)\,U^{-1}~&=~R_{ab}\,T_b\;,& \\ \label{eq:CPasOut3}
(\mathrm{iii})~&:~&\mathcal{C}\,(-{\gamma^\mu}^\mathrm{T})\,\mathcal{C}^{-1}~&=~{\gamma^\mu}\;.&
\end{align}
The first condition arises from the invariance of $\mathscr L_\mathrm{G}$, the second condition arises from 
the invariance of the fermion gauge coupling, and the last condition is required already by the invariance of the fermion kinetic term.
Furthermore, conservation of the terms quadratic in the gauge fields requires $R$ to be an orthogonal matrix, which can be traced back to
the orthogonality condition $\tr(T_aT_b)=k\,\delta_{ab}$.

The conditions (i)--(iii), however, are well--known from the previous sections. They imply that the CP transformation is an automorphism.
More specifically, it has to be an automorphism which maps all present symmetry representations
to their respective complex conjugate representation. This statement holds equally for gauge group and Lorentz group representations. 
It is clear that the CP automorphism is outer for the Lorentz group.
For semisimple compact Lie groups, however, this automorphism may be inner. 
This can be the case only if the corresponding group does not have complex representations. Whenever there are complex representations present
then the automorphism fulfilling (i) and (ii) has to be outer \cite{Grimus:1995zi}. 
Also in \cite{Grimus:1995zi}, it has been shown that the contragredient automorphism, corresponding to the root system inversion,
always fulfills (i) and (ii). Nevertheless, the conditions do not single out one particular automorphism. 
The only requirement is that it must be a consistent automorphism which maps representations to their complex conjugate, and so 
multiple automorphisms could in principle be qualified. The existence of a single automorphism which fulfills (i)--(iii)
is enough to warrant CP conservation for the gauge kinetic terms. 
As the contragredient automorphism always exists for semisimple compact gauge groups, CP is automatically conserved in the gauge kinetic terms.
In addition, it has been shown that for semisimple Lie groups one can always find 
a so--called CP basis in which $U=\mathbbm{1}$ \cite{Grimus:1995zi}. 
In the following it will, therefore, without loss of generality always be assumed that $U$ has been set to $\mathbbm{1}$, 
i.e.\ a CP basis has been chosen in the gauge group representation space of fermions.

Finally, note that if the field $\Psi_\mathrm{L}$ would, in addition, also transform in a representation of any other, say, global symmetry,
then it is imperative that also the corresponding representation of this global symmetry is mapped onto its own complex conjugate representation. 
That is, also for all additional groups the CP transformation should be a complex conjugation (outer) automorphism.\footnote{%
Strictly speaking this is true only for all fields which have some charge also under the SM gauge group, or couple to the SM charged fields in some way. 
} 
This statement is sufficiently general that one may actually use it as a definition of a physical CP transformation in the first place.

\section{Definition of CP as a special automorphism}
\label{sec:CPIsOut}

In the preceding section it has been demonstrated that the text--book CP transformation is
a complex conjugation (outer) automorphism of the space--time and gauge symmetry of a relativistic quantum field theory. 
In turn, any conserved complex conjugation outer automorphism warrants CP conservation. Therefore, it makes sense to identify these two
notions and define a physical CP transformation as a complex conjugation (outer) automorphism of all present symmetries. This includes space--time, gauge, and global symmetries. 

It should be noted that there have been advances to give a more precise definition of CP, for example,  
defining it as an automorphism which reverses certain quantum numbers (cf.\ \cite{Grimus:1995zi} and also \cite{Fallus:2015}).
Defining CP in such a way, however, is dependent on a specific choice of inner automorphism accompanying the complex conjugation outer automorphism. 
If there is any one inner automorphism for which the corresponding outer automorphism is conserved, however, then
it is also conserved for all other choices of accompanying inner automorphisms. 
Taking a specific inner automorphism, therefore, certainly gives a sanity check and an intuition for the action of the CP transformation, 
but it is not necessary to include it in the definition of the transformation in the first place.
In any case, the simple definition of CP as a complex conjugation automorphism of all present representations, as advertised here, 
includes all known examples \cite{Buchbinder:2000cq,Grimus:1995zi,Chen:2014tpa} without contradiction. In particular, it also holds for additional global (possibly finite) symmetry groups which are studied below.

The non--conservation of \textit{all possible}\footnote{%
While for gauge theories and the Poincar\'e group the complex conjugation (outer) automorphism is unique up to inner automorphisms this 
is not the case in general. For finite groups, for example, there may be multiple distinct outer automorphisms fulfilling the definition.
}
complex conjugation outer automorphisms is a necessary condition for CPV in a physical sense,
for example as prerequisite for baryogenesis. In contrast, if there are multiple possible and distinct CP transformations, then the conservation
of any of these transformations gives rise to physical CP conservation.

\section{Generalized CP transformations}
\label{sec:GeneralizedCP}

Despite the now clarified fact that a physical CP transformation always 
corresponds to a complex conjugation automorphism transformation of all present symmetries,
there is an additional subtlety worth mentioning. Many theories contain multiple identical 
copies of fields in equivalent symmetry representations. Most prominently this is the case in the SM, 
but it also happens, for example, in theories with multiple Higgs fields. 
In general, there is some degeneracy in the distinction of these fields, typically reflected by 
the freedom to perform a \U{n} basis transformation in the so--called ``horizontal'' space without 
changing physical observables. In the SM, the identical copies of representations correspond to the 
repetition of fermion generations and the corresponding horizontal rotations are simply the possible 
basis choices in flavor space. For multi--Higgs models, the corresponding horizontal rotations mix the multiple 
Higgs fields and are referred to as Higgs--basis rotations \cite{Botella:1994cs}, cf.\ also \cite{Barroso:2006pa, Branco:1999fs}. 

For a proper physical CP transformation every field in a symmetry representation should be mapped onto 
its own complex conjugate. Hence, all of the repeated fields spanning the horizontal space should, in principle, 
be mapped to their own complex conjugate fields. Nevertheless, due to the degeneracy in the horizontal space, 
it is always possible to amend the complex conjugation map by an additional rotation in the horizontal space. 
Recall that even for a single field there is an additional freedom in taking outer automorphisms corresponding to the rephasing of each field. 
This freedom has already been taken care of by amending the usual C, P, and T transformations by free phases 
$\eta_{\mathsf{C}}$, $\eta_{\mathsf{P}}$, and $\eta_{\mathsf{T}}$, cf. \Tabref{tab:classicalCPT}.
For the complex conjugation map of multiple fields in all identical representations, there can be more sophisticated transformations 
than just mapping each field onto its own complex conjugate.
Additionally to these ``canonical'' CP transformations, 
there can be non--trivial rotations in the horizontal space which are typically called ``generalized'' CP transformations. 
Generalized CP transformations have formally been introduced in \cite{Lee:1966ik} and firstly been used in the 
context of left--right symmetric models \cite{Ecker:1981wv, Ecker:1983hz}.
For a generalized CP transformation one can think of the $\eta$ phases as being promoted to matrices that act in the horizontal space.
Just as for the $\eta$ phases, the existence of a single set of matrices for which the 
corresponding generalized CP transformation is conserved is sufficient to warrant CP conservation.

For definiteness consider the quark sector of the SM (see \cite{Branco:1986gr} for an analogous discussion for the lepton sector). 
The most general possible CP transformation is given by \cite{Gronau:1986xb,Bernabeu:1986fc}%
\begin{align}\label{eq:generalizedCP}
Q~&\mapsto~U_\mathrm{L}\,\mathcal{C}\,Q^*\,,&                             \\
u_\mathrm{R}~&\mapsto~U^{u}_\mathrm{R}\,\mathcal{C}\,u_\mathrm{R}^*\,,&      \\
d_\mathrm{R}~&\mapsto~U^{d}_\mathrm{R}\,\mathcal{C}\,d_\mathrm{R}^*\,,&      
\end{align}
where $U_\mathrm{L}$, $U^{u}_\mathrm{R}$, and $U^{d}_\mathrm{R}$ are general $3\times3$ unitary matrices acting in flavor space
and $\mathcal{C}$ is the charge conjugation matrix for fermions defined in \Secref{sec:CPTDefinition}.\footnote{%
For clarity, it is remarked that $U$ of \eqref{eq:generalCP}, which acts in the gauge representation space of each fermion, has been set to $\mathbbm{1}$ here.}
In order for the generalized CP transformation to be 
conserved, the Yukawa coupling matrices, as introduced in \eqref{eq:SMYukawas}, have to fulfill
\begin{equation}
U^{\dagger}_\mathrm{L}\,y_u\,U^{u}_\mathrm{R}~=~y_u^*\;\qquad\text{and}\qquad U^{\dagger}_\mathrm{L}\,y_d\,U^{d}_\mathrm{R}~=~y_d^*\;,
\end{equation}
or equivalently
\begin{equation}\label{eq:gerneralizedCPYukawas}
U^{\dagger}_\mathrm{L}\,y_u\,y_u^\dagger\,\,U_\mathrm{L}~=~y_u^*\,y_u^\mathrm{T}\;\qquad\text{and}
\qquad U^{\dagger}_\mathrm{L}\,y_d\,y_d^\dagger\,\,U_\mathrm{L}~=~y_d^*\,y_d^\mathrm{T}\;.
\end{equation}
Using the Jarlskog invariant $J$ as defined in \eqref{eq:J} it is straightforward to check that $J=-J=0$, i.e.\ CP is conserved as a consequence of
these relations -- just as it would be for the particular ``canonical'' choice $U_\mathrm{L}=U^{u}_\mathrm{R}=U^{d}_\mathrm{R}=\mathbbm{1}$.

\subsection{New horizontal symmetries and exotic CP eigenstates}
\label{sec:generalizedCP}

Even though very tempting, one can in general not regard generalized CP transformations simply as canonical CP transformations amended 
by a basis rotation. This shall be detailed in the following. Consider, for example, the behavior of $U_\mathrm{L}$ under change 
of the left--handed quark flavor basis. That is, assuming that \eqref{eq:gerneralizedCPYukawas} is solved by $U_\mathrm{L}$ in one basis,
in a different basis $Q'=W_\mathrm{L} Q$ it is solved by 
\begin{equation}\label{eq:CPBasisChange}
U_\mathrm{L}'~=~W_\mathrm{L}\,U_\mathrm{L}\,W_\mathrm{L}^\mathrm{T}\;.
\end{equation}
Therefore, it becomes clear that $U_\mathrm{L}$ generally cannot be absorbed in a basis redefinition (because 
it rotates with $WUW^\mathrm{T}$, not with $WUW^\dagger$).
However, using the freedom to change the basis as in \eqref{eq:CPBasisChange}, it is possible to bring any unitary CP transformation matrix $U$ to a certain standard form 
which can be presented as \cite{Ecker:1987qp}(cf.\ also \cite[Ch.2, App.C]{Weinberg:1995mt2})%
\begin{equation}\label{eq:stdForm}
U~=~W^\dagger\,
	\begin{pmatrix}
	\Theta_1 & & & \\
	& \ddots & & \\
	& & \Theta_\ell & \\
	& & & \mathbbm{1}_m
	\end{pmatrix}\,
	W^*\;.
\end{equation}
Here, $\Theta_k$ are $2\times2$ orthogonal matrices 
\begin{equation}
\Theta_k~:=~\begin{pmatrix} \phantom{-}\cos{\theta_k} & \sin{\theta_k} \\ -\sin{\theta_k} & \cos{\theta_k} \end{pmatrix}\;,
\end{equation}
with angles $\theta_k$ that are given by the pairwise appearing eigenvalues $\e^{\pm2\,\I\,\theta_k}$ of the matrix $UU^*$.
The angles $\theta_k$ can be constrained to lie in the range $0\leq\theta_k\leq\pi/2$.%\frac{\pi}{2}$.

In contrast to the ubiquitous $\eta$ phases or the arbitrary phase of $U$, a generalized CP transformation does not automatically cancel if CP is applied twice.
Therefore, the requirement of a conserved generalized CP transformation will generally induce new linear horizontal symmetries \cite{Grimus:1987kn}.
Applying the generalized CP transformation \eqref{eq:generalizedCP} twice one finds that
\begin{equation}
Q~\xrightarrow{(\op{C\,P})^2}~U_\mathrm{L}\,U^*_\mathrm{L}\,Q~=:V_\mathrm{L}\,Q\;.    
\end{equation}
The matrix $V_\mathrm{L}$ then acts as the generator of a new linear symmetry in the horizontal space.

Typically, an enhancement of the linear symmetry is avoided by requiring that the CP
transformation acts as an involution on all fields, i.e.\ it squares to the identity $U\,U^*=V=\mathbbm{1}$. 
This requirement is equivalent to the statement that one can find a so--called ``CP basis'' in which $U=\mathbbm{1}$,
as can be seen from the discussion of the standard form of $U$ above and the fact that $V$ in this form only has unity eigenvalues.

Nevertheless, this requirement is somewhat arbitrary if one is just after CP conservation, 
as any non--involutory, i.e.\ higher--order CP transformation would also warrant CP conservation.
Higher--order CP transformations have, for example, been considered in two Higgs doublet models \cite{Maniatis:2007vn, Branco:2011iw}.
Another particularly interesting example is a three Higgs doublet model with a CP transformation of order 4 \cite{Ivanov:2015mwl}.
Furthermore, it has been shown that some discrete groups (of the so--called type~II~B below)
enforce higher--order CP transformations on the representations \cite{Chen:2014tpa}.

In general, one should note that for higher--order CP transformations it is not possible to attain a CP basis, 
as is clear from the standard form \eqref{eq:stdForm} of $U$ and the fact that $V=UU^*$ has non--unity eigenvalues. 
This opens up the phenomenologically unprecedented possibility that there are eigenstates of CP which are neither CP even nor CP odd, 
but CP ``half--odd'' \cite{Ivanov:2015mwl}. That is, finding the eigenstates of an order $2+2n$ $(n\in\mathbbm{N})$ CP operation $U$
one may find states not only with eigenvalues $\pm1$ but also with the ``half--odd'' (or even ``$1/2n$--odd'') eigenvalues $(-1)^{1/2n}$.

\subsection{Generalized CP and existing horizontal symmetries}

For the construction of generalized CP transformations in the previous section, the implicit assumption has been made 
that there is no pre--existing structure in the horizontal space.
There is, however, the possibility to have a symmetry $G$ acting in the horizontal space.
Sticking with the SM example of three fermion flavors for concreteness, $G$ would be a flavor or family symmetry.
If there is a horizontal symmetry, not every generalized CP transformation is admissible. In contrast, all 
possible CP transformations are given by the complex conjugation automorphisms of all symmetries, and in particular of $G$, 
as discussed in detail before. 
Of course, it would, in principle, always be possible to impose \textit{any} generalized CP symmetry --
just as it is possible to impose any additional linear symmetry.
Nevertheless, this would generally enhance the symmetry in the horizontal space, possibly up to the maximal \U{n} symmetry
of the gauge--kinetic interactions. In such a construction it would then be unreasonable to speak of a $G$--symmetric model, 
for what reason this possibility is discarded in the following. Therefore, the choice of possible generalized CP transformations
is limited to the explicit representation matrices of the complex conjugation automorphisms of $G$. 

The possible CP transformations, as automorphisms of all symmetries and in particular $G$, of course, will also come with explicit representation matrices.
For example, for the case of spinor representations of the Lorentz group, the representation matrix of the CP outer automorphism
is $\mathcal{C}$. For a generic horizontal space, this explicit representation matrix is usually called $U$. 
The basis transformation freedom, which acts on the explicit representation matrix $U$ as in \eqref{eq:CPBasisChange}, 
can be used to rotate $U$ to the standard form \eqref{eq:stdForm}. Depending on the order of $U$, this sometimes allows
to rotate $U$ to the identity matrix, i.e.\ find a CP basis, as discussed above. 
For example, this is always the case for all semisimple compact Lie groups \cite[App.\ F]{Grimus:1995zi} and for finite groups of the so--called type~II~A below. 
Nevertheless, for many models the CP basis is often not the most convenient choice to identify 
the physical states of a theory or to perform explicit higher--order computations \cite{Chen:2014tpa}.

For general symmetry groups it is by far not guaranteed that a CP basis can be found. Whether or not
this is possible in a given model crucially depends on the properties of the corresponding automorphisms of all the involved groups, 
and their representation matrices for the present representations. 
In fact, while it follows from the requirement $UU^*=V=\mathbbm{1}$ that the corresponding complex conjugation automorphism 
is an involutory automorphism of the corresponding group the reverse statement is not true. That is, even for complex conjugation automorphisms
which act as an involution on the level of the abstract symmetry groups, the 
explicit representation matrices can turn out to be such that $U\,U^*=V\neq\mathbbm{1}$.
In general it is true that the representation matrices $U$ have to be determined from the structure of all present symmetry groups 
and the corresponding complex conjugation automorphisms. If a complex conjugation automorphism is involutory one can show that the 
only non--trivial possibility besides $V=\mathbbm{1}$ is that $V=-\mathbbm{1}$. 
For higher order complex conjugation automorphisms necessarily more 
complicated forms of $V$ arise. 

For the case of finite discrete groups, for example, it will be shown in the following that 
there is a large class of groups for which $U$ cannot be rotated away. 
In fact, whether or not this is possible will be one of the criteria by which discrete 
groups are classified.
Whenever a model is such that the CP transformation matrix $U$ cannot be basis--rotated to the unity matrix, it is implicit that states with exotic CP properties,
such as the aforementioned ``half--odd'' states, exist in a model. In the following discussion of CP automorphisms in discrete groups it will be found
that this situation always arises for groups of the so--called type~II~B.

\chapter{CP and discrete groups}
\label{sec:CPAndFiniteGroups}
\enlargethispage{0.5cm}
In the previous sections it has been established that proper physical CP transformations correspond to automorphisms which map 
all of the present symmetry representations of a theory to their own respective complex conjugate representations.
If there are complex representations present, then the corresponding automorphism must be outer.

In this section, this situation shall be analyzed in depth for the case of finite groups. Finite groups find
applications in many models, in particle physics most prominently as flavor symmetries, cf.\ e.g.\ \cite{Altarelli:2010gt,Ishimori:2010au,King:2013eh,Feruglio:2015jfa} for reviews. 
If there are finite groups present, then CP transformations also have to be (outer) automorphisms of the finite groups.
This has firstly been shown in \cite{Holthausen:2012dk}, where it has been missed, however, 
that CP transformations are only a special subset of all outer automorphisms.
As will be shown in the following, not all finite groups allow for (outer) automorphisms that simultaneously 
map all representations of the group to their respective complex conjugate representations. 
Whenever a finite group with this property and a sufficient number of irreps is contained in a model, then there is no possible (outer) 
automorphism corresponding to a CP transformation. This implies that CP can never be a possible symmetry of such a model. 
In particular, there will be complex couplings, originating from the Clebsch--Gordan coefficients (CGs) of the group, which enter amplitudes in the form of 
CP violating weak phases.\footnote{%
``Weak'' here has nothing to do with the weak interaction but with the fact that the corresponding phase differs from the 
one of the CP conjugate process. This is in contrast to ``strong'' phases which do not change under CP and arise, for example, as the absorptive part of loop integrals
if a certain process is kinematically allowed \cite{Branco:1999fs}. The presence of weak phases is an unambiguous sign for CP violation.\label{fot:one}}
Interestingly, phases that originate from finite groups are calculable and assume fixed ``geometrical'' values, such as for example $\omega=\e^{2\pi \I/3}$.
An example model will be presented where this type of CP violating calculable weak phases are present, and a CP violating amplitude will be calculated explicitly.
 
That ``geometrical'' CP violation originating from complex CGs of the group $\mathrm{T'}$ 
could be a possibility has firstly been speculated on in \cite{Chen:2009gf}.
The group $\mathrm{T'}$, however, allows for a basis with real CGs \cite{Bickerstaff:1985jc}, 
thus, cannot lead to this form of CPV \cite{Chen:2014tpa}.
Nevertheless, explicit CPV from complex CGs, which nowadays is referred to as explicit geometrical CP violation \cite{Branco:2015hea}, 
is indeed possible as has firstly been demonstrated in \cite{Chen:2014tpa}.
There, also necessary and sufficient conditions for the occurrence of this form of CPV have been presented.

To proceed systematically, finite groups will first be classified according to their possible CP outer automorphisms or, reversely stated,
according to their ability to lead to CPV from group theory. Then, it will be discussed how the assumption of certain 
finite groups and their representations gives rise to explicit geometrical CP violation originating from complex CGs. 
Finally, there will be some remarks towards the use of explicit geometrical CP violation in possibly realistic flavor models.

\section{Classification of finite groups according to CP outer automorphisms}

This section provides a classification of finite groups according to their CP properties. That is, finite groups
shall be classified according to whether CP transformations are possible in general. 
Groups which generally do not allow for CP transformations allow for settings that give rise to calculable CP violating phases. 
It is possible to consider only scalar fields for the following classification of finite groups since the additional space--time transformation 
properties of a field do not matter for the discussion of its transformation properties under the discrete group. 
An extension of the argument to higher spin representations is straightforward, but it would not lead to new statements on the finite group. 
The classification of finite groups in this manner has firstly been brought forward in \cite{Chen:2014tpa}, where it is also discussed 
in somewhat more detail.

\subsection{Properties of CP outer automorphisms}
\label{Sec:CPautProps}

Assume that there is some scalar field $\phi$ in an irrep \rep[_i]{r} of a finite group $G$. 
Recall once again that a proper physical CP transformation for the finite group is given by a complex conjugation automorphism $u$ 
which maps every present irrep \rep[_i]{r} to its own complex conjugate representation $\rep[_i]{r}\mapsto\rep{r}^*_i\sim \rep[_i]{\bar{r}}$.
Therefore, the action on the field $\phi$ is given by
\begin{equation}
\label{eq:generalized_CP}
 \phi(x)  ~\mapsto~\UU[_i]{\rep{r}}\,\phi^*(\parityP\,x)\;,
\end{equation}
where $\UU[_i]{\rep{r}}$ is the unitary representation matrix of the automorphism $u$. Because of that $\UU[_i]{\rep{r}}$ fulfills the consistency condition 
\eqref{eq:consistencyEquation}, which here takes the form 
\begin{equation}\label{eq:CPConsistencyEquation}
\UU[_i]{\rep{r}}\,\rho_{\rep[_i]{r}}^*(\mathsf{g})\,\UU[_i]{\rep{r}}^\dagger~=~\rho_{\rep[_i]{r}}(u(\mathsf{g}))\;,\qquad\forall \mathsf{g}\in G\;.
\end{equation}
Without loss of generality, a basis for 
$\rep[_i]{\bar{r}}$ has been chosen such that $\rho_{\rep[_i]{\bar{r}}}(\mathsf{g})=\rho_{\rep[_i]{r}}^*(\mathsf{g})$. 

The integer $i$ enumerates all irreps of $G$.
In order to make model independent statements, it will be assumed in the following that $u$ is such that \eqref{eq:CPConsistencyEquation} holds for all $\rep[_i]{r}$ simultaneously.
Of course, it is not guaranteed that such an automorphism exist for a given group, and whether it does or not will be decisive for the classification.
In any case, the existence of such an automorphism $u$ allows to draw conclusions on its properties and the properties of the group $G$
which are elucidated in the following.

\paragraph{$\boldsymbol{u}$ must be class--inverting.}

Consider the characters of representations by taking the trace of \eqref{eq:CPConsistencyEquation}. One finds
\begin{equation}\label{eq:ClassInverting}
\chi_{\rep[_i]{r}}\!\left(u(\elm g)\right)~=~\tr\left[\rhoR{_i}\!\left(u(\elm g)\right)\right]~=~
\tr\left[\rhoR{_i}\!(\elm g)^*\right]~=~\chiR{_i}\!(\elm g)^*~=~\chiR{_i}\!(\elm g^{-1}).
\end{equation}
Since \eqref{eq:CPConsistencyEquation} and, therefore, also \eqref{eq:ClassInverting} has been required to be valid for all $i$,
one finds that $u$ must be a class--inverting automorphism.\footnote{%
A class--inverting automorphism $u$ maps every group element $g$ to another group element $u(g)$ which is part
of the same conjugacy class as $g^{-1}$, i.e.\ $u(g)=h\,g^{-1}\,h^{-1}$ for some $h\in G$.}

\paragraph{Remarks on the order of $\boldsymbol{u}$.} 

Applying the automorphism transformation $u$ twice, $\phi$ transforms as
 \begin{equation} 
  \phi~\stackrel{u^2}{\longmapsto}~\UU[_i]{\rep{r}}\,\left(
  \UU[_i]{\rep{r}}\,\phi^*(\parityP^2\,x)\right)^*
  ~=~\UU[_i]{\rep{r}}\,\UU[_i]{\rep{r}}^*\,\phi(x)~=:~V_{\rep[_i]{r}}\,\phi(x)\qquad \forall~i\;.
 \end{equation}
$V_{\rep[_i]{r}}$ is a unitary matrix that can be related to the automorphism $v=u^2$.
Therefore, imposing the CP transformation $u$ as a symmetry has the immediate 
consequence that also $\phi \mapsto V_{\rep[_i]{r}}\,\phi$ is imposed as a symmetry transformation.
Being the square of a class--inverting automorphism, $v=u^2$
is class--preserving. 

There are three logical possibilities for the square of $u$:
\begin{itemize}
 \item[(i)]~$u^2=v=\text{id}$, is the identity automorphism, or 
 \item[(ii)]~$u^2=v$ is a non--trivial inner automorphism, or
 \item[(iii)]~$u^2=v$ is a class--preserving outer automorphism.
\end{itemize}
The three cases will be examined in the following.

%\textbf{(i)}
\paragraph{(i).} 
The order of the automorphism $u$ is at most
two, i.e.\ $u$ squares to the identity and, therefore, is called an involutory automorphism.
Counterintuitively, this does not imply that $V_{\rep[_i]{r}}=\mathbbm1$.
In contrast, another possibility is that $V_{\rep[_i]{r}}=-\mathbbm1$.\footnote{%
That $u$ squares to the identity or to an inner automorphism has also been employed in \cite{Nishi:2013jqa}. 
However, that $u^2=\mathrm{id}$ can also imply $V_{\rep[_i]{r}}=-\mathbbm1$ has been missed.} 
Indeed, it is true that $V_{\rep[_i]{r}}=\pm\mathbbm1$ if and only if $u$ is involutory which will be shown in the following.

Applying the consistency condition \eqref{eq:CPConsistencyEquation} for
the group element $u(\elm g)$ while bringing all $\UU[_i]{\rep{r}}$'s to the other
side one finds 
\begin{equation}
  \rhoR{_i}\!(u(\elm g)) ~=~ \UU[_i]{\rep{r}}^\mathrm{T}\, \rhoR{_i}\!(u^2(\elm g))^* \,\UU[_i]{\rep{r}}^* 
  ~=~ \UU[_i]{\rep{r}}^\mathrm{T}\, \rhoR{_i}\!(\elm g)^* \,\UU[_i]{\rep{r}}^* \quad\forall~\elm g\in\DiscreteGroup~\text{and}~\forall~i\;,
\end{equation}
where in the last step it has been used that $u$ is involutory.
This reproduces the consistency condition \eqref{eq:CPConsistencyEquation} for $u$, but with the transposed matrices $\UU[_i]{\rep{r}}^\mathrm{T}$.
Due to the fact that $\rho_{\rep[_i]{r}}$ is an irrep one can then use Schur's lemma to show that
 \begin{equation}\label{eq:U(anti)symmetric}
   \UU[_i]{\rep{r}}^\mathrm{T} ~=~ \mathrm{e}^{\I\, \alpha} \, \UU[_i]{\rep{r}} \qquad \forall~i\;.
 \end{equation}
The only possible solutions for this are $\alpha=0$ or $\alpha=\pi$, meaning that 
$\UU[_i]{\rep{r}}$ is either a symmetric or an anti--symmetric unitary matrix, respectively.
Consequently, $V_{\rep[_i]{r}}=\UU[_i]{\rep{r}}\,\UU[_i]{\rep{r}}^*=\pm\mathbbm{1}$. 
 
To prove the reverse direction, assume that all $V_{\rep[_i]{r}}=\pm\mathbbm{1}$. 
Inserting \eqref{eq:CPConsistencyEquation} into itself one finds
\begin{equation}
 \rhoR{_i}\!(u^2(\elm g)) ~=~ \left(\UU[_i]{\rep{r}}\,\UU[_i]{\rep{r}}^*\right) \, 
 \rhoR{_i}\!(\elm g) \, \left(\UU[_i]{\rep{r}}\,\UU[_i]{\rep{r}}^*\right)^\dagger 
 ~=~ \rhoR{_i}\!(\elm g) \qquad \forall\,\elm g\in\DiscreteGroup~\text{and}~\forall~i\;.
\end{equation}
Being true for all irreps by assumption, it follows
that $u^2(\elm g)=\elm g$ for all $\elm g$ in \DiscreteGroup. Therefore, the order of $u$ can
only be one or two which shows that $u$ is involutory. This completes the proof
that $V_{\rep[_i]{r}}=\pm\mathbbm1$ if and only if $u$ is involutory.

This discussion shows that even though $u$ squares to the identity it is possible that 
\DiscreteGroup\ gets amended by an additional \Z2 symmetry upon imposing $u$ as a CP symmetry.
This is possible if and only if there is a representation $\rep[_i]{r}$ with $V_{\rep[_i]{r}}= -\mathbbm{1}$ present in the model.
The assignment of the \Z2 charges to the fields of a model then is uniquely fixed and given by the signs of the $V_{\rep[_i]{r}}$.

The case of an involutory automorphism $u$ is the most important case for the classification of finite groups 
and the consequences of $V_{\rep[_i]{r}}$ being $+\mathbbm1$ or $-\mathbbm1$ will be further discussed below. 

\paragraph{(ii).} 
The second possibility is that $u^2=v$ is an inner automorphism. As an illustration, note that this case may always be attained 
by amending an involutory automorphism $u$ (case (i)) by some inner automorphism $b$ such that $u\circ b$ does not square to the identity 
automorphism anymore. As the additional application of an inner automorphism corresponds to an already preserved symmetry transformation,
the automorphism $u\circ b$ can be regarded physically equivalent to the automorphism $u$.

Thus, applying this in reverse, the question whether case (ii) yields anything new in comparison to case (i) can be answered by checking whether one can always find
an inner automorphism which relates $u$ to an involutory automorphism $u'$ with $u'\circ b=u$. 
It has been proven explicitly for the majority of cases that any class--inverting automorphism that squares to an inner automorphism 
is related by an inner automorphism to a class--inverting involutory automorphism.
A proof exists for groups of odd order, automorphisms of odd order, and for automorphisms of order $\mathrm{ord}(u)=4n+2$
for some integer $n$ \cite{Chen:2014tpa}. The only case withstanding an explicit proof so far is for automorphisms of $\mathrm{ord}(u)=4n$.

Alternatively, one may also argue that any outer automorphism, by definition, actually corresponds to a coset of automorphisms, 
i.e.\ trivially contains all inner automorphisms. In this regard, any outer automorphism $u^2=v$, with $v$ inner, is by all means equivalent to 
the case that $u^2=\mathrm{id}$, due to the fact that the outer automorphism $\mathrm{id}$ already contains all inner automorphisms. 
Nevertheless, upon requiring $u$ as a symmetry, $G$ may still be enhanced by an Abelian factor in analogy to case (i).

\paragraph{(iii).} 
The last logical possibility is that $u^2=v$ is a non--trivial (necessarily class--preserving) outer
automorphism itself.
Then, there appears an additional generator $\elm h$ with
an explicit representation $\rhoR{_i}(\elm h)=V_{\rep[_i]{r}}$. It can be shown that
$\elm h$ does not commute with every group element of \DiscreteGroup, and, hence, extends $G$ to the 
larger semi--direct product group $H=\DiscreteGroup\SemiDirect_{v} \Z{\elm h}$, 
where $\Z{\elm h}$ is the cyclic group generated by $\elm h$.  Consequently, upon imposing $u$ as a CP symmetry, 
terms which are allowed by \DiscreteGroup\ but prohibited by $H$ are forced to be absent
from the Lagrangian. Since $v$ is class--preserving, it does not interrelate 
inequivalent representations such that the representation content of $H$ coincides with the one of
\DiscreteGroup. Nevertheless, upon imposing the CP transformation $u$,
complex conjugate representations are merged as usual. 

\enlargethispage{0.5cm}
As a remark, note that case (iii) seems to be rare among groups and, even though there is presently no general argument 
for its absence, no example group is known for this case. A \textsc{GAP} scan for class--inverting automorphism that square to a class--preserving outer automorphism 
yields a negative result for groups up to order 150 (with the exception of some groups of order 128 which have not been checked) \cite{Chen:2014tpa}.

\paragraph{}
In summary, a valid physical and model independent CP transformation 
is given by a class--inverting (outer, if $G$ has complex irreps) automorphism $u$ of $G$.

The requirement of $u$ being conserved, sometimes enforces other discrete symmetries as well. 
In case that $G$ needs to be extended by symmetries in addition to $u$, the corresponding symmetry actions
follow from the class--preserving automorphism $u^2$. The corresponding action on the representations of 
$G$, therefore, does not result in any new relations between inequivalent representations, besides, 
of course, the interrelation of $\rep[_i]{r}\leftrightarrow\rep[_i]{r}^*$ which is induced
by $u$ itself. 

One may argue that whenever a non--trivial enlargement of $G$ is necessary, the CP properties 
of a model should be studied from the ``top--down'', i.e.\ by the investigation of the 
enlarged symmetry group. This is, of course, justified whenever CP is an exact symmetry. 
However, since it is known that CP is violated in Nature, the presentation 
here has been oriented towards the ``bottom--up'' perspective. That is, possible CP transformation 
have been investigated without necessarily requiring them to be conserved. 
It has been shown that a group does not have to be extended by additional symmetries upon requiring $u$ to be a CP symmetry
if $u$ corresponds to a class--inverting automorphism of order 2 that can be represented by symmetric 
matrices $\UU[_i]{\rep{r}}$ for all $i$. In the next section the consequences of the existence of such an automorphism are further discussed.

\subsection{The Bickerstaff--Damhus automorphism}
This section establishes an interesting connection between the existence of proper physical CP transformations
and the possibility to find a basis for $G$ in which all CGs are real.

According to a theorem by Bickerstaff and Damhus \cite{Bickerstaff:1985jc}, all CGs of $G$ are real 
if and only if there exists an automorphism $u$ such that 
\begin{equation}\label{eq:BDAequation2}
 \rhoR{_i}\!\left(u(g)\right)~=~\rhoR{_i}\!(g)^*\quad\forall~g\in\DiscreteGroup ~\text{and}~\forall~i\;.
\end{equation}
With the methods of the preceding section it is straightforward to show that such an automorphism is
class--inverting and involutory. Note that both, \Eqref{eq:BDAequation2} and the 
fact that CGs are real, are basis dependent statements. 

By using the behavior of $U$ in \eqref{eq:CPConsistencyEquation} under basis rotations, 
i.e.\ $U\rightarrow VUV^\mathrm{T}$ cf.\ \Secref{sec:generalizedCP}, it is possible to rephrase the Bickerstaff--Damhus theorem 
in a basis independent manner. Namely, there exists a 
basis in which all CGs of $G$ are real if and only if there is an automorphism $u$ which fulfills
\begin{equation}\label{eq:BDAequation}
 \rhoR{_i}\!\left(u(g)\right)~=~\UU[_i]{\rep{r}}\,\rhoR{_i}\!(g)^*\,\UU[_i]{\rep{r}}^\dagger\,,\quad\text{with}~
 \UU[_i]{\rep{r}}~\text{unitary and symmetric,}\quad\forall~g\in\DiscreteGroup
 ~\text{and}~\forall~i\;.
\end{equation}
Interestingly, these are precisely the conditions that have been found above for the existence of a
proper physical CP transformation which does not lead to an extension of the finite group.
In what follows, an automorphism $u$ which
satisfies \Eqref{eq:BDAequation} will be referred to as a Bickerstaff--Damhus automorphism (BDA). 
To repeat, a BDA is a class--inverting involutory automorphism which fulfills the
consistency condition \eqref{eq:CPConsistencyEquation} with symmetric unitary
matrices $\UU[_i]{\rep{r}}$.

The basis in which the CGs can be chosen real is exactly the CP basis for which
all $\UU[_i]{\rep{r}}$ in \Eqref{eq:BDAequation} are unit matrices, i.e.\ for which
\Eqref{eq:BDAequation2} is achieved. Since orthogonal basis transformations
do not change the form of \eqref{eq:BDAequation2} this equation actually defines a whole set of bases
with real CGs.

As a remark, note that there can be several different BDAs which fulfill
\Eqref{eq:BDAequation2} for different bases. The different BDAs generally will 
not be related by inner automorphisms. For example, a group which has two, in this sense
distinct, BDAs is $\text{SG}(32,43)$ \cite{Chen:2014tpa}. Nevertheless,
when \eqref{eq:BDAequation2} is fulfilled in a certain basis, the corresponding 
automorphism $u$ is unique.

It can be shown that non--Abelian groups of odd order do not
admit BDAs \cite{Chen:2014tpa}. Therefore, odd order non--Abelian groups do not admit a basis with
completely real CGs.

\subsection{The twisted Frobenius--Schur indicator}

As a central part of the classification,
a basis independent method is presented in the following which allows to determine if a given finite group allows for a 
model independent physical CP transformation, i.e.\ whether $G$ admits a class--inverting involutory automorphism. 
On the side, a basis independent algorithm is given that allows one
to determine whether a group allows for real CGs.
All statements of this section can be proved by the use of well--known Schur orthogonality relations (see e.g.~\cite[p.~37]{Ramond:2010zz}),
and an explicit form of the proofs has been given in \cite{Chen:2014tpa}.

Recall the well--known Frobenius--Schur indicator (FSI)(c.f.\ e.g.~\cite[p.~48]{Ramond:2010zz}) 
which is defined by 
\begin{equation}
  \mathrm{FS}(\rep[_i]{r}) ~:=~ 
  \frac{1}{|\DiscreteGroup|}\,\sum_{g \in \DiscreteGroup} \, 
        \chiR{_i}\!(g^2) 
  ~=~ \frac{1}{|\DiscreteGroup|}\,\sum_{g \in \DiscreteGroup} \,
        \tr{\left[\rhoR{_i}\!(g)^2\right]}\; ,
\end{equation}
where $|\DiscreteGroup|$ is the order of the group $\DiscreteGroup$.
The FSI is used to determine whether a representation of a finite group is real, pseudo--real, or complex,
since it evaluates to
\begin{equation}
  \mathrm{FS}(\rep[_i]{r}) ~=~ 
  \left\{\begin{array}{ll}
    +1,& \text{~if \rep[_i]{r} is a real representation,}\\
    0, & \text{~if \rep[_i]{r} is a complex representation,}\\
    -1,& \text{~if \rep[_i]{r} is a pseudo--real representation.}
  \end{array}\right.
\end{equation}

Completely analogous to the FSI, there is 
the so--called twisted Frobenius--Schur indicator (\FSI) \cite{Bickerstaff:1985jc,Kawanaka1990} which additionally depends on an automorphism $u$.
The twisted Frobenius--Schur indicator for an irrep
\rep[_i]{r} and an automorphism $u$ is defined by
\begin{equation}
  \FSI(\rep[_i]{r}) ~:=~ 
  \frac{1}{|\DiscreteGroup|}\,\sum_{g \in \DiscreteGroup} \, 
        \chiR{_i}\!(g \, u(g)) 
   ~=~ \frac{1}{|\DiscreteGroup|}\,\sum_{g \in \DiscreteGroup} \,
        \tr{\left[\rhoR{_i}\!(g)\,\rhoR{_i}\!(u(g))\right]}\;.
\end{equation}
The definition of the \FSI is such that for $u\equiv\mathrm{id}$ one recovers the original FSI.
The \FSI then can be used to determine the nature of an automorphism $u$. In fact, for an automorphism $u$ one can show that \cite{Chen:2014tpa}%
\begin{equation}\label{eq:FSI}
  \FSI(\rep[_i]{r}) ~=~ 
  \left\{\begin{array}{ll}
    +1 \quad \forall~i, &~\text{if}~u\text{~is a Bickerstaff--Damhus automorphism,}\\
    +1~\text{or}~-1 \quad \forall~i,& ~\text{if}~u\text{~is class--inverting and involutory,}\\
    \text{$\neq\pm1$}~\text{for some $i$},& ~\text{if}~
        u\text{~is not class--inverting and/or not involutory.}
  \end{array}\right.
\end{equation}
The twisted Frobenius--Schur indicator \FSI vanishes
for at least one irrep if $u$ is not class--inverting.

On the other hand, if $u$ is class--inverting, one can show that \cite{Chen:2014tpa}%
\begin{equation}
  \FSI(\rep[_i]{r}) ~=~ \frac{1}{\dim \rep[_i]{r}}\,\tr{\left(\UU[_i]{\rep{r}} \, \UU[_i]{\rep{r}}^*\right)} 
  ~=~ \frac{1}{\dim \rep[_i]{r}}\,\tr{\left(V_{\rep[_i]{r}}\right)}\;.
\end{equation}

\begin{figure}[t]
\centerline{\includegraphics{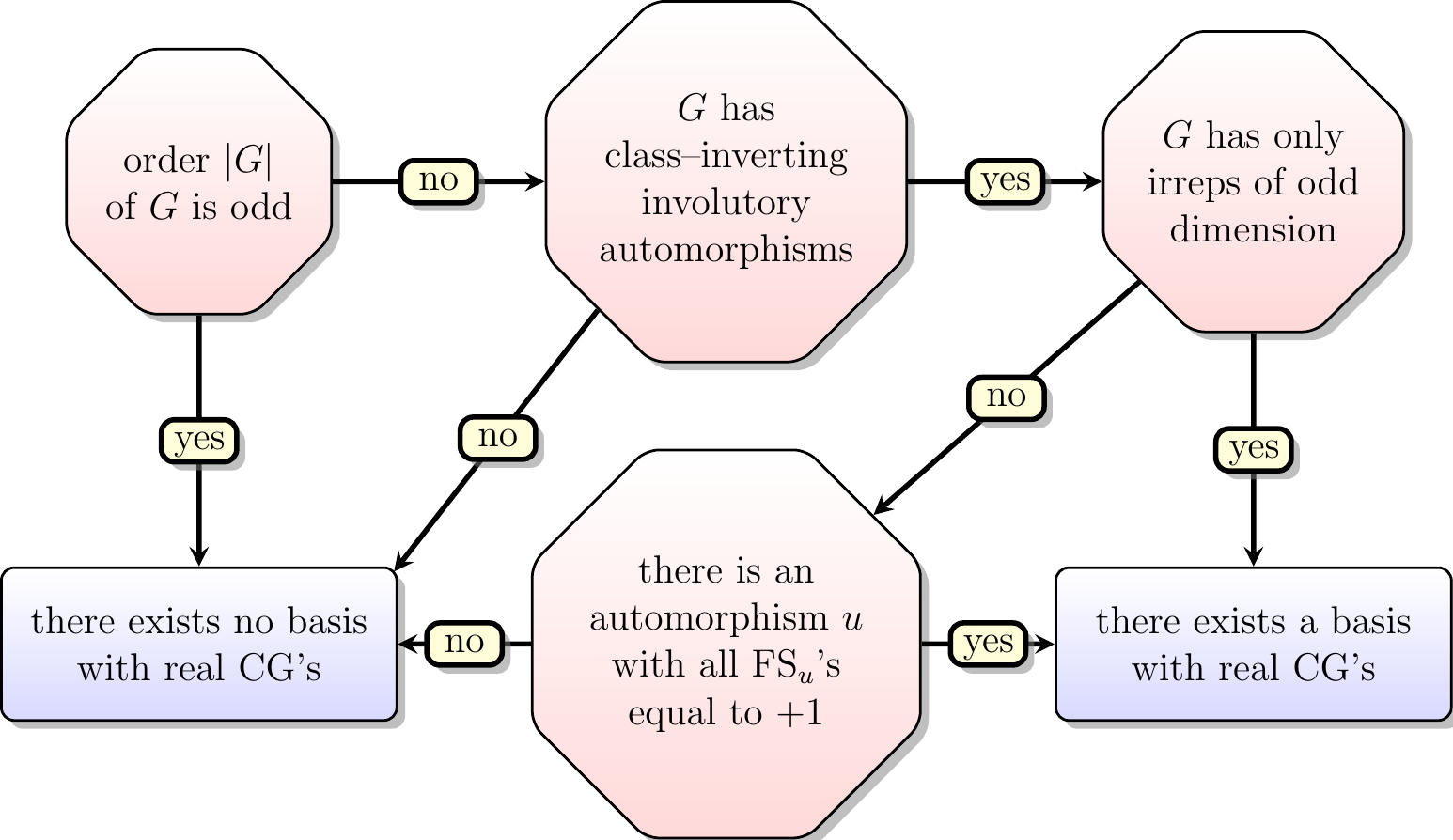}}
  \caption{A possible algorithm to determine whether a finite non--Abelian group $G$ allows for a basis with real
Clebsch--Gordan coefficients (figure taken from \cite{Chen:2014tpa}).}
  \label{fig:RealCGs} 
\end{figure}

It has been shown above that $V_{\rep[_i]{r}}=\pm \mathbbm{1}$ if and
only if $u$ is involutory, where plus(minus) signals a(n) (anti--)symmetric representation matrix $\UU[_i]{\rep{r}}$. 
Therefore, $\FSI=\pm1$ for all irreps \rep[_i]{r} if and only if $u$ is a class--inverting involutory automorphism.
Here, $\FSI(\rep[_i]{r})=+1$ applies if the corresponding transformation matrices $\UU[_i]{\rep{r}}$ are symmetric, while
$\FSI(\rep[_i]{r})=-1$ applies to the anti--symmetric case.

It is important to note that the \FSI can vanish for automorphisms of order larger than two even though they are class--inverting. 
In this case, it is possible to define an extended version of the indicator, which again
has the property to be $\pm1$ for all irreps in the class--inverting case and
$0$ for some irrep otherwise. The $n^\mathrm{th}$ extended twisted
Frobenius--Schur indicator is defined by
\begin{equation}\label{eq:FSIn}
  \FSI^{(n)}(\rep[_i]{r}) ~:=~ \frac{(\dim{\rep[_i]{r}})^{n-1}}{|\DiscreteGroup|^n}\,
  \sum_{g_1,\dots,g_n \in \DiscreteGroup} \, \chiR{_i}
  \bigl(g_1 \, u(g_1)\cdots g_n \, u(g_n)\bigr)\;,
\end{equation}
where $n=\ord{u}/2$ for even and $n=\ord{u}$ for odd--order automorphisms. 
The first extended twisted Frobenius--Schur
indicator $\FSI^{(n=1)}$ is identical to \FSI.

Due to the fact that a group allows for a basis with real CGs if and only if it has a BDA
it is possible to use the \FSI in order to develop an algorithm to determine whether this is
the case for a given (non--Abelian) group. A possible strategy to do this is shown in \Figref{fig:RealCGs}.
One should note here, that in principle only the very last step (mid, bottom) is necessary to make the decision.
The other steps, however, may be faster to compute. It should be remarked that Abelian groups always have a BDA, hence, always 
allow for real CGs.

A computer code to automatically compute the twisted Frobenius--Schur indicator with the aid of \textsc{GAP} is given in \Appref{app:FSI}.

\subsection{Classification of finite groups}
\label{sec:classification}

Finally, the insights of the previous sections can be used to categorize finite groups into three classes according to their CP properties. 
This task can be performed basis independently with the aid of the twisted Frobenius--Schur indicator.
To do this, the indicator must be calculated for all involutory automorphisms $u_\alpha$ of the specific finite group
\DiscreteGroup.\footnote{%
More precisely, one would have to calculate the $n^{\mathrm{th}}$
twisted $\FSI^{(n)}$ for all automorphisms. The difference, however, is only relevant for 
groups of the case (iii) of \Secref{Sec:CPautProps} for which there is no known example.}
A \textsc{GAP} code which automatizes this computation can be found in \cite{Chen:2014tpa}. 

\paragraph{}%
There are three types of groups:

\begin{description}
 \item[Type~I:] The group $G$ does not have a class--inverting automorphism. 
Therefore, not all irreps can simultaneously be mapped onto their respective complex conjugate irrep, implying that the group
does not allow for the definition of a model independent CP transformation.
Equivalently, for each automorphisms $u_\alpha$ of \DiscreteGroup\ there exists at least one representation $\rep[_i]{r}$ for which
 $\text{FS}^{(n)}_{u_\alpha}(\rep[_i]{r})=0$. Type~I groups do not allow for a basis in which all CGs are real. 

\item[Type~II:] There is at least one automorphism $u$ of
 \DiscreteGroup\ which is class--inverting, that is, it maps all irreps to their respective complex conjugate representations.
Based on this automorphism it is possible to define a model independent proper physical CP transformation.
There are two sub--cases:
 \begin{description}
  \item[Type~II~A:] There exists a Bickerstaff--Damhus automorphism, i.e.\ $G$ has a class--inverting involutory automorphism that
	can be represented by unitary and symmetric matrices $U$. Therefore, $G$ has a CP basis in which all CGs are real.
	For the BDA	all \FSI's are $+1$. 
  \item[Type~II~B:] Even though there exists a class--inverting automorphism it is either not involutory or it is involutory but cannot
	be represented by symmetric matrices $U$. This automorphism can be used to define a model independent proper physical CP transformation. 
	However, upon imposing CP as a symmetry the group $G$ is extended by additional transformations arising from $\UU[_i]{\rep{r}}\UU[_i]{\rep{r}}^*=V_{\rep[_i]{r}}$.\footnote{%
	It is not excluded that $V_{\rep[_i]{r}}$ is actually part of the group to begin with.
	In this case the group would not have to be extended. There is no known example for such a case, and it is presently not clear whether this case is possible at all. 
	All of the type~II~B groups investigated in this work have to be extended upon requiring CP conservation.}
	Hence, there is no CP basis and there exist, in general, ``half--odd'' or even more exotic CP eigenstates.  
	For the according class inverting automorphism some of the \FSI's are $-1$ while all others are $+1$.
	There is no BDA, and, hence, no basis in which all CGs are
  real.  
 \end{description}
\end{description}

\begin{figure}[!h!]
\centerline{\includegraphics{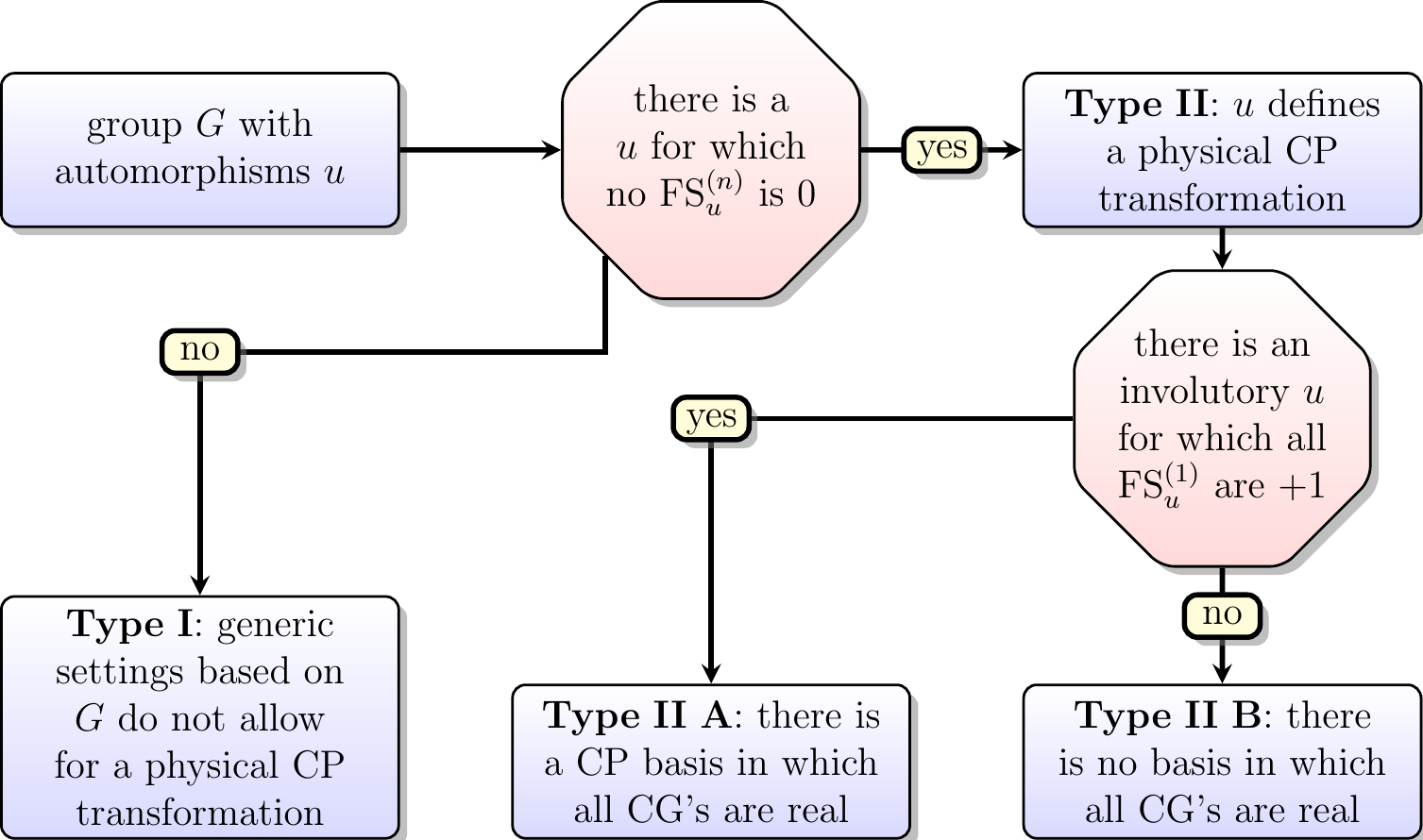}}
  \caption{Algorithm to distinguish between the three types of groups via the twisted 
   Frobenius--Schur indicator $\FSI^{(n)}$. The integer $n$ is $n=\ord{u}/2$ for even
   and $n=\ord{u}$ for odd order of $u$ (figure taken from \cite{Chen:2014tpa}).}
  \label{fig:3types} 
\end{figure}

An algorithm to classify a given group as one of the three types is illustrated in \Figref{fig:3types}. 
Some examples for groups of each type are listed in \Tabref{tab:groupTypes}.

There is a possible caveat due to the fact that these statements are made in the most general, model independent way.
Note, that in a specific model it may be possible to define a proper physical CP transformation even if the model features a 
discrete group of the type I, i.e.\ even though there is no class--inverting automorphism. 
This is the case whenever the representation content of a model is chosen such that there is an automorphism of 
$G$ which maps \textit{all present} representations to their respective complex conjugate representation.  
Whenever one defines a model based on a type I group in such a way that a CP outer automorphism is possible,
i.e.\ omits certain representations, this model is called non--generic. In contrast,
a setting is called generic, if the representation content is unconstrained in this way.
These statements will become clearer after studying the explicit example of a generic setting in \Secref{sec:Delta27}
and the example of a non--generic setting in \Secref{sec:3HDMExample}.

\renewcommand{\arraystretch}{1.2}
\begin{table}[t]
\begin{center}
\begin{subtable}{1.\textwidth}
\centering
\begin{tabular}{r|cccc}
  $G$& $\Z5 \SemiDirect \Z4$ & $\mathrm{T}_7$  & $\Delta(27)$ & $\Z9 \SemiDirect \Z3$ \\%& $A_9$ & $A_{11}$ & $A_{13}$ & $A_{n\geq 15}$ \\
  \hline
  SG & (20,3) & (21,1) & (27,3) & (27,4) \\
  \end{tabular}
\caption{Examples for type I groups.}
\label{tab:typeI}
\end{subtable}\vspace{0.4cm}
\begin{subtable}{1.\textwidth}
\centering
\begin{tabular}{r|cccccccc}
  $G$ 
  & $\mathrm{S}_3$ & $\mathrm{Q}_8$ & $\mathrm{A}_4$ & $\Z3\SemiDirect\Z8$ & $\mathrm{T}'$ &  $\mathrm{S}_4$ & $\mathrm{A}_5$\\
  \hline
  SG & (6,1) & (8,4) & (12,3) & (24,1) & (24,3) & (24,12) & (60,5)\\
  \end{tabular}
\caption{Examples for type~II~A groups. All Abelian groups are of this type.}
\label{tab:typeIIA}
\end{subtable}\vspace{0.4cm}
\begin{subtable}{1.\textwidth}
\centering
\begin{tabular}{r|cc}
  $G$ 
  & $\Sigma(72)$ & $\left((\Z3 \times \Z3) \SemiDirect \Z4\right) \SemiDirect \Z4$\\
  \hline
  SG & (72,41) & (144,120) \\
  \end{tabular}
\caption{Examples for type~II~B groups.}
\label{tab:typeIIB}
\end{subtable}
\end{center}  
\caption{Examples for each of the three types of groups: Type~I~\subref{tab:typeI}, type~II~A~\subref{tab:typeIIA}, and type~II~B~\subref{tab:typeIIB},
with their typical names and \textsc{GAP} SmallGroups library ID.}
\label{tab:groupTypes}
\end{table}
\renewcommand{\arraystretch}{1.0}%

The following sections give one example each for groups of the type~II~A, type~II~B, and type~I.
It will be shown how CP transformations can(not) be constructed for each of the groups and how CPV can arise. 
Type~II~A groups in this respect most closely resemble the well--known case of continuous Lie groups 
and, therefore, will be treated first. 
Secondly, the related type~II~B groups will be discussed where CP transformations exist but have to be generalized, 
i.e.\ CP transformations will always be accompanied by non--trivial representation matrices $U$.
Lastly, an example toy model will illustrate how explicit geometrical CP violation follows from the assumption of a type~I symmetry.
For the most interesting case of type~I groups, it will also be illustrated how spontaneous geometrical CP violation arises.
A different case of spontaneous geometrical CP violation will be treated in much more detail, also based on a much more interesting example model, in \Secref{sec:3HDMExample}.

\section{Type~II~A groups: ``Nothing special''}

\subsection{Explicit example:  \texorpdfstring{$\mathrm{T'}$}{}}
An example for a group of type~II~A is the group \Tprime.
\Tprime is listed in the SmallGroup catalogue of \textsc{GAP} as $\mathrm{SG}(24,3)$.
A presentation for \Tprime is given by
 \begin{equation}\label{eq:TprimePrsentation}
\Tprime~=~\Braket{\elm{S,T}~|~\mathsf{S}^4\,=\,\mathsf{T}^3\,=\,\left(\elm S\,\elm T\right)^3=\mathsf{e}~}\;.
\end{equation}
Besides the trivial singlet the group has two non--trivial one--dimensional,
three two--dimensional, and one three--dimensional irrep.
More details on the group can be found in \cite{Chen:2014tpa}
where also different basis conventions used in the literature are discussed and compared.

The group \Tprime has a unique involutory and class--inverting outer automorphism, which, therefore, swaps every representation
with its respective complex conjugate representation. 
A possible representation of this outer automorphism\footnote{%
As always, one particular choice of inner automorphism has been made to state the explicit action of the outer automorphism. 
While all other possible choices of inner automorphisms are admissible as well, they would only differ by a symmetry transformation and, therefore,
be physically equivalent.} 
is given by
\begin{equation}\label{eq:TprimeAutomorphism}
 u ~:~ (\elm S,\elm T)\,\mapsto\,(\elm S^3, \elm T^2)\;,
\end{equation}
and it acts on the irreps of \Tprime as
\begin{equation}
 u ~:~ 
 \rep[_i]{1}~\rightarrow~\UU[_i]{\rep{1}}\,\rep[_i]{1}^*\;,\quad
 \rep[_i]{2}~\rightarrow~\UU[_i]{\rep{2}}\,\rep[_i]{2}^*\;,\quad
 \rep{3}~\rightarrow~\UU{\rep{3}}\,\rep{3}^*\;.
\end{equation}
The explicit transformation matrices $\UU[_i]{\rep{r}}$ of $u$ can be deduced from the following arguments.
The twisted Frobenius--Schur indicators for $u$ are displayed in \Tabref{tab:TPrimeFSI}.
From the fact that all \FSI's are +1 one concludes that $u$ is a BDA. 
That is, $u$ is class--inverting, involutory, and
has symmetric representation matrices. 
Consequently, \Tprime admits a basis with real CGs.
This basis is also the CP basis in which all representation matrices of $u$ 
are unit matrices of dimension $\dim(\rep[_i]{r})$, i.e.\ $\UU[_i]{\rep{r}}=\mathbbm{1}_{\dim(\rep[_i]{r})}$.

\renewcommand{\arraystretch}{1.2}
\begin{table}[t]
\centering
\begin{tabular}{c|ccccccc}
 $\rep r$ & \rep[_0]{1} & \rep[_1]{1} & \rep[_2]{1} & \rep[_0]{2} & \rep[_1]{2} & \rep[_2]{2} & \rep{3} \\
\hline
$\TFS{u}(\rep r)$ & 1 & 1 & 1 & 1 & 1 & 1 & 1 
\end{tabular}
\caption{Twisted Frobenius--Schur indicators of the automorphism
\eqref{eq:TprimeAutomorphism} of $\Tprime$.}
\label{tab:TPrimeFSI}
\end{table}\renewcommand{\arraystretch}{1.0}%

An explicit form of the representation matrices in the CP basis is given by
\begin{equation}
S_{\rep{3}} ~=~
 \frac13\begin{pmatrix}
  -1 & 2 & 2 \\ 2 & -1 & 2 \\ 2 & 2 & -1
 \end{pmatrix}
 \quad \text{and} \quad
 T_{\rep{3}} ~=~
 \begin{pmatrix}
  1 & 0 & 0 \\ 0 & \omega  & 0 \\ 0 & 0 & \omega ^2
 \end{pmatrix}
\end{equation}
for the triplet representation, as well as by 
\begin{equation}
 S_{\rep[_i]{2}}~=~ -\frac{\I}{\sqrt{3}} 
 \begin{pmatrix}
  1 & \sqrt{2} \\ \sqrt{2} & -1 
 \end{pmatrix}
 \quad\text{and}\quad
 T_{\rep[_i]{2}}~=~\omega^i
 \begin{pmatrix}
  \omega^2 & 0 \\ 0 & \omega
 \end{pmatrix}
\end{equation}
for the doublet representations $\rep[_i]{2}$ $(i=0,1,2)$.
It is straightforward to check that in this basis
\begin{equation}
 S_{\rep{r}}^*~=~S_{\rep{r}}^3\quad\text{and}\quad T_{\rep{r}}^*~=~T_{\rep{r}}^2
\end{equation}
are fulfilled for every representation. Therefore, the Bickerstaff--Damhus condition \eqref{eq:BDAequation2}
 is fulfilled in this basis. 
The corresponding real CGs can be found in \cite{Ishimori:2010au} and will not be stated here.

Therefore, any setting based on the group \Tprime (and possibly other space--time and continuous internal symmetries)
allows for the definition of a CP transformation. 
In the \Tprime space, this CP transformation is based on the outer automorphism $u$ \eqref{eq:TprimeAutomorphism}.
Only in the CP basis this transformation acts trivially as $\rep[_i]{r}\mapsto\rep[_i]{r}^*$. In any other basis
the according, generally non--trivial, representation matrices of $u$ have to be taken into account. 
They can be obtained by basis--transforming $\UU[_i]{\rep{r}}$ as in \eqref{eq:CPBasisChange}. 
It will generically lead to inconsistencies if the representation matrices $\UU[_i]{\rep{r}}$ are not properly taken into account.  
For example, naively applying the map $\rep[_i]{r}\mapsto\rep[_i]{r}^*$ without taking into account the $\UU[_i]{\rep{r}}$'s
generically maps \Tprime invariants in the Lagrangian to non--invariants, cf.\ \cite{Holthausen:2012dk, Chen:2014tpa}.
However, the existence of $u$ as a consistent CP outer automorphism implies that a consistent CP transformation 
for \Tprime exists in any basis, as has just been demonstrated.

\subsection{CP violation for type~II~A groups}

Nevertheless, the mere existence of a consistent CP transformation does, of course, not imply that CP is conserved.
In this section, therefore, the CPV properties of models based on type~II~A groups shall be analyzed.
In addition to a discrete group $G$, the models under discussion can have the usual space--time 
and continuous internal symmetries without affecting any of the conclusions.

It will be demonstrated that models based on type~II~A groups behave, with respect to their CP transformation properties, 
just as models which are based on semisimple Lie groups such as, for example, \SU{n}.
This can be attributed to the fact that both, compact Lie groups as well as type~II~A groups, allow 
for involutory complex conjugation (CP) outer automorphisms which are, in the so--called CP basis, represented by unit matrices.

For definiteness, consider two fields $x$ and $y$ transforming in irreps $\rep{r}(x)=\rep[_x]{r}$ and  $\rep{r}(y)=\rep[_y]{r}$ of $G$
and assume that they can be contracted to the trivial singlet representation. This contraction can be written as
\begin{equation}\label{eq:Singletcontraction}
 \left(x\otimes y\right)_{\rep[_0]{1}}
 ~=~ 
 C_{\alpha\beta}\,x_\alpha\,y_\beta~=~x^\mathrm{T}\,C\,y\;,
\end{equation}
where $\alpha$ and $\beta$ are the vector indices of $x$ and
$y$, and $C_{\alpha\beta}$ denote the CGs of this contraction. 
For the last equality the vector indices of $x$, $y$, and $C$ have been suppressed, i.e.\  a matrix--vector notation has been introduced.
Requiring the action to be real, the presence of \eqref{eq:Singletcontraction} in a Lagrangian generically 
requires also the presence of the corresponding complex conjugate contraction, reading 
\begin{equation}\label{eq:contraction_conjugate}
 \left(x\otimes y\right)^*_{\rep[_0]{1}}
 ~=~ 
 C^*_{\alpha\beta}\,x^*_\alpha\,y^*_\beta~=~x^\dagger\,C^*\,y^*\;.
\end{equation}
Including arbitrary complex couplings $c$, a $G$ symmetric Lagrangian schematically would contain
\begin{equation}\label{eq:Linitial}
\mathscr{L}~\supset~ c \, \left(x^\mathrm{T}\,C\,y\right) + c^* \, \left(x^\dagger\,C^*\,y^*\right) \;.
\end{equation}
In general, the conjugation outer automorphism, i.e.\ the CP transformation will act on each, $x$ and $y$, according to 
\begin{equation}\label{eq:rep_trafo}
 \rep[_i]{r}~\mapsto~ 
 \UU[_i]{\rep{r}}\,\rep[_i]{r}^*\;.
\end{equation}
For groups of type~II it is guaranteed that such a transformation exists.
Under the action of this transformation the above Lagrangian is transformed to
\begin{equation}\label{eq:Ltrafo}
\mathscr{L}~\mapsto~\mathscr{L'}~\supset~ c \, \left(x^\dagger\,\UU[_x]{r}^\mathrm{T}\,C\,\UU[_y]{r}\,y^*\right) + c^* \, \left(x^\mathrm{T}\,\UU[_x]{r}^\dagger\,C^*\,\UU[_y]{r}^*\,y\right) \;.
\end{equation}
However, for groups of type~II~A there exists a CP basis in which $\UU[_i]{\rep{r}}=\mathbbm{1}_{\dim(\rep[_i]{r})}$, and in which the CGs are real numbers.
Working in this basis, the transformed Lagrangian reads
\begin{equation}
\mathscr{L'}~\supset~ c \, \left({x}^\dagger\,C\,{y}^*\right) + c^* \, \left({x}^\mathrm{T}\,{C}\,{y}\right) \;.
\end{equation}
Comparing this with the original form \eqref{eq:Linitial}, one concludes that CP is a symmetry of this setting if and only if
\begin{equation}
 c~\equiv~c^*\;.
\end{equation}
Therefore, a conserved involutory CP transformation, as always present in settings with type~II~A symmetry, generally requires real couplings. 
More precisely formulated, the requirement for CP conservation in type~II~A groups is that one can find a basis in which all couplings are real. 
Arbitrary basis choices and rephasings of field may give rise to so--called spurious phases \cite{Branco:1999fs}, which, 
however, can always be absorbed by basis transformations (i.e.\ field redefinitions) if CP is conserved. 
This argument is straightforwardly extended to an arbitrary number of fields, where any operator is mapped to its Hermitian conjugate operator as
long as the transformation \eqref{eq:rep_trafo} is applicable.

Consequently, CP can be violated in settings with type~II~A groups only if a sufficient number of field redefinitions is not possible, such that some of the complex phases of couplings become physical.
A very accessible presentation of criteria for when such a situation arises (that is, a systematic way of counting 
rephasing degrees of freedom vs.\ complex couplings) is given in \cite{Haber:2012np}. This type of CPV, for example, is present in the SM.
The situation in type~II~A groups, therefore, is very reminiscent to the well--known settings with continuous groups: 
CP transformations are always possible, they generically constrain the phases of couplings, and can be violated explicitly only if 
there are more complex phases in couplings than what can be absorbed by rephasings.
However, whether or not this mechanism really leads to CPV in possibly realistic theories cannot be decided from a theoretical point of view, but must be clarified by experiments.
In this sense, this type of CPV can never be predictive.

\section{Type~II~B groups: Non--trivial CPV and CP half--odd states}

\subsection{Explicit example: \texorpdfstring{$\Sigma(72)$}{}}
\label{sec:Sigma72Example}

An example for a group of the type~II~B is the non--Abelian group $\Sigma(72)$ which is listed
in the \textsc{GAP} SmallGroups library as $\mathrm{SG}(72,41)$. A minimal generating set for $\Sigma(72)$ is given by
\begin{equation}\label{eq:Sigma72Presentation}
\Sigma(72)~=~\Braket{ \elm{M,P}~|~ \elm M^4\,=\,\elm P^4\,=\,\left(\elm M\elm P\elm M\elm P\right)^3\,=\,\left(\elm P\elm M\elm P^2\elm M\right)^3\left(\elm M\elm P\right)^2\elm P\elm M^2\,=\,\elm e}\;.
\end{equation}
The group $\Sigma(72)$ has three one--dimensional (\rep[_{1-3}]{1}), a two--dimensional (\rep{2}), and an
eight--dimensional (\rep{8}) irrep. The character table of the group is shown in \Tabref{tab:Sigma72char}.
More details of $\Sigma(72)$ are given in \Appref{app:Sigma72}.

\begin{table}[t]
\centering
\begin{tabular}{c|*{6}{r}|}
                &  $C_{1a}$ & $C_{3a}$ & $C_{2a}$ & $C_{4a}$ & $C_{4b}$ & $C_{4c}$ \\
                &  1 &  8 &  9 &  18 &  18 &  18 \\
  $\Sigma(72)$   & $\elm e$ & $\elm M^2\elm P^2$ & $\elm M^2$ & $\elm M\elm P$ & $\elm P$ & $\elm M$ \\
  \hline
  $\rep[_0]{1}$ & $1$ & $1$  & $1$  & $1$  & $1$  & $1$ \\
  $\rep[_1]{1}$ & $1$ & $1$  & $1$  & $1$  & $-1$ & $-1$\\
  $\rep[_2]{1}$ & $1$ & $1$  & $1$  & $-1$ & $1$  & $-1$\\
  $\rep[_3]{1}$ & $1$ & $1$  & $1$  & $-1$ & $-1$ & $1$ \\
  $\rep{2}$     & $2$ & $2$  & $-2$ & $0$  & $0$  & $0$ \\
  $\rep{8}$     & $8$ & $-1$ & $0$  & $0$  & $0$  & $0$ \\
  \hline
\end{tabular}
\caption{Character table of $\Sigma(72)$. 
The second line gives the cardinality of the conjugacy class (c.c.)\ and
the third line gives a representative of the corresponding c.c.\ in the presentation specified
in \eqref{eq:Sigma72Presentation}.}
\label{tab:Sigma72char}
\end{table}

The group $\Sigma(72)$ is peculiar in the sense that every conjugacy class contains along with an element $\elm g$ also the inverse element $\elm g^{-1}$. This is equivalent to
the fact that all characters of the group are real.
Groups with this property are called ambivalent. 
Therefore, even though the outer automorphism group of $\Sigma(72)$ is the symmetric group\footnote{%
The outer automorphism group $\mathrm{Out}(\Sigma(72))=\mathrm{S}_3$ acts on the representations as permutation of the one--dimensional representations $\rep[_{1-3}]{1}$.}
$\mathrm{S}_3$, none of the outer automorphisms is class--inverting. 
On the other hand, for ambivalent groups, every inner automorphism is class--inverting. It is, thus, possible to use the identity automorphism to define
a consistent model--independent CP transformation. The corresponding twisted Frobenius--Schur indicators reduce to the ordinary Frobenius--Schur 
indicators and they are shown in \Tabref{tab:Sigma72FSI}.
The value $\TFS{\mathrm{id}}(\rep2)=-1$ signals that the two--dimensional representation
is pseudo--real, and therefore, transforms with an anti--symmetric matrix under this complex conjugation automorphism.
Altogether, this discussion shows that $\Sigma(72)$ does not have a BDA, and, therefore, also 
no basis in which all CGs can be chosen real. 

\renewcommand{\arraystretch}{1.2}
\begin{table}[t]
\centering
\begin{tabular}{c|*{6}{c}}
 $\rep r$ & \rep[_0]{1} & \rep[_1]{1} & \rep[_2]{1} & \rep[_3]{1} & \rep{2} & \rep{8} \\
\hline
$\TFS{\mathrm{id}}(\rep r)$ & $1$ & $1$ & $1$ & $1$ & $-1$ & $1$ \\
\end{tabular}
\caption{Twisted Frobenius--Schur indicators for the identity automorphisms of
$\Sigma(72)$.}
\label{tab:Sigma72FSI}
\end{table}\renewcommand{\arraystretch}{1.0}%

The CP transformation based on the class--inverting and involutory identity automorphism acts as
\begin{equation}\label{eq:S72CP}
 (\elm M,\elm P)~\mapsto~(\elm M, \elm P)   
 \quad\curvearrowright\quad 
 \rep[_i]{1}~\mapsto~\rep[_i]{1}^*\;,
 \quad
 \rep{2}~\mapsto~\UU{\rep{2}}\,\rep{2}^*\;,
 \quad
 \rep{8}~\mapsto~\UU{\rep{8}}\,\rep{8}^*\,.
\end{equation} 
As usual, the explicit representation matrices are found by solving the corresponding consistency 
condition \eqref{eq:CPConsistencyEquation}.
Using the basis specified in \Appref{app:Sigma72} as well as the identity automorphism one finds
\begin{equation}
  \UU{{\rep{2}}} ~=~ \begin{pmatrix}0 & 1\\ -1 & 0 \end{pmatrix}\;,\quad\text{and}\quad  \UU{{\rep{8}}} ~=~ \mathbbm{1}_8\;.
\end{equation}
From $\UU{{\rep{2}}}\UU{{\rep{2}}}^*=V_{\rep{2}}=-\mathbbm{1}_2$ it immediately follows that,
upon imposing CP to be conserved, a model based on $\Sigma(72)$ will pick up an additional \Z2 symmetry acting trivially on 
all representations besides the $\rep{2}$ on which it acts as $V_{\rep{2}}=-\mathbbm{1}$.\footnote{%
Strictly speaking, the simultaneous presence of \textit{both}, the faithful \rep{8} and the \rep{2}, is required that the
\Z2 extension appears. The resulting group after requiring CP then is $\mathrm{SG}(288,892)$, which is of type~II~B.}
Furthermore, the appearance of $\UU{{\rep{2}}}$ with $\UU{{\rep{2}}}\UU{{\rep{2}}}^*=-\mathbbm{1}_2$ immediately signals the presence of CP half--odd
states, cf.\ \Secref{sec:GeneralizedCP}.

There is another peculiarity related to type~II~B groups and their behavior under CP transformations. There are two possible ways to
contract the \rep{8} with itself to form a \rep{2}
\begin{equation}\label{eq:882}
\left(\rep{8} \otimes \rep{8}\right)_{\rep[_1]{2}} \quad\text{and}\quad \left(\rep{8} \otimes \rep{8}\right)_{\rep[_2]{2}}\;,
\end{equation}
where the respective CGs are given in \Appref{app:Sigma72}. Naively, 
one would expect that the two ``composite'' doublets $\rep[_1]{2}$ and
\rep[_2]{2} should transform in the same way under CP as the ``elementary'' doublet \rep{2}. But this is not the case.
While the transformation of the elementary \rep{2} is given in \eqref{eq:S72CP}, the composite doublets transform 
under the action of the complex conjugation automorphism as
\begin{equation}
\rep[_1]{2}~\mapsto~\UU{\rep{2}}\,\rep{2}^*_2 \quad\text{and}\quad \rep[_2]{2}~\mapsto~-\UU{\rep{2}}\,\rep{2}^*_1\;.
\end{equation}
That is, the two composite doublets are permuted under the action of the
complex conjugation automorphism. Furthermore, the composite doublets transform trivially under the additional \Z2 symmetry, 
in contrast to the elementary \rep{2} which picks up a sign. 

\subsection{CP violation for type~II~B groups}

As for groups of the type~II~A, the mere existence of a consistent CP transformation is, of course, not enough to warrant CP conservation.
In the following, the non--trivial consequences of requiring CP conservation in models with type~II~B groups shall be
investigated and contrasted to the type~II~A case. 

Consider, for example the $\Sigma(72)$ invariant Lagrangian
\begin{equation}\label{eq:S72Example}
  \mathscr{L} ~\supset~  c_1\,\left(\rep{2} \otimes \left(\rep{8} \otimes \rep{8}\right)_{\rep[_1]{2}} \right)_{\rep[_0]{1}}~+~
	c_2\,\left(\rep{2} \otimes \left(\rep{8} \otimes \rep{8}\right)_{\rep[_2]{2}} \right)_{\rep[_0]{1}}~+~\mathrm{h.c.}\;.
\end{equation}
Imposing CP here requires non--trivial relations amongst the previously unrelated couplings $c_1$ and $c_2$, i.e.\
operators are not necessarily mapped onto their own Hermitian conjugate but may be non--trivially permuted.
In fact, the relation on the couplings in \eqref{eq:S72Example} is such that all terms must identically vanish in order for CP to be conserved.
This can also be seen directly from the fact that the terms in \eqref{eq:S72Example} are odd under the additionally appearing \Z2 symmetry.

This shows the crucial difference between type~II~A and type~II~B groups.
For groups of the type~II~A one can always, that is independently of a specific model, find a CP basis in which field operators, and therefore couplings, are mapped
to their own complex conjugate. In contrast, for a generic model based on a type~II~B group such a basis does not exist. 
Therefore, CP conservation generally enforces non--trivial relations among otherwise unrelated couplings. 
This also implies that operators which are charged under the additionally appearing 
linear symmetry (generated by $V$) are forced to vanish if CP is to be conserved. 
Note that this does not show that the conservation of the additional linear symmetry is sufficient 
for CP conservation, and this is generally not expected to be the case.

Altogether, the origin of CPV in type~II~B groups $G$ can be different than in the 
continuous case. CP violation can be tied to certain operators which are permuted 
under the action of the CP transformation, and the conservation of an additional linear symmetry beyond $G$ is a necessary condition
for CP conservation. For all operators which are uncharged under the additional symmetry and mapped to their own Hermitian conjugate
the CP transformation acts as in the type~II~A or continuous symmetry case, that is, CP conservation restricts the phases of the corresponding couplings.
Therefore, also for type~II~B groups it can only be decided experimentally whether or not CP is violated, and if so, by what magnitude.

Deeply related to the peculiar effects which arise in type~II~B groups is the fact that some composite states transform differently under the CP transformation 
than elementary states with the same $G$ representation. This shall be discussed more generally in the following.

\subsection{Transformation of mesons and constituents}
\label{sec:Mesons}

Consider again two generic fields $x$ and $y$ transforming in irreps $\rep{r}(x)=\rep[_x]{r}$ and  $\rep{r}(y)=\rep[_y]{r}$ of $G$.
In contrast to \eqref{eq:Singletcontraction} above, take the contraction of $x$ and $y$ to an unspecified irrep $\rep[_z]{r}$.
In what follows, $\left(x\otimes y\right)_{\rep[_z]{r}}$ will be referred to as a ``meson'' and $x$
and $y$ are called the ``constituents''. This contraction can be written as
\begin{equation}\label{eq:contraction}
 \left[\left(\myvec{x}\otimes\myvec{y}\right)_{\rep[_z]{r}}\right]_\mu 
 ~=~ 
 C_{\mu,\alpha\beta}\,x_\alpha\,y_\beta~=~x^T\,C_{\mu}\,y\;,
\end{equation}
where $\alpha$ and $\beta$ denote the vector indices of $\myvec{x}$ and
$\myvec{y}$, and $C_{\mu, \alpha\beta}$ are the CGs for
the $\mu^\mathrm{th}$ component of the resulting representation vector of $z$. 
In the last step again a vector--matrix notation has been used.

The complex conjugation automorphism acts on representations as
\begin{equation}\label{eq:constituent_trafo}
 \rep[_i]{r}~\mapsto~ 
 \UU[_i]{\rep{r}}\,\rep[_i]{r}^*\;.
\end{equation}
As a result, the transformation of the meson can be derived from the transformation of its constituents,
\begin{equation}\label{eq:meson_trafo}
 \left[\left(x\otimes y\right)_{\rep[_z]{r}}\right]_\mu
 ~=~
 x^T\,C_{\mu}\,y
 ~\mapsto~ 
 x^\dagger\,\UU[_x]{r}^T\,C_{\mu}\,\UU[_y]{r}\,y^*\;.
\end{equation}
However, the representation $\rep[_z]{r}$ itself also transforms under the automorphism according to \eqref{eq:constituent_trafo},
such that one may consider the transformation of the meson as 
\begin{equation}\label{eq:bare_rep_trafo}
 \left[\left(x\otimes y\right)_{\rep[_z]{r}}\right]_\mu~\mapsto~ 
 (\UU[_z]{\rep{r}})_{\mu\nu}\,\left[\left(x\otimes y\right)^*_{\rep[_z]{r}}\right]_{\nu}~=~(\UU[_z]{\rep{r}})_{\mu\nu}\,\left[x^\dagger\,{C_{\nu}}^*\,y^*\right]\;,
\end{equation}
where in the last step the conjugate of \eqref{eq:contraction} has been used.

From the comparison of \eqref{eq:meson_trafo} and \eqref{eq:bare_rep_trafo}
one finds that a meson transforms in consistency with its constituents if and only if
\begin{equation}\label{eq:meson_consistency_condition}
 \UU[_x]{r}^T\,C_{\mu}\,\UU[_y]{r}~=~(\UU[_z]{r})_{\mu\nu}\,{C_{\nu}}^*\;.
\end{equation}
In general, this condition does not have to be fulfilled even if the
matrices $\UU[_x]{r}$, $\UU[_y]{r}$, and $\UU[_z]{r}$ are representations of a
class--inverting automorphism and solve the corresponding consistency condition \eqref{eq:CPConsistencyEquation}.
For example, in \Secref{sec:Sigma72Example} the contractions $\left(\rep{8}\otimes\rep{8}\right)_{\rep[_i]{2}}$ of $\Sigma(72)$ have been discussed which do 
not transform like an elementary $\rep{2}$ under the automorphism, i.e.\ they do not obey equation \eqref{eq:meson_consistency_condition}. 

The existence of an automorphism for which matrices which solve
\eqref{eq:CPConsistencyEquation} also satisfy \eqref{eq:meson_consistency_condition} is a
non--trivial property of a group.\footnote{A group is, up to isomorphism, defined by its CGs \cite{Feit:1967}.}
In the following it shall be investigated under what conditions \eqref{eq:meson_consistency_condition} can be solved.
For simplicity, the treatment here is restricted to class--inverting and involutory automorphisms, 
remarking that similar considerations of general automorphisms would certainly be a worthwhile pastime.

To start, it should be noted that there are many redundancies which can obscure possible solutions 
of \eqref{eq:meson_consistency_condition} in any basis. For example, there are arbitrary (unphysical)
global phase choices possible in the definition of 
\begin{itemize}
 \item the CGs (one global phase for each $\left(x\otimes y\right)_{\rep[_z]{r}}$);
 \item each of the explicit transformation matrices \UU[_x]{\rep{r}}, \UU[_y]{\rep{r}} and
  \UU[_z]{\rep{r}}.
\end{itemize}
Therefore, a specific basis choice will be made to analyze whether or not \eqref{eq:meson_consistency_condition} can be solved.
In particular, the standard form of generalized CP transformations by Grimus and Ecker \cite{Ecker:1987qp}, as already discussed in \Secref{sec:GeneralizedCP}, will be used.
Due to the restriction to class--inverting and involutory automorphisms the matrices
\UU[_i]{\rep{r}} are all either symmetric or anti--symmetric.
Therefore, all of the unitary transformation matrices can be written in the form
\begin{equation}
 U~=~W\,\Sigma\,W^T\;,
\end{equation}
with unitary $W$ and
\renewcommand{\arraystretch}{1.}
\begin{equation}\label{eq:Sigma}
 \Sigma ~=~\left\{\begin{array}{ll}
 \Sigma_+~=~\mathbbm{1}\;, & \text{if $U$ is symmetric,}\\
\Sigma_-~=~\left(\begin{array}{ccccc}
   & 1 & & & \\
	 -1 & & & & \\
   & & \ddots & & \\
   & & & & 1 \\ 
	 & & & -1 &
 \end{array}\right)
 \;, & \text{if $U$ is anti--symmetric.}
 \end{array}\right.
\end{equation}\renewcommand{\arraystretch}{1.}%
The matrices $W$ then can easily be absorbed by a unitary basis change
\begin{equation}\label{eq:CP_basis}
  \rep[_i]{r}~\to~ \WW[_i]{r}^\dagger\,\rep[_i]{r}
  \;,\quad 
  \rhoR{_i}(g) ~\to~ \WW[_i]{\rep{r}}^\dagger\,\rhoR{_i}(g)\,\WW[_i]{\rep{r}} 
  \quad \forall~g \in \DiscreteGroup\;.
\end{equation}
In the new basis \Eqref{eq:meson_consistency_condition} takes the simple form
\begin{equation}\label{eq:consistency_condition_prime_basis}
 \SIGMA[_x]{r}^T\,C'_{\mu}\,\SIGMA[_y]{r}
 ~=~
 (\SIGMA[_z]{r})_{\mu\nu}\,{(C'_{\nu})}^*\;,
\end{equation}
where $C'_\mu$ denotes the basis transformed CGs.

For type~II~A groups, where one has class--inverting involutory automorphisms with symmetric matrices (BDAs), 
all $\SIGMA[_i]{r}$ are equal to the identity matrix and the chosen basis is the CP basis
in which all CGs are real \cite{Bickerstaff:1985jc}.
Therefore, \eqref{eq:consistency_condition_prime_basis} is trivially fulfilled.
This statement then, of course, holds for all other bases as well. 
Therefore, for type~II~A groups, mesons always transform in consistency with their constituents under the BDA.

In contrast, for type~II~B groups it strictly depends on the symmetry properties of $\SIGMA[_x]{r}$, $\SIGMA[_y]{r}$, and
$\SIGMA[_z]{r}$ whether \eqref{eq:consistency_condition_prime_basis} can be solved. 
If both $\SIGMA[_x]{r}$ and $\SIGMA[_y]{r}$ are symmetric (S) or anti--symmetric (AS), $\SIGMA[_z]{r}$ has to be S as well,
while for the mixed case $\SIGMA[_z]{r}$ has to be AS in order for \eqref{eq:consistency_condition_prime_basis} to be fulfilled.
Whenever the group structure gives rise to other contractions, the corresponding mesons do not transform in consistency
with fields in the analogous elementary representation.
Put another way, this means that even though the meson transforms as the elementary representation $\rep[_z]{r}$ under the action of the group $G$,
the meson transforms differently than the elementary representation $\rep[_z]{r}$ under the action of automorphisms of $G$. 
The group $\Sigma(72)$ provides an explicit example for such representations. Namely, in the $\mathrm{S}\times\mathrm{S}$ contraction $\rep{8}\otimes\rep{8}$, 
the representation $\rep{2}$ appears twice. Since $\rep{2}$, however, transforms AS under the contraction it is impossible 
that $\left(\rep{8}\otimes\rep{8}\right)_{\rep[_i]{2}}$ transforms in the same way as the elementary $\rep{2}$ of $\Sigma(72)$. Indeed,
as pointed out in \Secref{sec:Sigma72Example}, the action of the automorphism is such that it permutes the representations $\rep[_1]{2}$ and $\rep[_2]{2}$, 
which is clearly distinct 
from the transformation behavior of the elementary $\rep{2}$.
 
The fact that there are representations which transform equally under $G$ but differently under automorphisms can be used to classify (composite) representations of
$G$ according to their transformation behavior under the automorphism. 
This is particularly relevant for composite trivial singlets, i.e.\ contractions which would appear in a $G$ invariant Lagrangian. 
This fact has so far went unnoticed but it could have far reaching, mostly unexplored, 
implications some of which will be touched in the following sections and chapters, and some of which will be commented on at the end of this work. 

To conclude the discussion of type~II groups, it is remarked once more that the relation 
between real CGs and the possibility of 
having a proper physical CP transformation is not one--to--one. That is,
there are groups (type~II~B) which do not allow for a basis with real CGs even though 
they allow for class--inverting and involutory automorphisms which can be used to define CP.
Typically, however, groups without real CGs have to be extended by additional symmetries upon requiring CP to be a symmetry.
It has not been investigated whether these additional symmetries prohibit all potentially complex coupling coefficients
and this certainly would be an interesting task to do. Also, it has been discussed that type~II~B groups generally feature 
composite states which transform differently under CP than elementary states in the same $G$ representation.
In the next section, finally, groups will be discussed which generally are inconsistent with physical CP transformations.
These groups never allow for a basis with real CGs.

\section{Type I groups: CP violation from a symmetry principle}

\subsection{Explicit example: \texorpdfstring{$\Delta(27)$}{}}
\label{sec:Delta27}

In the following sections it will be demonstrated that groups of the type~I generically give
rise to settings in which CP is violated by calculable complex phases originating 
from the CGs of the group. 

An example for a type~I group is $\Delta(27)$ which can be presented as
\begin{equation}\label{eq:Delta27Presentation}
\Delta(27)~=~\Braket{\elm{A,B}~|~\mathsf{A}^3\,=\,\mathsf{B}^3\,=\,\left(\elm A\,\elm B\right)^3=\mathsf{e}~}\;.
\end{equation}
The group has eleven irreps out of which nine are one--dimensional $(\rep[_0]{1},\rep[_{1-8}]{1})$ 
and the remaining two form a pair of complex conjugate triples $(\rep{3},\bar{\rep{3}})$.
Further details of the group can be found in \Appref{app:Delta27}, and also \cite{Chen:2014tpa}.
The outer automorphism group of $\Delta(27)$ is $\mathrm{GL}(2,3)\equiv\mathrm{SG}(48,29)$,
which is of order $48$, and therefore, bigger than the group itself. Nevertheless, there is no automorphism which 
simultaneously maps each representation of $\Delta(27)$ to its respective complex conjugate representation. 
This can easily be checked by computing the twisted \FSI for all automorphisms.
Nevertheless, there are automorphisms which map a subset of irreps to their complex conjugate representations, thereby 
allowing for model dependent physical CP transformations in so--called non--generic settings (cf.\ the discussion in \Secref{sec:classification}).
To provoke explicit geometrical CP violation it is, thus, crucial to include a sufficient amount of
irreps in a model such that no complex conjugation automorphism is possible.

\subsection{CP violation in a toy model based on \texorpdfstring{$\Delta(27)$}{}}
\label{sec:Delta27decay}

Let us consider a toy model based on the symmetry group $\Delta(27)$. 
The model contains three complex scalars $X$, $Y$ and $Z$ transforming as $\rep[_1]{1}$,
$\rep[_3]{1}$ and $\rep[_8]{1}$, as well as two fermion triplets $\Psi$ and
$\Sigma$, each transforming as $\rep{3}$ under $\Delta(27)$. 
Furthermore, in order to distinguish $\Psi$ and
$\Sigma$, a \U1 symmetry is introduced under which $Y$ is neutral, $\Psi$ has charge $q_\Psi$, $\Sigma$
has charge $q_\Sigma$, and $X$ and $Z$ both have charge
$q_X=q_Z=q_\Psi-q_\Sigma\neq0$. The renormalizable interaction Lagrangian is given by\footnote{%
A possible cubic coupling $Y^3$ is not displayed because it is irrelevant for this discussion.}
\begin{align}
 \mathscr{L}_\mathrm{toy}
 ~=~&
 g_X\,\left[
        X_{\rep[_1]{1}}\otimes \left(\overline{\Psi}\,\otimes\,\Sigma\right)_{\rep[_2]{1}}
 \right]_{\rep[_0]{1}}
 +g_Z\,\left[
        Z_{\rep[_8]{1}}\otimes \left(\overline{\Psi}\,\otimes\,\Sigma\right)_{\rep[_4]{1}}
 \right]_{\rep[_0]{1}}
 \nonumber\\
 &
 +h_\Psi\,\left[
        Y_{\rep[_3]{1}}\otimes \left(\overline{\Psi}\,\otimes\,\Psi\right)_{\rep[_6]{1}}
 \right]_{\rep[_0]{1}}
 +h_\Sigma\,\left[
        Y_{\rep[_3]{1}}\otimes \left(\overline{\Sigma}\,\otimes\,\Sigma\right)_{\rep[_6]{1}}
 \right]_{\rep[_0]{1}}
 +\text{h.c.}\;.
\end{align} 
In components the Lagrangian can be written as
\begin{align}\label{eq:Ltoy}
\mathscr{L}_\mathrm{toy}
 ~=~&
 G^{ij}_X\,X\,\overline{\Psi}_i\Sigma_j
 +G^{ij}_Z\,Z\,\overline{\Psi}_i\Sigma_j
 +H_\Psi^{ij}\,Y\,\overline{\Psi}_i\Psi_j
 +H_\Sigma^{ij}\,Y\,\overline{\Sigma}_i\Sigma_j
 +\text{h.c.}\;.
\end{align} 
In the basis specified in \Appref{app:Delta27}, where also the relevant CGs are given, 
the Yukawa coupling matrices take the form
\begin{align}
 G_X~&=~g_X\,\left(
\begin{array}{ccc}
 0 & 1 & 0 \\
 0 & 0 & 1 \\
 1 & 0 & 0 \\
\end{array}
\right),\quad
G_Z~=~g_Z\,
\left(
\begin{array}{ccc}
 0 & 0 & \omega \\
 \omega^2 & 0 & 0  \\
 0 & 1 & 0
\end{array}
\right)\;,\quad\text{and} \\
 H_{\Psi/\Sigma} ~&=~ h_{\Psi/\Sigma}\,\left(
\begin{array}{ccc}
 1 & 0 & 0 \\
 0 & \omega^2 & 0 \\
 0 & 0 & \omega \\
\end{array}
\right)\;.
\end{align}
Here $g_X$, $g_Z$, $h_{\Psi}$, and $h_{\Sigma}$ are complex couplings and
$\omega:=\mathrm{e}^{2\pi\,\I/3}$.

\begin{figure}[t]
%\begin{tabular}{ccc}
\centering
\begin{subfigure}[b]{.3\textwidth}
\centering
\includegraphics[width=0.66\textwidth]{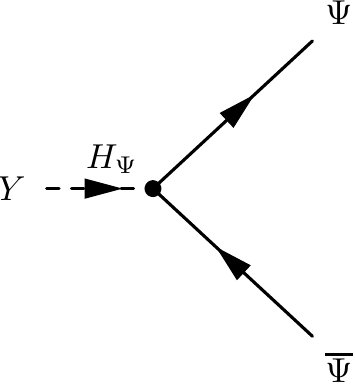}
\caption{}
\label{fig:YdecayTree1}
\end{subfigure}
\begin{subfigure}[b]{.3\textwidth}
\centering
\includegraphics[width=1\textwidth]{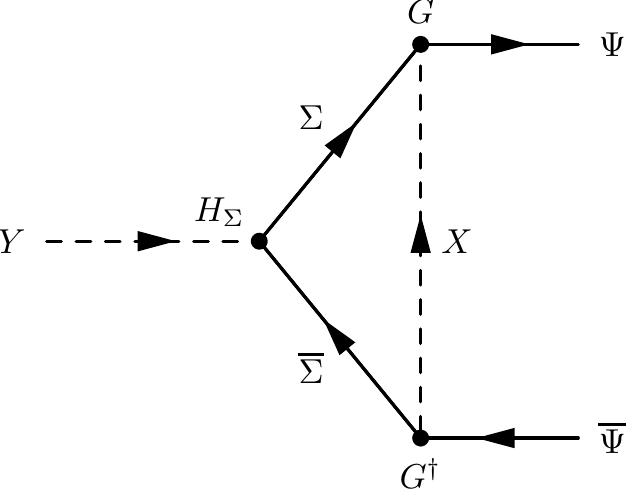}
\caption{}
\label{fig:Ydecay1Loop1}
\end{subfigure}
\begin{subfigure}[b]{.3\textwidth}
\centering
\includegraphics[width=1\textwidth]{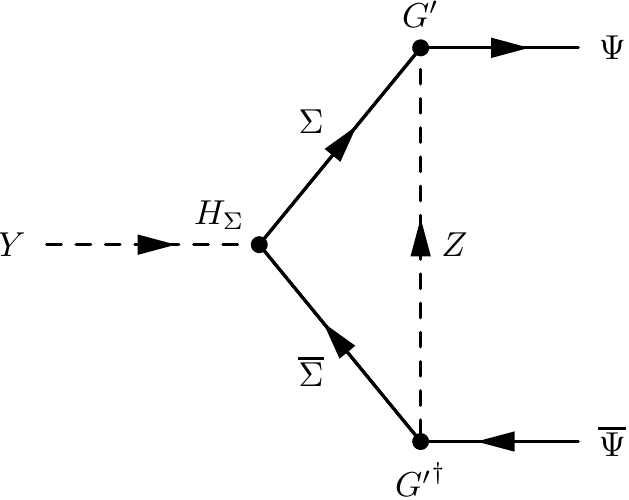}
\caption{}
\label{fig:Ydecay1Loop2}
\end{subfigure}
\caption{Tree level and one--loop diagrams contributing to the decay $Y\to\bar{\Psi}\Psi$.}
\label{fig:DecayAsymmetry}
\end{figure}

There are multiple ways to prove that CP is violated by geometrical phases in this model.
For example, consider the decay $Y\to\overline{\Psi}\Psi$. 
It is no contradiction to the appearance of CPV that neither a \U1 charge asymmetry nor a left--right asymmetry is produced by this decay, 
and distinguishing between $Y$ and $Y^*$ is possible by measuring their respective branching fractions to the final states $\overline{\Psi}\Psi$ and $\overline{\Sigma}\Sigma$.
Interference between
tree--level and one--loop diagrams (\Figref{fig:DecayAsymmetry}) gives rise to a CP asymmetry
\begin{equation}
 \varepsilon_{Y\to\overline{\Psi}\Psi}~:=~
 \frac{\left|\Gamma(Y\to\overline{\Psi}\Psi)\right|^2
 -\left|\Gamma(Y^*\to\overline{\Psi}\Psi)\right|^2 }{\left|\Gamma(Y\to\overline{\Psi}\Psi)\right|^2
 +\left|\Gamma(Y^*\to\overline{\Psi}\Psi)\right|^2}\;,
\end{equation}
which is proportional to
\begin{align}
 \varepsilon_{Y\to\overline{\Psi}\Psi}
 &~\propto~
 \im\left[I_X\right]\,\im\left[\tr\left(G_X^\dagger\,H_\Psi\,G_X\,H_\Sigma^\dagger\right)\right]
 +
 \im\left[I_Z\right]\,\im\left[\tr\left(G_Z^\dagger\,H_\Psi\,G_Z\,H_\Sigma^\dagger\right)\right]
 \nonumber\\
 &
 ~\propto~
 |g_X|^2\,\im\left[I_X\right]\,\im\left[\omega\,h_\Psi\,h_\Sigma^*\right]+
 |g_Z|^2\,\im\left[I_Z\right]\,\im\left[\omega^2\,h_\Psi\,h_\Sigma^*\right]\;.
 \label{eq:DecayAsymmetry}
\end{align}
Here $I_{X}=I(M_X,M_Y)$ and $I_{Z}=I(M_Z,M_Y)$ denote phase space
factors and the loop integral, which are non--trivial functions of the masses of
$X$ and $Y$, and $Z$ and $Y$, respectively. Assuming that the process 
is kinematically allowed, the imaginary parts of $I_{X}$ and $I_{Z}$ provide so--called strong phases (cf.\ the discussion in footnote \ref{fot:one}). 
Being a physical observable, $\varepsilon_{Y\to\overline{\Psi}\Psi}$ is of course invariant under rephasing 
of the fields and the particular basis choice. Indeed, the CP violating weak phases are governed by
\begin{align}
\im\left[\tr\left(G_X^\dagger\,H_\Psi\,G_X\,H_\Sigma^\dagger\right)\right]~&=~3\,|g_X|^2\,\im\left[I_X\right]\,\im\left[\omega\,h_\Psi\,h_\Sigma^*\right]\quad\text{and} \\
\im\left[\,\tr\left(G_Z^\dagger\,H_\Psi\,G_Z\,H_\Sigma^\dagger\right)\right]~&=~3\,|g_Z|^2\,\im\left[I_Z\right]\,\im\left[\omega^2\,h_\Psi\,h_\Sigma^*\right]\;,
\end{align}
which are CP odd basis invariants \cite{Varzielas:2015fxa} completely analogous to the Jarlskog invariant \eqref{eq:J}. 
Therefore, either one of them being non--zero is generally a sufficient condition for CPV. 

The geometrical phase $\omega$ in the Yukawa coupling coefficients originates from the complex CGs of $\Delta(27)$ and
enters the CP odd basis invariants as a CP violating phase. Therefore, this model and similar ones discussed in \cite{Chen:2014tpa}
are the first examples of explicit geometrical CP violation \cite{Chen:2009gf,Branco:2015hea}. Nevertheless, the predictivity of the geometrical phase $\omega$ is limited 
in this toy model by construction. This is because another CP odd phase $\varphi:=\arg(h_\Psi\,h_\Sigma^*)$
is present. In this sense, this model also has a ``plain old'' source of explicit CP violation simply due to the fact that there are not enough rephasing degrees of freedom
to render all couplings real. 

More recently, also models have been constructed whose only source of explicit CP violation is a geometrical phase \cite{Branco:2015hea, Branco:2015gna}.
These models, therefore, are predictive w.r.t.\ the explicitly CP violating weak phase. Since the discussion there is solely based on CP odd basis invariants, 
no CP violating process has been explicitly discussed so far. Simply due to the length of the corresponding invariants, however, it seems very likely that one has 
to go at least to the two--loop level to explicitly identify such a process. In any case, also these models are based on the type~I group $\Delta(27)$ and explicit geometrical CP violation
arises because there is no possible class--inverting automorphism \cite{Chen:2014tpa}.

\subsection{Spontaneous geometrical CP violation with calculable phases}
\label{sec:SSBandCPV}

There is one peculiar spot in the parameter space of the above model which deserves more attention. 
Note that the CP asymmetry of the $Y$ decay vanishes for the special choice of parameters (i) $M_Z=M_X$, (ii) $|g_X|=|g_Z|$, and (iii)
$\varphi=0$. In fact, it is not only for this particular process, but CP will globally be conserved in this model for this choice of parameters.
This can be understood by realizing that it is possible to enhance the ``flavor'' symmetry $\Delta(27)$ by a specific outer automorphism to the bigger group $\text{SG}(54,5)$,
thereby enforcing the specific values of parameters above. However, it is stressed that CP conservation is not imposed in this symmetry enhancement.
Nevertheless, CP is accidentally conserved at the level of the bigger group. This can be understood from the fact that $\text{SG}(54,5)$ is of type~II~A, 
and the fact that at the level of $\text{SG}(54,5)$ there are enough rephasing degrees of freedom to absorb all complex parameters.

The outer automorphism $w$ of \Dts which extends the group to $\text{SG}(54,5)$ (cf.\ \Appref{app:Delta27} for details) acts as 
\begin{equation}
 X~\stackrel{w}{\longleftrightarrow}~Z\;,\quad Y~\stackrel{w}{\longmapsto}~ Y\;,\quad 
 \Psi~\stackrel{w}{\longmapsto}~ U_{w}\,\Sigma^\mathrm{c}\quad\text{and}\quad
 \Sigma~\stackrel{w}{\longmapsto}~ U_{w}\,\Psi^\mathrm{c}\;,
\end{equation}
with $U_{w}$ stated in \Eqref{eq:Uw}.\footnote{%
The standard definition $\Psi^\mathrm{c}:=\mathcal{C}\,\bar{\Psi}^\mathrm{T}=\mathcal{C}\,\mathcal{\beta}^\mathrm{T}\,\Psi^*$ is used here.
}
This transformation is consistent with
the \U1 symmetry (for the choice $q_\Sigma=-q_\Psi$) and naturally ensures
relations (i)--(iii), thereby also granting the absence of CPV.

At the level of $\text{SG}(54,5)$, the previously separate fields $X$ and $Z$ are combined to a doublet, and $\Psi$ and
$\Sigma^\mathrm{c}$ are combined to a hexaplet. $Y$ still transforms in a non--trivial
one--dimensional representation. There are then enough field rephasings possible to
render all coupling phases unphysical, showing that the class--inverting involutory
automorphism of $\text{SG}(54,5)$ is an automatic CP symmetry of the setting.
Since relations (i)--(iii) are fulfilled due to an additional symmetry, they are 
also stable under renormalization group running \cite{Chen:2014tpa}. 

The stage is now set to demonstrate the spontaneous breaking of CP by calculable phases.
Introducing a \U1 neutral scalar field $\phi$ 
in the real non--trivial one--dimensional representation of
$\text{SG}(54,5)$, the symmetry is spontaneously broken to $\Delta(27)$ by the VEV of $\phi$.
The field $\phi$ couples to the scalars $X$ and $Z$ and gives rise to
a mass splitting after SSB,
\begin{equation}
  \mathscr{L}_\mathrm{toy}^\phi ~\supset~ M^2\,\left(|X|^2 + |Z|^2\right) 
  + \left[\frac{\mu}{\sqrt{2}}\,\langle \phi \rangle \, \left(|X|^2 - |Z|^2\right)
  +\text{h.c.}\right]\;,
\end{equation}
where $\mu$ denotes a parameter of mass dimension $1$. At the renormalizable level, 
$\phi$ does not change the Yukawa couplings of $X$, $Y$, and $Z$.
Furthermore, the VEV of $\phi$ will generally also split the
fermion masses, thereby making them distinguishable. Relations (ii) and (iii), i.e.\ the equalities
$|g_X|=|g_Z|$ and $h_{\Psi}=h_{\Sigma}$, are still valid after SSB while relation (i) 
is destroyed by the VEV--induced mass splitting of $M_X$ and $M_Z$.
Consequently, at the level of the residual \Dts symmetry there appears again the CPV decay asymmetry 
\begin{equation}
 \varepsilon_{Y\to\overline{\Psi}\Psi}~\propto~
 |g_X|^2\,|h_\Psi|^2\,\im\left[\omega\right]\,\left(\im\left[I_X\right]-\im\left[I_Z\right]\right)\;.
 \label{eq:XZDecayAsymmetry}
\end{equation}
In contrast to \eqref{eq:DecayAsymmetry}, however, the CP violating weak phase here is independent of the couplings and
fixed to a geometrical value, i.e.\ it is calculable.

This toy model exemplifies a simple recipe for the construction of models 
with predictable spontaneous CPV. Starting from a type II group
$\DiscreteGroup_\mathrm{II}$, which contains (and can be spontaneously broken
down to) a type I group $\DiscreteGroup_\mathrm{I}$, CP conservation is required.
By the spontaneous breaking $\DiscreteGroup_\mathrm{II}\to\DiscreteGroup_\mathrm{I}$, also CP will, at least
generically, be broken. The above example demonstrates that CPV phases then can be predicted.
Whether or not this is a general feature of such settings is unclear. 
Also note the peculiar nature of spontaneous CPV here:
The VEV by itself does not break CP directly but only gives rise to a mass splitting which in turn destroys the previously possible CP transformation.
Altogether CPV can again be related to the fact that there is generally no CP transformation possible at the level of \Dts. 
The SSB construction, here, merely served as a way to isolate the geometrical weak phase. 
To date, this toy example is the only known model where CPV arises in this way and it would certainly
be interesting to learn more about this particular type of spontaneous geometrical CPV. 

Note that due to their very similar CP transformation properties it would also be possible to start this discussion 
with a continuous group such as \SU{n} or \SO{n} instead of a discrete type~II group. 
The general breaking of continuous to discrete groups has been discussed in \cite{Adulpravitchai:2009kd, Luhn:2011ip, Merle:2011vy, Merle:2012xr}
and more specifically related to this situation in \cite{Fallbacher:2015pga}.
Most notably, for all investigated cases (including the breaking of \SU3 to \Dts or \Dff) it has been found that 
the branching of representations from the continuous to the discrete group is such that only highly non--generic settings arise 
in which CP is accidentally conserved \cite{Fallbacher:2015pga}. 
Nevertheless, so far there has been no general argument or even a no--go theorem presented which would show that 
this mechanism of spontaneous CPV is excluded in the breaking of continuous to finite groups and
it would certainly be interesting and worthwhile to explore this further.

Somewhat different models in which CP is violated spontaneously and geometrically directly by the VEV 
have been known for a long time \cite{Branco:1983tn} and will be discussed in detail in chapter \ref{sec:OtherOuts}.
For these models as well, geometrical CPV ultimately originates from the complex CGs of a type~I group.

\subsection{Action of other automorphisms, CP--like symmetries}
\label{sec:CPlike}
 
Having an explicit example at hand, it shall also be explicitly demonstrated that not every outer automorphism 
is suitable to define a physical CP transformation.
Consider for example the outer automorphism $c$ which is in every detail discussed in \ref{app:Delta27}.

The only way in which $c$ can act consistently with the symmetries of the \Dts example model is
\begin{equation}\label{eq:Delta27additionalTrafo}
 X~\stackrel{c}{\longmapsto}~X^*\,,\quad Z~\stackrel{c}{\longmapsto}~Z^*\,,\quad Y~\stackrel{c}{\longmapsto}~Y^*\,,\quad 
 \Psi~\stackrel{c}{\longmapsto}~ C\,\Sigma\,,\quad\text{and}\quad
 \Sigma~\stackrel{c}{\longmapsto}~ C\,\Psi\;,
\end{equation}
where $C$ is given in \eqref{eq:delta54gens}, and the fermion \U1 charges are fixed as $q_\Sigma=-q_\Psi$.
Clearly, this transformation acts like a charge conjugation on the \U1 because 
it maps fields with opposite \U1 charges onto each
other. Therefore, imposing $c$ cannot be viewed as an
enhancement of the flavor symmetry. Nonetheless, $c$ is not a physical CP symmetry
either, due to the fact that not all representations of $\Delta(27)$ are mapped
to their complex conjugate representations. In particular, $c$ maps $\rep{3}\mapsto\rep{3}$.

Requiring the transformation $c$ to be conserved, hence, does not entail physical CP conservation.
This can be proved explicitly by noting that none of the relations (i)--(iii) is
fulfilled due to $c$, implying that the physical CP asymmetry of the $Y$ decay, 
$\varepsilon_{Y\to\overline{\Psi}\Psi}$ is still non--vanishing.
Instead, imposing \eqref{eq:Delta27additionalTrafo} enforces equality between
the decay rates of $Y\to\overline{\Psi}\Psi$ and $Y^*\to\overline{\Sigma}\Sigma$.
This type of exotic transformation has been termed ``CP--like symmetry'' \cite{Chen:2014tpa},
as it acts in some but not all ways very similarly to a physical CP transformation.
This discussion explicitly shows that not every outer automorphism can serve as 
a physical CP transformation.

To conclude the discussion of the example model, 
it has been shown that CPV with calculable phases in type~I groups, in particular $\Delta(27)$, 
exists solely due to the properties of the symmetry group.
That is, the absence of class--inverting automorphisms in type~I groups signals that 
CP violating complex phases originate from the CGs of the group.
It has been demonstrated how these phases can give rise to so--called explicit geometrical CP violation.
Furthermore, it has also been shown that it is possible to have settings in which
a CP conserving group gets spontaneously broken down to a type~I group, for which CP then is violated by calculable phases.
Thus, for both cases -- explicit or spontaneous CP violation -- it is possible to predict CP violating phases from group theory.

\section{Towards realistic models with discrete flavor symmetries and drawbacks}

Discrete flavor symmetries are a well motivated way to address the flavor puzzle (cf.\ e.g.\ \cite{Altarelli:2010gt, Grimus:2011fk, King:2013eh} for reviews).
Besides reducing the number of parameters and thereby increasing the general predictivity, 
non--Abelian discrete groups are able to predict certain geometrical values for mixing angles \cite{Altarelli:2005yx,Altarelli:2005yp}.
Furthermore, in contrast to models with continuous symmetries, spontaneously broken discrete groups do not suffer from the 
appearance of massless Goldstone modes.
Attractive scenarios for the origin of non--Abelian discrete (flavor) symmetries include string theory \cite{Kobayashi:2006wq,Nilles:2012cy,Fischer:2012qj,Fischer:2013qza,Berasaluce-Gonzalez:2013bba} 
or the breaking of non--Abelian continuous gauge symmetries \cite{Adulpravitchai:2009kd, Luhn:2011ip, Merle:2011vy, Merle:2012xr,Fallbacher:2015pga}.
Adding to this list of attractive features, it has been shown in this work that discrete groups can predict geometrical values for complex phases that
violate CP explicitly and/or spontaneously.

Nevertheless, there are some drawbacks for models based on discrete groups. 
Namely, to explain the observed structure of mixing and masses, groups typically have to be broken 
completely in the quark sector \cite{Leurer:1992wg, Felipe:2014zka} and possibly to some very limited partial symmetries
in the charged lepton an neutrino sectors \cite{Lam:2008sh, Lam:2008rs, Hernandez:2012ra,Hernandez:2012sk}.
Such a breaking of discrete groups always gives rise to domain walls \cite{Zeldovich:1974uw} which are not observed 
and have to be argued away \cite{Preskill:1991kd} (cf.\ \cite{Kolb:1990vq}). Furthermore, it is technically difficult to achieve the
complete spontaneous breaking of a group in the first place, as global minima are typically located at symmetry enhanced points. 

A complete and elegant solution to the flavor puzzle based on discrete symmetries is
presently not available. Nevertheless, there are attractive candidates for an explanation of the lepton sector flavor structure 
based on the residual symmetry approach. The corresponding highly predictive models typically depend only on a single internal parameter
\cite{Feruglio:2012cw, Feruglio:2013hia, Hagedorn:2014wha, Ding:2014ora, Li:2015jxa, Ballett:2015wia, DiIura:2015kfa}.

Regarding the improved knowledge on outer automorphisms and CP transformations it also seems to be worthwhile to
revisit the strong CP problem. However, based on the most straightforward approaches, so far, only solutions have been found
which could be tracked back to variants of either left--right symmetric models (cf.\ e.g.\ \cite{Mohapatra:1997su}) or 
Nelson--Barr type constructions \cite{Nelson:1983zb,Barr:1984qx}. 
A detailed account of the approaches taken has already been given in \cite{Fallus:2015} and will not be repeated here.

\chapter{Outer automorphisms beyond CP}
\label{sec:OtherOuts}

Discussing (outer) automorphisms the focus so far has been on CP transformations, i.e.\ on complex conjugation automorphisms.
From the discussions in \Secref{sec:OutIntro}, however, 
it is clear that not all outer automorphisms are CP transformations. That is, CP transformations are only a subset of all possible outer automorphisms 
and there are automorphisms which are unrelated to CP. Examples demonstrating this have already appeared in \Secref{sec:OutIntro}, and in the context of a specific
toy model in \ref{sec:SSBandCPV} and \ref{sec:CPlike}.

In this section, the implications of outer automorphisms shall be studied in a wider context, without the restriction to complex conjugation transformations.
At the beginning, two logical possibilities are identified for the possible action of outer automorphisms on a given physical model.
The first possibility allows for a very general definition of what ``maximal'' breaking of a given transformation means (including C and P in the SM).
The second possibility for the action of outer automorphisms is more subtle and will subsequently be investigated in more detail.
The implications of possible outer automorphisms on the parameter space of a model, as well as for the multiplicity and 
appearance of stationary points of potentials, will be discussed in a general manner. 
As an example, a three Higgs doublet model (3HDM) with $\Delta(54)$ symmetry will be presented which features a rich structure of outer automorphisms.
It will be shown how the knowledge of outer automorphisms allows to identify physically redundant regions in the parameter space of this model.
Furthermore, it is demonstrated how the VEVs of the according, generally involved, Higgs potential can be calculated by solving only a homogeneous linear equation.
Finally, it is explained how the large set of outer automorphisms in this setting is related to spontaneous geometrical CP violation.

\section{Possible action of outer automorphism transformations}
\label{sec:ImplicationsOfOuts}

Form the discussion in \Secref{sec:OutIntro} it follows that outer automorphisms generally act as a permutation of representations of the same dimensionality.
That is, outer automorphisms map symmetry representations to other other representations, $\rep[_i]{r}\mapsto\rep[_j]{r}$.
Therefore, there are two logical possibilities for the action of an outer automorphism in a given model which contains $\rep[_i]{r}$:
\begin{itemize}
\item[(i)]  \rep[_j]{r} is not part of the model.
\item[(ii)]  Both, \rep[_i]{r} and \rep[_j]{r}, are included in the model. 
\end{itemize}
In the first case, operators are mapped to other operators which are not present in the model.
As a result, this type of transformations maps a model to an inherently different model. 
The corresponding transformation can never be a symmetry transformation because it is broken by the representation content of the model. 
This type of breaking is called ``maximal''.

In the second case, present operators are generally mapped to other, equally present operators.
Hence, such a transformation can also be described as acting in the space of couplings.
The corresponding transformation, as outer automorphism, is not a symmetry transformation by construction.
Consequently, this type of outer automorphism looks like a possible symmetry transformation which is, however, explicitly broken by the values of some couplings. 

There are two very well--known examples for each case:
\begin{itemize}
\item[(i)] The P transformation in the SM: $\rep[_{\mathrm{L}}]{r}\mapsto\rep[_{\mathrm{R}}]{r}$ for all chiral spinors.
\end{itemize}
This transformation maps all chiral Weyl fermions to chiral Weyl fermions of the opposite chirality without affecting the 
gauge group representation. For example, $Q_\mathrm{L}$ in $\left(\rep{3},\rep{2}\right)^{\mathrm{L}}_{1/6}$ would be mapped to
``$Q_\mathrm{R}$'' transforming as $\left(\rep{3},\rep{2}\right)^{\mathrm{R}}_{1/6}$. Such a field $Q_\mathrm{R}$, however, is not part of the model.
Therefore, parity is broken explicitly and maximally by the choice of representations of the model.
By the variety of outer automorphisms and representations, for example in discrete groups, it is straightforward to construct more examples for maximally broken transformations,
meaning that C and P are not special in this sense.

An example for the case (ii) is:
\begin{itemize}
\item[(ii)] The CP transformation in the SM: $\rep{r}\mapsto\rep{r}^*$ for all representations. 
\end{itemize}
This transformation maps all operators to their respective Hermitian conjugate operators.
Alternatively, this can be described as mapping all couplings to their respective complex conjugate couplings
and, therefore, $V_\mathrm{CKM}\mapsto \left(V_\mathrm{CKM}\right)^*$. 
Only experimentally, it is known that CP is not a symmetry because $\delta_\mathrm{CKM}\neq0,\pi$. 
Altogether, the SM CP transformation is possible in principle but broken explicitly by the values of couplings.

By the variety of outer automorphisms it is clear that also for case (ii) one can find many more examples in other models.
Nevertheless, the complex conjugation automorphism acting as CP transformation is somewhat special. 
This is because if a complex representation \rep{r} is present, then the presence of $\rep{r}^*$ cannot be avoided by the hermiticity of the Lagrangian.
Therefore, CP cannot be broken ``maximally'', i.e.\ by the absence of representations.\footnote{%
The way how CP is broken geometrically in models with type I symmetries, is not by the absence of representations but
by the absence of an adequate automorphism that maps $\rep[_i]{r}\mapsto\rep[_i]{r}^*$ (cf.\ \Secref{sec:Delta27}).}
Analogously to CP, also outer automorphisms which map 
$\rep[_i]{r}\mapsto\rep[_i]{r}$, i.e.\ each present representation to itself, cannot be broken maximally.

The possibility (i) of explicit and maximal breaking will not further be discussed here. 
In contrast, the focus will be on the case (ii) for which some general results shall be derived in the following.

\section{Redundancies in parameter space}
\label{sec:redundant}

It shall be shown how outer automorphism transformations allow to identify physically redundant regions in the parameter space of a model.
This discussion can be led in a very general manner including also gauge and fermion sectors. 
For clarity, however, let us focus on the particularly simple example of a pure scalar potential. 
Consider a model with multiple (Higgs) scalars which are charged under the $\SU2_\mathrm{L}\times\U1_\mathrm{Y}$ symmetry of the SM.\footnote{%
None of the arguments presented here depends on any of the chosen continuous internal symmetries.
}
The multiple copies of the Higgs field $\phi_a$ $(a=1,\dots,n)$ span a new $n$--dimensional ``horizontal'' space.
In the most general form, the renormalizable scalar potential can be written as
\begin{equation}
V\left(\Phi,\lambda\right)~=~Y_{ab}\left(\phi^\dagger_a\,\phi_b\right)~+~\frac12\,Z_{ab,cd}\left(\phi^\dagger_a\,\phi_b\right)\left(\phi^\dagger_c\,\phi_d\right)\;.
\end{equation}
Here it is summed over repeated indices, the contraction in the internal \SU2 space is implicit, $\Phi:=(\phi_1,\dots,\phi_n)$, and $\lambda$ collectively denotes all of the potential parameters.
Hermiticity of the potential requires that
\begin{equation}
Y_{ab}~=~\left(Y_{ba}\right)^*\;,\quad\text{and}\quad Z_{ab,cd}~=~\left(Z_{ba,dc}\right)^*\;.
\end{equation}
Furthermore, due to the internal \SU2 structure,
\begin{equation}\label{eq:Su2Req}
Z_{ab,cd}~=~Z_{cd,ab}\;.
\end{equation}

Consider now a symmetry $G$ acting in the horizontal space with $\Phi$ transforming in a representation \rep[_\Phi]{r}. 
That is, there is a set of explicit representation matrices $\rho_{\rep[_\Phi]{r}}(\elm g)$, $\elm g\in G$,
acting on $\Phi$ in the horizontal space while leaving the potential invariant. Having $G$ as a conserved symmetry, thus, amounts to requiring
\begin{equation}\label{eq:SymmetryReq}
V\left(\rho_{\rep[_\Phi]{r}}(\elm g)\,\Phi,\lambda\right)~=~V\left(\Phi,\lambda\right)\;,\quad\forall\elm g\in G\;.
\end{equation}
These requirements fix the functional form of the potential in the sense that, imposing \eqref{eq:SymmetryReq},
$Y$ and $Z$ have to fulfill the conditions 
\begin{align}\notag
Y_{ab}~&=~\left[\rho^*_{\rep[_\Phi]{r}}(\elm g)\right]_{aa'}\,Y_{a'b'}\,\left[\rho_{\rep[_\Phi]{r}}(\elm g)\right]_{b'b}\;,\quad\forall\elm g\in G\;,\quad\text{and}&\\ 
Z_{ab,cd}~&=~\left[\rho^*_{\rep[_\Phi]{r}}(\elm g)\right]_{aa'}\,\left[\rho^*_{\rep[_\Phi]{r}}(\elm g)\right]_{cc'}\,Z_{a'b',c'd'}\,
	\left[\rho_{\rep[_\Phi]{r}}(\elm g)\right]_{b'b}\,\left[\rho_{\rep[_\Phi]{r}}(\elm g)\right]_{d'd}\;,\quad\forall\elm g\in G\;.&
\label{eq:ParameterReq}
\end{align}
Therefore, some $Y_{ab}$ and $Z_{ab,cd}$ generally will be forced to vanish while
others are related to one another, analogous to \eqref{eq:Su2Req}. 
Illustratively, one should really imagine $Y$ and $Z$ as $n\times n$ and $n\times n\times n\times n$ tensors, respectively, with most of their entries vanishing while certain 
symmetric patterns of entries are non--vanishing and partly interrelated. 
The list of scalar parameters $\lambda$ then contains one entry for each independent non--zero component of $Y$ and $Z$.

Note that it is always possible to perform a $\U{n}$ basis change in the $n$--dimensional horizontal space without any 
physical consequences. That is, physical observables must be independent of the Higgs basis choice and can only depend 
on basis invariant quantities derived from $Y$ and $Z$, cf.\ e.g.\ \cite{Botella:1994cs, Branco:1999fs, Davidson:2005cw, Haber:2006ue}.
Settling to one specific basis, the scalar parameters $\lambda$ generally can take values within some domain, the so--called parameter space of the model. 
Each of the distinct points in the parameter space generally gives rise to different physical predictions of a model. 
Consequently, the parameter space is typically restricted by physical requirements
such as boundedness of the potential, the presence of charge--conserving minima and so on.

Consider now the action of an outer automorphism $u$ which maps the representation $\rep{r}_\Phi$ to itself with an explicit representation
matrix $U$. That is, $U$ fulfills the consistency condition \eqref{eq:consistencyEquation} in the form 
\begin{equation}\label{eq:consistencyEquationHiggs}
U\,\rho_{\rep{r}_\Phi}(\mathsf{g})\,U^{-1}~=~\rho_{\rep{r}_\Phi}(u(\mathsf{g}))\;,\qquad\forall \mathsf{g}\in G\;.
\end{equation}
Acting with the outer automorphism on $\Phi$ in the potential, one finds that it transforms
\begin{equation}
V\left(\Phi,\lambda\right)~\mapsto~V\left(U\,\Phi,\lambda\right)~=~V\left(\Phi,\lambda'\right)\;,
\end{equation}
where $\lambda'$ collectively denotes the transformed parameters
\begin{align}\notag
Y'_{ab}~&=~U^*_{aa'}\,Y_{a'b'}\,U_{b'b}\;,\quad\text{and}& \\
Z'_{ab,cd}~&=~U^*_{aa'}\,U^*_{cc'}\,Z_{a'b',c'd'}\,U_{b'b}\,U_{d'd}\;.&
\label{eq:ParameterTrafo}
\end{align}
Due to the property \eqref{eq:consistencyEquationHiggs} of $U$, it is straightforward to show that\footnote{%
For example: $Y'=U^\dagger Y U\stackrel{\eqref{eq:ParameterReq}}{=}U^\dagger\rho^\dagger Y\rho U=U^\dagger\rho^\dagger UU^\dagger Y UU^\dagger\rho U
\stackrel{\eqref{eq:consistencyEquationHiggs}}{=}\rho'^\dagger Y'\rho'$ has to hold for all $\rho'$, that is, for all $\rho$.
}
\begin{align}\notag
Y'_{ab}~&=~\left[\rho^*_{\rep[_\Phi]{r}}(\elm g)\right]_{aa'}\,Y'_{a'b'}\,\left[\rho_{\rep[_\Phi]{r}}(\elm g)\right]_{b'b}\;,\quad\forall\elm g\in G\;,\quad\text{and}&\\ 
Z'_{ab,cd}~&=~\left[\rho^*_{\rep[_\Phi]{r}}(\elm g)\right]_{aa'}\,\left[\rho^*_{\rep[_\Phi]{r}}(\elm g)\right]_{cc'}\,Z'_{a'b',c'd'}\,
	\left[\rho_{\rep[_\Phi]{r}}(\elm g)\right]_{b'b}\,\left[\rho_{\rep[_\Phi]{r}}(\elm g)\right]_{d'd}\;,\quad\forall\elm g\in G\;.&
\label{eq:ParameterReqPrime}
\end{align}
That is, $Y'$ and $Z'$ have to fulfill \eqref{eq:ParameterReq} in exactly the same way as $Y$ and $Z$.
This can be understood as a consequence of the fact that $u$ 
leaves the set of all symmetry transformations invariant. In other words, $u$ may permute all available symmetry transformations 
but writing the symmetry transformations as a list $\{\rho_{\rep[_\Phi]{r}}\}$, $u$ would not change the content of the list.
This implies that the functional form of the potential, i.e.\ which entries of $Y$ and $Z$ compared to $Y'$ and $Z'$ are non--zero,
is unchanged under the action of the automorphism $u$. 
Nevertheless, the non--zero entries cannot all coincide. That is, $Y'\neq Y$ and/or $Z'\neq Z$ must hold. 
Otherwise $u$ would not be an outer automorphism but an inner automorphism, i.e.\ a symmetry transformation to begin with.

As a result of this discussion one notes that the outer automorphism is equivalent to a mapping in the parameter space of the theory.
That is, by the application of $u$ one moves the theory from a set of scalar parameters $\lambda$ to a different set of scalar parameters $\lambda'$ without changing the functional form of the potential.
This, however, implies that the two a priori physically distinct spots in the parameter space with values $\lambda$ and $\lambda'$ are related by a basis transformation.
Namely, 
\begin{equation}\label{eq:OutTrafo}
V(\Phi,\lambda')~=~V(U\Phi,\lambda)~=~V(\Phi',\lambda)\;,
\end{equation}
describes exactly the same physics for $\Phi':=U\Phi$, as $V(\Phi,\lambda)$ does for $\Phi$. The a priori physically distinct spots in parameter space
$\lambda$ and $\lambda'$, thus, merely differ by a physically meaningless relabeling of fields \cite{Fallbacher:2015rea}.
%In other terms, since a basis transformation cannot matter physically, one concludes that the parameter values $\lambda$ and $\lambda'$ describe equivalent physical scenarios. 

For clarity, this should be contrasted to the behavior of the potential under general Higgs--basis changes, which are sometimes also called 
reparametrization transformations \cite{Botella:1994cs,Ivanov:2006yq,Ivanov:2010ww}. The physical predictions of a theory do, of course, not depend on the specific way the Lagrangian is expressed. 
Therefore, it is always possible to perform a field redefinition, i.e.\ to rewrite the Lagrangian in terms of new fields $\widetilde{\Phi} = U\Phi$ with an arbitrary unitary matrix $U$.
The resulting potential
\begin{equation}
  V(U^{-1}\widetilde{\Phi},\lambda)~=:~\widetilde{V}(\widetilde{\Phi},\lambda)
  \;,
\end{equation}
however, is in general a different function of its arguments than $V$. Consequently it is, in general, 
impossible to pass on the difference in the functional dependence of $\widetilde{V}$ in comparison to $V$ to the scalar couplings $\lambda$.
This is possible if and only if $U$ is the representation matrix of an outer automorphism transformation, in which case the functional form of the potential is unchanged and one has
\begin{equation}
 \widetilde{V}(\widetilde{\Phi},\lambda)~=~V(\widetilde{\Phi},\widetilde{\lambda})\;.
\end{equation}
To sum up, outer automorphism transformations can be used to relate different regions of the
parameter space. For the complete discussion of the physical phenomenology of a model
it is, thus, sufficient to consider only a restricted region of the parameter space from which the complete 
parameter space can be reached by outer automorphisms.

So far, only transformations $u:\rep[_\Phi]{r}\mapsto\rep[_\Phi]{r}$ have been discussed. 
The same argument, however, also holds for any complex conjugation (CP) outer automorphism 
$u:\rep[_\Phi]{r}\mapsto\rep{r}^*_\Phi$. If such a CP transformation is not a symmetry, 
it is actually well--known that it maps the theory to a different 
spot in the parameter space. In the simplest cases this implies a mapping of all parameters to their respective complex conjugate parameters. 
The resulting theory is physically equivalent to its pre--image in the sense that it describes the same dynamics as before but for the CP conjugate set of fields.
Whether one describes the underlying physics with fields or their respective conjugates, however, is completely arbitrary. 

As a final remark, note that this discussion extends to gauge and fermion sectors of a theory in a straightforward way.
Furthermore, if there are otherwise indistinguishable fields in representations \rep{r} and \rep{r'}, then this discussion also
applies to outer automorphisms $u$ which map $\rep{r}$ to $\rep{r'}$.

\section{Outer automorphisms and VEVs}
\label{sec:OutsAndVEVs}

In the previous section it has been established that settings which allow for outer automorphisms have physical degeneracies in the parameter space.
In this section the presence of outer automorphism transformations shall further be used to establish an interesting relation between different stationary 
points of potentials.
For this, assume that the potential $V(\Phi,\lambda$) has a VEV $\Phi_0(\lambda):=\langle \Phi \rangle$ which is, in general, a continuous function of the couplings $\lambda$.
Therefore, for all values of $\lambda$ one has
\begin{equation}\label{eq:VEVexistence}
\left.\nabla_\Phi\,V(\Phi,\lambda)\right|_{\Phi_0(\lambda)}~=~\left.\nabla_{\Phi^*}\,V(\Phi,\lambda)\right|_{\Phi_0(\lambda)}~=~0\;,
\end{equation}
where $\nabla_{\Phi^{(*)}}$ denotes the differentiation with respect to $(\phi_1,\dots,\phi_n)$ or the complex conjugate fields, respectively. Furthermore, 
assume that there is an outer automorphism of the setting mapping $\Phi\mapsto \Phi'=U\Phi$ such that the potential fulfills \eqref{eq:OutTrafo}, i.e.\
$V(\Phi',\lambda)=V(\Phi,\lambda')$. 

The claim is that the potential $V(\Phi,\lambda)$ then has, besides $\Phi_0(\lambda)$, also the stationary point $U\Phi_0(\lambda')$.
This assertion is proven by noting that \cite{Fallbacher:2015rea}%
\begin{equation}
\left.\nabla_\Phi\,V(\Phi,\lambda)\right|_{U\,\Phi_0(\lambda')}~=~\left.U^{-1}\,\nabla_\Phi\,V(U\,\Phi,\lambda)\right|_{\Phi_0(\lambda')}
~=~\left.U^{-1}\,\nabla_\Phi\,V(\Phi,\lambda')\right|_{\Phi_0(\lambda')}~=~0\;,
\end{equation}
and analogously for the derivative w.r.t.\ $\Phi^*$.
Here it has been used that $U$ is invertible, the second equality follows from \eqref{eq:OutTrafo},
and the last equality is a direct consequence of \eqref{eq:VEVexistence}. In case the outer automorphism does not map $\rep[_\Phi]{r}\mapsto\rep[_\Phi]{r}$ but
$\rep[_\Phi]{r}\mapsto\rep{r}^*_\Phi$, i.e.\ $\Phi\mapsto U\Phi^*$, a completely analogous argument holds for the new stationary point $U\Phi^*_0(\lambda')$.

In summary, one concludes that if there is an outer automorphism transformation $u$ acting consistently with the symmetries and representations of a model
then it is possible to obtain new VEVs from a known one $\langle\Phi(\lambda)\rangle$ simply by taking
\begin{equation}\label{eq:additionalVEV}
  \langle\Phi(\lambda)\rangle_\mathrm{new}~=~
	\left\{\begin{array}{ll}
    U\,\langle\Phi\left( \lambda\rightarrow\lambda' \right)\rangle\;,& \text{~if}\quad u~:~\rep[_\Phi]{r}~\mapsto~U\,\rep{r}_\Phi\;,~\text{or}\\[0.2cm]
    U\,\langle\Phi\left( \lambda\rightarrow\lambda' \right)\rangle^*\;, & \text{~if}\quad u~:~\rep[_\Phi]{r}~\mapsto~U\,\rep{r}^*_\Phi\;.
  \end{array}\right.
\end{equation}

This implies that stationary points of potentials always appear in complete multiplets of the available group of outer automorphisms.

Note that there is a very close similarity to the well--known so--called group orbits of VEVs. 
That is, in close analogy to above one can prove that if $\langle\Phi(\lambda)\rangle$ is a VEV of a potential, 
then so is $\rho_{\rep[_\Phi]{r}}(\elm g)\langle\Phi(\lambda)\rangle,~\forall \elm g\in G$.
All stationary points which are obtained by the action of all group elements on a given VEV form a so--called group orbit.
Stationary points, thus, always appear in complete permutation representation multiplets of the symmetry group.

In complete analogy to the symmetry group orbit, an orbit of VEVs is defined by the action of all possible outer automorphism transformations on a given VEV à la \eqref{eq:additionalVEV}. 
Since non--trivial outer automorphism transformations are always distinct from the symmetry transformations this new orbit
is, in fact, `perpendicular' to the group orbit.

There is a close relation of the symmetry group orbit to the pattern of spontaneous symmetry breaking. That is, $G$ is spontaneously broken to a subgroup $H\subset G$, 
if and only if the VEV is invariant under the action of $H$, i.e. $\rho(\elm g)\langle\Phi\rangle=\langle\Phi\rangle~\forall \elm g\in H$, and non--invariant under elements of the coset $G/H$. 
The orbit stabilizer theorem (cf.\ e.g.\ \cite[p.\ 80]{CeccheriniSilberstein2008}) 
then yields the number of distinct but physically equivalent VEVs contained in the group orbit as $|G|/|H|$. 
The VEVs in each orbit are physically equivalent because they are only distinguished by a symmetry action, also implying that they break to isomorphic subgroups.
It is this presence of several isolated but equivalent minima which typically gives rise to domain walls in 
the spontaneous breaking of discrete groups \cite{Zeldovich:1974uw,Kolb:1990vq}.

In complete analogy, there are now also the `perpendicular' orbits of VEVs resulting from the action of the outer automorphism group.
Since physics does not change under the application of a globally available outer automorphism transformation, 
also VEVs belonging to the same outer automorphism orbit deserve to be called physically equivalent. 
If, however, the corresponding outer automorphisms are only available at the level of the potential while being explicitly and maximally broken by other sectors of the theory, 
then the stationary points will no longer be physically equivalent. Nevertheless, in this case one may still use 
the available outer automorphisms of the potential to analyze the stationary points.

Just as for the case of symmetry transformations, also the length of orbits under outer automorphism transformations generally does not exhaust all outer automorphisms. 
That is, just as for $H\subset G$, there is some subgroup $O_\Phi\subset\mathrm{Out}(G)$ of the outer automorphism group which leaves a given VEV invariant. 
This is very interesting because it implies that if a theory allows for outer automorphisms then the vacuum of the theory 
may have additional (emergent) symmetries -- recruited 
from the outer automorphism group. By construction, these symmetries were not realized in the original Lagrangian, and they are, therefore, not exact.
Nevertheless, it is possible to construct scenarios in which the VEV establishes the emergent symmetry in a complete sector of a theory,
for which symmetry breaking effects then only arise at higher order. This point has never been noted before and it is conceivable that there are interesting 
applications of this mechanism of emerging symmetries.

Altogether, in this section it has been shown that stationary points of a potential only appear in multiplets which are permutation 
representations of the available outer automorphism group. That is, with a given VEV one may simply apply \eqref{eq:additionalVEV} to
obtain others. Furthermore, as there generally are fixed points in the action of $\mathrm{Out}(G)$ on a given stationary point, the vacua 
of theories with outer automorphisms can give rise to emergent symmetries. 
In combination, both facts can be employed in order to compute stationary points of potentials in a novel way. 
For instance, in \Secref{sec:VEVcomputation} it will be shown how the VEVs of an example model can be computed by solving only a simple homogeneous linear equation
instead of a complicated system of coupled polynomial equations.

\section{Explicit example: 3HDM with \texorpdfstring{$\Delta(54)$}{} symmetry}
\label{sec:3HDMExample}

\subsection{The model}
\label{Sec:TheModel}

In order to illustrate and substantiate the findings of the last sections, this section provides an explicit example for 
the presence of outer automorphisms beyond CP in quantum field theories. The model under discussion is a 3HDM
with $\Delta(54)$ symmetry.
That is, in a new horizontal space three copies of the SM Higgs field are assumed to transform like a triplet representation of 
the group $\Delta(54)$. In foresight of this discussion, 
all possible outer automorphisms of $\Delta(54)$ have already been derived in \Secref{sec:OutIntro}.

The model under discussion is actually well--known in the literature as the 3HDM with \Dts symmetry introduced by Branco, Gerard, and Grimus \cite{Branco:1983tn}.
It has been extensively studied in the literature 
\cite{deMedeirosVarzielas:2011zw,Ivanov:2012ry,Ivanov:2012fp,Varzielas:2012nn,Varzielas:2012pd,Degee:2012sk,Holthausen:2012dk,Bhattacharyya:2012pi,Varzielas:2013sla,Ivanov:2013nla,Ma:2013xqa, Varzielas:2013eta}, 
because it is the prime example for the occurrence of so--called spontaneous geometrical CP violation. 
The term spontaneous geometrical CP violation refers to spontaneous CPV by relative complex phases of several VEVs, which are independent of the exact values of couplings and, therefore,
calculable directly from the underlying symmetry of the model. Several efforts have been undertaken to improve the understanding of geometrical CP violation in the original model 
\cite{deMedeirosVarzielas:2011zw,Degee:2012sk,Holthausen:2012dk}, but also for potentials of higher order \cite{Varzielas:2012nn} and in multi--Higgs models \cite{Varzielas:2012pd,Ivanov:2013nla}.
Nevertheless, a complete understanding of geometrical CPV and the origin of calculable phases has not been achieved to date.

The approach which is used here closely follows \cite{Fallbacher:2015rea}.
It is different from previous treatments in the literature and strongly motivated by the unavoidably present outer automorphisms of the model.

To simplify the discussion, the internal $\SU2_\mathrm{L}$ structure will not be displayed and $H:=(H_1,H_2,H_3)^\mathrm{T}$ stands for 
a vector of three EW doublets transforming like a triplet \rep{3} under $\Delta(54)$.\footnote{%
Due to the fact that $C$ and $-C$ only differ by a global phase it is completely irrelevant for the whole discussion whether $H$ is assumed to 
transform in \rep[_1]{3} or \rep[_2]{3} of \Dff.
}
The basis choice for the triplet representation
can be found in \Secref{Sec:Delta54}, along with all other necessary group theoretical details. Even if just \Dts is required 
as discrete symmetry, the actual discrete symmetry group of the Higgs potential turns out to be \Dff \cite{Branco:1983tn, deMedeirosVarzielas:2011zw,
Varzielas:2012nn,Ivanov:2012ry,Ivanov:2012fp}. This is because the continuous symmetries and the 
representation content of the model are such that the \Dts potential has an accidental symmetry corresponding to $H\mapsto CH$,  which automatically 
enlarges the discrete symmetry group from \Dts to \Dff (cf.\ \Appref{app:Delta27}).

The scalar potential can be written as
\begin{equation}\label{eq:Invariants}
\begin{split}
 V(H,\vec{a})~=~&-m^2\,H_i^\dagger H_i + a_0\, I_0(H^\dagger,H) +& \\
                &+a_1\,I_1(H^\dagger,H)+ a_2\,I_2(H^\dagger,H)+ a_3\,I_3(H^\dagger,H)+ a_4\,I_4(H^\dagger,H)\;,
\end{split}
\end{equation}
where $m$ denotes the mass term and $a_k$ $(k=0,..,4)$ are five real quartic couplings corresponding to the five real quartic invariants $I_k(H^\dagger,H)$.
The quartic invariants are defined as the five possible singlet contractions of \mbox{$(\rep{\bar{3}}\otimes\rep{3})\otimes(\rep{\bar{3}}\otimes\rep{3})$} in $\Delta(54)$:
\begin{equation}\label{eq:invariants}
 \begin{split}
  I_0(H^\dagger,H)~&:=~\left[\left(H^\dagger\otimes H\right)_{\rep[_0]{1}}\otimes\left(H^\dagger\otimes H\right)_{\rep[_0]{1}}\right]\;, \\
  I_1(H^\dagger,H)~&:=~\frac{1}{\sqrt{2}}\,\left[\left(H^\dagger\otimes H\right)_{\rep[_1]{2}}\otimes\left(H^\dagger\otimes H\right)_{\rep[_1]{2}}\right]_{\rep[_0]{1}}\;,\\
 I_2(H^\dagger,H)~&:=~\frac{1}{\sqrt{2}}\, \left[\left(H^\dagger\otimes H\right)_{\rep[_3]{2}}\otimes\left(H^\dagger\otimes H\right)_{\rep[_3]{2}}\right]_{\rep[_0]{1}}\;,\\
 I_3(H^\dagger,H)~&:=~\frac{1}{\sqrt{2}}\,\left[\left(H^\dagger\otimes H\right)_{\rep[_4]{2}}\otimes\left(H^\dagger\otimes H\right)_{\rep[_4]{2}}\right]_{\rep[_0]{1}}\;,\\
 I_4(H^\dagger,H)~&:=~\frac{1}{\sqrt{2}}\, \left[\left(H^\dagger\otimes H\right)_{\rep[_2]{2}}\otimes\left(H^\dagger\otimes H\right)_{\rep[_2]{2}}\right]_{\rep[_0]{1}}\;.					      
\end{split}
\end{equation}
The necessary CGs as well as the explicit form of the invariants is given in \Appref{app:Delta54}.
This way of writing the potential differs from the originally chosen form \cite{Branco:1983tn} or more recently used forms \cite[Eq.\ (14)]{Ivanov:2012ry} (cf.\ \cite{Fallbacher:2015rea} 
for a detailed translation of parameters).
While the basis used for the $\Delta(54)$ (and \Dts) triplet generators is exactly the same in all approaches, the difference lies in how the Lagrangian
is written in terms of linear combination of the quartic $\Delta(54)$ symmetry invariants. While, of course, any choice for the basis of invariants is admissible, 
there is somehow no clear motivation for the chosen basis in the space of invariants in \cite{Branco:1983tn} and \cite{Ivanov:2012ry}.
In the formulation used here, the linearly independent symmetry invariants are derived directly from the direct product decompositions of the group. 
The benefit of this ``derived'' basis of symmetry invariants will become clear upon considering the action of outer automorphisms.

In this parametrization the potential is bounded below if and only if \cite{Fallbacher:2015rea}%
\begin{equation}\label{eq:physicalityA}
 0~<~a_0+a_\ell\;, \qquad\text{for}~\;\ell=1,..,4\;.
\end{equation}

Assuming that the vacuum preserves the electric charge\footnote{%
In fact, this assumption is automatically fulfilled by any global minimum of this potential, cf.\ \cite[App.\ B.1]{Fallbacher:2015rea}.
} 
the doublet VEVs can be parametrized as 
\begin{equation}\label{eq:VEVparametrization}
 \Braket{0\!|H_i|\!0}~\equiv~\langle H_i \rangle ~:=~\begin{pmatrix} 0 \\ \mathrm{v}_i\,\e^{\I\,\varphi_i} \end{pmatrix}\qquad\text{for}~\;i=1,..,3\;,
\end{equation}
with $\mathrm{v}_i>0$ and $0\leq\varphi_i<2\pi$. In a compact notation, the triplet of EW doublet VEVs will be denoted by 
\begin{equation}
 \langle H \rangle~=~(\mathrm{v}_1\,\e^{\I\,\varphi_1} ,\mathrm{v}_2\,\e^{\I\,\varphi_2},\mathrm{v}_3\,\e^{\I\,\varphi_3})^\mathrm{T}\;.
\end{equation}
A complete discussion of the analytical minimization of the potential can be found in \cite{Fallbacher:2015rea}.
Earlier analyses have already shown that this potential gives rise to very specific stationary points with discrete physical phases \cite{Branco:1983tn,deMedeirosVarzielas:2011zw}.
A careful analysis reveals that the minima of the potential can be grouped into four distinct classes $\mathrm{I}-\mathrm{IV}$, 
with a representative of each class given by \cite{Varzielas:2012nn,Ivanov:2014doa,Fallbacher:2015rea}%
\begin{equation}\label{eq:VEVs}
\langle H \rangle_\mathrm{I} ~=~ v_\mathrm{1} \begin{pmatrix} 1 \\ 1 \\ 1 \end{pmatrix},\;
   \langle H \rangle_\mathrm{II} ~=~ v_\mathrm{2} \begin{pmatrix} \omega \\ 1 \\ 1 \end{pmatrix},\;
   \langle H \rangle_\mathrm{III} ~=~ v_\mathrm{3} \begin{pmatrix} \omega^2 \\ 1 \\ 1 \end{pmatrix},\;
   \langle H \rangle_\mathrm{IV} ~=~ v_\mathrm{4} \begin{pmatrix} \sqrt{3} \\ 0 \\ 0 \end{pmatrix},
\end{equation}
where as usual $\omega:=\e^{2\pi\,\I/3}$. Note that only the depth of each minimum depends on the parameters of the potential as
\begin{equation}\label{eq:VEVMagnitude}
\left| v_\ell \right|~:=~\frac{m}{\sqrt{2 \left(a_0+a_\ell\right)}}\;,\qquad\text{for}~\;\ell=1,..,4\;,
\end{equation}
while the directions, including the relative phases, of the VEVs are fixed independently of the potential parameters.
Therefore, the direction of each of the VEVs, including the relative phases, is stable under renormalization group (RGE) running \cite{Branco:1983tn}.
Which of the stationary points in \Eqref{eq:VEVs} actually is the global minimum depends on the values of the couplings $a_\ell$.
The stationary point with the smallest $a_\ell$ hosts the global minimum, as the value of the potential 
at the different stationary points is $V^{(\ell)}_\mathrm{min}=-\frac34m^4(a_0+a_\ell)^{-1}$. 

For completeness, note that there is one more class of stationary points which can never host the global minimum. These are given by
\begin{equation}\label{eq:Saddle}
 \langle H \rangle_\mathrm{V}~=~v_5\,\left(0,-\I,+\I \right)^\mathrm{T}\;,
\end{equation}
with 
\begin{equation}\label{eq:SaddleDepth}
 \left| v_{5} \right|~=~\frac{m\sqrt{3}}{\sqrt{4\,a_0+a_1+a_2+a_3+a_4}}\;.
\end{equation}
These types of stationary points typically correspond to saddle points, at
which the potential has the value $V^{(5)}_\mathrm{sad}=-3m^4(4a_0+a_1+a_2+a_3+a_4)^{-1}$.

Each of the distinct classes of stationary points $\mathrm{I}-\mathrm{V}$ actually corresponds a whole group orbit of physically equivalent
stationary points which can be reached by acting on any of the given vectors with all available symmetry transformations.
Furthermore, the overall global phase of each of the VEV triplets is physically meaningless 
since it can always be shifted by a global hypercharge rotation.

As has been remarked before, physical quantities cannot depend on the chosen basis.
It is, therefore, useful to consider basis--invariant quantities. 
The lowest order CP odd basis invariant quantity in this model can be written as \cite{Nishi:2016}%
\begin{equation}\label{eq:CPInvariant}
 I_6~=~-9\,\sqrt{3}\,(a_1-a_2)(a_1-a_3)(a_1-a_4)(a_2-a_3)(a_2-a_4)(a_3-a_4)\;.
\end{equation}

\subsection{Action of outer automorphisms}
\label{sec:OutAction}

The stage is now set to perform a complete analysis of the \Dff Higgs potential with respect to the action of outer automorphism transformations.
The complete outer automorphism group of \Dff is \Sfour, the permutation group of four elements. 
All the necessary details to understand the following section have been derived in \Secref{Sec:Delta54}.
In particular, recall the outer automorphism transformation of \Dff triplets in \eqref{eq:Delta54AutTriplet}, 
and the corresponding transformation of doublets in \eqref{eq:STActionDoublets}.
All elements of the outer automorphism group can be obtained as compositions of the generating outer automorphisms $s$ and $t$.

The only fundamental (i.e.\ non--composite) \Dff representation present in the 3HDM example model
is the triplet $H$ in \rep[_i]{3}. With respect to triplet representations the outer automorphism group of \Dff splits into two kinds of transformations:
\begin{enumerate}[label=(\roman*)]
  \item \label{enum:equiv} Transformations which map $\rep[_i]3~\mapsto~U\,\rep[_i]3$,\quad and 
  \item \label{enum:CP} Transformations which map $\rep[_i]3~\mapsto~U\,\rep{3}^*_i$.
\end{enumerate}
$U$ here stands for an arbitrary representation matrix of the respective outer automorphism transformation which can for 
each specific case be obtained as solution to \eqref{eq:consistencyEquation}.
All representations which can be reached as an image of \rep{3} under outer automorphism transformations are automatically present in the model. 
Therefore, none of the transformations is broken maximally (cf.\ \Secref{sec:ImplicationsOfOuts}) and the complete set of outer automorphisms is available for the 3HDM example model.

One finds that there are in total $12$ possible transformations of the first category \ref{enum:equiv} corresponding 
to all even permutations of four elements in \Sfour (the identity, three transformations of order two, and eight of order three). 
In the second category \ref{enum:CP}, there are $12$ possible transformations corresponding to all odd permutations of four elements in \Sfour (six of order two and six of order four).
For the second category this counting has been done before and is consistent with the results of \cite{Nishi:2013jqa}.

Consider now the action of the outer automorphisms on the triplet of Higgs doublets.
If a transformation \ref{enum:equiv} would be conserved, this would increase the linear symmetry of the theory (which may accidentally also lead to CP conservation). 
In contrast, any transformation \ref{enum:CP} would warrant CP conservation (and, if not involutory, increases the linear symmetry in addition).
However, as they are outer automorphisms, none of the above transformations is a symmetry of the \Dff potential to begin with.
Both types of transformations generally map the theory to different spots in the parameter space and, hence, can be treated on an equal footing.

\paragraph{Action on the parameters.}
In \Secref{sec:redundant} it has been shown that outer automorphism generally act non--trivially on the couplings of a model, thereby allowing
for the identification of physically equivalent regions in the parameter space. 
Specifically for this model, it is straightforward to infer the corresponding transformations of the couplings 
along the lines discussed around \eqref{eq:OutTrafo}.  

This shall be illustrated with a few examples. The parameters $m$ and $a_0$ stay inert under the action of all outer automorphisms.
The transformation $t$, for instance, is of category \ref{enum:equiv}.
Acting with $t$ on the triplet $H$, i.e.\ mapping $H\mapsto U_{t}H$ in \eqref{eq:Invariants} with $U_{t}$ as given in \eqref{eq:Delta54Aut}, %(and of course also $H^\dagger\mapsto H^\dagger U^\dagger_h$)
is equivalent to the parameter mapping 
\begin{equation}\label{eq:TAction}
 (a_1,a_2,a_3,a_4)~\mapsto~(a_1,a_3,a_4,a_2)\;.
\end{equation}
Following the discussion in \Secref{sec:redundant} this shows that a model with parameters 
$(a_1,a_2,a_3,a_4)$ is physically equivalent to a theory with parameters $(a_1,a_3,a_4,a_2)$.
That is, a model with the second set of parameters makes w.r.t.\ the fields $(U_t H)$ exactly the same physical predictions 
as a model with the first set of parameters w.r.t.\ the fields $H$.

Another example for an outer automorphism of the category \ref{enum:equiv} is given by the transformation $s\circ t^{-1}\circ s\circ t$.
This transformation is equivalent to the parameter mapping%
\begin{equation}\label{eq:T2Action}
 (a_1,a_2,a_3,a_4)~\mapsto~(a_2,a_3,a_1,a_4)\;.
\end{equation}
Again, this transformation identifies parameter regions which are physically equivalent.

Let us now discuss outer automorphisms of the category \ref{enum:CP}. 
A priori, all outer automorphisms which map $\rep3$ to $\rep3^*$ are possible physical CP transformations of this model. 
That is, any of these transformations, if conserved, warrants the vanishing of all CP odd basis invariants.
Consider, for example, the outer automorphism $s$. It is straightforwardly confirmed that this transformation 
is equivalent to the mapping
\begin{equation}\label{eq:SAction}
 (a_1,a_2,a_3,a_4)~\mapsto~(a_3,a_2,a_1,a_4)\;.
\end{equation}
In this case this implies that a theory with parameters $(a_1,a_2,a_3,a_4)$ describes, with respect to $H$, precisely the same dynamics as a theory with parameters $(a_3,a_2,a_1,a_4)$ with respect to $(U_{s}H^*)$.  
Furthermore, $s$ is a CP symmetry of the theory if and only if the couplings fulfill the relation $a_1=a_3$.

A supposedly very particular example is what can be called the canonical CP transformation in the chosen basis.\footnote{%
In fact, this transformation is nothing special. None of the order two CP transformations is distinguished with 
respect to any other order two CP transformation. In particular, it only depends on the chosen basis whether a given CP transformation
would be called generalized or canonical.}
On the triplets this transformation acts as $\rep{3}\mapsto U\,\rep{3}^*$ with $U=\mathbbm{1}$. The corresponding outer automorphism 
transformation is given by $s\circ t^{-1}\circ s\circ t\circ s$. 
In parameter space this transformation corresponds to the mapping
\begin{equation}\label{eq:S2Action}
 (a_1,a_2,a_3,a_4)~\mapsto~(a_1,a_3,a_2,a_4)\;.
\end{equation}
Also here, this implies that both parameter regions are physically degenerate and 
this CP transformation is conserved if and only if $a_2=a_3$. 
Indeed, this already hints at a more general principle, namely, CP is conserved whenever (at least) two of the four parameters $(a_1,a_2,a_3,a_4)$ are equal.

Inspecting the contractions \eqref{eq:invariants},
one realizes that one could also have used the transformation of the intermediate (composite) doublet representations, cf.\ \Eqref{eq:STActionDoublets}, 
involved in the construction of the quartic invariants in order to arrive at the results \eqref{eq:TAction}--\eqref{eq:S2Action}.
The complete set of outer automorphisms act as all possible permutations of the four doublets \rep[_{1-4}]{2} of \Dff,
corresponding also to all possible permutations of the four individual \Dff invariants $I_{1-4}(H^\dagger,H)$.
Consequently, the outer automorphism group \Sfour, being the full permutation group of four elements, also describes all possible 
permutation of the four couplings $a_{1-4}$. 

The behavior of the symmetry invariants $I_{1-4}(H^\dagger,H)$ is also a prime example for the remarks at the end of \Secref{sec:Mesons}. 
Even though composites here are transforming like the elementary states, the derived $\Delta(54)$ symmetry invariants are clearly distinguished by, and 
hence may be classified according to, their transformation behavior under the outer automorphisms.

This shows the advantage of parametrizing the Lagrangian in terms of the derived invariants: The independent invariants and, hence, also the couplings,
are nicely permuted under the action of the outer automorphism.
In contrast, in the conventional forms of the Lagrangian \cite{Branco:1983tn,Ivanov:2012ry} separate invariants mix under the action of the outer
automorphism. This unnecessarily complicates the action on the couplings \cite{Fallbacher:2015rea} and obscures the underlying structure.

Altogether, the possible outer automorphism transformations indicate that the three Higgs doublet potential with $\Delta(54)$ symmetry and a given set of parameters is physically equivalent 
to every potential which can be obtained by any permutation of the four parameters $a_{1-4}$.
This equivalence can be made explicit by a field redefinition for all even permutations or by a complex field redefinition, 
i.e.\ a CP transformation, for all odd permutations of the $a_{1-4}$, respectively.

\paragraph{Action on the VEVs.}
From the discussion in \ref{sec:OutsAndVEVs} it is clear that the outer automorphisms also act non--trivially on the stationary points of the potential.
In fact, one can even use outer automorphism to constrain -- and in this specific case even fully compute --
the form of all stationary points of the potential. This will be detailed in \Secref{sec:VEVcomputation} below.
For now, the plain observation of how outer automorphisms act on the VEVs is sufficient.

Consider, for instance, the outer automorphism transformation $t$. Acting with the explicit representation matrix $U_t$ in \eqref{eq:Delta54AutTriplet} 
on the triplets of stationary points in \eqref{eq:VEVs}, one finds that the VEVs are permuted as 
\begin{equation}\label{eq:TVEVaction}
 \mathrm{I}~\mapsto~\mathrm{I}\quad\text{and}\quad\mathrm{II}~\mapsto~\mathrm{III}~\mapsto~\mathrm{IV}~\mapsto~\mathrm{II}\;.
\end{equation}
The capital roman numerals here denote the different types of VEVs as defined in \eqref{eq:VEVs}. 
Note, however, that the resulting vectors are only valid stationary points of the potential 
if the arguments of their absolute values $v_\ell(a_\ell)$ are also permuted in the same way.
For example, transforming
\begin{equation}
 \langle H \rangle_\mathrm{II}~\mapsto \langle H \rangle'_\mathrm{II}~=~
 \left.U_t\,\langle H \rangle_\mathrm{II}\right|_{a_2\,\rightarrow\,a_3}~=~
  \omega^2\,\langle H \rangle_\mathrm{III}\;,
\end{equation}
only results in an actual stationary point of the potential if the argument of the 
modulus of the VEV is also transformed consistently with the outer automorphism, i.e.\ $a_2\mapsto a_3$.
For $t$, the complete transformation of parameters is stated in \eqref{eq:TAction}. 
This nicely agrees with the permutations of the categories of VEVs in \eqref{eq:TVEVaction}, corresponding to the general rule \eqref{eq:additionalVEV}.

It should be remarked, however, that \eqref{eq:TVEVaction} here only holds up to a global rephasing of each of the VEVs in \eqref{eq:VEVs}.
This is not a general feature but merely an artifact of the inital global phase choice for each to the VEVs in \eqref{eq:VEVs} to begin with.
The arbitrary global phases in \eqref{eq:VEVs} have been chosen such as to make the connection to the previous discussions of this model in the literature.
This is not necessary and in \eqref{eq:Block_representants72} below, for example, global phases of the VEVs are chosen such that the transformations $U_s$ and $U_t$ and, therefore, all outer automorphisms,
act as a permutation of the VEVs without the need of an additional rephasing. 

As another example, consider the transformation $s$. Because $s$ maps $\rep[_i]{3}$ to $\rep{3}^*_i$, the VEVs in this case have to be conjugated in addition to the multiplication with $U_s$
and the formal replacement of their moduli's arguments. The action on a VEV then is given by
\begin{equation}\label{eq:SVEVaction}
 \langle H \rangle_\mathrm{I}~\mapsto \langle H \rangle'_\mathrm{I}~=~
\left.U_s\,\langle H \rangle^*_\mathrm{I}\right|_{a_1\,\rightarrow\,a_3}~=~-C\,\langle H \rangle_\mathrm{III}\;.
\end{equation}
This transformation permutes VEVs of the different types according to
\begin{equation}\label{eq:SVEVaction2}
 \mathrm{I}~\longleftrightarrow~\mathrm{III}\;,\quad\mathrm{II}~\mapsto~\mathrm{II}\quad\text{and}\quad\mathrm{IV}~\mapsto~\mathrm{IV}\;,
\end{equation}
again completely consistent with the transformation of the parameters in \eqref{eq:SAction}.

As can be seen from \eqref{eq:SVEVaction}, $s$ again permutes the VEVs only up to a global phase.
In the same way as for $t$, this can be avoided by making a globally consistent phase choice for all VEVs such as the one given in \eqref{eq:Block_representants72} below.
In addition, however, the transformation $s$ permutes some of the VEVs only up to an inner automorphism $c$ as apparent in \eqref{eq:SVEVaction}.
As any of the categories of VEVs in \eqref{eq:VEVs} is, strictly speaking, only defined modulo inner automorphisms to begin with, this is not an issue.
In the systematic construction presented in \Secref{sec:VEVcomputation} the correct inner automorphisms appear automatically at the right places.
In a manual construction the appropriate inner automorphisms can, for example, be inferred from the appearance of inner automorphisms in 
the doublet transformations \eqref{eq:STActionDoublets} while carefully taking into account the definition of the invariants \eqref{eq:invariants}.

Altogether, the two transformations $s$ and $t$ generate all possible permutations of the four classes of VEVs $\mathrm{I}-\mathrm{IV}$.
That is, completely analogous to the four invariants $I_{1-4}(H^\dagger,H)$ and the four corresponding potential parameters $a_{1-4}$, also
the four categories of stationary points $\mathrm{I}-\mathrm{IV}$ transform as a \rep{4}--plet under the outer automorphism group \Sfour. 
This shows that each of the classes of stationary points $\mathrm{I}-\mathrm{IV}$ is physically equivalent as they conserve isomorphic subgroups.

In an analogous way one finds that the category $\mathrm{V}$ of stationary points (cf.\ \eqref{eq:Saddle}) transforms as a trivial singlet 
under the action of the outer automorphisms. That is, all stationary points within $\mathrm{V}$ are always mapped back to a stationary points
which are also part of $\mathrm{V}$. This is consistent with the transformation of \eqref{eq:SaddleDepth} under the action of the outer automorphism on the couplings.
Since \eqref{eq:SaddleDepth} contains $a_{1-4}$ in a completely symmetric manner, it is invariant under the permutations induced by the outer automorphism.

\subsection{Physical implications}

\paragraph{Parameter space degeneracies.}
As a direct consequence of the preceding discussion the complete physical phenomenology of the potential is already contained in a restricted region of the parameter space. 
Outer automorphisms, here, correspond to all possible permutations of the parameters $a_{1-4}$ implying that
a possible choice for a set of physically non--degenerate parameters is \mbox{$a_1\leq a_2\leq a_3\leq a_4$}.
All other regions of the parameter space are physically equivalent because they can be related to this wedge by outer automorphisms. 

If and only if two or more parameters are equal, the conserved symmetry group of the model is larger than \Dff.
Therefore, first consider the interior of the wedge, \mbox{$a_1<a_2<a_3<a_4$}.
For this choice of parameters the symmetry of the model is exactly \Dff and CP is explicitly broken. 
If an order two CP symmetry should initially be conserved 
two out of the four parameters $a_{1-4}$ have to be equal.
This statement follows from the fact that CP transformations correspond to odd permutations of the four parameters,
and it can nicely be understood also by the means of the CP odd basis invariant \eqref{eq:CPInvariant}.

If more than two out of the four parameters $a_{1-4}$ are equal this automatically ensures explicit CP conservation but also 
implies an enhancement of the linear symmetry of the model. 
If there are two pairs of equal parameters the discrete symmetry of the potential is enhanced from \Dff to 
$\widetilde{G}:=\left(\left(\Z3\times\Z3\right)\rtimes\Z3\right)\rtimes\Z4\cong\mathrm{SG}(108,15)$.
In case three (or more) out of the four parameters are equal, then the symmetry is enhanced to a continuous group. 
This is completely in agreement with the maximal ``realizable'' symmetry $\Sigma(36)\cong \widetilde{G}/\Z3$ of a 3HDM as found in \cite{Ivanov:2012ry,Ivanov:2012fp}.

\paragraph{Spontaneous geometrical CP violation.}
The possibility of spontaneous (geometrical) CP violation in this model shall be discussed in the following. 
In general, CP is spontaneously violated by the vacuum of the Higgs potential if and only if there is no $U$ fulfilling
\begin{equation}\label{eq:VEVcondition}
\langle H \rangle ~=~ U\,\langle H \rangle^*\;,
\end{equation}
while at the same time the transformation $H\mapsto U H^*$ is a CP symmetry of the Lagrangian (cf.\ e.g.\ \cite{Branco:1983tn}).
In contrast, if there is a $U$ which fulfills \eqref{eq:VEVcondition} without being a symmetry of the Lagrangian,
this would be an emergent CP symmetry of the vacuum (cf.\ the discussion in \Secref{sec:OutsAndVEVs}).

Two important results from the preceding discussion are worth to be stressed again. 
Firstly, each of the classes of stationary points $\mathrm{I}-\mathrm{IV}$ is physically equivalent. 
Secondly, one may without loss of generality focus on a wedge of the parameter space with hierarchical ordering of the parameters $a_{1-4}$, as 
all other orderings of parameters are physically equivalent.

In order to obtain a setting with spontaneous CP violation, CP, of course, has to be a conserved symmetry to begin with. 
According to the previous discussion, this implies that a pair of the parameters $a_{1-4}$ must be equal. For definiteness,
consider the CP transformation induced by the outer automorphism $s$. 
All other order two CP transformations can be obtained from $s$ by the action of outer automorphisms, 
implying that they are physically equivalent. Therefore, it is possible to focus on $s$ without loss of generality.
The according CP transformation is a symmetry of the Lagrangian if and only if $a_1=a_3$, cf.\ \eqref{eq:SAction}.
Furthermore, from \eqref{eq:SVEVaction2} it follows that $s$ permutes the VEVs of category $\mathrm{I}$ with those of category $\mathrm{III}$, 
while it leaves the VEVs of category $\mathrm{II}$ and $\mathrm{IV}$ invariant.
From the discussion after \eqref{eq:VEVMagnitude} it is clear 
that the global minimum of the potential is given by the category of VEVs in \eqref{eq:VEVs} which features the smallest $a_\ell$.
Considering the phenomenology of the global minimum of the theory in case the CP transformation $s$ is conserved, 
there are then two possible physically distinct scenarios: 
either \mbox{$a_2<a_4<a_1=a_3$}, or \mbox{$a_1=a_3<a_2<a_4$}.\footnote{%
In principle, one could also have \mbox{$a_2<a_1=a_3<a_4$} but concerning the behavior of the global minimum this would be equivalent to the case \mbox{$a_2<a_4<a_1=a_3$}.
Also note that the relative ordering of $a_2$ and $a_4$ is completely irrelevant for the discussion here.}

In the first case, the global minima are given by VEVs of the category $\mathrm{II}$ (or $\mathrm{IV}$, which would not make a difference). 
As this category is invariant under the action of $s$, it is always possible to find a transformation which solves \eqref{eq:VEVcondition} 
while at the same time being also a symmetry of the Lagrangian.
Consequently, CP is conserved by the Higgs VEV in this case. 

In the second case, the global minima are by VEVs of the categories $\mathrm{I}$ and $\mathrm{III}$ which 
are not invariant under $s$ implying that \eqref{eq:VEVcondition} cannot be fulfilled and 
this transformation is spontaneously broken by the Higgs VEV. 
In order to claim that this also implies the spontaneous violation of CP, however, one has to assure that there is no other CP symmetry of the Lagrangian which fulfills \eqref{eq:VEVcondition}.
Regarding the choice of parameters, this is trivially achieved. Given that there is no equal pair of parameters besides $a_1$ and $a_3$ by assumption, 
the only conserved CP symmetry at the level of the Lagrangian is $s$. Thus, CP in this case is indeed spontaneously violated by
the VEVs of categories $\mathrm{I}$ and $\mathrm{III}$. Spontaneous CP violation, here, is geometrical in the sense that all CP violating phases are independent of the potential parameters and can be calculated from the 
necessarily complex CGs of the symmetry group $\Dff$ \cite{Chen:2014tpa}. 

It may appear surprising that a triplet VEV of category $\mathrm{I}$ with only real entries, i.e.\ no relative phases between the individual Higgs VEVs, can give rise to spontaneous CP violation.
Equally surprising may be the fact that a triplet VEV which does have fixed relative phases between the Higgs VEVs, such as $\mathrm{II}$, is CP conserving.
However, as it is clear from this discussion, spontaneous CPV always appears in a combination of an initial CP transformation and a VEV. 
In case $s$ is spontaneously broken by a VEV of the category $\mathrm{I}$, the geometrical phases are carried within the explicit transformation matrix $U_s$.
For any specific CP transformation in this model, one may always perform a basis transformation such that it is represented by $U=\mathbbm{1}$. 
In the new basis, the geometrical CPV phases would then reside in the VEV, whereas VEVs which do not lead to CPV would be all real.\footnote{%
Alternatively, one may also consider the canonical CP transformation in the given basis which transforms the parameters and VEVs according to \eqref{eq:S2Action}.
Fully consistent with our claims, this transformation is conserved by the ``all--real'' types of VEVs $\mathrm{I}$ and $\mathrm{IV}$, and broken by the VEVs with 
relative phases $\mathrm{II}$ and $\mathrm{III}$.}

As an aside, note that from the relation $a_1\equiv a_3$ it follows directly that vacua of the categories $\mathrm{I}$ and $\mathrm{III}$ are energy degenerate.
This also follows from the fact that $s$ has been added to the symmetry transformations, because the vacua then are part of one and the same group orbit.
However, as the two vacua are, in principle, distinguishable, there will be domain walls present after the spontaneous breaking \cite{Zeldovich:1974uw}.
This, of course, is always the case if there are multiple isolated vacua related by discrete group operations. The delicacy in this case is that
the different domains have different properties also with respect to CP.\footnote{%
This may have interesting consequences for baryogenesis via tunneling processes as discussed in \cite{Hook:2015foa}.
}

Another remark concerns the emergence of approximate symmetries in the vacuum of the potential. The previous discussion shows that for each of the VEVs, 
there are always non--trivial stabilizers originating from the group of outer automorphisms. 
Consider, for example, a VEV of category $\mathrm{I}$. While VEVs of this category spontaneously break the transformation $s$, they are invariant under 
other outer automorphisms whose action only involves the permutation of stationary points of the other categories.
In this way also CP could appear as an emergent symmetry.

\paragraph{Outer automorphisms and other sectors.}
In order to investigate outer automorphisms in possibly realistic theories it is, of course, not sufficient to limit the focus only to the Higgs potential.
Therefore, it should be commented on the validity of the whole discussion
once the Higgs fields are coupled to other sectors of a model, such as the Yukawa couplings to fermions.

Firstly, it is clear that if the new sector does not obey the full symmetry group of the Higgs potential but only a smaller group, 
then the whole discussion can only be led based on the outer automorphisms of that group.
However, even if the symmetry is not reduced by the addition of a new sector, the new sector generically will have fields in representations other than the Higgses'.
This typically reduces the number of available outer automorphism transformations. That is, outer automorphisms which are available at the level of the Higgs potential 
might be explicitly and maximally broken by the additional representations (cf.\ \Secref{sec:ImplicationsOfOuts}).

In particular, it may occur that parameter regions or VEVs, which are related by outer automorphisms and, thus, seem to be physically equivalent at the level of the Higgs potential, 
in fact give rise to distinct physical predictions of masses, mixings, and CPV, etc.\ in the additional sectors.
This, for example, is the case in models which employ the \Dts symmetric potential for the explanation of 
fermion masses and mixing patterns \cite{Bhattacharyya:2012pi,Varzielas:2013sla,Ma:2013xqa, Varzielas:2013eta}.
Even if the complete set of outer automorphisms then cannot be used to identify necessarily physically equivalent parameter regions, 
it is still a powerful tool to analyze the Higgs potential in order to find all possible VEVs and possible emergent symmetries.
Furthermore, all globally (i.e.\ for all sectors) available outer automorphism transformations will still point to parameter space redundancies.

\subsection{Computation of VEVs with outer automorphisms}
\label{sec:VEVcomputation}

Finally, to finish off the discussion of the 3HDM example with \Dff symmetry, it shall be demonstrated how the 
knowledge about outer automorphisms of this setting can be used in order to compute the stationary points of the potential.

\paragraph{Prequel.}
Stationary points of a potential are typically found directly by solving for points of the potential with vanishing gradient. 
The plain three Higgs doublet potential has $3\times4$ real degrees of freedom. Limiting the discussion to charge conserving minima, 
one may use the parametrization of VEVs given in \eqref{eq:VEVparametrization}, thereby reducing the
relevant number of real degrees of freedom to $3\times2$. Differentiating the potential with respect to each of these, results 
in a system of six coupled cubic polynomial equations which has to be solved in order to find the stationary points. 
The number of solutions, i.e.\ the number of possible VEVs then is strictly bounded above by \mbox{$3^6=729$} \cite{Schmid1995}. 
In practice, the actual bound will be more restrictive due to additional symmetries which have not been taken into account 
in this counting. A complete analytical minimization of the 3HDM potential with $\Delta(54)$ 
symmetry performed along these lines can be found in \cite[App.\ B]{Fallbacher:2015rea}.

Complementary to the traditional minimization, the direction and relative phases of the stationary points in the 3HDM with \Dff symmetry 
can also be computed by solving only a homogeneous linear equation, as will be shown in the following.
The modulus of a specific stationary point is not fixed by the outlined procedure. 
Nevertheless, the moduli can easily be computed subsequently by plugging one of the resulting stationary points of fixed direction 
into the gradient of the potential. 

Before starting the actual calculation, the underlying idea shall be outlined.
In \Secref{sec:OutsAndVEVs} it has been established that VEVs do not only form orbits under the symmetry group $G$ 
but, `perpendicular' to that, also under the group of outer automorphism transformations of the specific setting.
This has explicitly been demonstrated for the 3HDM example model in \Secref{sec:OutAction}, for which the group
of available outer automorphisms exhausts the full outer automorphism group $\Out(G)$.
The crucial link to understand that the four categories of VEVs $\mathrm{I}-\mathrm{IV}$ in \eqref{eq:VEVs} form a single entity and are, at least on the level of the potential,
all physically equivalent, is the fact that VEVs of all types are part of a single orbit under the action of the outer automorphism group.

If the corresponding orbits under the group and outer automorphism group are shorter than $|G|$ or $|\Out(G)|$, respectively,
then it is imperative that the corresponding VEVs have stabilizers in $G$ and $\Out(G)$.
That is, each of the stationary points is an eigenvector, also called fixed point, to one or more elements of $G$ and $\Out(G)$.
For the case of $G$ this is not unusual, and it is well--known that the corresponding stabilizers in $G$ 
form precisely the subgroup which is left unbroken by the VEV. For the case of $\Out(G)$, however, this is a new and 
non--trivial insight. Stabilizers in $\Out(G)$ do not correspond to unbroken symmetries, as the operations 
in $\Out(G)$ are no symmetries of the Lagrangian to begin with. Nevertheless, the VEVs 
have to be eigenvectors to some operations in $\Out(G)$. This generally gives additional non--trivial restrictions 
on the possible form of the VEVs. In the example discussed here, these conditions are so restrictive 
as to completely fix the form and direction of the VEVs. In combination with the appearance of calculable phases, this explains
the origin of spontaneous geometrical CP violation. 

The discussion here closely follows the analysis in \cite{Fallbacher:2015rea}.
Since it is the first analysis of this kind, details have been taken into account probably a bit more thoroughly than necessary. 
In particular, all outer and inner automorphisms are taken into account for completeness, resulting in a very big group of transformations.
However, this may not be necessary in principle, as equivalent results also have been obtained by just working with the much smaller outer automorphism
group alone. Another possible simplification for the future might be not to work with the whole set of stationary points and their permutations, as here, 
but to exploit the fact that VEVs must be eigenvectors of elements in $G$ and $\Out(G)$ directly.

The $\Delta(54)$ example model will be used to illustrate this method in the following.
The presented method can straightforwardly be adapted to any other potential and it will be commented on this below.

\paragraph{Computation of stationary points in the 3HDM $\Delta(54)$ example.} 
Put aside, for the moment, the existence of the global $\U{1}_\text{Y}$, i.e.\ the fact that VEVs can be re--phased continuously,
and focus on the orbits of stationary points only under discrete transformations. 
Together, symmetry and outer automorphism transformations are referred to as equivalence transformations of the potential \cite{Fallbacher:2015rea}.
The group of equivalence transformations, therefore, describes the complete orbit of a stationary point. It can be constructed as $\Equi\cong G\SemiDirect\Out(G)$.

For the $\Delta(54)$ example, $E$ can be constructed by the combination of \eqref{eq:DffPrsentation}, \eqref{eq:S4algebra}, and \eqref{eq:SfourAction},
implying that a possible presentation of the group is
\renewcommand{\arraystretch}{1.45}
\begin{equation}\label{eq:EPresentation}
\begin{split}
E~=~\bigl\langle\elm{A,B,C,S,T}\,\big|\,&\mathsf{A}^3\,=\,\mathsf{B}^3\,=\,\mathsf{C}^2\,=\,\left(\mathsf{A}\,\mathsf{B}\right)^3\,=\,\left(\mathsf{A}\,\mathsf{C}\right)^2\,=\,
\left(\mathsf{B}\,\mathsf{C}\right)^2\,=\,\mathsf{e}\;,\bigr. \\
&\mathsf{S}^2\,=\,\mathsf{T}^3\,=\,\left(\mathsf{T}^2\mathsf{S}\right)^4\,=\,\mathsf{e}\;, \\ 
&\hspace{-5pt}\begin{array}{ll} 
\mathsf{T\,A\,}\mathsf{T}^{-1}~=~\mathsf{A}\;, & \mathsf{S\,A\,S}^{-1}~=~\mathsf{AB^2A}\;, \\ 
\mathsf{T\,B\,T}^{-1}~=~\mathsf{ABA}\;, & \mathsf{S\,B\,S}^{-1}~=~\mathsf{B}\;,  \\
\mathsf{T\,C\,T}^{-1}~=~\mathsf{C}\;, & \mathsf{S\,C\,S}^{-1}~=\mathsf{C}\;\bigr\rangle\;.
\end{array}
\end{split}
\end{equation}\renewcommand{\arraystretch}{1.0}%
\Equi has order $|\Equi|=|G|\times|\Out(G)|=1296$ as expected, and is contained in the SmallGroup library of GAP as $\mathrm{SG}(1296,2891)$.
The orbit of a given stationary point $\vev\equiv\langle H \rangle$ is denoted by $\VEV$, 
and is obtained from the \mbox{(left--)action} of all elements of \Equi, i.e.\ \mbox{$\VEV:=\{\elm p\vev \,|\, \elm p \in \Equi\}\equiv\Equi \vev$}.

The theoretically maximal total orbit length, i.e.\ the maximal number of distinct VEVs which can be obtained from a given VEV by equivalence transformations, is
given by the total number of possible equivalence transformations $|\Equi|=1296$.
However, the actual orbit length can, of course, not exceed the upper bound on the total number of stationary points which is here given by $729$.
Consequently, each VEV has to be a fixed point (i.e.\ an eigenvector) of typically several equivalence transformations. 
That is, for each $\vev$ there are several elements in \Equi that leave $\vev$ invariant.
Indeed, it is easy to show that for \emph{any} VEV $\vev$, it is always a subgroup $\Equi_\vev\subset\Equi$ which leaves the VEV invariant.
The crucial factor which determines the size and shape of an orbit is $\Equi_\vev$, which is a priori unknown.
In the subsequent discussion everything is derived for general $\Equi_\vev$, and specific cases will be treated in the end.
If there are several distinct orbits of VEVs they are disjoint and can be considered separately.

Note that $G$ is by construction a normal subgroup of \Equi. Consequently, the orbits of \Equi have a very special structure (e.g.\ \cite[p.\ 12]{Bogopolskij2008}). 
Namely, $\VEV$ splits under the action of the normal subgroup $G$ and can be written in the form 
\begin{equation}
 \VEV~=~\begin{pmatrix} \,\boxed{\leftarrow ~ G\,\vev_1 ~ \rightarrow} \,, \,\boxed{\leftarrow ~ G\,\vev_2 ~ \rightarrow}\,, \qquad \cdots \qquad ,  \,\boxed{\leftarrow ~ G\,\vev_n ~ \rightarrow} ~~\end{pmatrix}^\mathrm{T}\;.
\end{equation}
Here, the boxes correspond to equally--sized blocks which themselves contain $G$--orbits of VEVs \mbox{$G\vev_i\equiv \{\elm g\vev_i \,|\, \elm g \in G\}$},
which are obviously disjoint under the action of $G$.
The individual blocks have size \mbox{$r:=|G|/|G\cap \Equi_\vev|$} and the number of blocks is given by $n:=|\Equi||G\cap \Equi_\vev|/\left(|G||\Equi_\vev|\right)$. 
By the orbit stabilizer theorem $|\VEV|=|\Equi|/|\Equi_\vev|=r\cdot n\,$.

Under the action of elements in $G$, the VEVs are permuted transitively\footnote{%
The action of a group $G$ on a set $X$ is called transitive if for any pair of elements $x,y\in X$, there is a $\elm g\in G$ such that $\elm{g}x=y$.
}
only within the individual blocks. In contrast, under the action of elements in \Equi which are not in $G$, i.e.\ under the action of $\Equi/G\cong\Out(G)$, the blocks themselves are permuted transitively.
This is the precise mathematical formulation of the statement in \Secref{sec:OutsAndVEVs} that orbits 
obtained from the outer automorphism group are `perpendicular' to the $G$--orbits.

In the following, the transformation of $\VEV$ under \Equi shall be investigated in detail.
For this, it is more practical to switch the presentation of \Equi from \eqref{eq:EPresentation} to a minimal generating set which is
given by 
\begin{equation}
 \mathsf{P}~:=~\mathsf{T}\;, \quad\text{and}\quad 
 \mathsf{Q}~:=~(\mathsf{T\,S})^2\,(\mathsf{T}^{-1}\,\mathsf{S})^2\, \mathsf{C}
 \,(\mathsf{T}^{-1}\,\mathsf{S})^2\, \mathsf{C}\, \mathsf{A}\, (\mathsf{T}^{-1}\,\mathsf{B}^{-1}\,\mathsf{T\,B\,A})^4\;.
\end{equation}
The explicit action of $\elm P$ and $\elm Q$ on the triplet representation then is given by
\begin{equation}\label{eq:PQTripletReps}
  P~=~\frac{\I}{\sqrt{3}}\,\begin{pmatrix} 1 & \omega^2 & \omega^2\\ \omega^2 & 1 & \omega^2\\ \omega^2 & \omega^2 & 1\end{pmatrix}\, \qquad \text{and} 
  \qquad Q~=~ \frac{\I}{\sqrt{3}}\,\begin{pmatrix} \omega^2 & \omega & \omega^2\\ \omega & \omega & 1\\ 1 & \omega & \omega\end{pmatrix}\,.
\end{equation}
The transformation $\elm Q$ corresponds to a complex conjugation mapping of $\rep[_i]{3}\mapsto Q\,\rep[_i]{3}^*$. 
Therefore, it is more convenient to work with the representation $\rep[_i]{6}=\rep[_i]{3}\oplus\rep[_i]{\bar{3}}$, 
as for example also discussed above \Eqref{eq:Rep6}. The representation matrices of the minimal generating set for the \rep{6}--plet representation then are given by
\begin{equation}\label{eq:PQHexpletReps}
  P_{\rep{6}}  ~=~ \begin{pmatrix} P & \boldsymbol{0}\\ \boldsymbol{0} & P^*\end{pmatrix} \qquad \text{and} \qquad Q_{\rep{6}}~=~\begin{pmatrix} \boldsymbol{0} & Q\\ Q^* & \boldsymbol{0}\end{pmatrix}\,. 
\end{equation}
The explicit action of \Equi on a given triplet VEV $\vev$ can be obtained by letting all combinations of the explicit representation matrices $P_{\rep{6}}$ and $Q_{\rep{6}}$
act on the vector $(\vev,\vev^*)^\mathrm{T}$. 

However, letting a given transformation act on all possible VEVs $\vev$ simultaneously, each $\vev\in\VEV$ must be mapped bijectively to another VEV $\vev'\in\VEV$.
As a result, all possible transformations must correspond to a permutation of the components of $\VEV$. 
Due to the fact that \Equi acts transitively on $\VEV$, this permutation is equivalent 
to the permutation of elements of the coset space $\Equi/\Equi_\vev$ under 
the action of \Equi by left--multiplication (e.g.\ \cite[p.\ 80]{CeccheriniSilberstein2008}).\footnote{%
This equivalence is also used in the general construction of effective Lagrangians for spontaneously broken continuous symmetries 
\cite{Coleman:1969sm,Callan:1969sn}. There, however, only the action of the symmetry group is considered in order to parametrize the vacua, whereas
here, important additional information from the outer automorphism group is taken into account.}
Therefore, any set of VEVs $\VEV$ must correspond to a possible permutation representation of $\Equi/\Equi_\vev$ (under left--action of \Equi) for 
a given subgroup $\Equi_\vev$. The fact that the explicit action of \Equi on $\VEV$ has to be equivalent to one of these possible permutations 
is a necessary condition on all VEVs with non--trivial stabilizer. 
These necessary conditions shall be explicitly derived in the following.
 
The explicit action on every single VEV in a set of VEVs \VEV is given by $P_{\rep{6}}$ and $Q_{\rep{6}}$ as stated in \eqref{eq:PQHexpletReps} above.
The action on the whole set \VEV with $|\VEV|=r\cdot n$ then is simply given by the $|\VEV|$--fold direct sum $(\oplus)$ of these matrices
\begin{equation}\label{eq:DirectPQ}
 P_\VEV~:=~\bigoplus_{i=1}^{r\cdot n}\, P_{\rep{6}}\;, \qquad \text{and} \qquad Q_\VEV~:=~\bigoplus_{i=1}^{r\cdot n}\, Q_{\rep{6}}\;.
\end{equation}
This action has to be consistent with a permutation of the VEVs in $\VEV$. 
As noted above, this permutation is equivalent to the permutation of elements of the coset space $\Equi/\Equi_\vev$.
For a given subgroup $\Equi_\vev$ this permutation representation can easily be obtained via GAP \cite{GAP4}, and
a computer code which performs this task is given in \Appref{App:VEVCalculation}.
The matrices corresponding to the minimal generating set of this permutation representation are denoted by $\Pi_{\mathsf{P}}$ and $\Pi_{\mathsf{Q}}$.
Altogether, thus, the permutation acts on \VEV as
\begin{equation}\label{eq:PermutationPQ}
 \Pi_{\mathsf{P}}^\VEV~:=~\Pi_{\mathsf{P}}\otimes\Id_6\;, \qquad \text{and} \qquad \Pi_{\mathsf{Q}}^\VEV~:=~\Pi_{\mathsf{Q}}\otimes\Id_6\;,
\end{equation}
where $\otimes$ denotes the Kronecker product of matrices. The additional six--dimensional space here corresponds to of each of the VEVs $(\vev,\vev^*)^\mathrm{T}$
which are being permuted. 

For consistency, acting with either of the two transformations \eqref{eq:DirectPQ} or \eqref{eq:PermutationPQ} on \VEV has to yield the same result,
i.e.\ 
\begin{equation}\label{eq:consistency}
 \left(P_\VEV-\Pi_{\mathsf{P}}^\VEV\right)\,\VEV~=~0\;, \qquad \text{and} \qquad  \left(Q_\VEV-\Pi_{\mathsf{Q}}^\VEV\right)\,\VEV~=~0\;.
\end{equation}
These two homogeneous linear equations \emph{must} be fulfilled by the orbit $\VEV$ of any admissible VEV $\vev$.
That is, fulfilling \eqref{eq:consistency} is a necessary condition for any VEV.

Note that the derivation of \eqref{eq:consistency} requires to know the corresponding stabilizer subgroup $\Equi_\vev$.
However, $\Equi_\vev$ is, in principle, only known in consequence of a given VEV.
Nevertheless, simply assuming a certain subgroup $\Equi_\vev\subset E$ one may check whether a solution to \eqref{eq:consistency} is even possible. 
Scanning over all subgroups in this way, one finds candidates for VEVs.\footnote{%
If the trivial stabilizer subgroup, i.e.\ a complete breaking of the discrete group, is admissible then \eqref{eq:consistency} does not 
impose any constraint on the corresponding VEVs.} 
Depending on the specific subgroup under assumption, the combined rectangular matrix
\begin{equation}\label{eq:Mconsitency}
M~:=~\begin{pmatrix}P_\VEV-\Pi_{\mathsf{P}}^\VEV\\ Q_\VEV-\Pi_{\mathsf{Q}}^\VEV\end{pmatrix}\,
\end{equation}
either has $\mathrm{rank}(M) = 6\,|\VEV|$, implying that there is none but the trivial solution for $\VEV$, or
$\mathrm{rank}(M)<6\,|\VEV|$, implying that there is a non--trivial solution for $\VEV$.
In the first case, VEVs that conserve the assumed subgroup $\Equi_\vev$ cannot exist, whereas in the second case
the solutions of \eqref{eq:consistency} are candidates for orbits of non--trivial VEVs.

The only information used in order to arrive at \eqref{eq:consistency} is the discrete symmetry group of the potential as well as
the group of available outer automorphism transformations, which implicitly also contains information about the representation content of the model.
The derived constraints on the VEVs, hence, are independent of the precise form of the potential and 
solving \eqref{eq:consistency} simply reveals what (orbits of) VEVs are possible in principle.
In order to check whether a non--trivial solution of \eqref{eq:consistency} really is a stationary point of the potential 
one would still have to plug an element of $\VEV$ into the gradient of the potential. This then also fixes any remaining free parameters,
such as the modulus of the VEVs in the case under discussion. 

For the 3HDM example, scanning over the subgroups of \Equi reveals that the largest subgroups which allow for a non--trivial solution to \eqref{eq:consistency}
are given by $\mathrm{SG}(18,4)$, $\mathrm{SG}(18,3)$, and $\mathrm{SG}(48,29)$.\footnote{%
The scan can be limited to conjugacy classes of subgroups, as stabilizer subgroups of points on the same orbit are conjugate to each other.}
All other possible solutions are given by subgroups of these subgroups which unavoidably allow for non--trivial solutions to \eqref{eq:consistency} by construction,
as they are less or equally restrictive on $\VEV$. 
The permutation representations corresponding to the largest subgroups are labeled as $\rep{72}_1$, $\rep{72}_2$, and $\rep{27}$, respectively, 
and their minimal set of generators is given in \Appref{App:VEVCalculation}.

Solving \eqref{eq:consistency} explicitly for the representation $\rep{72}_1$ one finds that 
\begin{equation}
\VEV_{\rep{72}}~=~ \begin{pmatrix} ~\boxed{ G\,\vev_1 } \,, \,\boxed{ G\,\vev_2 }\,, \,\boxed{ G\,\vev_3 } \,, \,\boxed{ G\,\vev_4 }  ~~\end{pmatrix}^\mathrm{T}\;,
\end{equation}
where $\vev_{1-4}$ are representatives of the different blocks, which are given by
\begin{equation}\label{eq:Block_representants72}
 \left( \vev_1, \vev_2, \vev_3, \vev_4 \right)~=~\left( \begin{pmatrix}-\omega \\ -\omega \\ -\omega \end{pmatrix}v_1, \begin{pmatrix}-\omega \\ -1 \\ -1 \end{pmatrix}v_2, 
 \begin{pmatrix}\omega \\ \omega^2 \\ \omega^2 \end{pmatrix}v_3, \begin{pmatrix} \I\omega\sqrt{3} \\ 0 \\ 0 \end{pmatrix}v_4\right)\;.
\end{equation}
Modulo the ubiquitous global $\U{1}_\text{Y}$ rephasing, this exactly reproduces the four categories of VEVs $\mathrm{I}-\mathrm{IV}$ \eqref{eq:VEVs} found in the conventional way.
In contrast to \eqref{eq:VEVs}, the global phase choice for the VEVs in \eqref{eq:Block_representants72} is consistent by construction, in the sense 
that the VEVs permute under the outer automorphisms $t$ and $s$ as in \eqref{eq:TVEVaction} and \eqref{eq:SVEVaction2} without the need of any additional rephasing. 
The remaining free parameters are the moduli $v_{\ell}$, which can easily be obtained by plugging $\vev_{\ell}$ into the gradient of the potential. 
The results, of course, are the same as stated in \eqref{eq:VEVMagnitude}.

An analogous computation for the stabilizer subgroup $\mathrm{SG}(18,3)$ corresponding to the permutation representation $\rep{72}_2$ yields 
a result which differs from \eqref{eq:Block_representants72} only by a global phase. Therefore, this does not give rise to any new VEVs
in the presence of a global $\U{1}_\text{Y}$.

Instead, solving \eqref{eq:consistency} for the permutation representation $\rep{27}$ results in 
\begin{equation}
\VEV_{\rep{27}}^\mathsf{T}~=~ \begin{pmatrix} \,\boxed{ G\,\vev_{27} }  ~~\end{pmatrix}\;,
\end{equation}
which only has a single block under the action of the outer automorphisms. A representative of this class 
of stationary points is given by
\begin{equation}\label{eq:Block_representants27}
 \vev_{27}~=~ v_{27}\begin{pmatrix} 0 \\ -\I \\ +\I \end{pmatrix}\;.
\end{equation}
Here, $v_{27}$ is a free parameter which can again be fixed by plugging $\vev_{27}$ into the gradient of the potential, resulting
in \eqref{eq:SaddleDepth}. 

Therefore, all VEVs have been found and the result is completely in agreement with the analytical minimization of the potential.
This completes the computation of VEVs by means of the outer automorphism group.

Some comments are in order. 
There are other subgroups of $E$ which also solve \eqref{eq:consistency}.
However, as mentioned above, all of them are also subgroups of the maximally allowed subgroups $\mathrm{SG}(18,4)$, $\mathrm{SG}(18,3)$, and $\mathrm{SG}(48,29)$.
Therefore, the corresponding \Eqref{eq:consistency} can only be equally or less restrictive on the possible form of \VEV. 
Indeed, by performing the explicit computations, one finds that the form of \VEV is less constrained if and only if the corresponding subgroup $E_\vev$ 
is in the intersection of two or more of the maximally allowed subgroups of $E$ stated above.
If, instead, the subgroup is only contained in one of the maximally allowed subgroups, then
the solution for \VEV derived from $E_\vev$ is identical to the solution of the parent group, implying that any $\vev\in\VEV$ automatically conserves the parent group.
The fact that solutions for subgroups in the intersection of the maximally allowed subgroups are less constrained must, of course, be expected 
due to the fact that the corresponding solution $\VEV_\text{sub}$ has to accommodate all 
otherwise mutually exclusive solutions $\VEV_\text{parent}$ by fixing additional free parameters. 

Due to the upper bound on the total number of stationary points, one can be sure that there 
are no stationary points with trivial stabilizer. Thus, after having scanned over all non--trivial subgroups of $E$ 
one can be sure that all possible VEVs have been found. Interestingly, the \Dff Higgs potential allows for all of the stationary 
points which conserve the maximally allowed subgroups of $E$. 
Note that the method presented here does not provide an explanation for \textit{why} the potential realizes all those VEVs.

Nevertheless, one may speculate that the information on the available outer automorphism group,
which, in fact, also contains information on the representation content, is already enough to uniquely determine the VEVs in general.
This conjecture is supported by recalling that all, the invariants in the potential, the parameters of the potential,
as well as the stationary points transform in the same representation, namely as a \rep{4}, under the outer automorphism group \Sfour (cf.\ \Secref{sec:OutAction}).
This \rep{4}--plet structure is a bit obscured here, due to the fact that the inner automorphism have explicitly been taken into account.
Nevertheless, note that the \mbox{$\rep{72}_1$--plet} $\VEV_{\rep{72}}$ 
(and equivalently $\rep{72}_2$) of $E$ decomposes under $G$ as $\rep{72}_1=\rep{18}_1\oplus\rep{18}_2\oplus\rep{18}_3\oplus\rep{18}_4$.
Here, the $\rep{18}_\ell$ correspond to the $G$ orbits, while the set of $\rep{18}$--plets transforms as a $\rep{4}$--plet under the action of the outer automorphism of $G$.

The conjecture that VEVs under outer automorphisms either transform in the same representation as the couplings or are invariant \cite{Fallbacher:2015rea} 
holds true for all cases that have been investigated. For example, it has been checked and confirmed that this method allows one to find the also VEVs of the pure \Dts potential 
without any continuous symmetries. Furthermore, applying this method to the 3HDM potential with \A4 symmetry \cite{Degee:2012sk,Ivanov:2014doa} rough bounds 
on the form of the VEVs were be obtained from the present \Z2 outer automorphism. Finally, for settings with no outer automorphism, 
as for instance in the 3HDM with $\Sfour$ symmetry, no additional constraints on the VEVs could be obtained in agreement with expectation. 
Altogether, there is confidence that the method of computing, or at least constraining, the form of stationary points by the use of outer automorphisms works in general. 
After all, it is clear that VEVs of any potential which allows for outer automorphism necessarily must be 
solutions to consistency equations analogous to \eqref{eq:consistency}.

Specifically for the example 3HDM with $\Delta(54)$ symmetry, the derived necessary conditions on the stationary points are so restrictive as to completely fix their directions and relative phases.
The geometrical phases of the VEVs, hence, can be tracked back to 
the complex CGs of the group.
This is because the corresponding \Eqref{eq:consistency} involves the representation matrices of the outer automorphisms, which themselves carry discrete complex phases from
the necessarily discrete complex phases of the CGs of $\Delta(54)$.

Even though there are presently no generally known sufficient conditions for the appearance of spontaneous geometrical CP violation,
two conditions seem to be necessary for the appearance of VEVs with calculable phases.
Firstly, it seems to be required that the VEVs only depend on a small number of potential parameters. 
This is equivalent to saying that $M$ in equation \eqref{eq:Mconsitency} should have close to maximal rank.
By this requirement it shall be guaranteed that any VEV can be brought to the form $(v,0,..,0)$ by a Higgs--basis rotation which is independent of the couplings.
Secondly, in this new basis there must be a CP transformation with fixed complex phases which is broken by this VEV.
Clearly, both of these conditions favor a large outer automorphism group. The appearance of complex entries in the representation matrices of 
the CP outer automorphism, furthermore, is deeply related to the complexity of the CGs. This last consideration, therefore, 
favors groups of type I. 

Altogether it should be noted that the understanding of spontaneous geometrical CP violation generally is not yet as 
mature as the understanding of explicit geometrical CP violation. The latter can be fully understood by the absence of a 
mutual complex conjugation outer automorphism for all representations \cite{Chen:2014tpa}.

\section{Future applications of outer automorphisms}

The concept of outer automorphisms is as general as the concept of symmetry. 
Therefore, whenever dealing with a physical setting based on a symmetry, it might be beneficial to investigate whether this setting allows for outer 
automorphisms and consider their physical implications.

In this work it has been shown that outer automorphisms correspond to mappings in the parameter space of a theory,
and that outer automorphisms can give rise to emergent symmetries. 
The first property arises from the fact that operators which are symmetry invariants generally are not
outer automorphism invariants. Therefore, operators can be classified, i.e.\ grouped into multiplets, according to their outer automorphism properties.
The second feature arises from the fact that outer automorphisms permute the one--point correlation functions.
Regarding these observations, it is not far--fetched to conjecture that outer automorphisms could have wide--ranging applications beyond 
what is considered in this thesis. Some very preliminary and naive thoughts are gathered here, as
a possible guideline for future explorations of outer automorphisms. 

The classification of invariants in the SM effective field theory (EFT) according to helicity counting 
methods \cite{Cheung:2015aba} explains certain non--renormalization features \cite{Alonso:2014rga}. 
Classifying symmetry invariant operators by helicity counting, however, is analogous to classifying them according to the Lorentz group representations of their constituents.
As these representations are permuted by outer automorphisms one ought to think that also the proposed classification of composite operators can be generalized 
according to the transformation behavior of composite operators under outer automorphisms. 
Just as in the 3HDM example, outer automorphisms can be a decisive criterion for the selection of a convenient operator basis, for example, in recent 
systematic constructions of the SM EFT \cite{Lehman:2015via,Henning:2015alf}.

From the observation that outer automorphisms give relations between the solutions of a coupled system of polynomial equations, 
it is tempting to speculate that a similar situation could arise in coupled systems of differential (renormalization group) equations.
Together with the fact that outer automorphisms are acting in the parameter space of theories,
there could be a deep relation between RGE flows, the anomalous dimension matrix, and the outer automorphism structure.
At least in the 3HDM example at hand, it is clear that the outer automorphisms define the fixed boundaries of the RGE flow.
On these grounds, it would not be surprising if the construction of RGE invariants (cf.\ e.g.\ \cite{Santamaria:1993ah, Chang:2002yr, Liu:2009vh, Chiu:2015ega, Feldmann:2015nia, Chiu:2016qra}) could be systematized 
by the use of outer automorphisms. Furthermore, it has also been noted that dilatations are outer automorphisms \cite{Buchbinder:2000cq}.
In this respect outer automorphisms also seem to be a promising tool to address formal questions in scale or conformally invariant theories \cite{Fortin:2014}.

From the fact that they permute one--point amplitudes, it seems likely that also multi--point amplitudes are related by outer automorphisms.
In this respect, reconsidering the recent developments in on--shell scattering amplitudes in the light of outer automorphisms could be a worthwhile pastime (cf.\ e.g.\ \cite{Elvang:2013cua,Dixon:2013uaa}).

\enlargethispage{1cm}
Finally, note that emergent symmetries are an active field of study in modern condensed matter physics (cf.\ e.g.\ \cite{Batista:2004, Hassan:2012, Gomes:2015eea}).
In this context, it seems to be worthwhile to further investigate the emergence of symmetries from the outer automorphism group of a model, as proposed in this work.

\chapter{Summary and conclusion}

Flavor mixing, masses, and explicit CP violation in the Standard Model all originate from the Higgs Yukawa couplings.
In addition, phases of the Yukawa couplings also enter $\bar{\theta}$, thereby relating the flavor puzzle to the strong CP problem.
This suggests that the origin of flavor and the origin of CP violation could be closely related.
Understanding the possible origin of CP violation, thus, could give invaluable hints for the understanding of the experimentally observed
pattern of parameters in the flavor sector, including the strong CP problem.

After reviewing the SM flavor puzzle and the strong CP problem, the standard definitions of C, P, and T have been recapitulated. 
It has been discussed how C and P are explicitly and maximally violated by the absence of representations, 
while CP is explicitly violated by the non--vanishing CKM phase. 
Then, outer automorphism have been introduced in a pedagogical manner, based on the discrete group $\Delta(54)$ and the compact simple Lie group \SU3.
A very general consistency condition, \Eqref{eq:consistencyEquation}, has been proposed which determines the representation matrices of automorphisms.

Following references \cite{Buchbinder:2000cq} and \cite{Grimus:1995zi} is has been shown that C, P, and T transformations 
correspond to outer automorphisms of the space--time and gauge symmetries. It has been argued that CP is the complex conjugation 
automorphism of the Lorentz group. Subsequently, a physical CP transformation has been defined as \emph{a
complex conjugation (outer) automorphisms which maps all present representations of all (space--time, local, global) symmetries to their respective complex conjugate representations}.
Whenever a group has complex representations, then a corresponding physical CP transformation necessarily is an outer automorphism.
This definition of a physical CP transformation is new, but a straightforward generalization of the definitions of CP as the contragredient automorphism in simple Lie groups \cite{Grimus:1995zi},
or as a class--inverting automorphism in finite groups \cite{Chen:2014tpa} (cf.\ also \cite{Fallus:2015}).
Any transformation which fulfills the above condition equally qualifies as a CP transformation, meaning that there is, in general, no unique CP transformation.
For example, in the 3HDM discussed in \Secref{sec:3HDMExample} there are 12 possible distinct CP transformations.
A sufficient condition for physical CP conservation is that there is a single unbroken physical CP transformation. 
Any such transformation, if conserved, causes all CP odd basis invariants to vanish. 
Conversely, to have CP violated it is enough if there is a single non--vanishing CP odd basis invariant, meaning that there 
must not be a single conserved CP transformation. 

\enlargethispage{0.5cm}
The subtle difference between generalized CP transformations in additional horizontal spaces and CP transformations as outer automorphisms has been clarified.
If there is a symmetry acting in a horizontal space, and one insists on not or only minimally extending this symmetry, then CP transformations 
in the horizontal space have to be automorphisms of that symmetry. In contrast, if the horizontal space is unconstrained or it is not insisted on 
keeping the horizontal symmetry minimal, then any generalized CP transformation can be imposed in the horizontal space. 
In general, if the corresponding CP transformation does
neither square to the identity nor to a symmetry element then the linear symmetry in the horizontal space is enhanced by requiring the CP transformation \cite{Grimus:1987kn}. 
As a consequence, mass degenerate states \cite{Weinberg:1995mt2} and CP half--odd \cite{Ivanov:2015mwl} or even more exotic eigenstates of CP appear.

While the possible automorphisms of the Poincaré group and compact semisimple Lie groups have been known for some time, 
the systematic study of automorphisms of finite (discrete) groups in a physical context is rather new \cite{Holthausen:2012dk}. 
In this work, it has been shown that discrete groups can be classified into three disjoint types according to their automorphism properties \cite{Chen:2014tpa}.
Systematically, this classification can be done via the twisted Frobenius--Schur indicator \eqref{eq:FSI}.

Groups of type~II~A have a Bickerstaff--Damhus automorphism \cite{Bickerstaff:1985jc}, which is a class--inverting, involutory automorphism 
with only symmetric representation matrices. The BDA simultaneously maps every representation to its own
complex conjugate and squares to the identity. Therefore, the BDA corresponds to a model independent physical CP transformation which squares to the identity.
Groups of type~II~A, hence, most closely resemble the case of semisimple Lie groups: 
It is always possible to find a physical CP transformation, and this transformation is broken explicitly if and only if there are 
complex phases which cannot be absorbed by a rephasing of fields. 
Due to the fact that the corresponding CPV phases have to be determined by experiment, 
this way of explicit CP violation can never be predictive.
A group is of type~II~A if and only if it allows for a basis with purely real Clebsch--Gordan coefficients. This basis is also the CP basis.

Groups of type~II~B do have a class--inverting automorphism which suits to define a physical CP transformation. 
Nevertheless, for generic settings this automorphism can never be represented by only symmetric matrices and, therefore, never squares to the identity for explicit representations.
CP violation then can be tied to the presence of certain operators which are charged under the additionally appearing linear symmetry. 
Furthermore, CP half--odd or even more exotic states necessarily appear in generic models based on type~II~B groups.
Type~II~B groups do not allow for a basis with real Clebsch--Gordan coefficients.

Presumably the most interesting category of discrete groups are those of type~I. 
These groups do not allow for automorphisms which simultaneously map 
all representations to their complex conjugate representations. Therefore, if there are sufficiently many different representations present, 
then CP is violated as a consequence of the type~I group, simply because there is no possible physical CP transformation.
This has been demonstrated by the computation of a decay asymmetry in an explicit example model based on the group $\Delta(27)$. 
CP violation in groups of type~I can be tracked back to the necessarily complex Clebsch--Gordan coefficients, which enter
CP odd basis invariants in the form of CP violating weak phases \cite{Chen:2014tpa}.
The existence of this type of CP violation has first been conjectured in \cite{Chen:2009gf}, 
and it has been termed explicit geometrical CP violation in \cite{Branco:2015hea}.
Necessary and sufficient criteria for the occurrence of explicit geometrical CP violation, together with an explicit example model, have firstly been
discussed in \cite{Chen:2014tpa}.

The physical relevance of CP outer automorphisms also motivates the general study of outer automorphisms beyond C, P, or T. 
It has been argued that there are only two possibilities for the action of a general outer automorphism.
Either, the outer automorphism is broken explicitly and maximally by the absence of representations. Or, the
outer automorphism is broken explicitly by the values of already present couplings. This allows for a clear 
and precise definition of maximal breaking of a transformation (in agreement with C and P in the SM). 
According to this notion of maximal breaking, CP can never be maximally broken.
Studying non--maximally broken outer automorphisms in general, it has been argued that these transformations allow to identify 
physically equivalent regions in the parameter space of models. 
Furthermore, it has been shown that VEVs form orbits under the outer automorphism group, 
just as they do under the coset of broken symmetry transformations.
Consequently, VEVs are generally invariant under a subset of outer automorphism, meaning that the
vacuum of theories with outer automorphisms can be more symmetric than the Lagrangian itself. 
This is a novel mechanism for the origin of emergent symmetries. 

A three Higgs doublet model with $\Delta(54)$ symmetry has been presented as an example for a model with
a rich outer automorphism structure. It has been explicitly shown how the outer automorphisms 
permute couplings, and thereby allow to identify physically equivalent regions in the parameter space. 
Furthermore, also the stationary points are permuted under the action of outer automorphisms. 
This has been used in order to set up necessary conditions on the VEVs which constrain their relative directions including 
the relative phases. In the example model, these constraints are so restrictive as to completely fix the directions and phases of VEVs, 
thereby providing a reason for spontaneous geometrical CP violation in this model \cite{Branco:1983tn}.

In summary, in this thesis, the notion of CP transformations as particular outer automorphism has been introduced in a coherent way.
This insight has been used in order to demonstrate that certain discrete groups allow to predict geometrical CP violating phases from group theory.
Even though a fully realistic model has not been fleshed out, this work paves the way for model building with predictive CP violation.
This could be intimately related to a solution of the flavor puzzle, the origin of the baryon asymmetry, 
or towards understanding the microscopic arrow of time.
In addition, outer automorphisms beyond C, P, or T have firstly been studied.
Besides simplifying the computation of stationary points, it has been demonstrated that outer automorphisms are conceptually and phenomenologically relevant 
to understand emergent symmetry including the origin of spontaneous geometrical CP violation. 
As a consequence of this work, further studies of outer automorphisms are appreciable and worthwhile. 

\appendix

\chapter{Dirac spinor representation matrices}
\label{app:DiracMatrices}

For completeness, this appendix lists the explicit form of matrices for the Dirac spinor representation of the Lorentz group.
The notation of \cite{Srednicki_2007} is followed loosely, all statements are made in the so--called chiral or Weyl basis.

A generic Dirac spinor is given by 
\begin{equation}
\Psi~:=~\begin{pmatrix} \chi_a \\ \xi^{\dagger\dot{a}} \end{pmatrix}\;,
\end{equation}
with a left--handed Weyl spinor $\chi_a$ and a right handed Weyl spinor $\xi^{\dagger\dot{a}}$ $(a,\dot{a}=1,2)$.

In this basis, the $\gamma$--matrices are given by
\begin{equation}
\gamma^\mu~=~\begin{pmatrix} & \sigma^\mu_{a\dot{c}}\\ \bar{\sigma}^{\mu,\dot{a}c} &  \end{pmatrix}\;,
\end{equation}
where $\sigma^\mu:=(\mathbbm{1},\vec{\sigma})$ and $\bar{\sigma}^\mu:=(\mathbbm{1},-\vec{\sigma})$, with the Pauli matrices
\begin{equation}
\vec{\sigma}~=~\begin{pmatrix} \begin{pmatrix} & 1 \\ 1 &  \end{pmatrix} , \begin{pmatrix} & -\I \\ \I &  \end{pmatrix} , \begin{pmatrix} 1 &  \\  & -1 \end{pmatrix} \end{pmatrix}\;.
\end{equation}
Furthermore, there appear the matrices  
\begin{equation}
\gamma_5~:=~\I\,\gamma^0\,\gamma^1\,\gamma^2\,\gamma^3~=~ \left(\begin{array}{cc} -\delta^{~c}_{a} & \\ & \delta^{\dot{a}}_{~\dot{c}} \end{array}\right)\;,
\end{equation}
\begin{equation}
 \mathcal{\beta}~=~\left(\begin{array}{cc} & \delta^{\dot{a}}_{~\dot{c}} \\ \delta^{~c}_{a} & \end{array}\right)\,,\quad\text{and}\quad
\mathcal{C}~=~\left(\begin{array}{cc} \varepsilon_{ac} & \\ & \varepsilon^{\dot{a}\dot{c}} \end{array}\right)~=~
\left(\begin{array}{cc} -\varepsilon^{ac} & \\ & -\varepsilon_{\dot{a}\dot{c}} \end{array}\right)\;,
\end{equation}
with $\varepsilon_{ac}$ ($\varepsilon_{12}=-1$) being the total antisymmetric tensor in two dimensions.

\chapter{Proof of the consistency condition}
\label{app:CCProof}

It shall be proven that 
\begin{equation}\label{eq:consistencyEquation2}
U\,\rho_{\rep{r'}}(\mathsf{g})\,U^{-1}~=~\rho_{\rep{r}}(u(\mathsf{g}))\;,\qquad\forall \mathsf{g}\in G\;,
\end{equation}
with unitary matrices $U$ holds for all irreps $\rep{r}$ of $G$ if and only if $u(\elm g)$ is an automorphism of the group $G$. Here it is assumed
that $\rho_{\rep{r}}(\elm g)$ are matrix representations.

For the first direction, it must be shown that \eqref{eq:consistencyEquation2} holds if $u:G\rightarrow G$ is an automorphism of the group $G$.
Therefore, assume that $u$ is an automorphism, i.e.\ $u$ is a bijective homomorphism, $u(\elm g\elm h)=u(\elm g)u(\elm h)$.
In order to deduce \eqref{eq:consistencyEquation2}, one has to show that 
\begin{equation}
\rho_{\rep{r}}(u(\mathsf{g}))~=:~\Gamma(\elm g)\;
\end{equation}
is a matrix representation itself, i.e.\ $\Gamma(\elm g)$ fulfills the group algebra. 
This is easily verified by
\begin{equation}
\Gamma(\elm{g\,h})~=~\rho_{\rep{r}}(u(\mathsf{g\,h}))~=~\rho_{\rep{r}}(u(\elm g)\,u(\elm h))~=~\rho_{\rep{r}}(u(\elm g))\,\rho_{\rep{r}}(u(\elm h))~=~\Gamma(\elm g)\,\Gamma(\elm h)\;.
\end{equation}
In the second step it has been used that $u$ is a homomorphism, and in the third step it has been used that $\rho_{\rep{r}}:G\rightarrow\mathrm{GL}(V)$ as a representation is
also a homomorphic map.
Therefore, one can write $\Gamma(\elm g)\equiv U\rho_{\rep{r'}}(\elm g)U^{-1}$, and the unitary matrices $U$ are understood as the freedom to chose a
basis for the representation $\rho_{\rep{r'}}(\elm g)$. 

To prove the reverse direction it needs to be shown that if \eqref{eq:consistencyEquation2} holds, then $u$ is an automorphism of $G$, i.e.\ a bijective homomorphism $G\to G$.
This is shown in three steps. 
First it will be shown that $u$ is injective, then it will be shown that $u$ is surjective. Lastly, the homomorphism property will be shown.
In order to show that $u$ is injective one needs to show that 
\begin{equation}
 u(\elm g)~=~u(\elm g')\quad\Rightarrow\quad\elm g~=~\elm g'\quad\forall \elm g,\elm g'\in G\;.
\end{equation}
Applying the representation map to both sides of the equation $u(\elm g)=u(\elm g')$, then applying \eqref{eq:consistencyEquation2} for 
$\elm g$ and for $\elm g'$ on the two sides, one obtains
\begin{equation}\label{eq:injective}
\rho_{\rep{r'}}(\mathsf{g})~=~\rho_{\rep{r'}}(\mathsf{g'})\;.
\end{equation}
Since \eqref{eq:consistencyEquation2} is assumed to hold for all irreps of $G$ it will hold for at least one faithful representation.
It is enough that there is at least one faithful irrep $\rep{r'}$ for which the representation map $\rho_{\rep{r'}}$ in \eqref{eq:injective} can be inverted. 
Performing the inversion on both sides results in $\elm g=\elm g'$ thereby proving that $u$ is injective. 
In order to prove that $u$ is surjective one needs to show that
\begin{equation}
\forall\elm g\in G~\exists~\elm g'\in G:u(\elm g')~=~\elm g\;.
\end{equation}
This is most easily shown by constructing the element $\elm g'$ explicitly. 
Taking \eqref{eq:consistencyEquation2} for $\elm g'$ and a faithful (therefore, invertible) representation and
requiring that $u(\elm g')=\elm g$ one finds that 
\begin{equation}
\elm g'=~\rho^{-1}_{\rep{r'}}\left(U^{-1}\,\rho_{\rep{r}}(\mathsf{g})\,U\right)\;.
\end{equation}
Lastly, the homomorphism property can be shown to be fulfilled by taking
\begin{equation}
\begin{split}
\rho_{\rep{r}}(u(\mathsf{g\,h}))~=&~U\,\rho_{\rep{r'}}(\elm g\,\elm h)\,U^{-1}~=~U\,\rho_{\rep{r'}}(\elm g)\,\rho_{\rep{r'}}(\elm h)\,U^{-1}~=~\\
~=&~U\,\rho_{\rep{r'}}(\elm g)\,U^{-1}\,U\,\rho_{\rep{r'}}(\elm h)\,U^{-1}~=~\rho_{\rep{r}}(u(\elm g))\,\rho_{\rep{r}}(u(\elm h))~=~ \\
~=&~\rho_{\rep{r}}(u(\elm g)\,u(\elm h))\;.
\end{split}
\end{equation}
Taken again for a faithful representation, it is possible to invert $\rho_{\rep{r}}$ on both sides thereby showing that $u$ is a 
homomorphic map. Together this shows that $u$ is a bijective homomorphic map $G\rightarrow G$ and, therefore, an automorphism. \hfill {\large\changefont{pzc}{m}{n} q.e.d.}

\chapter{Computer codes}

\section{Outer automorphism structure of finite groups}
\label{app:Out}

A \textsc{GAP} \cite{GAP4} code to compute the automorphism group (\texttt{AutG}) of a finite group \texttt{G} (for example $\Delta(54)$), as well as its inner (\texttt{InnG}) and outer 
automorphism group (\texttt{OutG}) is given by:
\begin{verbatim}
  G:=SmallGroup(54,8);;
  AutG:=AutomorphismGroup(G);;
  InnG:=InnerAutomorphismsAutomorphismGroup(AutG);;
  OutG:=AutG/InnG;;
  IdGroup(OutG);
  StructureDescription(OutG);
\end{verbatim}

\section{Twisted Frobenius--Schur indicator}% with \textsc{GAP}}
\label{app:FSI}

A \textsc{GAP} code which computes the (first) twisted
Frobenius--Schur indicator (\FSI) for all irreps with respect to a given automorphism \texttt{aut} of 
of a finite group \texttt{G} is given by:
\begin{verbatim}
  twistedFS:=function(G,aut)
    local elG,tbl,irr,fsList;
    elG:=Elements(G);
    tbl:=CharacterTable(G);
    irr:=Irr(tbl);
    fsList:=List(elG,x->x*x^aut);
    return List(irr,y->Sum(fsList,x->x^y))/Size(G);
  end;
\end{verbatim}
For example, to print out all \FSI's for the group \Tprime ($\mathrm{SG}(24,3)$) one can use:
\begin{verbatim}
  G:=SmallGroup(24,3);;
  autG:=AutomorphismGroup(G);;
  elAutG:=Elements(autG);;
  for i in elAutG do Print(twistedFS(G,i)); od;
\end{verbatim}
The fact that there is an automorphism $u$ with $\FSI(\rep{r})=1$ for all irreps shows that \Tprime
is of type~II. For a code to compute the $n^\mathrm{th}$ twisted \FSI see \cite{Fallus:2015}.

\section{Permutation representations}
\label{App:VEVCalculation}

Assume $\Equi$ is a group with a subgroup $\Equi_\vev$. The action of the group via left--multiplication on the coset $E/E_\vev$ 
defines a permutation representation. The following GAP code computes the explicit permutation matrix $(\Pi_{\mathsf{P}})^{-1}$ of a group element 
$\mathsf{P}\in E$ in this representation:

\small
\begin{verbatim}
action:=ActionHomomorphism(E,RightCosets(E,E_lambda),OnRight);;
Pi_P_inverse:=Image(action,P);
\end{verbatim}
\normalsize

For the group $E=\mathrm{SG}(1296,2891)$ and the subgroup $E_\vev=\mathrm{SG}(18,4)$ this permutation representation 
is denoted by $\rep{72}_1$ and a minimal generating (in cycles) given by
\begin{equation}
\begin{split}
 (\Pi_{\mathsf{P}}^{\rep{72}_1})^{-1}~:=&~(2,9,5)(4,13,33)(6,17,14)(7,19,15)(8,22,47)(10,26,24)(11,28,53) \\ 
 &~ (12,31,55)(16,38,58)(18,40,59)(20,42,60)(21,44,62)(23,46,65) \\ 
 &~ (25,49,66)(27,51,67)(29,45,64)(30,54,69)(32,57,35)(37,52,68) \\
 &~ (43,61,71)(48,56,70)(50,63,72)\;, \\
 (\Pi_{\mathsf{Q}}^{\rep{72}_1})^{-1}~:=&~(1,59, 8,56,26,72,37,44)( 2,33,18,63,36,68,30,46) \\ 
 &~ ( 3,60,25,38, 9,64,35,61)( 4,42,19,70,11,40,17,58) \\
 &~ ( 5,62,12,28,41,67,48,22)(6,65,29,52, 7,69,50,32) \\
 &~ (10,53,20,45,34,57,23,54)(13,39,55,16,49,24,71,27) \\
 &~ (14,47,21,51,15,66,43,31)\;.
\end{split}
\end{equation}
For the subgroup $E_\vev=\mathrm{SG}(18,3)$ the permutation representation 
is denoted by $\rep{72}_2$ and a minimal generating set is given by
\begin{equation}
\begin{split}
 (\Pi_{\mathsf{P}}^{\rep{72}_2})^{-1}~:=&~(2,9,5)(4,13,33)( 6,17,14)( 7,19,15)( 8,22,47)(10,26,24)(11,28,54)\\ 
 &~(12,31,55)(16,38,58)(18,40,59)(20,42,60)(21,44,63)(23,48,66)\\ 
 &~(25,50,67)(27,52,68)(29,45,64)(30,46,65)(32,57,35)(37,53,69)\\ 
 &~(43,61,71)(49,56,70)(51,62,72)\;, \\
 (\Pi_{\mathsf{Q}}^{\rep{72}_2})^{-1}~:=&~( 1,58,29,53,23,54,42,18)( 2,63,22,55,33,62,31,26)\\
 &~( 3,47,17,52, 4,71,37,49)( 5,60,21,68,24,38,27,61)\\
 &~( 6,56,13,57,15,64,48,41)( 7,59,30,36, 8,72,35,32)\\
 &~( 9,50,12,70,34,46,20,51)(10,40,25,44,14,67,45,43)\\
 &~(11,66,19,65,28,69,16,39)\;.
\end{split}
\end{equation}
For the subgroup $E_\vev=\mathrm{SG}(48,29)$ the permutation representation 
is denoted by $\rep{27}$ and has a minimal generating set
\begin{equation}
\begin{split}
 (\Pi_{\mathsf{P}}^{\rep{27}})^{-1}~:=&~( 1, 4,12)( 2, 8,20)( 3, 9,21)( 5,15,24)( 6,16,11) \\
 &~ ( 7,17,19)(10,23,25)(13,27,14)(18,22,26)\;, \\
 (\Pi_{\mathsf{Q}}^{\rep{27}})^{-1}~:=&~( 1, 8,25,24,14,27,26,13)( 2,22,15, 9,11,19,16, 4) \\
 &~( 3,10,21,12,20,17, 5, 6)(18,23)\;.
\end{split}
\end{equation}

\chapter{Group theory}

\section{On the group \SU3}
\label{app:SU3}

All elements $A\in\SU3$ can be written in the form $A=\exp[\I\, \theta_a T_a]$ where
$\theta_a$ $(a=1,..,8)$ are real parameters and $T_a$ are the eight $3\times3$ traceless matrix generators of the group.
A conventional basis choice for the generators of \SU3 is given by $T_a=\lambda_a/2$ with the Hermitian Gell--Mann matrices \cite{GellMann:1962xb} 
\begin{equation}
\begin{split}\label{eq:SU3GensGM}
\lambda_1~&=~\begin{pmatrix} 0 & 1 & 0 \\ 1 & 0 & 0  \\ 0 & 0 & 0 \end{pmatrix},&
\lambda_2~&=~\begin{pmatrix} 0 & -\I & 0 \\ \I & 0 & 0  \\ 0 & 0 & 0 \end{pmatrix},&
\lambda_3~&=~\begin{pmatrix} 1 & 0 & 0 \\ 0 & -1 & 0  \\ 0 & 0 & 0 \end{pmatrix}, \\
\lambda_4~&=~\begin{pmatrix} 0 & 0 & 1 \\ 0 & 0 & 0  \\ 1 & 0 & 0 \end{pmatrix},&
\lambda_5~&=~\begin{pmatrix} 0 & 0 & -\I \\ 0 & 0 & 0  \\ \I & 0 & 0 \end{pmatrix},&
\lambda_6~&=~\begin{pmatrix} 0 & 0 & 0 \\ 0 & 0 & 1  \\ 0 & 1 & 0 \end{pmatrix}, \\
\lambda_7~&=~\begin{pmatrix} 0 & 0 & 0 \\ 0 & 0 & -\I  \\ 0 & \I & 0 \end{pmatrix},&
\lambda_8~&=~\frac{1}{\sqrt{3}}\begin{pmatrix} 1 & 0 & 0 \\ 0 & 1 & 0  \\ 0 & 0 & -2 \end{pmatrix}.
\end{split}
\end{equation}
The non--zero structure constants in the basis $T_a=\lambda_a/2$ are given by
\begin{equation}
\begin{split}
 f^{\mathrm{GM}}_{123}~&=~1\,,\quad\,f^{\mathrm{GM}}_{147}~=~f^{\mathrm{GM}}_{165}~=~f^{\mathrm{GM}}_{246}~=~f^{\mathrm{GM}}_{257}~=~f^{\mathrm{GM}}_{345}~=~f^{\mathrm{GM}}_{376}~=\frac12\,,\\
 f^{\mathrm{GM}}_{458}~&=~f^{\mathrm{GM}}_{678}~=~\frac{\sqrt{3}}{2}\,,
\end{split}
\end{equation}
and all other $f^{\mathrm{GM}}_{abc}$ which can be obtained from these by a (completely anti--symmetric) permutations of the indices.\enlargethispage{0.5cm}

The generators used in \Secref{sec:SU3example} are obtained from the Gell--Mann matrices as
\begin{equation}\label{eq:basisChange}
\begin{split}
 \HI~&=~\frac12\lambda_3\,,&\quad\HY~&=~\frac12\lambda_8\,,& \\
 \EEpm~&=~\frac{1}{2\sqrt{2}}\left(\lambda_1\pm\I\,\lambda_2\right)\,&\quad
 \ETpm~&=~\frac{1}{2\sqrt{2}}\left(\lambda_4\pm\I\,\lambda_5\right)\,&\quad
 \EZpm~&=~\frac{1}{2\sqrt{2}}\left(\lambda_6\pm\I\,\lambda_7\right)\,.
\end{split}
\end{equation}
The non--zero structure constants $f^{c}_{ab}$ in this basis are given by
\begin{equation}
\begin{split}
 f^1_{13}~&=f^2_{32}~=~f^3_{21}~=~\I\,, \\
 f^1_{74}~&=f^2_{56}~=~f^4_{61}~=~f^5_{27}~=~f^6_{62}~=~f^7_{15}~=~\frac{\I}{\sqrt{2}}\,,\\
 f^3_{54}~&=f^3_{67}~=~f^4_{43}~=~f^5_{35}~=~f^6_{36}~=~f^7_{73}~=~\frac{\I}{2}\,,\\
 f^8_{54}~&=f^8_{76}~=~f^4_{48}~=~f^5_{85}~=~f^6_{68}~=~f^7_{87}~=~\I\frac{\sqrt{3}}{2}\,.\\
\end{split}
\end{equation}
This is not an orthonormal basis and the structure constants are anti--symmetric 
only w.r.t.\ permutations of the lower two indices.

The outer automorphism $u_\Delta$ (as defined in \eqref{eq:SU3Out}) 
in the Gell--Mann basis acts as
\begin{equation}
R^{\mathrm{GM}}_\Delta~=~\diag(-1,+1,-1,-1,+1,-1,+1,-1)
\end{equation}
in the adjoint space. The consistency condition, therefore, simply reads  
\begin{equation}
-\lambda_a^\mathrm{T}~=~\eta_a\,\lambda_a\;,
\end{equation}
with $\eta_{2,5,7}=1$ for the complex, and $\eta_{1,3,4,6,8}=-1$ for the real symmetric Gell--Mann matrices.
This is precisely what is naively expected from a transformations which maps $A=\exp[\I\, \theta_a \lambda_a/2]\mapsto A^*$.
The Gell--Mann basis as well as the non--orthogonal bases for the generators are both CP bases because $U=\mathbbm{1}$.

\section{On the group \texorpdfstring{$\Sigma(72)$}{}}
\label{app:Sigma72}

The group $\Sigma(72)$ is contained in the SmallGroups library of \textsc{GAP} as $\mathrm{SG}(72,41)$.
A possible minimal generating set and the corresponding presentation have been given in \eqref{eq:Sigma72Presentation}.
The character table of the group is shown in \Tabref{tab:Sigma72char}.

Explicit matrix representations for the generators for the two--dimensional representation can be chosen as 
\begin{equation}
  M_{\rep{2}} ~=~ \begin{pmatrix}0 & 1\\ -1 & 0 \end{pmatrix}\;, \quad\text{and}\quad
  P_{\rep{2}} ~=~ \begin{pmatrix}-\I & 0\\ 0 & \I \end{pmatrix}\;,
\end{equation}
and the generators of the eight--dimensional representation can be chosen as
\begin{align}\notag
  M_{\rep{8}} ~=&~ \begin{pmatrix} 
    0 & 0 & 1 & 0 & 0 & 0 & 0 & 0 \\
    0 & 0 & 0 & 1 & 0 & 0 & 0 & 0 \\
    1 & 0 & 0 & 0 & 0 & 0 & 0 & 0 \\
    0 & -1 & 0 & 0 & 0 & 0 & 0 & 0 \\
    0 & 0 & 0 & 0 & 0 & 0 & 1 & 0 \\
    0 & 0 & 0 & 0 & 0 & 0 & 0 & 1 \\
    0 & 0 & 0 & 0 & 1 & 0 & 0 & 0 \\
    0 & 0 & 0 & 0 & 0 & -1 & 0 & 0
  \end{pmatrix}\;, \quad \text{and}~ \\
  P_{\rep{8}} ~=&~ \frac{1}{2}\,\begin{pmatrix}
    0 & 0 & 0 & 0 & -1 &  -\sqrt{3} & 0 & 0 \\
    0 & 0 & 0 & 0 & \sqrt{3} &  -1 & 0 & 0 \\
    0 & 0 & 0 & 0 & 0 & 0 & -1 &  \sqrt{3} \\ 
    0 & 0 & 0 & 0 & 0 & 0 & \sqrt{3} &  1 \\
    2 & 0 & 0 & 0 & 0 & 0 & 0 & 0 \\
    0 & -2 & 0 & 0 & 0 & 0 & 0 & 0 \\
    0 & 0 &  -1 &  -\sqrt{3} & 0 & 0 & 0 & 0 \\
    0 & 0 &  \sqrt{3} & -1 & 0 & 0 & 0 & 0
  \end{pmatrix}\;.
\end{align}
Some more details on $\Sigma(72)$ can be found in \cite{Damhus:1981} and \cite{Chen:2014tpa} where, however, 
a different (non--minimal) presentation has been used.
The relation between the generators $\elm{M,P}$ used in this work and the generators $\elm{m,n,p}$ which are used in \cite{Chen:2014tpa} is given by
\begin{align}
 \elm M~&=~\elm m\;,& \elm P~&=~\elm m\,\elm n\,\elm p^2\,\elm m\;,& \\
 \elm m~&=~\elm M\;,& \elm n~&=~\elm P\,\elm M\,\elm P^2\,\elm M\;,& \elm p~&=~\elm M^2\,\elm P^2\;.&
\end{align}

The non--trivial direct product rules for the $\Sigma(72)$ irreps are
\begin{align}
\rep{2}\otimes\rep{2}~&=~\rep[_0]{1}\oplus\rep[_1]{1}\oplus\rep[_2]{1}\oplus\rep[_3]{1}\;, \\
\rep{2}\otimes\rep{8}~&=~\rep[_1]{8}\oplus\rep[_2]{8}\;, \\
\rep{8}\otimes\rep{8}~&=~\rep[_0]{1}\oplus\rep[_1]{1}\oplus\rep[_2]{1}\oplus\rep[_3]{1}\oplus\rep[_1]{2}\oplus\rep[_2]{2}\oplus\rep[_1]{8}\oplus
\rep[_2]{8}\oplus\rep[_3]{8}\oplus\rep[_4]{8}\oplus\rep[_5]{8}\oplus\rep[_6]{8}\oplus\rep[_7]{8}\;,
\end{align}
The subscript on the resulting higher--dimensional representations corresponds the fact that even though they transform like the 
elementary irreps \rep{8} and \rep{2} under $\Sigma(72)$,
the composite representations \rep[_i]{8} and \rep[_i]{2} can be distinguished by their different 
transformation behavior under the outer automorphisms of the group.

The Clebsch--Gordan coefficients needed for the discussion in this work are given by
\begin{align}\notag
  \left(x_{\rep{2}}\otimes y_{\rep{2}}\right)_{\rep[_0]{1}} & ~=~ \frac{1}{\sqrt{2}}\left(x_1\, y_2 - x_2\, y_1 \right)\;,\\ \notag
	\left(x_{\rep{8}}\otimes y_{\rep{8}}\right)_{\rep{2}^1} & ~=~ \frac{1}{2}\,
	\begin{pmatrix}\I\, x_2\, y_1 - \I\, x_1\, y_2 - x_6\, y_5 + x_5\, y_6 \\ \I\, x_4\, y_3 - \I\, x_3\, y_4 - x_8\, y_7 + x_7\, y_8\end{pmatrix}\;,\\
  \left(x_{\rep{8}}\otimes y_{\rep{8}}\right)_{\rep{2}^2} & ~=~ \frac{1}{2}\,
  \begin{pmatrix}\I\, x_4\, y_3 - \I\, x_3\, y_4 + x_8\, y_7 - x_7\, y_8 \\ -\I\, x_2\, y_1 + \I\, x_1\, y_2 - x_6\, y_5 + x_5\, y_6\end{pmatrix}\;.
\end{align}
Other CGs can be found in \cite{Chen:2014tpa}.

\section{On the group \texorpdfstring{$\Delta(27)$}{}}
\label{app:Delta27}

The group $\Delta(27)$ is listed in the \textsc{GAP} SmallGroups library as $\mathrm{SG(27,3)}$. It can be presented by the two generators $\elm A$ and $\elm B$ fulfilling \eqref{eq:Delta27Presentation}.
The character table is given in \ref{tab:Delta27char}.
\begin{table}[t!]
\centering
\resizebox{\textwidth}{!}{\begin{tabular}{c|ccccccccccc|}
                      &  $C_{1a}$ & $C_{3a}$ & $C_{3b}$ & $C_{3c}$ & $C_{3d}$ & $C_{3e}$ & $C_{3f}$ & $C_{3g}$ & $C_{3h}$ & $C_{3i}$ & $C_{3j}$ \\
                      &  1 &  3 &  3 &  3 &  3 &  3 &  3 &  3 &   3 &  1 &   1 \\
$\Delta(27)$       & $e$ & $A$ & $A^2$ & $B$    & $B^2$      & $ABA$      & $BAB$      & $AB$       & $A^2B^2$   & $AB^2ABA$ & $BA^2BAB$ \\
\hline
 $\rep[_0]{1}$ & $1$ & $1$        & $1$        & $1$        & $1$        & $1$        & $1$        & $1$        & $1$        & $1$       & $1$ \\
 $\rep[_1]{1}$ & $1$ & $1$        & $1$        & $\omega^2\hspace{-5pt}$ & $\omega$   & $\omega^2\hspace{-5pt}$ & $\omega$   & $\omega^2\hspace{-5pt}$ & $\omega$   & $1$       & $1$ \\
 $\rep[_2]{1}$ & $1$ & $1$        & $1$        & $\omega$   & $\omega^2\hspace{-5pt}$ & $\omega$   & $\omega^2\hspace{-5pt}$ & $\omega$   & $\omega^2\hspace{-5pt}$ & $1$       & $1$ \\
 $\rep[_3]{1}$ & $1$ & $\omega^2\hspace{-5pt}$ & $\omega$   & $1$        & $1$        & $\omega$   & $\omega^2\hspace{-5pt}$ & $\omega^2\hspace{-5pt}$ & $\omega$   & $1$       & $1$ \\
 $\rep[_4]{1}$ & $1$ & $\omega^2\hspace{-5pt}$ & $\omega$   & $\omega^2\hspace{-5pt}$ & $\omega$   & $1$        & $1$        & $\omega$   & $\omega^2\hspace{-5pt}$ & $1$       & $1$ \\
 $\rep[_5]{1}$ & $1$ & $\omega^2\hspace{-5pt}$ & $\omega$   & $\omega$   & $\omega^2\hspace{-5pt}$ & $\omega^2\hspace{-5pt}$ & $\omega$   & $1$        & $1$        & $1$       & $1$ \\
 $\rep[_6]{1}$ & $1$ & $\omega$   & $\omega^2\hspace{-5pt}$ & $1$        & $1$        & $\omega^2\hspace{-5pt}$ & $\omega$   & $\omega$   & $\omega^2\hspace{-5pt}$ & $1$       & $1$ \\
 $\rep[_7]{1}$ & $1$ & $\omega$   & $\omega^2\hspace{-5pt}$ & $\omega^2\hspace{-5pt}$ & $\omega$   & $\omega$   & $\omega^2\hspace{-5pt}$ & $1$        & $1$        & $1$       & $1$ \\
 $\rep[_8]{1}$ & $1$ & $\omega$   & $\omega^2\hspace{-5pt}$ & $\omega$   & $\omega^2\hspace{-5pt}$ & $1$        & $1$        & $\omega^2\hspace{-5pt}$ & $\omega$   & $1$       & $1$ \\
 $\rep{3}$       & $3$ &  $0$   & $0$ & $0$ & $0$ & $0$ & $0$ & $0$ & $0$ & $3\omega$ & $3\omega^2\hspace{-5pt}$   \\
 $\rep{\bar{3}}$ & $3$ &  $0$   & $0$ & $0$ & $0$ & $0$ & $0$ & $0$ & $0$ & $3\omega^2\hspace{-5pt}$   & $3\omega$ \\
\hline
\end{tabular}}
\caption{The character table of $\Delta(27)$. As usual, $\omega=\mathrm{e}^{2\pi\,\I/3}$. 
The second line gives the cardinality of the conjugacy class (c.c.)\ and
the third line gives a representative of the corresponding c.c.\ in the presentation specified
in \eqref{eq:Delta27Presentation}. An error appearing in the last two columns of the analogous table in \cite{Chen:2014tpa} has been corrected.}
\label{tab:Delta27char}
\end{table}

By computing the \FSI's for all possible automorphisms, it is readily confirmed that 
\Dts is of type~I, because it does not allow for any class--inverting automorphism. 
Therefore, $\Delta(27)$ does in general not allow for a consistent model independent physical CP transformation. 
However, since one can find outer automorphisms 
which simultaneously map the triplet and at most two non--trivial
one--dimensional representations to their respective complex conjugate,
consistent CP transformations are possible in non--generic models with such a constrained field content.

The outer automorphism group of $\Delta(27)$ can be generated by the operations 
\begin{equation}\label{eq:OutDelta27Gens}
s:~(\mathsf{A,B})\mapsto(\mathsf{AB^2A,B})\quad\text{and}\quad t:~(\mathsf{A,B})\mapsto(\mathsf{A,ABA})\,,
\end{equation}
and is $\mathrm{GL}(2,3)$, a group of order $48$. Note the striking similarity to \eqref{eq:SfourAction}.
Obviously, there is a close relation between the groups \Dts and \Dff. 
Namely, \Dts can be extended by the outer automorphism $c:=\left(t^2\circ s\right)^4\equiv\mathrm{conj}(\elm C)$ which acts as
\begin{equation}
 c:~(\mathsf{A,B})\mapsto(\mathsf{A^2,B^2})\;,
\end{equation}
resulting in the group \Dff (cf.\ the action of $\elm C$ in the explicit presentation of \Dff in \eqref{eq:DffPrsentation}).

An explicit matrix representation of the triplet of \Dts is given by the matrices $A$ and $B$ in \Eqref{eq:delta54gens}.
The explicit action of $c$ on the triplet representations of \Dts is given by 
\begin{equation}
 \rep{3}~\mapsto~C\,\rep{3}\;,~\quad\bar{\rep{3}}~\mapsto~C\,\bar{\rep{3}}\;,
\end{equation}
where the matrix $C$ has been stated already in \eqref{eq:delta54gens}. Furthermore, 
the action of $c$ exchanges all mutually complex conjugate one--dimensional representations of \Dts,
thereby combining them to the real doublet representations of \Dff as 
$\rep[_1]{2}=(\rep[_1]{1},\rep[_2]{1})$, $\rep[_2]{2}=(\rep[_3]{1},\rep[_6]{1})$, $\rep[_3]{2}=(\rep[_4]{1},\rep[_8]{1})$, and $\rep[_4]{2}=(\rep[_5]{1},\rep[_7]{1})$.
Therefore, $\Dts$ is a normal subgroup of \Dff and all outer automorphisms of \Dts are also available at the level of \Dff, 
where $c$ becomes an inner automorphisms.

Another outer automorphism used in this work is $w\equiv t^2\circ s\circ t$ which acts as
\begin{equation}
w : (A,B)\,\rightarrow\,(BAB, B^2) \curvearrowright\, \rep[_1]{1}\leftrightarrow\rep[_8]{1}\;,\;\rep[_2]{1}\leftrightarrow\rep[_4]{1}\;,\;\rep[_5]{1}\leftrightarrow\rep[_7]{1}\;,\;\rep3\rightarrow U_{w}\,\rep3^*\;,
\end{equation}
with the explicit representation matrix for the triplets given by
\begin{equation}\label{eq:Uw}
  U_{w} ~=~ \begin{pmatrix}-\omega^2 & 0 & 0\\ 0 & -\omega & 0 \\ 0 & 0 & -\omega \end{pmatrix}\;.
\end{equation}
Note that $w$ has been called $u_3$ in \cite{Chen:2014tpa}.

The only non--trivial direct product of \Dts irreps relevant for this work is
\begin{equation}
 x_{\rep{3}}\otimes \bar{y}_{\rep{\bar3}}~=~\sum^9_{i=1}\,\rep1_i\;.
\end{equation}
The corresponding CGs in the basis \eqref{eq:delta54gens} are
\begin{subequations}
\begin{align}
 \rep1_0 & ~=~\frac{\left(x_{1}\,\bar{y}_{1}+        x_{2}\,\bar{y}_{2}+         x_{3}\,\bar{y}_{3}\right)}{\sqrt{3}}\;,
 \\ \notag
 \rep1_1 & ~=~\frac{\left(x_{1}\,\bar{y}_{2}+        x_{2}\,\bar{y}_{3}+        
 x_{3}\,\bar{y}_{1}\right)}{\sqrt{3}}\;,&
 \rep1_2 & ~=~\frac{\left(x_{2}\,\bar{y}_{1}+        x_{3}\,\bar{y}_{2}+         x_{1}\,\bar{y}_{3}\right)}{\sqrt{3}}\;,
 \\ \notag
 \rep1_3 & ~=~\frac{\left(x_{1}\,\bar{y}_{1}+\omega\,  x_{2}\,\bar{y}_{2}
 +\omega^2\,x_{3}\,\bar{y}_{3}\right)}{\sqrt{3}}\;,&   
 \rep1_6 & ~=~\frac{\left(x_{1}\,\bar{y}_{1}+\omega^2\,x_{2}\,\bar{y}_{2}+\omega\,  
 x_{3}\,\bar{y}_{3}\right)}{\sqrt{3}}\;,
 \\ \notag
 \rep1_4 & ~=~\frac{\left(x_{2}\,\bar{y}_{3}+\omega\,  x_{3}\,\bar{y}_{1}
 +\omega^2\, x_{1}\,\bar{y}_{2}\right)}{\sqrt{3}}\;,&    
 \rep1_8 & ~=~\frac{\left(x_{3}\,\bar{y}_{2}+\omega^2\,x_{1}\,\bar{y}_{3}
 +\omega\,   x_{2}\,\bar{y}_{1}\right)}{\sqrt{3}}\;,
 \\ \notag
 \rep1_5 & ~=~\frac{\left(x_{3}\,\bar{y}_{2}+\omega\,  x_{1}\,\bar{y}_{3}
 +\omega^2\, x_{2}\,\bar{y}_{1}\right)}{\sqrt{3}}\;,& 
 \rep1_7 & ~=~\frac{\left(x_{2}\,\bar{y}_{3}+\omega^2\,x_{3}\,\bar{y}_{1}
 +\omega\,   x_{1}\,\bar{y}_{2}\right)}{\sqrt{3}}\;. 
\end{align}
\end{subequations}
The a priori free global phases here have been adjusted such that both, $s$ and $t$ act
as permutation of the given contractions without the need of any additional phase multiplication.
This phase choice also ensures consistency with \eqref{eq:Delta54CGs} and the corresponding 
transformation behavior of the \Dff doublets under $s$ and $t$, cf.\ \Eqref{eq:STActionDoublets}.

\section{On the group \texorpdfstring{$\Delta(54)$}{}}
\label{app:Delta54}

The group $\Delta(54)$ is listed in the \textsc{GAP} SmallGroups library as $\mathrm{SG(54,8)}$.
A possible minimal generating set and the corresponding presentation have been given in \eqref{eq:DffPrsentation}.
The character table of the group is shown in \Tabref{tab:Delta54char}.
Explicit representation matrices for singlet and doublet representations can be found in \eqref{eq:DoubletDefs} and \eqref{eq:delta54gens}, respectively.

The CGs of \Dff relevant to this work are given by
\begin{subequations}\label{eq:Delta54CGs}
\begin{align}
  \left(x_{\rep[_i]{2}}\otimes y_{\rep[_i]{2}}\right)_{\rep[_0]{1}} & ~=~ \frac{1}{\sqrt{2}}\left(x_1\, y_2 + x_2\, y_1 \right)\;,\nonumber\\
	\left(x_{\rep[_i]{3}}\otimes y_{\rep[_i]{\bar{3}}}\right)_{\rep[_0]{1}} & ~=~ \frac{1}{\sqrt{3}}\left(x_1\, \bar{y}_1 + x_2\, \bar{y}_2 + x_3\, \bar{y}_3 \right)\;,\nonumber\\
  \left(x_{\rep[_i]{3}}\otimes y_{\rep[_i]{\bar{3}}}\right)_{\rep[_1]{2}} & ~=~ 
  \frac{1}{\sqrt{3}}\begin{pmatrix}x_1\, \bar{y}_2 + x_3\, \bar{y}_1 + x_2\, \bar{y}_3 \\ x_2\, \bar{y}_1 + x_1\, \bar{y}_3 + x_3\, \bar{y}_2 \end{pmatrix}\;,\nonumber\\
  \left(x_{\rep[_i]{3}}\otimes y_{\rep[_i]{\bar{3}}}\right)_{\rep[_2]{2}} & ~=~ 
  \frac{1}{\sqrt{3}}\begin{pmatrix}x_1\, \bar{y}_1 + \omega\, x_2\, \bar{y}_2 + \omega^2\, x_3\, \bar{y}_3 \\ x_1\, \bar{y}_1 + \omega^2\, x_2\, \bar{y}_2 + \omega\, x_3\, \bar{y}_3 \end{pmatrix}\;,\nonumber\\
  \left(x_{\rep[_i]{3}}\otimes y_{\rep[_i]{\bar{3}}}\right)_{\rep[_3]{2}} & ~=~ 
  \frac{1}{\sqrt{3}}\begin{pmatrix}x_2\, \bar{y}_3 + \omega\, x_3\, \bar{y}_1 + \omega^2\, x_1\, \bar{y}_2 \\ \omega\, x_2\, \bar{y}_1 + x_3\, \bar{y}_2 + \omega^2\, x_1\, \bar{y}_3 \end{pmatrix}\;,\nonumber\\
  \left(x_{\rep[_i]{3}}\otimes y_{\rep[_i]{\bar{3}}}\right)_{\rep[_4]{2}} & ~=~ 
  \frac{1}{\sqrt{3}}\begin{pmatrix}\omega^2\, x_2\, \bar{y}_1 + x_3\, \bar{y}_2 + \omega\, x_1\, \bar{y}_3 \\ x_2\, \bar{y}_3 + \omega^2\, x_3\, \bar{y}_1 + \omega\, x_1\, \bar{y}_2 \end{pmatrix}\;.
\end{align}
\end{subequations}
CGs for other contractions are listed in \cite{Escobar:2011mq}, where, however, a different labeling for the representations is used.
By computing the \FSI's for all automorphisms, it is readily confirmed that \Dff is of type I according to the classification in \Secref{sec:classification}.

The CGs can be used in order to compute the invariants of the direct product $\rep{\bar{3}}\otimes\rep{3}\otimes\rep{\bar{3}}\otimes\rep{3}$,
as needed in section \Secref{Sec:TheModel}. Using $H=(H_1,H_2,H_3)$ for the triplet, as well as the Hermitian conjugate for \rep{\bar{3}} one finds
\begin{equation}
 \begin{split}
  I_0(H^\dagger,H)~&=~\frac13\left(H_1^\dagger H_1^{\phantom{\dagger}}+H_2^\dagger H_2^{\phantom{\dagger}}+H_3^\dagger H_3^{\phantom{\dagger}}\right)^2\;, \\
  I_1(H^\dagger,H)~&=~\frac{1}{3}\left[\left(H_1^\dagger H_2^{\phantom{\dagger}}H_1^\dagger H_3^{\phantom{\dagger}}+
					      H_2^\dagger H_1^{\phantom{\dagger}}H_2^\dagger H_3^{\phantom{\dagger}}+
					      H_3^\dagger H_1^{\phantom{\dagger}}H_3^\dagger H_2^{\phantom{\dagger}}+\mathrm{h.c.}\right) + \right.\\
					      &\,\phantom{=~\left[\frac{\sqrt{2}}{3}\right.}\left. 
					      H_1^\dagger H_2^{\phantom{\dagger}}H_2^\dagger H_1^{\phantom{\dagger}}+
					      H_1^\dagger H_3^{\phantom{\dagger}}H_3^\dagger H_1^{\phantom{\dagger}}+
					      H_2^\dagger H_3^{\phantom{\dagger}}H_3^\dagger H_2^{\phantom{\dagger}}
					      \right]\;,\\
 I_2(H^\dagger,H)~&=~\frac{1}{3}\left[\left(\omega^2H_1^\dagger H_2^{\phantom{\dagger}}H_1^\dagger H_3^{\phantom{\dagger}}+
					      \omega^2H_2^\dagger H_1^{\phantom{\dagger}}H_2^\dagger H_3^{\phantom{\dagger}}+
					      \omega^2H_3^\dagger H_1^{\phantom{\dagger}}H_3^\dagger H_2^{\phantom{\dagger}}+\mathrm{h.c.}\right) + \right.\\
					      &\,\phantom{=~\left[\frac{\sqrt{2}}{3}\right.}\left. 
					      H_1^\dagger H_2^{\phantom{\dagger}}H_2^\dagger H_1^{\phantom{\dagger}}+
					      H_1^\dagger H_3^{\phantom{\dagger}}H_3^\dagger H_1^{\phantom{\dagger}}+
					      H_2^\dagger H_3^{\phantom{\dagger}}H_3^\dagger H_2^{\phantom{\dagger}}
					      \right]\;,\\
I_3(H^\dagger,H)~&=~\frac{1}{3}\left[\left(\omega H_1^\dagger H_2^{\phantom{\dagger}}H_1^\dagger H_3^{\phantom{\dagger}}+
					      \omega H_2^\dagger H_1^{\phantom{\dagger}}H_2^\dagger H_3^{\phantom{\dagger}}+
					      \omega H_3^\dagger H_1^{\phantom{\dagger}}H_3^\dagger H_2^{\phantom{\dagger}}+\mathrm{h.c.}\right) + \right.\\
					      &\,\phantom{=~\left[\frac{\sqrt{2}}{3}\right.}\left. 
					      H_1^\dagger H_2^{\phantom{\dagger}}H_2^\dagger H_1^{\phantom{\dagger}}+
					      H_1^\dagger H_3^{\phantom{\dagger}}H_3^\dagger H_1^{\phantom{\dagger}}+
					      H_2^\dagger H_3^{\phantom{\dagger}}H_3^\dagger H_2^{\phantom{\dagger}}
					      \right]\;,\\
I_4(H^\dagger,H)~&=~\frac{1}{3}\left[H_1^\dagger H_1^{\phantom{\dagger}}H_1^\dagger H_1^{\phantom{\dagger}}+
					     H_2^\dagger H_2^{\phantom{\dagger}}H_2^\dagger H_2^{\phantom{\dagger}}+
					     H_3^\dagger H_3^{\phantom{\dagger}}H_3^\dagger H_3^{\phantom{\dagger}}  \right.\\
					     &\,\phantom{=~\left[\frac{\sqrt{2}}{3}\right.}\left. 
					     -H_1^\dagger H_1^{\phantom{\dagger}}H_2^\dagger H_2^{\phantom{\dagger}}
					     -H_1^\dagger H_1^{\phantom{\dagger}}H_3^\dagger H_3^{\phantom{\dagger}}
					     -H_2^\dagger H_2^{\phantom{\dagger}}H_3^\dagger H_3^{\phantom{\dagger}}
					      \right]\;.
 \end{split}
\end{equation}
The definition of the $I_k(H^\dagger,H)$ is given in \eqref{eq:invariants}.

\clearpage
\backmatter

%%%%%%%%%%%%%%%%%%%%%%%%%%%%%%%%%%%%%%%%%%%%%%%%%%%%%%%%%%%%%%%%%%%%%%%%%%%%%%%%%%%%%%%%%%%%%%%%%%%%%%%%%%%%%%%
\newpage
\thispagestyle{empty}

\vspace*{2cm}
\noindent\textsf{\textbf{\LARGE Acknowledgements}} \\

This research was supported in parts by the DFG cluster of excellence ``Origin and Structure of the Universe''.
It is a pleasure to thank Michael Ratz for his supervision, support, advice, 
and countless hours of interesting physical discussions.
Thanks to Karin Ramm for being the humanly center of the institute, and for the support on countless occasions.
Thanks to all colleagues from T30e and T30d for the great working atmosphere, 
help, encouragement, support whenever needed, and all the good times we had. 
Thanks also to all other groups, fellow students, professors, and staff, whom I have interacted with during the past 9 years here at the Physics Department of TUM, for providing 
the working atmosphere which I will certainly miss and nostalgically look back to on many occasions in the future.
Thanks to S.\ Biermann, D.\ Chakravorty, S.\ Ingenhütt, A.\ Mütter, and A. Solaguren--Beascoa for carefully proofreading parts of the manuscript and valuable comments.
Thanks to M.\ Fallbacher for many fruitful and insightful physical discussions and the good collaborations.
Thanks to M.\--C.\ Chen for collaborations, support, and the great hospitality in Irvine.
Thanks to P.\ Vaudrevange for collaboration and many interesting coffee discussions. Thanks to I.\ Ivanov for useful discussions.
Thanks to A.\ Buras for support, as well as for being a mentor and role model for how much fun and satisfaction physics can be.
Thanks to D.B.\ Kaplan for introducing me to group theory.
Thanks to all my various office mates during these years, including S.\ Ingenhütt, A.\ Lamperstorfer, D.\ Meindl, A.\ Mütter, C.\ Niehoff, P.\ Stangl, and M.\ Totzauer, 
for willingness to discuss physics and providing Mathematica support at all times. 
Thanks to all my former teachers and professors for passing the knowledge and bearing with a questioning student.
Thanks to all my friends who are the basis for what I am doing.
Thanks to Ana for many hearty discussions.
Last but not least I am eternally grateful to my parents, my sister, and the rest of my family 
for the everlasting support at all times and for letting me freely unfold to what I am now. 
Without you this would not have been possible. 

\vspace*{\fill}
\begin{center}
 {\large\changefont{pzc}{m}{n} If I have seen further it is by standing on the shoulders of Giants.}
\end{center}
\begin{flushright}
 \small\textsc{Isaac Newton} \\
\end{flushright}

\vspace*{\fill}
%%%%%%%%%%%%%%%%%%%%%%%%%%%%%%%%%%%%%%%%%%%%%%%%%%%%%%%%%%%%%%%%%%%%%%%%%%%%%%%%%%%%%%%%%%%%

\addcontentsline{toc}{chapter}{Bibliography}
\bibliography{Orbifold}
\bibliographystyle{utphys} %ONLINE VERSION
\clearpage

\end{document}

%% file: Titel_arxiv.tex
\makeatletter 
\newcommand{\semiHuge}{\@setfontsize\semiHuge{23.6}{27.38}}
\newcommand{\TitleHuge}{\@setfontsize\TitleHuge{30}{40}}
\makeatother

\newcommand{\Title}{ \TitleHuge{CP and other symmetries of symmetries}}

\begin{titlepage}

\begin{center}

\par
\raisebox{-.5\height}{ \includegraphics[width=.2\linewidth]{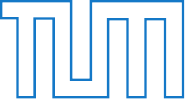}}%
\hfill
\raisebox{-.5\height}{\includegraphics[width=.25\linewidth]{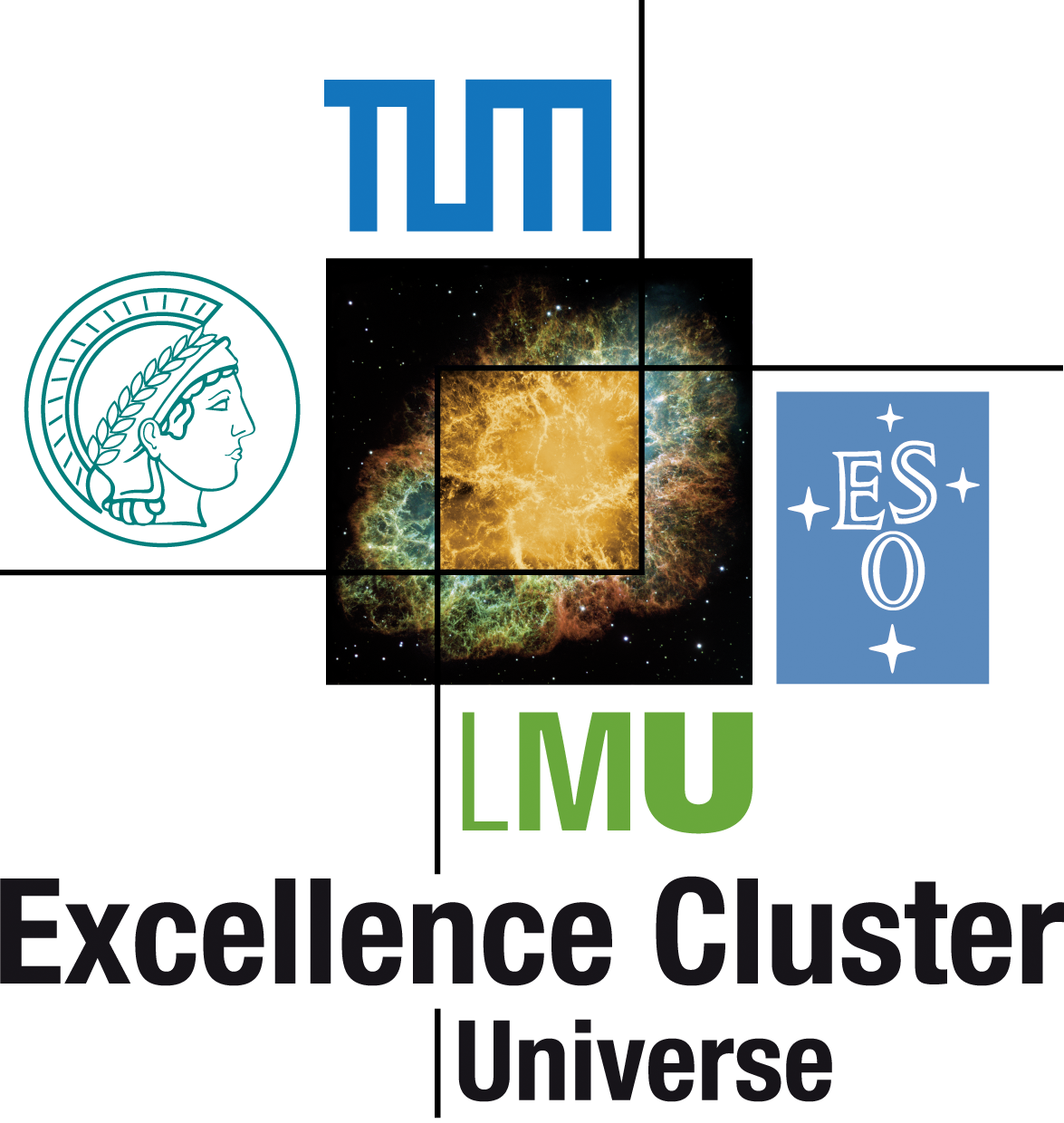}}\\[1cm]%
\par

\textsc{\Large Technische Universität München}\\[0.1cm]
\Large  \textrm{\&}\\[0.1cm] 
\textsc{\Large Excellence Cluster Universe}\\[1.5cm]

\linespread{1.2}
    \HRule\\
    {\Huge\bfseries CP and other \\ Symmetries of Symmetries\par}
      \HRule\\[1.5cm]

\textsc{\Large Dissertation}\\[0.5cm]

{\Large  \textrm{by}\\[0.1cm] \textsc{Andreas Trautner}\\[1.0cm]}

\vfill

\includegraphics[width=0.15\textwidth]{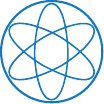}\\[0.6cm]

\textsc{\Large Physik Department T30e}\\[0.1cm]

\end{center}

\end{titlepage}

\begin{titlepage}

\begin{center}
  
\includegraphics[width=0.15\textwidth]{tumlogo.pdf}\\[0.6cm]

\textsc{\LARGE Technische Universität München}\\[0.6cm]

\textsc{\Large Physik Department T30e}\\[2.0cm]

\vspace{1.5cm}

\linespread{1.2}
    {\Huge\bfseries CP and other \\ Symmetries of Symmetries\par}
\vspace{2.0cm}
			
\textsc{\large Andreas Trautner}\\[1.5cm]
\vfil

\begin{minipage}{1.0\textwidth}

Vollständiger Abdruck der von der Fakultät für Physik der Technischen Universität München zur Erlangung des akademischen Grades eines

\begin{center}
\textbf{Doktors der Naturwissenschaften} 
\end{center}

genehmigten Dissertation.\\

\begin{center}
\begin{tabular}{lrl}
Vorsitzende: & & Prof.\ Dr.\ Elisa Resconi \\
Prüfer der Dissertation: & 1. & Prof.\ Dr.\ Michael Ratz\\
& 2. &  Prof.\ Dr.\ Andrzej J.\ Buras \\
& 3. &  Prof.\ Pierre Ramond, \\
&    &  University of Florida, USA (nur schriftliche Beurteilung)
\end{tabular}
\end{center}

\end{minipage}

\vspace{1.0cm}
\enlargethispage{1.0cm}
\begin{minipage}{1.0\textwidth}

Die Dissertation wurde am 08.06.2016 bei der Technischen Universität München \mbox{eingereicht} und durch die Fakultät für Physik am 15.07.2016 angenommen.

\end{minipage}

\end{center}

\end{titlepage}

%% file: Dissertation_arxiv_upload.bbl
\providecommand{\href}[2]{#2}\begingroup\raggedright\begin{thebibliography}{100}

\bibitem{Weyl:1952}
H.~Weyl, {\em Symmetry}.
\newblock Princeton University Press, 1952.

\bibitem{Chen:2013dpa}
M.-C. Chen, M.~Ratz, and A.~Trautner, ``{Non-Abelian discrete $R$
  symmetries},'' \href{http://dx.doi.org/10.1007/JHEP09(2013)096}{{\em JHEP}
  {\bfseries 1309} (2013) 096},
\href{http://arxiv.org/abs/1306.5112}{{\ttfamily arXiv:1306.5112 [hep-ph]}}.
%%CITATION = ARXIV:1306.5112;%%.

\bibitem{Chen:2014tpa}
M.-C. Chen, M.~Fallbacher, K.~Mahanthappa, M.~Ratz, and A.~Trautner, ``{CP
  Violation from Finite Groups},'' {\em Nucl. Phys.} {\bfseries B883} (2014)
  267,
\href{http://arxiv.org/abs/1402.0507}{{\ttfamily arXiv:1402.0507 [hep-ph]}}.
%%CITATION = ARXIV:1402.0507;%%.

\bibitem{Fallbacher:2015rea}
M.~Fallbacher and A.~Trautner, ``{Symmetries of symmetries and geometrical CP
  violation},'' \href{http://dx.doi.org/10.1016/j.nuclphysb.2015.03.003}{{\em
  Nucl. Phys.} {\bfseries B894} (2015) 136--160},
\href{http://arxiv.org/abs/1502.01829}{{\ttfamily arXiv:1502.01829 [hep-ph]}}.
%%CITATION = ARXIV:1502.01829;%%.

\bibitem{Chen:2015aba}
M.-C. Chen, M.~Fallbacher, M.~Ratz, A.~Trautner, and P.~K.~S. Vaudrevange,
  ``{Anomaly-safe discrete groups},''
  \href{http://dx.doi.org/10.1016/j.physletb.2015.05.047}{{\em Phys. Lett.}
  {\bfseries B747} (2015) 22--26},
\href{http://arxiv.org/abs/1504.03470}{{\ttfamily arXiv:1504.03470 [hep-ph]}}.
%%CITATION = ARXIV:1504.03470;%%.

\bibitem{Chen:2015dka}
M.-C. Chen, M.~Ratz, and A.~Trautner, ``{Nonthermal cosmic neutrino
  background},'' \href{http://dx.doi.org/10.1103/PhysRevD.92.123006}{{\em Phys.
  Rev.} {\bfseries D92} no.~12, (2015) 123006},
\href{http://arxiv.org/abs/1509.00481}{{\ttfamily arXiv:1509.00481 [hep-ph]}}.
%%CITATION = ARXIV:1509.00481;%%.

\bibitem{Sakharov:1967dj}
A.~Sakharov, ``{Violation of CP Invariance, c Asymmetry, and Baryon Asymmetry
  of the Universe},''
\href{http://dx.doi.org/10.1070/PU1991v034n05ABEH002497}{{\em Pisma
  Zh.Eksp.Teor.Fiz.} {\bfseries 5} (1967) 32--35}.
%%CITATION = ZFPRA,5,32;%%.

\bibitem{Cohen:1987vi}
A.~G. Cohen and D.~B. Kaplan, ``{Thermodynamic Generation of the Baryon
  Asymmetry},''
\href{http://dx.doi.org/10.1016/0370-2693(87)91369-4}{{\em Phys. Lett.}
  {\bfseries B199} (1987) 251--258}.
%%CITATION = PHLTA,B199,251;%%.

\bibitem{Hook:2015foa}
A.~Hook, ``{Baryogenesis in a CP invariant theory},''
  \href{http://dx.doi.org/10.1007/JHEP11(2015)143}{{\em JHEP} {\bfseries 11}
  (2015) 143},
\href{http://arxiv.org/abs/1508.05094}{{\ttfamily arXiv:1508.05094 [hep-ph]}}.
%%CITATION = ARXIV:1508.05094;%%.

\bibitem{Kobayashi:1973fv}
M.~Kobayashi and T.~Maskawa, ``{CP Violation in the Renormalizable Theory of
  Weak Interaction},''
\href{http://dx.doi.org/10.1143/PTP.49.652}{{\em Prog.Theor.Phys.} {\bfseries
  49} (1973) 652--657}.
%%CITATION = PTPKA,49,652;%%.

\bibitem{NPP:2008}
``{The 2008 Nobel Prize in Physics - Press Release},''.
  \url{http://www.nobelprize.org/nobel_prizes/physics/laureates/2008/press.html}.

\bibitem{Kuzmin:1985mm}
V.~Kuzmin, V.~Rubakov, and M.~Shaposhnikov, ``{On the Anomalous Electroweak
  Baryon Number Nonconservation in the Early Universe},''
\href{http://dx.doi.org/10.1016/0370-2693(85)91028-7}{{\em Phys.Lett.}
  {\bfseries B155} (1985) 36}.
%%CITATION = PHLTA,B155,36;%%.

\bibitem{Riotto:1998bt}
A.~Riotto, ``{Theories of baryogenesis},''
  \href{http://arxiv.org/abs/hep-ph/9807454}{{\ttfamily arXiv:hep-ph/9807454
  [hep-ph]}}.
\url{http://inspirehep.net/record/473645/files/arXiv:hep-ph_9807454.pdf}.
%%CITATION = HEP-PH/9807454;%%.

\bibitem{Bernreuther:2002uj}
W.~Bernreuther, ``{CP violation and baryogenesis},'' {\em Lect. Notes Phys.}
  {\bfseries 591} (2002) 237--293,
  \href{http://arxiv.org/abs/hep-ph/0205279}{{\ttfamily arXiv:hep-ph/0205279
  [hep-ph]}}.
[,237(2002)].
%%CITATION = HEP-PH/0205279;%%.

\bibitem{Branco:1999fs}
G.~C. Branco, L.~Lavoura, and J.~P. Silva, ``{CP Violation},''
{\em Int.Ser.Monogr.Phys.} {\bfseries 103} (1999) 1--536.
%%CITATION = IMPHA,103,1;%%.

\bibitem{Baker:2006ts}
C.~A. Baker {\em et~al.}, ``{An Improved experimental limit on the electric
  dipole moment of the neutron},''
  \href{http://dx.doi.org/10.1103/PhysRevLett.97.131801}{{\em Phys. Rev. Lett.}
  {\bfseries 97} (2006) 131801},
\href{http://arxiv.org/abs/hep-ex/0602020}{{\ttfamily arXiv:hep-ex/0602020
  [hep-ex]}}.
%%CITATION = HEP-EX/0602020;%%.

\bibitem{Agashe:2014kda}
{\bfseries Particle Data Group} Collaboration, K.~A. Olive {\em et~al.},
  ``{Review of Particle Physics},''
\href{http://dx.doi.org/10.1088/1674-1137/38/9/090001}{{\em Chin. Phys.}
  {\bfseries C38} (2014) 090001}.
%%CITATION = CHPHD,C38,090001;%%.

\bibitem{Weinberg:1977hb}
S.~Weinberg, ``{The Problem of Mass},''
\href{http://dx.doi.org/10.1111/j.2164-0947.1977.tb02958.x}{{\em Trans. New
  York Acad. Sci.} {\bfseries 38} (1977) 185--201}.
%%CITATION = TNYAA,38,185;%%.

\bibitem{Peccei:1997mz}
R.~D. Peccei, ``{The Mystery of flavor},'' {\em AIP Conf. Proc.} {\bfseries
  424} (1997) 354--364, \href{http://arxiv.org/abs/hep-ph/9712422}{{\ttfamily
  arXiv:hep-ph/9712422 [hep-ph]}}.
[,354(1997)].
%%CITATION = HEP-PH/9712422;%%.

\bibitem{Fritzsch:1999ee}
H.~Fritzsch and Z.-z. Xing, ``{Mass and flavor mixing schemes of quarks and
  leptons},'' \href{http://dx.doi.org/10.1016/S0146-6410(00)00102-2}{{\em Prog.
  Part. Nucl. Phys.} {\bfseries 45} (2000) 1--81},
\href{http://arxiv.org/abs/hep-ph/9912358}{{\ttfamily arXiv:hep-ph/9912358
  [hep-ph]}}.
%%CITATION = HEP-PH/9912358;%%.

\bibitem{Xing:2014sja}
Z.-z. Xing, ``{Quark Mass Hierarchy and Flavor Mixing Puzzles},''
  \href{http://dx.doi.org/10.1142/S0217751X14300671}{{\em Int. J. Mod. Phys.}
  {\bfseries A29} (2014) 1430067},
\href{http://arxiv.org/abs/1411.2713}{{\ttfamily arXiv:1411.2713 [hep-ph]}}.
%%CITATION = ARXIV:1411.2713;%%.

\bibitem{Feruglio:2015jfa}
F.~Feruglio, ``{Pieces of the Flavour Puzzle},''
  \href{http://dx.doi.org/10.1140/epjc/s10052-015-3576-5}{{\em Eur. Phys. J.}
  {\bfseries C75} no.~8, (2015) 373},
\href{http://arxiv.org/abs/1503.04071}{{\ttfamily arXiv:1503.04071 [hep-ph]}}.
%%CITATION = ARXIV:1503.04071;%%.

\bibitem{Buchbinder:2000cq}
I.~L. Buchbinder, D.~M. Gitman, and A.~L. Shelepin, ``{Discrete symmetries as
  automorphisms of the proper Poincare group},''
  \href{http://dx.doi.org/10.1023/A:1015244830241}{{\em Int. J. Theor. Phys.}
  {\bfseries 41} (2002) 753--790},
\href{http://arxiv.org/abs/hep-th/0010035}{{\ttfamily arXiv:hep-th/0010035
  [hep-th]}}.
%%CITATION = HEP-TH/0010035;%%.

\bibitem{Grimus:1995zi}
W.~Grimus and M.~Rebelo, ``{Automorphisms in gauge theories and the definition
  of CP and P},'' \href{http://dx.doi.org/10.1016/S0370-1573(96)00030-0}{{\em
  Phys.Rept.} {\bfseries 281} (1997) 239--308},
\href{http://arxiv.org/abs/hep-ph/9506272}{{\ttfamily arXiv:hep-ph/9506272
  [hep-ph]}}.
%%CITATION = HEP-PH/9506272;%%.

\bibitem{Holthausen:2012dk}
M.~Holthausen, M.~Lindner, and M.~A. Schmidt, ``{CP and Discrete Flavour
  Symmetries},'' \href{http://dx.doi.org/10.1007/JHEP04(2013)122}{{\em JHEP}
  {\bfseries 04} (2013) 122},
\href{http://arxiv.org/abs/1211.6953}{{\ttfamily arXiv:1211.6953 [hep-ph]}}.
%%CITATION = ARXIV:1211.6953;%%.

\bibitem{Chen:2009gf}
M.-C. Chen and K.~Mahanthappa, ``{Group Theoretical Origin of CP Violation},''
  \href{http://dx.doi.org/10.1016/j.physletb.2009.10.059}{{\em Phys.Lett.}
  {\bfseries B681} (2009) 444--447},
\href{http://arxiv.org/abs/0904.1721}{{\ttfamily arXiv:0904.1721 [hep-ph]}}.
%%CITATION = ARXIV:0904.1721;%%.

\bibitem{Branco:2015hea}
G.~C. Branco, I.~de~Medeiros~Varzielas, and S.~F. King, ``{Invariant approach
  to CP in family symmetry models},''
  \href{http://dx.doi.org/10.1103/PhysRevD.92.036007}{{\em Phys. Rev.}
  {\bfseries D92} no.~3, (2015) 036007},
\href{http://arxiv.org/abs/1502.03105}{{\ttfamily arXiv:1502.03105 [hep-ph]}}.
%%CITATION = ARXIV:1502.03105;%%.

\bibitem{Lee:1973iz}
T.~Lee, ``{A Theory of Spontaneous T Violation},''
\href{http://dx.doi.org/10.1103/PhysRevD.8.1226}{{\em Phys.Rev.} {\bfseries D8}
  (1973) 1226--1239}.
%%CITATION = PHRVA,D8,1226;%%.

\bibitem{Branco:1983tn}
G.~Branco, J.~Gerard, and W.~Grimus, ``{Geometrical T Violation},''
\href{http://dx.doi.org/10.1016/0370-2693(84)92024-0}{{\em Phys. Lett.}
  {\bfseries B136} (1984) 383}.
%%CITATION = PHLTA,B136,383;%%.

\bibitem{Ivanov:2015mwl}
I.~P. Ivanov and J.~P. Silva, ``{A CP-conserving multi-Higgs model without real
  basis},'' \href{http://dx.doi.org/10.1103/PhysRevD.93.095014}{{\em Phys.
  Rev.} {\bfseries D93} no.~9, (2016) 095014},
\href{http://arxiv.org/abs/1512.09276}{{\ttfamily arXiv:1512.09276 [hep-ph]}}.
%%CITATION = ARXIV:1512.09276;%%.

\bibitem{Aad:2012tfa}
{\bfseries ATLAS Collaboration} Collaboration, G.~Aad {\em et~al.},
  ``{Observation of a new particle in the search for the Standard Model Higgs
  boson with the ATLAS detector at the LHC},''
  \href{http://dx.doi.org/10.1016/j.physletb.2012.08.020}{{\em Phys. Lett.}
  {\bfseries B716} (2012) 1--29},
\href{http://arxiv.org/abs/1207.7214}{{\ttfamily arXiv:1207.7214 [hep-ex]}}.
%%CITATION = ARXIV:1207.7214;%%.

\bibitem{Chatrchyan:2012xdj}
{\bfseries CMS} Collaboration, S.~Chatrchyan {\em et~al.}, ``{Observation of a
  new boson at a mass of 125 GeV with the CMS experiment at the LHC},''
  \href{http://dx.doi.org/10.1016/j.physletb.2012.08.021}{{\em Phys. Lett.}
  {\bfseries B716} (2012) 30--61},
\href{http://arxiv.org/abs/1207.7235}{{\ttfamily arXiv:1207.7235 [hep-ex]}}.
%%CITATION = ARXIV:1207.7235;%%.

\bibitem{NPP:2015}
``{The 2015 Nobel Prize in Physics - Press Release},''.
  \url{http://www.nobelprize.org/nobel_prizes/physics/laureates/2015/press.html}.

\bibitem{Maki:1962mu}
Z.~Maki, M.~Nakagawa, and S.~Sakata, ``{Remarks on the unified model of
  elementary particles},''
\href{http://dx.doi.org/10.1143/PTP.28.870}{{\em Prog. Theor. Phys.} {\bfseries
  28} (1962) 870--880}.
%%CITATION = PTPKA,28,870;%%.

\bibitem{Chau:1984fp}
L.-L. Chau and W.-Y. Keung, ``{Comments on the Parametrization of the
  Kobayashi-Maskawa Matrix},''
\href{http://dx.doi.org/10.1103/PhysRevLett.53.1802}{{\em Phys. Rev. Lett.}
  {\bfseries 53} (1984) 1802}.
%%CITATION = PRLTA,53,1802;%%.

\bibitem{Bona:2006ah}
{\bfseries UTfit} Collaboration, M.~Bona {\em et~al.}, ``{The Unitarity
  Triangle Fit in the Standard Model and Hadronic Parameters from Lattice QCD:
  A Reappraisal after the Measurements of Delta m(s) and BR(B to tau
  nu(tau))},'' \href{http://dx.doi.org/10.1088/1126-6708/2006/10/081}{{\em
  JHEP} {\bfseries 10} (2006) 081},
  \href{http://arxiv.org/abs/hep-ph/0606167}{{\ttfamily arXiv:hep-ph/0606167
  [hep-ph]}}.
{Results taken from
  \url{http://www.utfit.org/UTfit/ResultsSummer2014PostMoriondSM}}.
%%CITATION = HEP-PH/0606167;%%.

\bibitem{Baak:2014ora}
{\bfseries Gfitter Group} Collaboration, M.~Baak, J.~Cúth, J.~Haller,
  A.~Hoecker, R.~Kogler, K.~Mönig, M.~Schott, and J.~Stelzer, ``{The global
  electroweak fit at NNLO and prospects for the LHC and ILC},''
  \href{http://dx.doi.org/10.1140/epjc/s10052-014-3046-5}{{\em Eur. Phys. J.}
  {\bfseries C74} (2014) 3046},
\href{http://arxiv.org/abs/1407.3792}{{\ttfamily arXiv:1407.3792 [hep-ph]}}.
%%CITATION = ARXIV:1407.3792;%%.

\bibitem{Charles:2015gya}
J.~Charles {\em et~al.}, ``{Current status of the Standard Model CKM fit and
  constraints on $\Delta F=2$ New Physics},''
  \href{http://dx.doi.org/10.1103/PhysRevD.91.073007}{{\em Phys. Rev.}
  {\bfseries D91} no.~7, (2015) 073007},
\href{http://arxiv.org/abs/1501.05013}{{\ttfamily arXiv:1501.05013 [hep-ph]}}.
%%CITATION = ARXIV:1501.05013;%%.

\bibitem{Gonzalez-Garcia:2014bfa}
M.~C. Gonzalez-Garcia, M.~Maltoni, and T.~Schwetz, ``{Updated fit to three
  neutrino mixing: status of leptonic CP violation},''
  \href{http://dx.doi.org/10.1007/JHEP11(2014)052}{{\em JHEP} {\bfseries 11}
  (2014) 052},
\href{http://arxiv.org/abs/1409.5439}{{\ttfamily arXiv:1409.5439 [hep-ph]}}.
%%CITATION = ARXIV:1409.5439;%%.

\bibitem{Thomas:2009ae}
S.~A. Thomas, F.~B. Abdalla, and O.~Lahav, ``{Upper Bound of 0.28eV on the
  Neutrino Masses from the Largest Photometric Redshift Survey},''
  \href{http://dx.doi.org/10.1103/PhysRevLett.105.031301}{{\em Phys. Rev.
  Lett.} {\bfseries 105} (2010) 031301},
\href{http://arxiv.org/abs/0911.5291}{{\ttfamily arXiv:0911.5291
  [astro-ph.CO]}}.
%%CITATION = ARXIV:0911.5291;%%.

\bibitem{Riemer-Sorensen:2013jsa}
S.~Riemer-Sørensen, D.~Parkinson, and T.~M. Davis, ``{Combining Planck data
  with large scale structure information gives a strong neutrino mass
  constraint},'' \href{http://dx.doi.org/10.1103/PhysRevD.89.103505}{{\em Phys.
  Rev.} {\bfseries D89} (2014) 103505},
\href{http://arxiv.org/abs/1306.4153}{{\ttfamily arXiv:1306.4153
  [astro-ph.CO]}}.
%%CITATION = ARXIV:1306.4153;%%.

\bibitem{Ade:2015xua}
{\bfseries Planck} Collaboration, P.~A.~R. Ade {\em et~al.}, ``{Planck 2015
  results. XIII. Cosmological parameters},''
\href{http://arxiv.org/abs/1502.01589}{{\ttfamily arXiv:1502.01589
  [astro-ph.CO]}}.
%%CITATION = ARXIV:1502.01589;%%.

\bibitem{Nir15}
Y.~Nir, ``Lecture notes invisibles15 school, la cristalera in miraflores de la
  sierra, spain,'' June, 2015.

\bibitem{Mohapatra:1998rq}
R.~N. Mohapatra and P.~B. Pal, ``{Massive neutrinos in physics and
  astrophysics. Second edition},''
{\em World Sci. Lect. Notes Phys.} {\bfseries 60} (1998) 1--397.
%%CITATION = 00327,60,1;%%.

\bibitem{Delepine:2009qg}
D.~Delepine, V.~G. Macias, S.~Khalil, and G.~L. Castro, ``{Probing Majorana
  neutrino CP phases and masses in neutrino-antineutrino conversion},''
\href{http://dx.doi.org/10.1016/j.physletb.2010.08.068}{{\em Phys. Lett.}
  {\bfseries B693} (2010) 438--442}.
%%CITATION = PHLTA,B693,438;%%.

\bibitem{Simkovic:2012hq}
F.~Simkovic, S.~M. Bilenky, A.~Faessler, and T.~Gutsche, ``{Possibility of
  measuring the CP Majorana phases in 0$\nu\beta\beta$ decay},''
  \href{http://dx.doi.org/10.1103/PhysRevD.87.073002}{{\em Phys. Rev.}
  {\bfseries D87} no.~7, (2013) 073002},
\href{http://arxiv.org/abs/1210.1306}{{\ttfamily arXiv:1210.1306 [hep-ph]}}.
%%CITATION = ARXIV:1210.1306;%%.

\bibitem{Xing:2013woa}
Z.-z. Xing and Y.-L. Zhou, ``{Majorana CP-violating phases in
  neutrino-antineutrino oscillations and other lepton-number-violating
  processes},'' \href{http://dx.doi.org/10.1103/PhysRevD.88.033002}{{\em Phys.
  Rev.} {\bfseries D88} (2013) 033002},
\href{http://arxiv.org/abs/1305.5718}{{\ttfamily arXiv:1305.5718 [hep-ph]}}.
%%CITATION = ARXIV:1305.5718;%%.

\bibitem{Minakata:2014jba}
H.~Minakata, H.~Nunokawa, and A.~A. Quiroga, ``{Constraining Majorana CP phase
  in the precision era of cosmology and the double beta decay experiment},''
  \href{http://dx.doi.org/10.1093/ptep/ptv010}{{\em PTEP} {\bfseries 2015}
  (2015) 033B03},
\href{http://arxiv.org/abs/1402.6014}{{\ttfamily arXiv:1402.6014 [hep-ph]}}.
%%CITATION = ARXIV:1402.6014;%%.

\bibitem{Fujikawa:1979ay}
K.~Fujikawa, ``Path integral measure for gauge invariant fermion theories,''
{\em Phys. Rev. Lett.} {\bfseries 42} (1979) 1195.
%%CITATION = PRLTA,42,1195;%%.

\bibitem{Fujikawa:1980eg}
K.~Fujikawa, ``Path integral for gauge theories with fermions,''
{\em Phys. Rev.} {\bfseries D21} (1980) 2848.
%%CITATION = PHRVA,D21,2848;%%.

\bibitem{Srednicki_2007}
M.~Srednicki, {\em Quantum Field Theory}.
\newblock Cambridge University Press, 1~ed., February, 2007.

\bibitem{Perez:2014fja}
P.~Fileviez~Perez and H.~H. Patel, ``{The Electroweak Vacuum Angle},''
  \href{http://dx.doi.org/10.1016/j.physletb.2014.03.064}{{\em Phys. Lett.}
  {\bfseries B732} (2014) 241--243},
\href{http://arxiv.org/abs/1402.6340}{{\ttfamily arXiv:1402.6340 [hep-ph]}}.
%%CITATION = ARXIV:1402.6340;%%.

\bibitem{Bell:1996nh}
J.~S. Bell, ``{Time reversal in field theory},''. [Proc. Roy. Soc.
  Lond.A231,479(1955)].

\bibitem{Wigner1932}
E.~Wigner, ``Ueber die operation der zeitumkehr in der quantenmechanik,'' {\em
  Nachrichten von der Gesellschaft der Wissenschaften zu Göttingen,
  Mathematisch-Physikalische Klasse} {\bfseries 1932} (1932) 546--559.
  \url{http://eudml.org/doc/59401}.

\bibitem{Weinberg:1995mt2}
S.~Weinberg, {\em {The Quantum theory of fields. Vol. 1: Foundations}}.
\newblock Cambridge University Press, 2005.

\bibitem{Dreiner:2008tw}
H.~K. Dreiner, H.~E. Haber, and S.~P. Martin, ``{Two-component spinor
  techniques and Feynman rules for quantum field theory and supersymmetry},''
  \href{http://dx.doi.org/10.1016/j.physrep.2010.05.002}{{\em Phys. Rept.}
  {\bfseries 494} (2010) 1--196},
\href{http://arxiv.org/abs/0812.1594}{{\ttfamily arXiv:0812.1594 [hep-ph]}}.
%%CITATION = ARXIV:0812.1594;%%.

\bibitem{Streater:1989vi}
R.~F. Streater and A.~S. Wightman, {\em {PCT}, spin and statistics, and all
  that}.
\newblock Redwood City, USA: Addison-Wesley, 1989.
\newblock 207 p. (Advanced book classics).

\bibitem{Christenson:1964fg}
J.~H. Christenson, J.~W. Cronin, V.~L. Fitch, and R.~Turlay, ``{Evidence for
  the 2 pi Decay of the k(2)0 Meson},''
\href{http://dx.doi.org/10.1103/PhysRevLett.13.138}{{\em Phys. Rev. Lett.}
  {\bfseries 13} (1964) 138--140}.
%%CITATION = PRLTA,13,138;%%.

\bibitem{NPP:1980}
``{The 1980 Nobel Prize in Physics - Press Release},''.
  \url{http://www.nobelprize.org/nobel_prizes/physics/laureates/1980/press.html}.

\bibitem{Lees:2012}
{\bfseries The BABAR Collaboration} Collaboration, J.~P. Lees {\em et~al.},
  ``Observation of time-reversal violation in the ${B}^{0}$ meson system,''
  \href{http://dx.doi.org/10.1103/PhysRevLett.109.211801}{{\em Phys. Rev.
  Lett.} {\bfseries 109} (Nov, 2012) 211801}.
  \url{http://link.aps.org/doi/10.1103/PhysRevLett.109.211801}.

\bibitem{Jarlskog:1985ht}
C.~Jarlskog, ``{Commutator of the Quark Mass Matrices in the Standard
  Electroweak Model and a Measure of Maximal CP Violation},''
\href{http://dx.doi.org/10.1103/PhysRevLett.55.1039}{{\em Phys. Rev. Lett.}
  {\bfseries 55} (1985) 1039}.
%%CITATION = PRLTA,55,1039;%%.

\bibitem{Gronau:1984nm}
M.~Gronau and J.~Schechter, ``{A Physical {CP} Phase and Maximal {CP}
  Violation},''
\href{http://dx.doi.org/10.1103/PhysRevLett.54.385}{{\em Phys. Rev. Lett.}
  {\bfseries 54} (1985) 385}.
%%CITATION = PRLTA,54,385;%%.

\bibitem{Bernabeu:1986fc}
J.~Bernabeu, G.~Branco, and M.~Gronau, ``{CP RESTRICTIONS ON QUARK MASS
  MATRICES},''
\href{http://dx.doi.org/10.1016/0370-2693(86)90659-3}{{\em Phys. Lett.}
  {\bfseries B169} (1986) 243--247}.
%%CITATION = PHLTA,B169,243;%%.

\bibitem{Botella:1994cs}
F.~J. Botella and J.~P. Silva, ``{Jarlskog - like invariants for theories with
  scalars and fermions},''
  \href{http://dx.doi.org/10.1103/PhysRevD.51.3870}{{\em Phys. Rev.} {\bfseries
  D51} (1995) 3870--3875},
\href{http://arxiv.org/abs/hep-ph/9411288}{{\ttfamily arXiv:hep-ph/9411288
  [hep-ph]}}.
%%CITATION = HEP-PH/9411288;%%.

\bibitem{Cornwell:1985xs}
J.~F. Cornwell, {\em {GROUP THEORY IN PHYSICS. VOL. 1}}.
\newblock 1985.

\bibitem{Georgi:1999jb}
H.~Georgi, {\em LIE ALGEBRAS IN PARTICLE PHYSICS. {F}ROM ISOSPIN TO UNIFIED
  THEORIES, 2ND EDITION}, vol.~54.
\newblock Perseus Books, 1999.

\bibitem{Fuchs:1997jv}
J.~Fuchs and C.~Schweigert, {\em Symmetries, {Lie} algebras and
  representations: A graduate course for physicists}.
\newblock Cambridge, UK: University Press, 1997.
\newblock 438 p.

\bibitem{Ramond:2010zz}
P.~Ramond, {\em Group theory: A physicist's survey}.
\newblock 2010.

\bibitem{GAP4}
The GAP~Group, {\em {GAP -- Groups, Algorithms, and Programming, Version
  4.5.5}}, 2012.
\newblock \url{http://www.gap-system.org}.

\bibitem{Holthausen:2011vd}
M.~Holthausen and M.~A. Schmidt, ``{Natural Vacuum Alignment from Group Theory:
  The Minimal Case},'' \href{http://dx.doi.org/10.1007/JHEP01(2012)126}{{\em
  JHEP} {\bfseries 1201} (2012) 126},
\href{http://arxiv.org/abs/1111.1730}{{\ttfamily arXiv:1111.1730 [hep-ph]}}.
%%CITATION = ARXIV:1111.1730;%%.

\bibitem{Fallus:2015}
M.~Fallbacher, {\em Discrete Groups in Model Building and the Definition of
  CP}.
\newblock Dissertation, Technische Universität München, München, 2015.

\bibitem{Gantmacher:1939}
F.~{Gantmacher}, ``{Canonical representation of automorphisms of a complex
  semisimple Lie group.},'' {\em {Rec. Math. Moscou, n. Ser.}} {\bfseries 5}
  (1939) 101--144.

\bibitem{Cornwell:1985xt}
J.~F. Cornwell, {\em GROUP THEORY IN PHYSICS. VOL. 2}.
\newblock Academic, 1985.
\newblock 589 p.

\bibitem{Hall:2000}
B.~C. {Hall}, ``{An Elementary Introduction to Groups and Representations},''
  {\em ArXiv Mathematical Physics e-prints} (May, 2000) ,
  \href{http://arxiv.org/abs/math-ph/0005032}{{\ttfamily math-ph/0005032}}.

\bibitem{Hall:2015}
B.~C.~H. (auth.), {\em Lie Groups, Lie Algebras, and Representations: An
  Elementary Introduction}.
\newblock Graduate Texts in Mathematics 222. Springer International Publishing,
  2~ed., 2015.

\bibitem{Jacobson:1979}
N.~Jacobson, {\em Lie Algebras}.
\newblock Dover Publications, 1979.
\newblock {p. 202}.

\bibitem{Dynkin:2016}
 by Tomruen [under the license CC BY-SA 3.0
  (\url{http://creativecommons.org/licenses/by-sa/3.0})], via Wikimedia
  Commons,
  \url{https://commons.wikimedia.org/wiki/File\%3AFinite_Dynkin_diagrams.svg},
  last visited 05/10/16.

\bibitem{GellMann:1962xb}
M.~Gell-Mann, ``{Symmetries of baryons and mesons},''
\href{http://dx.doi.org/10.1103/PhysRev.125.1067}{{\em Phys. Rev.} {\bfseries
  125} (1962) 1067--1084}.
%%CITATION = PHRVA,125,1067;%%.

\bibitem{Barroso:2006pa}
A.~Barroso, P.~Ferreira, R.~Santos, and J.~P. Silva, ``{Stability of the normal
  vacuum in multi-Higgs-doublet models},''
  \href{http://dx.doi.org/10.1103/PhysRevD.74.085016}{{\em Phys. Rev.}
  {\bfseries D74} (2006) 085016},
\href{http://arxiv.org/abs/hep-ph/0608282}{{\ttfamily arXiv:hep-ph/0608282
  [hep-ph]}}.
%%CITATION = HEP-PH/0608282;%%.

\bibitem{Lee:1966ik}
T.~D. Lee and G.~C. Wick, ``{Space Inversion, Time Reversal, and Other Discrete
  Symmetries in Local Field Theories},''
  \href{http://dx.doi.org/10.1103/PhysRev.148.1385}{{\em Phys. Rev.} {\bfseries
  148} (1966) 1385--1404}.
[,445(1966)].
%%CITATION = PHRVA,148,1385;%%.

\bibitem{Ecker:1981wv}
G.~Ecker, W.~Grimus, and W.~Konetschny, ``{Quark Mass Matrices in Left-right
  Symmetric Gauge Theories},''
\href{http://dx.doi.org/10.1016/0550-3213(81)90309-6}{{\em Nucl. Phys.}
  {\bfseries B191} (1981) 465}.
%%CITATION = NUPHA,B191,465;%%.

\bibitem{Ecker:1983hz}
G.~Ecker, W.~Grimus, and H.~Neufeld, ``Spontaneous cp violation in left-right
  symmetric gauge theories,''
\href{http://dx.doi.org/10.1016/0550-3213(84)90373-0}{{\em Nucl. Phys.}
  {\bfseries B247} (1984) 70--82}.
%%CITATION = NUPHA,B247,70;%%.

\bibitem{Branco:1986gr}
G.~Branco, L.~Lavoura, and M.~Rebelo, ``{Majorana Neutrinos and {CP} Violation
  in the Leptonic Sector},''
\href{http://dx.doi.org/10.1016/0370-2693(86)90307-2}{{\em Phys. Lett.}
  {\bfseries B180} (1986) 264}.
%%CITATION = PHLTA,B180,264;%%.

\bibitem{Gronau:1986xb}
M.~Gronau, A.~Kfir, and R.~Loewy, ``{Basis Independent Tests of {CP} Violation
  in Fermion Mass Matrices},''
\href{http://dx.doi.org/10.1103/PhysRevLett.56.1538}{{\em Phys. Rev. Lett.}
  {\bfseries 56} (1986) 1538}.
%%CITATION = PRLTA,56,1538;%%.

\bibitem{Ecker:1987qp}
G.~Ecker, W.~Grimus, and H.~Neufeld, ``{A Standard Form for Generalized {CP}
  Transformations},''
\href{http://dx.doi.org/10.1088/0305-4470/20/12/010}{{\em J.Phys.} {\bfseries
  A20} (1987) L807}.
%%CITATION = JPHGB,A20,L807;%%.

\bibitem{Grimus:1987kn}
W.~Grimus, ``{{CP} Violating Phenomena and Theoretical Results},''
{\em Fortsch. Phys.} {\bfseries 36} (1988) 201.
%%CITATION = FPYKA,36,201;%%.

\bibitem{Maniatis:2007vn}
M.~Maniatis, A.~von Manteuffel, and O.~Nachtmann, ``{CP violation in the
  general two-Higgs-doublet model: A Geometric view},''
  \href{http://dx.doi.org/10.1140/epjc/s10052-008-0712-5}{{\em Eur. Phys. J.}
  {\bfseries C57} (2008) 719--738},
\href{http://arxiv.org/abs/0707.3344}{{\ttfamily arXiv:0707.3344 [hep-ph]}}.
%%CITATION = ARXIV:0707.3344;%%.

\bibitem{Branco:2011iw}
G.~C. Branco, P.~M. Ferreira, L.~Lavoura, M.~N. Rebelo, M.~Sher, and J.~P.
  Silva, ``{Theory and phenomenology of two-Higgs-doublet models},''
  \href{http://dx.doi.org/10.1016/j.physrep.2012.02.002}{{\em Phys. Rept.}
  {\bfseries 516} (2012) 1--102},
\href{http://arxiv.org/abs/1106.0034}{{\ttfamily arXiv:1106.0034 [hep-ph]}}.
%%CITATION = ARXIV:1106.0034;%%.

\bibitem{Altarelli:2010gt}
G.~Altarelli and F.~Feruglio, ``{Discrete Flavor Symmetries and Models of
  Neutrino Mixing},'' \href{http://dx.doi.org/10.1103/RevModPhys.82.2701}{{\em
  Rev.Mod.Phys.} {\bfseries 82} (2010) 2701--2729},
\href{http://arxiv.org/abs/1002.0211}{{\ttfamily arXiv:1002.0211 [hep-ph]}}.
%%CITATION = ARXIV:1002.0211;%%.

\bibitem{Ishimori:2010au}
H.~Ishimori, T.~Kobayashi, H.~Ohki, Y.~Shimizu, H.~Okada, {\em et~al.},
  ``{Non-Abelian Discrete Symmetries in Particle Physics},''
  \href{http://dx.doi.org/10.1143/PTPS.183.1}{{\em Prog.Theor.Phys.Suppl.}
  {\bfseries 183} (2010) 1--163},
\href{http://arxiv.org/abs/1003.3552}{{\ttfamily arXiv:1003.3552 [hep-th]}}.
%%CITATION = ARXIV:1003.3552;%%.

\bibitem{King:2013eh}
S.~F. King and C.~Luhn, ``{Neutrino Mass and Mixing with Discrete Symmetry},''
  \href{http://dx.doi.org/10.1088/0034-4885/76/5/056201}{{\em Rept. Prog.
  Phys.} {\bfseries 76} (2013) 056201},
\href{http://arxiv.org/abs/1301.1340}{{\ttfamily arXiv:1301.1340 [hep-ph]}}.
%%CITATION = ARXIV:1301.1340;%%.

\bibitem{Bickerstaff:1985jc}
R.~Bickerstaff and T.~Damhus, ``{A necessary and sufficient condition for the
  existence of real coupling coefficients for a finite group},'' {\em
  International Journal of Quantum Chemistry} {\bfseries XXVII} (1985)
  381--391.

\bibitem{Nishi:2013jqa}
C.~Nishi, ``{Generalized CP symmetries in Delta(27) flavor models},''
  \href{http://dx.doi.org/10.1103/PhysRevD.88.033010}{{\em Phys. Rev.}
  {\bfseries D88} (2013) 033010},
\href{http://arxiv.org/abs/1306.0877}{{\ttfamily arXiv:1306.0877 [hep-ph]}}.
%%CITATION = ARXIV:1306.0877;%%.

\bibitem{Kawanaka1990}
N.~Kawanaka and H.~Matsuyama, ``A twisted version of the {F}robenius-{S}chur
  indicator and multiplicity-free permutation representations,'' {\em Hokkaido
  Math.J.} {\bfseries 19} (1990) 495--508.
  \url{http://projecteuclid.org/euclid.hokmj/1381517495}.

\bibitem{Haber:2012np}
H.~E. Haber and Z.~Surujon, ``{A Group-theoretic Condition for Spontaneous CP
  Violation},'' \href{http://dx.doi.org/10.1103/PhysRevD.86.075007}{{\em Phys.
  Rev.} {\bfseries D86} (2012) 075007},
\href{http://arxiv.org/abs/1201.1730}{{\ttfamily arXiv:1201.1730 [hep-ph]}}.
%%CITATION = ARXIV:1201.1730;%%.

\bibitem{Feit:1967}
W.~Feit, {\em {Characters of finite groups}}.
\newblock Benjamin, 1967.

\bibitem{Varzielas:2015fxa}
I.~d.~M. Varzielas, ``{Adding CP to flavour symmetries},''
  \href{http://dx.doi.org/10.1088/1742-6596/631/1/012020}{{\em J. Phys. Conf.
  Ser.} {\bfseries 631} no.~1, (2015) 012020},
\href{http://arxiv.org/abs/1503.02633}{{\ttfamily arXiv:1503.02633 [hep-ph]}}.
%%CITATION = ARXIV:1503.02633;%%.

\bibitem{Branco:2015gna}
G.~C. Branco, I.~de~Medeiros~Varzielas, and S.~F. King, ``{Invariant approach
  to $\mathcal {CP}$ in unbroken $\Delta(27)$},''
  \href{http://dx.doi.org/10.1016/j.nuclphysb.2015.07.024}{{\em Nucl. Phys.}
  {\bfseries B899} (2015) 14--36},
\href{http://arxiv.org/abs/1505.06165}{{\ttfamily arXiv:1505.06165 [hep-ph]}}.
%%CITATION = ARXIV:1505.06165;%%.

\bibitem{Adulpravitchai:2009kd}
A.~Adulpravitchai, A.~Blum, and M.~Lindner, ``{Non-Abelian Discrete Groups from
  the Breaking of Continuous Flavor Symmetries},''
  \href{http://dx.doi.org/10.1088/1126-6708/2009/09/018}{{\em JHEP} {\bfseries
  0909} (2009) 018},
\href{http://arxiv.org/abs/0907.2332}{{\ttfamily arXiv:0907.2332 [hep-ph]}}.
%%CITATION = ARXIV:0907.2332;%%.

\bibitem{Luhn:2011ip}
C.~Luhn, ``{Spontaneous breaking of SU(3) to finite family symmetries: a
  pedestrian's approach},''
  \href{http://dx.doi.org/10.1007/JHEP03(2011)108}{{\em JHEP} {\bfseries 1103}
  (2011) 108},
\href{http://arxiv.org/abs/1101.2417}{{\ttfamily arXiv:1101.2417 [hep-ph]}}.
%%CITATION = ARXIV:1101.2417;%%.

\bibitem{Merle:2011vy}
A.~Merle and R.~Zwicky, ``{Explicit and spontaneous breaking of SU(3) into its
  finite subgroups},'' \href{http://dx.doi.org/10.1007/JHEP02(2012)128}{{\em
  JHEP} {\bfseries 1202} (2012) 128},
\href{http://arxiv.org/abs/1110.4891}{{\ttfamily arXiv:1110.4891 [hep-ph]}}.
%%CITATION = ARXIV:1110.4891;%%.

\bibitem{Merle:2012xr}
A.~Merle and R.~Zwicky, ``{On explicit and spontaneous symmetry breaking in
  regard to SU(3) and its finite subgroups},'' in {\em {Proceedings, 2nd
  Workshop on Flavor Symmetries and Consequences in Accelerators and Cosmology,
  FLASY 12, Dortmund, Germany, 30 Jun - 4 Jul 2012}}, pp.~191--198.
\newblock 2012.
\newblock \url{https://inspirehep.net/record/1207710/files/FLASY12-25.pdf}.

\bibitem{Fallbacher:2015pga}
M.~Fallbacher, ``{Breaking classical Lie groups to finite subgroups – an
  automated approach},''
  \href{http://dx.doi.org/10.1016/j.nuclphysb.2015.07.004}{{\em Nucl. Phys.}
  {\bfseries B898} (2015) 229--247},
\href{http://arxiv.org/abs/1506.03677}{{\ttfamily arXiv:1506.03677 [hep-th]}}.
%%CITATION = ARXIV:1506.03677;%%.

\bibitem{Grimus:2011fk}
W.~Grimus and P.~O. Ludl, ``{Finite flavour groups of fermions},''
  \href{http://dx.doi.org/10.1088/1751-8113/45/23/233001}{{\em J. Phys.}
  {\bfseries A45} (2012) 233001},
\href{http://arxiv.org/abs/1110.6376}{{\ttfamily arXiv:1110.6376 [hep-ph]}}.
%%CITATION = ARXIV:1110.6376;%%.

\bibitem{Altarelli:2005yx}
G.~Altarelli and F.~Feruglio, ``{Tri-bimaximal neutrino mixing, A(4) and the
  modular symmetry},''
  \href{http://dx.doi.org/10.1016/j.nuclphysb.2006.02.015}{{\em Nucl.Phys.}
  {\bfseries B741} (2006) 215--235},
\href{http://arxiv.org/abs/hep-ph/0512103}{{\ttfamily arXiv:hep-ph/0512103
  [hep-ph]}}.
%%CITATION = HEP-PH/0512103;%%.

\bibitem{Altarelli:2005yp}
G.~Altarelli and F.~Feruglio, ``{Tri-bimaximal neutrino mixing from discrete
  symmetry in extra dimensions},''
  \href{http://dx.doi.org/10.1016/j.nuclphysb.2005.05.005}{{\em Nucl.Phys.}
  {\bfseries B720} (2005) 64--88},
\href{http://arxiv.org/abs/hep-ph/0504165}{{\ttfamily arXiv:hep-ph/0504165
  [hep-ph]}}.
%%CITATION = HEP-PH/0504165;%%.

\bibitem{Kobayashi:2006wq}
T.~Kobayashi, H.~P. Nilles, F.~Pl{\"o}ger, S.~Raby, and M.~Ratz, ``Stringy
  origin of non-{A}belian discrete flavor symmetries,'' {\em Nucl. Phys.}
  {\bfseries B768} (2007) 135--156,
\href{http://arxiv.org/abs/hep-ph/0611020}{{\ttfamily hep-ph/0611020}}.
%%CITATION = HEP-PH/0611020;%%.

\bibitem{Nilles:2012cy}
H.~P. Nilles, M.~Ratz, and P.~K.~S. Vaudrevange, ``{Origin of Family
  Symmetries},'' \href{http://dx.doi.org/10.1002/prop.201200120}{{\em Fortsch.
  Phys.} {\bfseries 61} (2013) 493--506},
\href{http://arxiv.org/abs/1204.2206}{{\ttfamily arXiv:1204.2206 [hep-ph]}}.
%%CITATION = ARXIV:1204.2206;%%.

\bibitem{Fischer:2012qj}
M.~Fischer, M.~Ratz, J.~Torrado, and P.~K.~S. Vaudrevange, ``{Classification of
  symmetric toroidal orbifolds},''
  \href{http://dx.doi.org/10.1007/JHEP01(2013)084}{{\em JHEP} {\bfseries 01}
  (2013) 084},
\href{http://arxiv.org/abs/1209.3906}{{\ttfamily arXiv:1209.3906 [hep-th]}}.
%%CITATION = ARXIV:1209.3906;%%.

\bibitem{Fischer:2013qza}
M.~Fischer, S.~Ramos-Sanchez, and P.~K.~S. Vaudrevange, ``{Heterotic
  non-Abelian orbifolds},''
  \href{http://dx.doi.org/10.1007/JHEP07(2013)080}{{\em JHEP} {\bfseries 1307}
  (2013) 080},
\href{http://arxiv.org/abs/1304.7742}{{\ttfamily arXiv:1304.7742 [hep-th]}}.
%%CITATION = ARXIV:1304.7742;%%.

\bibitem{Berasaluce-Gonzalez:2013bba}
M.~Berasaluce-González, G.~Ramírez, and A.~M. Uranga, ``{Antisymmetric tensor
  $Z_p$ gauge symmetries in field theory and string theory},''
  \href{http://dx.doi.org/10.1007/JHEP01(2014)059}{{\em JHEP} {\bfseries 01}
  (2014) 059},
\href{http://arxiv.org/abs/1310.5582}{{\ttfamily arXiv:1310.5582 [hep-th]}}.
%%CITATION = ARXIV:1310.5582;%%.

\bibitem{Leurer:1992wg}
M.~Leurer, Y.~Nir, and N.~Seiberg, ``{Mass matrix models},''
  \href{http://dx.doi.org/10.1016/0550-3213(93)90112-3}{{\em Nucl. Phys.}
  {\bfseries B398} (1993) 319--342},
\href{http://arxiv.org/abs/hep-ph/9212278}{{\ttfamily arXiv:hep-ph/9212278
  [hep-ph]}}.
%%CITATION = HEP-PH/9212278;%%.

\bibitem{Felipe:2014zka}
R.~G. Felipe, I.~Ivanov, C.~Nishi, H.~Serodio, and J.~P. Silva, ``{Constraining
  multi-Higgs flavour models},''
  \href{http://dx.doi.org/10.1140/epjc/s10052-014-2953-9}{{\em Eur.Phys.J.}
  {\bfseries C74} (2014) 2953},
\href{http://arxiv.org/abs/1401.5807}{{\ttfamily arXiv:1401.5807 [hep-ph]}}.
%%CITATION = ARXIV:1401.5807;%%.

\bibitem{Lam:2008sh}
C.~S. Lam, ``{The Unique Horizontal Symmetry of Leptons},''
  \href{http://dx.doi.org/10.1103/PhysRevD.78.073015}{{\em Phys. Rev.}
  {\bfseries D78} (2008) 073015},
\href{http://arxiv.org/abs/0809.1185}{{\ttfamily arXiv:0809.1185 [hep-ph]}}.
%%CITATION = ARXIV:0809.1185;%%.

\bibitem{Lam:2008rs}
C.~S. Lam, ``{Determining Horizontal Symmetry from Neutrino Mixing},''
  \href{http://dx.doi.org/10.1103/PhysRevLett.101.121602}{{\em Phys. Rev.
  Lett.} {\bfseries 101} (2008) 121602},
\href{http://arxiv.org/abs/0804.2622}{{\ttfamily arXiv:0804.2622 [hep-ph]}}.
%%CITATION = ARXIV:0804.2622;%%.

\bibitem{Hernandez:2012ra}
D.~Hernandez and A.~{\relax Yu}. Smirnov, ``{Lepton mixing and discrete
  symmetries},'' \href{http://dx.doi.org/10.1103/PhysRevD.86.053014}{{\em Phys.
  Rev.} {\bfseries D86} (2012) 053014},
\href{http://arxiv.org/abs/1204.0445}{{\ttfamily arXiv:1204.0445 [hep-ph]}}.
%%CITATION = ARXIV:1204.0445;%%.

\bibitem{Hernandez:2012sk}
D.~Hernandez and A.~{\relax Yu}. Smirnov, ``{Discrete symmetries and
  model-independent patterns of lepton mixing},''
  \href{http://dx.doi.org/10.1103/PhysRevD.87.053005}{{\em Phys. Rev.}
  {\bfseries D87} no.~5, (2013) 053005},
\href{http://arxiv.org/abs/1212.2149}{{\ttfamily arXiv:1212.2149 [hep-ph]}}.
%%CITATION = ARXIV:1212.2149;%%.

\bibitem{Zeldovich:1974uw}
{\relax Ya}.~B. Zeldovich, I.~{\relax Yu}. Kobzarev, and L.~B. Okun,
  ``{Cosmological Consequences of the Spontaneous Breakdown of Discrete
  Symmetry},'' {\em Zh. Eksp. Teor. Fiz.} {\bfseries 67} (1974) 3--11. [Sov.
  Phys. JETP40,1(1974)].

\bibitem{Preskill:1991kd}
J.~Preskill, S.~P. Trivedi, F.~Wilczek, and M.~B. Wise, ``Cosmology and broken
  discrete symmetry,''
{\em Nucl. Phys.} {\bfseries B363} (1991) 207--220.
%%CITATION = NUPHA,B363,207;%%.

\bibitem{Kolb:1990vq}
E.~W. Kolb and M.~S. Turner, ``{The Early Universe},''
{\em Front. Phys.} {\bfseries 69} (1990) 1--547.
%%CITATION = FRPHA,69,1;%%.

\bibitem{Feruglio:2012cw}
F.~Feruglio, C.~Hagedorn, and R.~Ziegler, ``{Lepton Mixing Parameters from
  Discrete and CP Symmetries},''
  \href{http://dx.doi.org/10.1007/JHEP07(2013)027}{{\em JHEP} {\bfseries 1307}
  (2013) 027},
\href{http://arxiv.org/abs/1211.5560}{{\ttfamily arXiv:1211.5560 [hep-ph]}}.
%%CITATION = ARXIV:1211.5560;%%.

\bibitem{Feruglio:2013hia}
F.~Feruglio, C.~Hagedorn, and R.~Ziegler, ``{A realistic pattern of lepton
  mixing and masses from $S_4$ and CP},''
  \href{http://dx.doi.org/10.1140/epjc/s10052-014-2753-2}{{\em Eur. Phys. J.}
  {\bfseries C74} (2014) 2753},
\href{http://arxiv.org/abs/1303.7178}{{\ttfamily arXiv:1303.7178 [hep-ph]}}.
%%CITATION = ARXIV:1303.7178;%%.

\bibitem{Hagedorn:2014wha}
C.~Hagedorn, A.~Meroni, and E.~Molinaro, ``{Lepton mixing from $\Delta(3n^2)$
  and $\Delta(6n^2)$ and CP},''
  \href{http://dx.doi.org/10.1016/j.nuclphysb.2014.12.013}{{\em Nucl. Phys.}
  {\bfseries B891} (2015) 499--557},
\href{http://arxiv.org/abs/1408.7118}{{\ttfamily arXiv:1408.7118 [hep-ph]}}.
%%CITATION = ARXIV:1408.7118;%%.

\bibitem{Ding:2014ora}
G.-J. Ding, S.~F. King, and T.~Neder, ``{Generalised CP and $\Delta(6n^2)$
  family symmetry in semi-direct models of leptons},''
  \href{http://dx.doi.org/10.1007/JHEP12(2014)007}{{\em JHEP} {\bfseries 12}
  (2014) 007},
\href{http://arxiv.org/abs/1409.8005}{{\ttfamily arXiv:1409.8005 [hep-ph]}}.
%%CITATION = ARXIV:1409.8005;%%.

\bibitem{Li:2015jxa}
C.-C. Li and G.-J. Ding, ``{Lepton Mixing in $A_5$ Family Symmetry and
  Generalized CP},'' \href{http://dx.doi.org/10.1007/JHEP05(2015)100}{{\em
  JHEP} {\bfseries 05} (2015) 100},
\href{http://arxiv.org/abs/1503.03711}{{\ttfamily arXiv:1503.03711 [hep-ph]}}.
%%CITATION = ARXIV:1503.03711;%%.

\bibitem{Ballett:2015wia}
P.~Ballett, S.~Pascoli, and J.~Turner, ``{Mixing angle and phase correlations
  from A5 with generalized CP and their prospects for discovery},''
  \href{http://dx.doi.org/10.1103/PhysRevD.92.093008}{{\em Phys. Rev.}
  {\bfseries D92} no.~9, (2015) 093008},
\href{http://arxiv.org/abs/1503.07543}{{\ttfamily arXiv:1503.07543 [hep-ph]}}.
%%CITATION = ARXIV:1503.07543;%%.

\bibitem{DiIura:2015kfa}
A.~Di~Iura, C.~Hagedorn, and D.~Meloni, ``{Lepton mixing from the interplay of
  the alternating group A$_{5}$ and CP},''
  \href{http://dx.doi.org/10.1007/JHEP08(2015)037}{{\em JHEP} {\bfseries 08}
  (2015) 037},
\href{http://arxiv.org/abs/1503.04140}{{\ttfamily arXiv:1503.04140 [hep-ph]}}.
%%CITATION = ARXIV:1503.04140;%%.

\bibitem{Mohapatra:1997su}
R.~N. Mohapatra, A.~Rasin, and G.~Senjanovic, ``{P, C and strong CP in
  left-right supersymmetric models},''
  \href{http://dx.doi.org/10.1103/PhysRevLett.79.4744}{{\em Phys. Rev. Lett.}
  {\bfseries 79} (1997) 4744--4747},
\href{http://arxiv.org/abs/hep-ph/9707281}{{\ttfamily arXiv:hep-ph/9707281
  [hep-ph]}}.
%%CITATION = HEP-PH/9707281;%%.

\bibitem{Nelson:1983zb}
A.~E. Nelson, ``{Naturally Weak CP Violation},''
\href{http://dx.doi.org/10.1016/0370-2693(84)92025-2}{{\em Phys. Lett.}
  {\bfseries B136} (1984) 387}.
%%CITATION = PHLTA,B136,387;%%.

\bibitem{Barr:1984qx}
S.~M. Barr, ``{Solving the Strong CP Problem Without the Peccei-Quinn
  Symmetry},''
\href{http://dx.doi.org/10.1103/PhysRevLett.53.329}{{\em Phys. Rev. Lett.}
  {\bfseries 53} (1984) 329}.
%%CITATION = PRLTA,53,329;%%.

\bibitem{Davidson:2005cw}
S.~Davidson and H.~E. Haber, ``{Basis-independent methods for the
  two-Higgs-doublet model},''
  \href{http://dx.doi.org/10.1103/PhysRevD.72.099902,
  10.1103/PhysRevD.72.035004}{{\em Phys. Rev.} {\bfseries D72} (2005) 035004},
  \href{http://arxiv.org/abs/hep-ph/0504050}{{\ttfamily arXiv:hep-ph/0504050
  [hep-ph]}}.
[Erratum: Phys. Rev.D72,099902(2005)].
%%CITATION = HEP-PH/0504050;%%.

\bibitem{Haber:2006ue}
H.~E. Haber and D.~O'Neil, ``{Basis-independent methods for the
  two-Higgs-doublet model. II. The Significance of tan beta},''
  \href{http://dx.doi.org/10.1103/PhysRevD.74.015018}{{\em Phys. Rev.}
  {\bfseries D74} (2006) 015018},
\href{http://arxiv.org/abs/hep-ph/0602242}{{\ttfamily arXiv:hep-ph/0602242
  [hep-ph]}}.
%%CITATION = HEP-PH/0602242;%%.

\bibitem{Ivanov:2006yq}
I.~P. Ivanov, ``{Minkowski space structure of the Higgs potential in 2HDM},''
  \href{http://dx.doi.org/10.1103/PhysRevD.76.039902,
  10.1103/PhysRevD.75.035001}{{\em Phys. Rev.} {\bfseries D75} (2007) 035001},
  \href{http://arxiv.org/abs/hep-ph/0609018}{{\ttfamily arXiv:hep-ph/0609018
  [hep-ph]}}.
[Erratum: Phys. Rev.D76,039902(2007)].
%%CITATION = HEP-PH/0609018;%%.

\bibitem{Ivanov:2010ww}
I.~P. Ivanov and C.~C. Nishi, ``{Properties of the general NHDM. I. The Orbit
  space},'' \href{http://dx.doi.org/10.1103/PhysRevD.82.015014}{{\em Phys.
  Rev.} {\bfseries D82} (2010) 015014},
\href{http://arxiv.org/abs/1004.1799}{{\ttfamily arXiv:1004.1799 [hep-th]}}.
%%CITATION = ARXIV:1004.1799;%%.

\bibitem{CeccheriniSilberstein2008}
T.~Ceccherini-Silberstein, F.~Scarabotti, and F.~Tolli, {\em Harmonic analysis
  on finite groups}.
\newblock Cambridge studies in advanced mathematics ; 108. Cambridge Univ.
  Press, Cambridge, 2008.

\bibitem{deMedeirosVarzielas:2011zw}
I.~de~Medeiros~Varzielas and D.~Emmanuel-Costa, ``{Geometrical CP Violation},''
  \href{http://dx.doi.org/10.1103/PhysRevD.84.117901}{{\em Phys. Rev.}
  {\bfseries D84} (2011) 117901},
\href{http://arxiv.org/abs/1106.5477}{{\ttfamily arXiv:1106.5477 [hep-ph]}}.
%%CITATION = ARXIV:1106.5477;%%.

\bibitem{Ivanov:2012ry}
I.~Ivanov and E.~Vdovin, ``{Discrete symmetries in the three-Higgs-doublet
  model},'' \href{http://dx.doi.org/10.1103/PhysRevD.86.095030}{{\em Phys.
  Rev.} {\bfseries D86} (2012) 095030},
\href{http://arxiv.org/abs/1206.7108}{{\ttfamily arXiv:1206.7108 [hep-ph]}}.
%%CITATION = ARXIV:1206.7108;%%.

\bibitem{Ivanov:2012fp}
I.~Ivanov and E.~Vdovin, ``{Classification of finite reparametrization symmetry
  groups in the three-Higgs-doublet model},''
  \href{http://dx.doi.org/10.1140/epjc/s10052-013-2309-x}{{\em Eur.Phys.J.}
  {\bfseries C73} (2013) 2309},
\href{http://arxiv.org/abs/1210.6553}{{\ttfamily arXiv:1210.6553 [hep-ph]}}.
%%CITATION = ARXIV:1210.6553;%%.

\bibitem{Varzielas:2012nn}
I.~de~Medeiros~Varzielas, D.~Emmanuel-Costa, and P.~Leser, ``{Geometrical CP
  Violation from Non-Renormalisable Scalar Potentials},''
  \href{http://dx.doi.org/10.1016/j.physletb.2012.08.008}{{\em Phys.Lett.}
  {\bfseries B716} (2012) 193--196},
\href{http://arxiv.org/abs/1204.3633}{{\ttfamily arXiv:1204.3633 [hep-ph]}}.
%%CITATION = ARXIV:1204.3633;%%.

\bibitem{Varzielas:2012pd}
I.~de~Medeiros~Varzielas, ``{Geometrical CP violation in multi-Higgs models},''
  \href{http://dx.doi.org/10.1007/JHEP08(2012)055}{{\em JHEP} {\bfseries 1208}
  (2012) 055},
\href{http://arxiv.org/abs/1205.3780}{{\ttfamily arXiv:1205.3780 [hep-ph]}}.
%%CITATION = ARXIV:1205.3780;%%.

\bibitem{Degee:2012sk}
A.~Degee, I.~Ivanov, and V.~Keus, ``{Geometric minimization of highly symmetric
  potentials},'' \href{http://dx.doi.org/10.1007/JHEP02(2013)125}{{\em JHEP}
  {\bfseries 1302} (2013) 125},
\href{http://arxiv.org/abs/1211.4989}{{\ttfamily arXiv:1211.4989 [hep-ph]}}.
%%CITATION = ARXIV:1211.4989;%%.

\bibitem{Bhattacharyya:2012pi}
G.~Bhattacharyya, I.~de~Medeiros~Varzielas, and P.~Leser, ``{A common origin of
  fermion mixing and geometrical CP violation, and its test through Higgs
  physics at the LHC},''
  \href{http://dx.doi.org/10.1103/PhysRevLett.109.241603}{{\em Phys. Rev.
  Lett.} {\bfseries 109} (2012) 241603},
\href{http://arxiv.org/abs/1210.0545}{{\ttfamily arXiv:1210.0545 [hep-ph]}}.
%%CITATION = ARXIV:1210.0545;%%.

\bibitem{Varzielas:2013sla}
I.~de~Medeiros~Varzielas and D.~Pidt, ``{Towards realistic models of quark
  masses with geometrical CP violation},''
  \href{http://dx.doi.org/10.1088/0954-3899/41/2/025004}{{\em J.Phys.}
  {\bfseries G41} (2014) 025004},
\href{http://arxiv.org/abs/1307.0711}{{\ttfamily arXiv:1307.0711 [hep-ph]}}.
%%CITATION = ARXIV:1307.0711;%%.

\bibitem{Ivanov:2013nla}
I.~Ivanov and L.~Lavoura, ``{Geometrical CP violation in the N-Higgs-doublet
  model},'' \href{http://dx.doi.org/10.1140/epjc/s10052-013-2416-8}{{\em
  Eur.Phys.J.} {\bfseries C73} no.~4, (2013) 2416},
\href{http://arxiv.org/abs/1302.3656}{{\ttfamily arXiv:1302.3656 [hep-ph]}}.
%%CITATION = ARXIV:1302.3656;%%.

\bibitem{Ma:2013xqa}
E.~Ma, ``{Neutrino Mixing and Geometric CP Violation with $\Delta(27)$
  Symmetry},'' \href{http://dx.doi.org/10.1016/j.physletb.2013.05.011}{{\em
  Phys.Lett.} {\bfseries B723} (2013) 161--163},
\href{http://arxiv.org/abs/1304.1603}{{\ttfamily arXiv:1304.1603 [hep-ph]}}.
%%CITATION = ARXIV:1304.1603;%%.

\bibitem{Varzielas:2013eta}
I.~Medeiros~Varzielas and D.~Pidt, ``{Geometrical CP violation with a complete
  fermion sector},'' \href{http://dx.doi.org/10.1007/JHEP11(2013)206}{{\em
  JHEP} {\bfseries 1311} (2013) 206},
\href{http://arxiv.org/abs/1307.6545}{{\ttfamily arXiv:1307.6545 [hep-ph]}}.
%%CITATION = ARXIV:1307.6545;%%.

\bibitem{Ivanov:2014doa}
I.~P. Ivanov and C.~C. Nishi, ``{Symmetry breaking patterns in 3HDM},''
  \href{http://dx.doi.org/10.1007/JHEP01(2015)021}{{\em JHEP} {\bfseries 01}
  (2015) 021},
\href{http://arxiv.org/abs/1410.6139}{{\ttfamily arXiv:1410.6139 [hep-ph]}}.
%%CITATION = ARXIV:1410.6139;%%.

\bibitem{Nishi:2016}
C.~Nishi, ``{In preparation},''. private communication.

\bibitem{Schmid1995}
J.~Schmid, ``On the affine bézout inequality,''
  \href{http://dx.doi.org/10.1007/BF02567819}{{\em manuscripta mathematica}
  {\bfseries 88} (1995) 225--232}.

\bibitem{Bogopolskij2008}
O.~V. Bogopol'skij, {\em Introduction to group theory}.
\newblock EMS textbooks in mathematics. European Mathematical Society, Zürich,
  2008.

\bibitem{Coleman:1969sm}
S.~R. Coleman, J.~Wess, and B.~Zumino, ``{Structure of phenomenological
  Lagrangians. 1.},''
\href{http://dx.doi.org/10.1103/PhysRev.177.2239}{{\em Phys.Rev.} {\bfseries
  177} (1969) 2239--2247}.
%%CITATION = PHRVA,177,2239;%%.

\bibitem{Callan:1969sn}
C.~G. Callan, Jr., S.~R. Coleman, J.~Wess, and B.~Zumino, ``{Structure of
  phenomenological Lagrangians. 2.},''
\href{http://dx.doi.org/10.1103/PhysRev.177.2247}{{\em Phys.Rev.} {\bfseries
  177} (1969) 2247--2250}.
%%CITATION = PHRVA,177,2247;%%.

\bibitem{Cheung:2015aba}
C.~Cheung and C.-H. Shen, ``{Nonrenormalization Theorems without
  Supersymmetry},''
  \href{http://dx.doi.org/10.1103/PhysRevLett.115.071601}{{\em Phys. Rev.
  Lett.} {\bfseries 115} no.~7, (2015) 071601},
\href{http://arxiv.org/abs/1505.01844}{{\ttfamily arXiv:1505.01844 [hep-ph]}}.
%%CITATION = ARXIV:1505.01844;%%.

\bibitem{Alonso:2014rga}
R.~Alonso, E.~E. Jenkins, and A.~V. Manohar, ``{Holomorphy without
  Supersymmetry in the Standard Model Effective Field Theory},''
  \href{http://dx.doi.org/10.1016/j.physletb.2014.10.045}{{\em Phys. Lett.}
  {\bfseries B739} (2014) 95--98},
\href{http://arxiv.org/abs/1409.0868}{{\ttfamily arXiv:1409.0868 [hep-ph]}}.
%%CITATION = ARXIV:1409.0868;%%.

\bibitem{Lehman:2015via}
L.~Lehman and A.~Martin, ``{Hilbert Series for Constructing Lagrangians:
  expanding the phenomenologist's toolbox},''
  \href{http://dx.doi.org/10.1103/PhysRevD.91.105014}{{\em Phys. Rev.}
  {\bfseries D91} (2015) 105014},
\href{http://arxiv.org/abs/1503.07537}{{\ttfamily arXiv:1503.07537 [hep-ph]}}.
%%CITATION = ARXIV:1503.07537;%%.

\bibitem{Henning:2015alf}
B.~Henning, X.~Lu, T.~Melia, and H.~Murayama, ``{2, 84, 30, 993, 560, 15456,
  11962, 261485, ...: Higher dimension operators in the SM EFT},''
\href{http://arxiv.org/abs/1512.03433}{{\ttfamily arXiv:1512.03433 [hep-ph]}}.
%%CITATION = ARXIV:1512.03433;%%.

\bibitem{Santamaria:1993ah}
A.~Santamaria, ``{Masses, mixings, Yukawa couplings and their symmetries},''
  \href{http://dx.doi.org/10.1016/0370-2693(93)91110-9}{{\em Phys. Lett.}
  {\bfseries B305} (1993) 90--97},
\href{http://arxiv.org/abs/hep-ph/9302301}{{\ttfamily arXiv:hep-ph/9302301
  [hep-ph]}}.
%%CITATION = HEP-PH/9302301;%%.

\bibitem{Chang:2002yr}
S.~Chang and T.-K. Kuo, ``{Renormalization invariants of the neutrino mass
  matrix},'' \href{http://dx.doi.org/10.1103/PhysRevD.66.111302}{{\em Phys.
  Rev.} {\bfseries D66} (2002) 111302},
\href{http://arxiv.org/abs/hep-ph/0205147}{{\ttfamily arXiv:hep-ph/0205147
  [hep-ph]}}.
%%CITATION = HEP-PH/0205147;%%.

\bibitem{Liu:2009vh}
L.-X. Liu, ``{Renormalization Invariants and Quark Flavor Mixings},''
  \href{http://dx.doi.org/10.1142/S0217751X10050640}{{\em Int. J. Mod. Phys.}
  {\bfseries A25} (2010) 4975--4991},
\href{http://arxiv.org/abs/0910.1326}{{\ttfamily arXiv:0910.1326 [hep-ph]}}.
%%CITATION = ARXIV:0910.1326;%%.

\bibitem{Chiu:2015ega}
S.~H. Chiu and T.~K. Kuo, ``{Renormalization of the Neutrino Mass Matrix},''
\href{http://arxiv.org/abs/1510.07368}{{\ttfamily arXiv:1510.07368 [hep-ph]}}.
%%CITATION = ARXIV:1510.07368;%%.

\bibitem{Feldmann:2015nia}
T.~Feldmann, T.~Mannel, and S.~Schwertfeger, ``{Renormalization Group Evolution
  of Flavour Invariants},''
  \href{http://dx.doi.org/10.1007/JHEP10(2015)007}{{\em JHEP} {\bfseries 10}
  (2015) 007},
\href{http://arxiv.org/abs/1507.00328}{{\ttfamily arXiv:1507.00328 [hep-ph]}}.
%%CITATION = ARXIV:1507.00328;%%.

\bibitem{Chiu:2016qra}
S.~H. Chiu and T.~K. Kuo, ``{Renormalization of the quark mass matrix},''
  \href{http://dx.doi.org/10.1103/PhysRevD.93.093006}{{\em Phys. Rev.}
  {\bfseries D93} no.~9, (2016) 093006},
\href{http://arxiv.org/abs/1603.04568}{{\ttfamily arXiv:1603.04568 [hep-ph]}}.
%%CITATION = ARXIV:1603.04568;%%.

\bibitem{Fortin:2014}
J.-F. Fortin, B.~Grinstein, and A.~Stergiou, {\em RG Cycles, Scale vs Conformal
  Invariance, and All That…},
  \href{http://dx.doi.org/10.1142/9789814566254_0026}{ch.~26, pp.~247--261}.
\newblock WORLD SCIENTIFIC, 2014.
\newblock
  \url{http://www.worldscientific.com/doi/abs/10.1142/9789814566254_0026}.

\bibitem{Elvang:2013cua}
H.~Elvang and Y.-t. Huang, ``{Scattering Amplitudes},''
\href{http://arxiv.org/abs/1308.1697}{{\ttfamily arXiv:1308.1697 [hep-th]}}.
%%CITATION = ARXIV:1308.1697;%%.

\bibitem{Dixon:2013uaa}
L.~J. Dixon, \href{http://dx.doi.org/10.5170/CERN-2014-008.31}{``{A brief
  introduction to modern amplitude methods},''} in {\em {Proceedings, 2012
  European School of High-Energy Physics (ESHEP 2012): La Pommeraye, Anjou,
  France, June 06-19, 2012}}, pp.~31--67.
\newblock 2014.
\newblock \href{http://arxiv.org/abs/1310.5353}{{\ttfamily arXiv:1310.5353
  [hep-ph]}}.
\newblock
  \url{https://inspirehep.net/record/1261436/files/arXiv:1310.5353.pdf}.

\bibitem{Batista:2004}
C.~D. {Batista} and G.~{Ortiz}, ``{Algebraic approach to interacting quantum
  systems},'' \href{http://dx.doi.org/10.1080/00018730310001642086}{{\em
  Advances in Physics} {\bfseries 53} (Feb., 2004) 1--82},
  \href{http://arxiv.org/abs/cond-mat/0207106}{{\ttfamily cond-mat/0207106}}.

\bibitem{Hassan:2012}
S.~R. {Hassan}, P.~V. {Sriluckshmy}, S.~K. {Goyal}, R.~{Shankar}, and
  D.~{S{\'e}n{\'e}chal}, ``{Stable Algebraic Spin Liquid in a Hubbard Model},''
  \href{http://dx.doi.org/10.1103/PhysRevLett.110.037201}{{\em Physical Review
  Letters} {\bfseries 110} no.~3, (Jan., 2013) 037201},
  \href{http://arxiv.org/abs/1208.5240}{{\ttfamily arXiv:1208.5240
  [cond-mat.str-el]}}.

\bibitem{Gomes:2015eea}
P.~R.~S. Gomes, ``{Aspects of Emergent Symmetries},''
  \href{http://dx.doi.org/10.1142/S0217751X1630009X}{{\em Int. J. Mod. Phys.}
  {\bfseries A31} no.~10, (2016) 1630009},
\href{http://arxiv.org/abs/1510.04492}{{\ttfamily arXiv:1510.04492 [hep-th]}}.
%%CITATION = ARXIV:1510.04492;%%.

\bibitem{Damhus:1981}
T.~Damhus, ``On the existence of real clebsch–gordan coefficients,''
  \href{http://dx.doi.org/http://dx.doi.org/10.1063/1.524757}{{\em Journal of
  Mathematical Physics} {\bfseries 22} no.~1, (1981) 7--14}.
  \url{http://scitation.aip.org/content/aip/journal/jmp/22/1/10.1063/1.524757}.

\bibitem{Escobar:2011mq}
J.~Escobar, ``{Flavor $\Delta$(54) in SU(5) SUSY Model},''
  \href{http://dx.doi.org/10.1103/PhysRevD.84.073009}{{\em Phys. Rev.}
  {\bfseries D84} (2011) 073009},
\href{http://arxiv.org/abs/1102.1649}{{\ttfamily arXiv:1102.1649 [hep-ph]}}.
%%CITATION = ARXIV:1102.1649;%%.

\end{thebibliography}\endgroup
